\documentclass{article}%
\usepackage{lscape}
\usepackage{makecell}
\usepackage[utf8]{inputenc}
\usepackage[nosort]{cite}
\usepackage{graphicx}
\usepackage{multicol}
\usepackage{amsfonts}
\usepackage{amssymb}
\usepackage{amsmath}
\usepackage{heck}
\usepackage{tabularx}
\usepackage{afterpage}
\usepackage{setspace}
\usepackage{verbatim}
\usepackage{color}
\usepackage{longtable}
\usepackage{float}
\usepackage{subcaption}
\usepackage{caption}
\usepackage{epsfig}
\usepackage{enumerate}
\usepackage{epstopdf}
\usepackage[enableskew, vcentermath]{youngtab}
\usepackage{adjustbox}
\usepackage{multirow}
\usepackage{tikz}
\usepackage[margin=1in]{geometry}
\usepackage{titletoc}
\usepackage{hyperref}
\usepackage{mathdots}
\usepackage{changepage}
\usepackage{xfrac}
\usepackage{color, colortbl}
\usepackage{xcolor}
\usepackage[utf8]{inputenc}
\usepackage{numprint}
\usepackage[CaptionAfterwards]{fltpage}
\usepackage{booktabs}
\npdecimalsign{.}
\nprounddigits{3}
\usepackage[mathscr]{euscript}
\usepackage{yfonts}
\usetikzlibrary{external}
\usetikzlibrary{shapes.geometric}
\usetikzlibrary{shapes.misc}
\usetikzlibrary{calc}
\usepackage{calc}
\usepackage{ifthen}
\usepackage{placeins}
\usetikzlibrary{arrows}

\providecommand{\U}[1]{\protect\rule{.1in}{.1in}}
\pdfoutput=1
\newsavebox{\mysavebox}

\hypersetup{colorlinks,citecolor=black,filecolor=black,linkcolor=black,urlcolor=black}
\usetikzlibrary{decorations.markings}

\numberwithin{equation}{section}

\usetikzlibrary{chains}
\usetikzlibrary{shapes,snakes}
\allowdisplaybreaks
\tikzset{node distance=2em, ch/.style={circle,draw,on chain,inner sep=2pt},chj/.style={ch,join},every path/.style={shorten >=0pt,shorten <=0pt},line width=1pt,baseline=-1ex}

\def\op#1{\displaystyle{\mathop{\mbox{\large $\oplus$}}_{#1}}}

\newcommand{\ksu}{\mathfrak{su}}

\newcommand{\kso}{\mathfrak{so}}
\newcommand{\ba}{\begin{eqnarray}}
\newcommand{\ea}{\end{eqnarray}}

\newcommand{\mf}{\mathfrak}

\DeclareMathOperator{\su}{\mathit{su}}

\newcommand{\be}{\begin{equation}}
\newcommand{\ee}{\end{equation}}
\tikzstyle{startstop} = [rectangle, rounded corners, minimum width=3cm, minimum height=1cm,text centered, draw=black, fill=blue!10]
\tikzstyle{startstop} = [rectangle, rounded corners, minimum width=3cm, minimum height=1cm,text centered, draw=black, fill=blue!10]
\tikzstyle{io} = [trapezium, trapezium left angle=70, trapezium right angle=110, minimum width=3cm, minimum height=1cm, text centered, draw=black, fill=blue!30]
\tikzstyle{process} = [rectangle, minimum width=3cm, minimum height=1cm, text centered, draw=black, fill=orange!30]
\tikzstyle{decision} = [diamond, minimum width=3cm, minimum height=1cm, text centered, draw=black, fill=green!30]
\tikzstyle{arrow} = [thick,->,>=stealth]
\tikzset{->-/.style={decoration={
  markings, mark=at position #1 with {\arrow[scale=1.5]{stealth}}},postaction={decorate}}}
\tikzset{rootArrow/.style={-,decoration={markings,mark=at position 0.7 with {\arrow[scale=1,black]{>}}}, postaction={decorate}, line width=1pt, shorten <=0, shorten >=0}}
\tikzset{quivArrow/.style={-,decoration={markings,mark=at position 0.6 with {\arrow[scale=.7,black]{>}}}, postaction={decorate}, line width=1pt, shorten <=0, shorten >=0}}
\tikzset{doubleArrowa/.style={-,decoration={markings,mark=at position 0.5 with {\arrow[scale=.7,black]{>}}}, postaction={decorate}, line width=1pt, shorten <=0, shorten >=0}}
\tikzset{doubleArrowb/.style={-,decoration={markings,mark=at position 0.4 with {\arrow[scale=.7,black]{>}}}, postaction={decorate}, line width=1pt, shorten <=0, shorten >=0}}
\tikzset{paramsAbranes/.style={circle,draw=black, fill=black, inner sep=0pt,minimum size=5pt}}
\tikzset{paramsBbranes/.style={circle,draw=black, fill=white, inner sep=0pt,minimum size=5pt}}
\tikzset{paramsbranes/.style={circle,draw=black, fill=black, inner sep=0pt,minimum size=2.5pt}}
\tikzset{whitebranes/.style={circle,draw=black, fill=white, inner sep=0pt,minimum size=2.5pt}}
\tikzset{blankbranes/.style={circle,draw=white, fill=white, inner sep=0pt,minimum size=2.5pt}}
\makeatletter \@addtoreset{equation}{section} \makeatother

\definecolor{LightCyan}{rgb}{0.88,1,1}

\date{July 2019}
\title{T-Branes, String Junctions, and 6D SCFTs}

\institution{UO}{\centerline{${}^{1}$Department of Physics, University of Oviedo, Avda. Federico Garc\'ia Lorca 18, 33007 Oviedo, Spain}}
\institution{PENN}{\centerline{${}^{2}$Department of Physics and Astronomy, University of Pennsylvania, Philadelphia, PA 19104, USA}}
\institution{PRINCETON}{\centerline{${}^{3}$School of Natural Sciences, Institute for Advanced Study, Princeton, NJ 08540, USA}}
\authors{Falk Hassler\worksat{\UO}\footnote{e-mail: {\tt falk@fhassler.de}},
Jonathan J. Heckman\worksat{\PENN}\footnote{e-mail: {\tt jheckman@sas.upenn.edu}},\\[4mm]
Thomas B. Rochais\worksat{\PENN}\footnote{e-mail: {\tt thb@sas.upenn.edu}},
Tom Rudelius\worksat{\PRINCETON}\footnote{e-mail: {\tt rudelius@ias.edu}},
and Hao Y. Zhang\worksat{\PENN}\footnote{e-mail: {\tt zhangphy@sas.upenn.edu}}}

\begin{document}
\abstract{Recent work on 6D superconformal field theories (SCFTs) has established an intricate correspondence
between certain Higgs branch deformations and nilpotent orbits of flavor symmetry algebras associated with
T-branes. In this paper, we return to the stringy origin of these theories and show that many aspects of these deformations can be understood in terms of simple combinatorial data associated with multi-pronged strings
stretched between stacks of intersecting $7$-branes in F-theory. This data lets us determine the
full structure of the nilpotent cone for each semi-simple flavor symmetry algebra, and it further allows us to
 characterize symmetry breaking patterns in quiver-like theories with classical gauge groups. An especially
 helpful feature of this analysis is that it extends to ``short quivers'' in which the breaking
patterns from different flavor symmetry factors are correlated.}
\maketitle
\setcounter{tocdepth}{2}
\tableofcontents
\newpage

\section{Introduction}

\label{sec:INTRO}

One of the surprises from string theory is the
prediction of whole new classes of quantum field theories decoupled from
gravity. Central examples of this sort are 6D superconformal field
theories (SCFTs). The only known way to reliably engineer examples of such theories is to start with a
background geometry in string / M- / F-theory, and to consider a singular
limit in which all length scales are sent to zero or infinity (for early work in this
direction see e.g. \cite{Witten:1995zh, Strominger:1995ac, Seiberg:1996qx}).
Since small deformations away from these scaling limits have a sensible coupling to
higher-dimensional gravity, there is strong evidence that this leads to an
interacting conformal fixed point.

The most flexible method known for constructing such theories is via F-theory on a non-compact, elliptically-fibered Calabi-Yau threefold. SCFTs are generated by simultaneously contracting a configuration of curves in the base geometry. There is now a classification of
all elliptic threefolds which can generate a 6D SCFT, and in fact, each known
6D SCFT can be associated with some such threefold \cite{Heckman:2013pva, Heckman:2015bfa} (see also \cite{Bhardwaj:2015xxa, Bhardwaj:2019hhd}).\footnote{The caveat to
this statement is that in all known constructions, there is a non-trivial tensor branch. Additionally,
in F-theory there can be ``frozen'' singularities \cite{Witten:1997bs, Tachikawa:2015wka, Bhardwaj:2018jgp}.
We note that all such models still are described by elliptic threefolds with
collapsing curves in the base.} For a recent review,
see reference \cite{Heckman:2018jxk}.

In these sorts of constructions, one
begins away from the fixed point of interest and then tunes to zero various
operator vevs in the low energy effective field theory. In this UV limit, the
effective field theory description breaks down, but the stringy description
still remains well-behaved. From this perspective, the main question is to
better understand the microscopic structure of these 6D SCFTs.

The F-theory realization of 6D SCFTs provides insight into the corresponding
structure of these theories as well as their moduli spaces (see \cite{Heckman:2018jxk}).
Perhaps surprisingly, all known 6D SCFTs resemble generalizations of quiver gauge
theories in which (on a partial tensor branch) the theory involves ADE gauge
groups linked together by 6D conformal matter \cite{DelZotto:2014hpa, Heckman:2014qba}.
The topology of these quivers is rather simple, and consists of a single spine of such gauge groups. The
space of tensor branch deformations translates in the geometry to the moduli
space of volumes for the contractible curves in the base of the elliptic
threefolds. Additionally, Higgs branch deformations translate to complex
structure deformations of the corresponding elliptic threefolds.

The quiver-like description of 6D SCFTs also suggests that
Higgs branch deformations can be understood in terms of breaking patterns
associated with the flavor symmetries of these theories. For example, in the
$7$-brane gauge theory, nilpotent
elements of the flavor symmetry algebra correspond to ``T-brane
configurations'' of $7$-branes. For a partial list of references to the T-brane literature, see references
\cite{Aspinwall:1998he, Donagi:2003hh,Cecotti:2009zf,Cecotti:2010bp,Donagi:2011jy,Anderson:2013rka,
Collinucci:2014qfa,Cicoli:2015ylx,Heckman:2016ssk,Collinucci:2016hpz,Bena:2016oqr,
Marchesano:2016cqg,Anderson:2017rpr,Collinucci:2017bwv,Cicoli:2017shd,Marchesano:2017kke,
Heckman:2018pqx, Apruzzi:2018xkw, Cvetic:2018xaq, Carta:2018qke, Marchesano:2019azf, Bena:2019rth, Barbosa:2019bgh}.

A pleasant aspect of nilpotent elements is that they
come equipped with a partial ordering, as dictated by the symmetry breaking
pattern in the original UV theory. Indeed, the orbit of each nilpotent element
under the adjoint action specifies (under Zariski closure) a partially ordered set.
This partial ordering determines fine-grained structure for Higgs branch
flows between different 6D SCFTs \cite{Heckman:2016ssk, Mekareeya:2016yal} and points the way to a possible
classification of RG flows between 6D SCFTs \cite{Heckman:2018pqx}.\footnote{See also references
\cite{Heckman:2010qv, Apruzzi:2018xkw} for a related discussion of partial ordering in
the case of certain 4D SCFTs.}

This has been established in the case of 6D SCFTs with a sufficient number of gauge group factors in the quiver-like description,
i.e., ``long quivers,'' where Higgsing of the different flavor symmetries is uncorrelated,
and there are also hints that it extends to the case of ``short quivers'' in which the structure of
Higgsing is correlated.

One feature which is somewhat obscure in this characterization of Higgs
branch flows is the actual breaking pattern taking place in the quiver-like
gauge theory. Indeed, in the case of a weakly-coupled quiver gauge theory, the
appearance of matter transforming in representations of different gauge groups
means that the corresponding D-flatness conditions for one vector multiplet
will automatically be correlated with those of neighboring gauge group nodes.
This means that each breaking pattern defined on the exterior of a quiver will
necessarily propagate towards the interior of the quiver. Even in the case of
quiver gauge theories with classical algebras, the resulting combinatorics for
tracking the breaking pattern of a Higgs branch deformation can be quite intricate.

To address these issues, in this paper we use the physics of
brane recombination to extract the combinatorics of Higgs branch flows in 6D SCFTs. In stringy
terms, brane recombination is associated with the condensation of strings
stretched between different branes. In the context of F-theory, strings can be
bound states of F1- and D1- strings, and they can have multiple ends. Our task,
then, will be to show how such multi-pronged strings attach between different
stacks of branes, and moreover, how this leads to a natural characterization
of brane recombination for Higgs branch flows in 6D SCFTs.

Since we will be primarily interested in flows driven by nilpotent orbits,
we first spell out how a given configuration of multi-pronged strings attached to bound states of $[p,q]$
$7$-branes maps on to the breaking pattern associated with a particular nilpotent orbit of an algebra. Separating
these branes from one another corresponds to a choice of Cartan subalgebra,
and strings stretched between these separated branes correspond to Lie algebra
elements associated with roots of the Lie algebra, defining a directed graph
between the nodes spanned by these branes. In particular, we show that we can
always generate a nilpotent element of the (complexified) Lie algebra by
working in terms of a directed graph which points in one direction. We also
show that, starting from such a directed graph, appending additional strings
always leads to a nilpotent element with a strictly larger nilpotent orbit.
We thus construct the entire nilpotent cone of each Lie algebra of type ABCDEFG using such multi-pronged string junctions.

With this result in place, we next turn to an analysis of Higgs branch flows
in quiver-like 6D SCFTs, as generated by T-brane deformations. We primarily
focus on 6D SCFTs generated by M5-branes probing an ADE singularity
with flavor symmetry $G_{ADE} \times G_{ADE}$, as well as tensor branch
deformations of these cases to non-simply laced flavor symmetry algebras. As found in
\cite{Heckman:2018pqx}, these are progenitor theories for many
6D SCFTs (the other being E-string probes of ADE singularities \cite{Heckman:2014qba, Heckman:2015bfa, Mekareeya:2017jgc, Frey:2018vpw, Cabrera:2019izd}). The partial
tensor branch of these parent UV theories are all of the form:
\begin{equation}
\label{quivo}[G_{0}] - G_{1} - ... - G_{k} - [G_{k+1}]
\end{equation}
with $G_{0}, G_{k+1}$ flavor symmetries and $G_1,...,G_k$ gauge symmetries. We show that Higgs
branch flows are determined by a system of coupled D-term constraints, one for each
node of such a quiver gauge theory. This in turn means that the ``links''
between gauge nodes behave as a generalization of matter, as suggested by the
structure of these quivers. We also show that condensing these strings leads
to a sequence of brane recombinations, relying on a parallel with Hanany-Witten moves
\cite{Hanany:1996ie} seen in the type IIA framework to derive the type IIB recombination
moves. We present a complete characterization of quiver-like theories with classical algebras,
and briefly discuss what would be needed to extend this analysis to quiver-like theories with
exceptional gauge group factors.

The explicit characterization of nilpotent orbits in terms of string junctions
also allows us to study Higgs branch flows in which the number of gauge groups
is small. This case is especially interesting because
there are non-trivial correlations on the symmetry breaking patterns,
one emanating from the left flavor symmetry $G_{0}$ and the subsequent D-term constraints on
its gauged neighbors and one emanating from
the right flavor symmetry $G_{k+1}$ and its gauged neighbors in the quiver of line (\ref{quivo}).
This sort of phenomenon occurs whenever the size of the nilpotent orbit
of the flavor algebras is sufficiently
large, and the number of gauge groups $k$ is sufficiently small. We study these ``overlapping T-branes'' in detail in the case of the classical
algebras. In particular, we show how to extract the resulting IR SCFT using
our picture in terms of brane recombination. We leave the case of short
quivers with exceptional gauge groups / flavor symmetries to future work.

The rest of this paper is organized as follows. First, in section \ref{sec:QUIVER}, we
review in general terms the structure of 6D SCFTs as quiver-like gauge
theories, and we explain how the worldvolume theory on $7$-branes leads to a
direct link between Higgs branch flows and nilpotent orbits of flavor
symmetries. In section \ref{sec:NILPJUNC}, we show how to reconstruct the nilpotent cone of a
flavor symmetry algebra in terms of the combinatorial data of strings
stretched between stacks of $[p,q]$ $7$-branes. Section \ref{sec:RECOMBO} uses
this combinatorial data to provide a systematic method for analyzing Higgs
branch flows in quiver-like theories with classical gauge groups, including cases with
6D conformal matter. In section \ref{sec:GETSHORTY}, we study Higgs branch flows
from overlapping nilpotent orbits in short quivers, and in section \ref{sec:CONC} we
present our conclusions. A number of additional detailed computations are
included in the Appendices.

\section{6D SCFTs as Quiver-Like Gauge Theories \label{sec:QUIVER}}

In this section, we briefly review the relevant aspects of 6D\ SCFTs which we
shall be studying in the remainder of this paper. The main item of interest
for us will be the quiver-like structure of all such theories, and the
corresponding Higgs branch flows associated with nilpotent orbits of the flavor
symmetry algebra.

To begin, we recall that the F-theory realization of 6D\ SCFTs involves
specifying a non-compact elliptically-fibered Calabi-Yau threefold
$X\rightarrow B$, where the base $B$ of the elliptic fibration is
a non-compact K\"{a}hler surface. In
minimal Weierstrass form, these elliptic threefolds can be viewed as a
hypersurface:%
\begin{equation}
y^{2}=x^{3}+fx+g.
\end{equation}
The order of vanishing for the coefficients $f$, $g$ and the discriminant
$\Delta=4f^{3}+27g^{2}$ dictate the structure of possible gauge groups, flavor
symmetries and matter content in the 6D\ effective field theory. We are
particularly interested in the construction of 6D\ SCFTs, which requires us to simultaneously collapse a collection of curves in the base to zero size at
finite distance in the Calabi-Yau metric moduli space. This can occur for
curves with negative self-intersection, and compatibility with the condition
that we maintain an elliptic fibration over generic points of each curve
imposes further restrictions \cite{Heckman:2013pva}. Each such configuration can be viewed as being
built up from intersections of non-Higgsable clusters (NHCs) \cite{Morrison:2012np} and possible
enhancements in the singularity type over each such curve. The tensor branch
of the 6D\ SCFT corresponds to resolving the collapsing curves in the base to
finite size, and the Higgs branch of the 6D\ SCFT corresponds to blow-downs and smoothing
deformations of the Weierstrass model such as \cite{Heckman:2015ola}:
\begin{equation}
y^{2}=x^{3}+(f+\delta f)x+(g+\delta g).
\end{equation}

In references \cite{Heckman:2013pva, Heckman:2015bfa}, the full list of possible F-theory geometries
which could support a 6D\ SCFT was determined. Quite remarkably, all of these
theories have the structure of a quiver-like gauge theory with a single spine
of gauge group nodes and only small amounts of decoration by (generalized) matter on the left and right of each quiver. In this description,
$7$-branes with ADE\ gauge groups intersect at points where additional
curves have collapsed. These points are often referred to as \textquotedblleft
conformal matter\textquotedblright\ since they localize at points just as in the case of ordinary matter
in F-theory \cite{DelZotto:2014hpa, Heckman:2014qba}. These configurations indicate
the presence of additional operators in the 6D\ SCFT and, like ordinary matter, can have non-trivial
vevs, leading to a deformation onto the Higgs branch. A
streamlined approach to understanding the vast majority of 6D\ SCFTs was
obtained in \cite{Heckman:2018pqx} where it was found that any 6D\ SCFT can be viewed as
\textquotedblleft fission products,\textquotedblright\ namely as deformations
of a quiver-like theory with partial tensor branch such as:%
\begin{equation}
\lbrack E_{8}]\overset{\mathfrak{g}_{ADE}}{1}\overset{\mathfrak{g}_{ADE}%
}{2}...\overset{\mathfrak{g}_{ADE}}{2}[G_{ADE}] \label{hetinst}%
\end{equation}
or:%
\begin{equation}
\lbrack G_{ADE}]\overset{\mathfrak{g}_{ADE}}{2}...\overset{\mathfrak{g}%
_{ADE}}{2}[G_{ADE}], \label{m5probe}%
\end{equation}
where the few SCFTs which cannot be understood in this way can be obtained by
adding a tensor multiplet and weakly gauging a common flavor symmetry of these fission products through a
process known as fusion. In the above, each compact curve of self-intersection
$-n$ with a $7$-brane gauge group of ADE\ type is denoted as
$\overset{\mathfrak{g}_{ADE}}{n}$. The full tensor branch of these theories is
obtained by performing further blowups at the collision points between the
compact curves (in the D- and E-type cases). To emphasize this quiver-like
structure, we shall often write:%
\begin{equation}
\lbrack G_{0}]-G_{1}-...G_{k}-[G_{k+1}],
\end{equation}
to emphasize that there are two flavor symmetry factors (indicated by square
brackets), and the rest are gauge symmetries.

The 6D SCFTs given by lines (\ref{hetinst}) and (\ref{m5probe}) can also be
realized in M-theory. The theories of line (\ref{hetinst}) arise from an
M5-brane probing an ADE\ singularity which is wrapped by an $E_{8}$
nine-brane. The theories of line (\ref{m5probe}) arise from M5-branes probing
an ADE\ singularity. In what follows, we shall primarily be interested in
understanding Higgs branch flows associated with the theories of line (\ref{m5probe}).

For $G_{ADE}$ of A or D type, the IR SCFTs of these Higgs branch flows can also be realized
in type IIA. $SU$ gauge algebras are obtained from the worldvolume of D6-branes suspended between
spacetime-filling NS5-branes, while $SO$ algebras and $Sp$ gauge algebras also require $O6^-$ and
$O6^+$ branes, respectively, stretched between $\frac{1}{2}$ NS5-branes. These constructions
will prove especially useful in section \ref{sec:RECOMBO}, where we discuss Hanany-Witten moves of
the branes of the type IIA construction.

One of the main ways to cross-check the structure of proposed RG flows is through anomaly
matching constraints. The anomaly polynomial of a 6D SCFT is calculable because the tensor branch description of each such theory is available
from the F-theory description, and the anomaly polynomial obtained on this branch of moduli space can be
matched to that of the conformal fixed point \cite{Ohmori:2014pca, Ohmori:2014kda,
Heckman:2015ola, Cordova:2015fha, Cordova:2018cvg}. To fix conventions, we often write this as a formal
eight-form with conventions (as in reference \cite{Heckman:2018jxk}):
\begin{align}
I_{8}  &  = \alpha c_{2}(R)^{2} + \beta c_{2}(R) p_{1}(T) + \gamma
p_{1}(T)^{2} + \delta p_{2}(T)\nonumber\\
&  + \sum_{i} \left[  \mu_{i} \, \mathrm{Tr} F_{i}^{4}
+ \, \mathrm{Tr} F_{i}^{2} \left(  \rho_{i}
p_{1}(T) + \sigma_{i} c_{2}(R) + \sum_{j} \eta_{ij} \, \mathrm{Tr} F_{j}^{2}
\right)  \right] , \label{anomalypoly}%
\end{align}
where in the above, $c_{2}(R)$ is the second Chern class of the $SU(2)_{R}$ symmetry,
$p_{1}(T)$ is the first Pontryagin class of the tangent bundle, $p_{2}(T)$ is
the second Pontryagin class of the tangent bundle, and $F_{i}$ is the field
strength of the $i^{th}$ symmetry, where $i$ and $j$ run over the
flavor symmetries of the theory. See the review article
\cite{Heckman:2018jxk} as well as the Appendices for
additional details on how to calculate the anomaly polynomial in specific
6D SCFTs.

Returning to the F-theory realization of the 6D\ SCFTs of line (\ref{m5probe}),
there is a large class of Higgs branch deformations associated with
nilpotent orbits of the flavor symmetry algebras.\footnote{We note that
although a T-brane deformation has vanishing Casimirs and may thus appear to
be \textquotedblleft invisible\textquotedblright\ to the geometry, we can
consider a small perturbation away from a T-brane which then would register as
a complex structure deformation. Since we are dealing with the limiting case
of an SCFT, all associated mass scales (as well as fluxes localized on
$7$-branes)\ will necessarily scale away. This also means that each nilpotent
element can be associated with an elliptic threefold \cite{DelZotto:2014hpa}.}
Moreover, nilpotent elements admit a partial ordering which also dictates a
partial ordering of 6D\ fixed points. We say that a nilpotent element
$\mu\preceq\nu$ when there is an inclusion of the orbits under the adjoint
action: Orbit$(\mu)\subseteq$ $\overline{\text{Orbit}(\nu)}$.

In the 6D\ SCFT, there is a triplet of adjoint valued moment maps $D_{\text{adj}}^{1},$
$D_{\text{adj}}^{2},$ $D_{\text{adj}}^{3}$ which couple to the flavor symmetry
current supermultiplet. The nilpotent element can be identified with the
complexified combination $D_{\text{adj}}^{\mathbb{C}}=D_{\text{adj}}%
^{1}+iD_{\text{adj}}^{2}$. Closely related to this triplet of moment maps are
the triplet of D-term constraints for each gauge group factor $G_{j}$ for
$j=1,...,k$. Labeling these as a three-component vector taking values in the
adjoint of each such group $\overrightarrow{D}_{j}$, supersymmetric vacua are
specified in part by the conditions:%
\begin{equation}
\overrightarrow{D}_{j}=0\text{ for all }j,
\end{equation}
modulo unitary gauge transformations. We note that in the weakly coupled
context, the D-term constraints for each gauge group factor are in fact
correlated with one another. In particular, if we specify a choice of moment
map $\overrightarrow{D}_{0}\neq0$ and $\overrightarrow{D}_{k+1}\neq0$ on the
left and right of the quiver, respectively, this propagates to a non-trivial
breaking pattern in the interior of the quiver.

That being said, the actual description of this breaking pattern using
6D\ conformal matter is poorly understood because there is no weakly
coupled description available for these degrees of freedom. So, while we
expect there to be a correlated breaking pattern for gauge groups in the
interior of a quiver, the precise structure of these terms is unclear due to
the unknown structure of the microscopic degrees of freedom in the field theory.

In spite of this, it is often possible to extract the resulting IR fixed point after such a deformation,
even in the absence of a Lagrangian description. The main reason this
is possible is because in the context of an F-theory compactification, we
already have a classification of all possible outcomes which could have
resulted from a Higgs branch flow (since we have a classification of
6D\ SCFTs). In many cases, this leads to a unique candidate
theory after Higgsing, and this has been used to directly determine the
Higgsed theory. Even so, this derivation of the
theory obtained after Higgsing involves a number of steps which are not
entirely systematic, thus leading to potential ambiguities in cases where the number of gauge group
factors in the quiver is sufficiently small that there is a non-trivial correlation in the
symmetry breaking pattern obtained from a pair of nilpotent orbits (one on the left and one on the
right of the quiver). We refer to such quivers as being ``short,'' and the case where there is no correlation between
breaking patterns from different nilpotent orbits as ``long.''

One of our aims in the present
paper will be to determine the condensation of strings stretched between
different stacks of branes. Our general strategy for analyzing
Higgs branch flows will therefore split into two parts:

\begin{itemize}
\item First, we determine the particular configuration of multi-pronged strings
associated with each nilpotent orbit.

\item Second, we determine how to consistently condense these multi-pronged
string states to trigger brane recombination in the quiver-like gauge theory.
\end{itemize}

\section{Nilpotent Orbits from String Junctions \label{sec:NILPJUNC}}

One of our aims in this paper is to better understand the combinatorial
structure associated with symmetry breaking patterns for 6D\ SCFTs. In
this section we show how to construct all of the nilpotent
orbits of a semi-simple Lie algebra of type ABCDEFG from the structure of
multi-pronged string junctions. The general idea follows earlier
work on the construction of such algebras, as in \cite{Gaberdiel:1997ud, DeWolfe:1998zf, Bonora:2010bu} (see also \cite{Grassi:2013kha, Grassi:2014ffa, Grassi:2018wfy}). We refer the interested reader to Appendix \ref{sec:Nilpotents}
for additional details and terminology on nilpotent orbits which we shall
reference throughout this paper.

Recall that in type IIB, we engineer such algebras using $[p,q]$ $7$-branes, namely
a bound state of $p$ D7-branes and $q$ S-dual
$7$-branes. Labeling the monodromy of the axio-dilaton around a source of
$7$-branes by a general element of $SL(2,\mathbb{Z})$:
\begin{equation}
\tau\mapsto\frac{a\tau+b}{c\tau+d}\text{ \ \ for \ \ }\left[
\begin{array}
[c]{cc}%
a & b\\
c & d
\end{array}
\right]  \in SL(2,\mathbb{Z})\text{,}%
\end{equation}
a $[p,q]$ $7$-brane determines a conjugacy class in $SL(2,\mathbb{Z})$ as specified
by the orbit of:\footnote{A note on conventions: One can either consider this matrix or its inverse depending on
whether we pass a branch cut counterclockwise or clockwise. This will not affect our discussion in any material way.}
\begin{equation}
M_{[p,q]} = \left[
\begin{array}
[c]{cc}%
1 + pq & - p^2 \\
q^2 & 1 - pq
\end{array}
\right].
\end{equation}
\\
The relevant structure for realizing the different ADE\ algebras are the monodromies:%
\begin{align}
A= M_{[1,0]} = \left[
\begin{array}
[c]{cc}%
1 & -1\\
0 & 1
\end{array}
\right]  \text{, \ \ }B= M_{[1,-1]} = \left[
\begin{array}
[c]{cc}%
0 & -1\\
1 & 2
\end{array}
\right]  \text{, \ \ } \nonumber \\
C= M_{[1,1]} = \left[
\begin{array}
[c]{cc}%
2 & -1\\
1 & 0
\end{array}
\right]  \text{, \ \ }
X= M_{[2,-1]} = \left[
\begin{array}
[c]{cc}%
-1 & -4\\
1 & 3
\end{array}
\right]  .
\end{align}
The $7$-branes necessary to engineer various
Lie algebras follow directly from the Kodaira classification of possible
singular elliptic fibers at real codimension two in the base of an F-theory
model \cite{Vafa:1996xn, Morrison:1996na, Morrison:1996pp}. They can also
be directly related to a set of basic building blocks in the string junction picture
worked out in \cite{Gaberdiel:1997ud} which we label as in reference \cite{DeWolfe:1998pr}:
\begin{align}
A_{N}  &  :A^{N+1}\\
H_{N}  &  :A^{N+1}C\text{ \ \ (for }N=0,1,2\text{)}\\
D_{N}  &  :A^{N}BC\\
E_{N}  &  :A^{N - 1}BC^{2}\text{ \ \ (for }N=6,7,8\text{)}\\
\widetilde{E}_{N}  &  :A^{N}XC\text{ \ \ (for }N=6,7,8\text{)}.
\end{align}
The $H_{N}$ series in the second line represents an alternative way to realize low
rank $SU$ type algebras. We also note that in the case of the A- and D-
series, it is possible to remain at weak string coupling, while the H- and
E-series require order one values for the string coupling. Here, we have indicated two alternate presentations
of the $E$-type algebras (see reference \cite{DeWolfe:1998pr}). It will prove convenient in what follows to use the
$\widetilde{E}_N$ realization with an $X$-brane. The non-simply laced algebras have the
same $SL(2,\mathbb{Z})$ monodromy type. In the string junction description, this involves further
identifications of some of the generators of the algebra by a suitable outer automorphism. Some aspects of this
case are discussed in \cite{Bonora:2010bu}.

We would like to understand the specific way that nilpotent generators of the
Lie algebra are encoded in this physical description. In all these cases, the
main idea is to first separate the $7$-branes so that we have a physical
realization of the Cartan subalgebra. Then, a string which stretches from one
brane to another will correspond to an 8D vector boson with mass dictated by
the length of the path taken to go from one stack to the other:%
\begin{equation}
\text{mass}\sim\frac{\text{length}}{\ell_{\ast}^{2}},
\end{equation}
with $\ell_{\ast}$ a short distance cutoff. In the limit where
all the $7$-branes are coincident, we get a massless state.

With this in mind, let us recall how we engineer the gauge algebra $\mathfrak{su}(N)$
using D7-branes. All we are required to do in this case is introduce $N$
D7-branes, which are $[p,q]$ $7$-branes with $p=1$ and $q=0$. Labeling the
$7$-branes as $A_{1},...,A_{N}$, we can consider an open string which
stretches from brane $A_{i}$ to brane $A_{j}$. Since this string comes with an
orientation, we can write:%
\begin{equation}
A_{i}\rightarrow A_{j}\text{,}%
\end{equation}
and introduce a corresponding nilpotent $N\times N$ matrix with a single entry
in the $i^{th}$ row and $j^{th}$ column. We denote by $E_{i,j}$ the matrix
with a one in this single entry so that the corresponding nilpotent element is
written as $v_{i,j} E_{i,j}$ with no summation on indices. Conjugation by an $SL(n,\mathbb{C})$
element reveals that the actual entry does not affect the orbit. We will, however, be
interested in RG flows generated by adding perturbations away from a single
entry, so we will often view $v_{i,j}$ as indicating a vev / energy scale. In this manner, we can represent an RG flow triggered by moving onto the Higgs branch of the theory, which is labeled by a nilpotent orbit of a Lie algebra, in terms of a collection of strings stretched between the $7$-branes.

\begin{landscape}
\begin{table}
\centering
\begin{tabular}{|>{\centering}m{.25cm} |>{\centering}m{2.2cm} |>{\centering}m{5.8cm} |>{\centering}m{4cm} |>{\centering}m{5cm}|m{3.5cm}|}
\hline
& Dynkin diagram & IIB with mirror plane & Physical picture from \cite{DeWolfe:1998zf} & Branching rule to $\mathfrak{su}(4) \times \mathfrak{u}(1)$ & \begin{center}Positive roots \end{center} \\
\hline
$A_4$
&\includegraphics{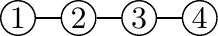}
&\includegraphics{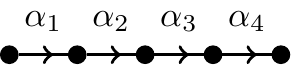}
&\includegraphics{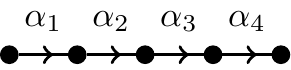}
& $24 \rightarrow 15_0 + 4_1 + \overline{4}_{\scalebox{0.75}[1.0]{-}1} + 1_0$ & \makecell{10 one-pronged strings: \\ $a_i-a_j$} \\ \hline
$B_4$
&\includegraphics{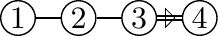}
&\includegraphics{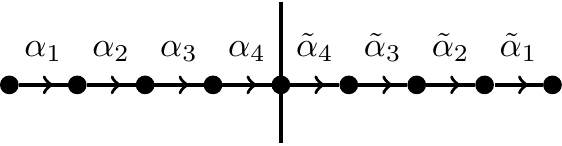}
&\includegraphics{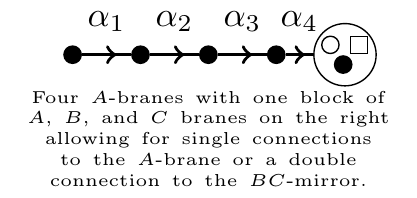}
& $36 \rightarrow 15_0 + 6_2 + \overline{6}_{\scalebox{0.75}[1.0]{-}2}+4_1+\overline{4}_{\scalebox{0.75}[1.0]{-}1} + 1_0$
& \makecell{10 one-pronged strings: \\ $a_i-a_j, \ \tilde{a}_j-\tilde{a}_i$ \\
  6 two-pronged strings: \\ $a_i-\tilde{a}_j, \ a_j-\tilde{a}_i$} \\
  \hline
$C_4$
&\includegraphics{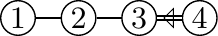}
&\includegraphics{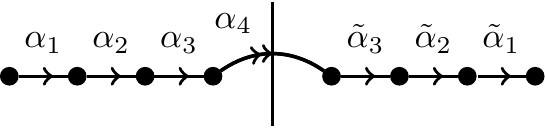}
&\includegraphics{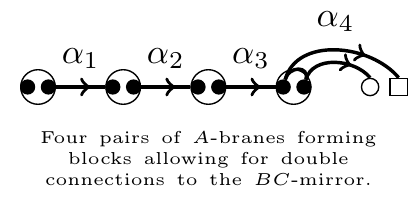}
  & $36 \rightarrow 10_2 + \overline{10}_{\scalebox{0.75}[1.0]{-}2} + 15_0 +1_0$ & \makecell{6 one-pronged strings: \\ $a_i-a_j, \ \tilde{a}_j-\tilde{a}_i$ \\
  4 double strings: \\ $a_i-\tilde{a}_i$ \\
6 two-pronged strings: \\ $a_i-\tilde{a}_j, \ a_j-\tilde{a}_i$} \\ \hline
$D_4$
&\includegraphics{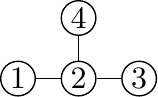}
&\includegraphics{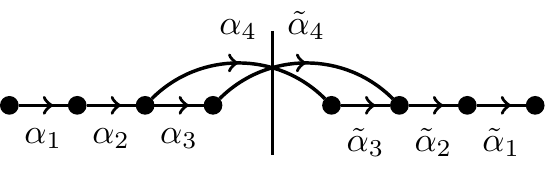}
&\includegraphics{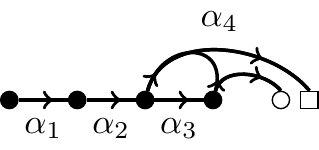}
& $28 \rightarrow 6_2 + \overline{6}_{\scalebox{0.75}[1.0]{-}2}+15_0 + 1_0$
& \makecell{6 one-pronged strings: \\ $a_i-a_j, \ \tilde{a}_j-\tilde{a}_i$ \\
6 two-pronged strings: \\ $a_i-\tilde{a}_j, \ a_j-\tilde{a}_i$} \\ \hline
\end{tabular}
\caption{Summary of basic properties for the string junction realization of the classical Lie algebras $A_4$, $B_4$, $C_4$, $D_4$. The columns from left to right are: Dynkin diagrams, IIB brane picture, string junction picture from \cite{DeWolfe:1998zf}, branching rule of adjoint decomposition in $\mathfrak{su}(4) \times \mathfrak{u}(1)$, explicit expression of groups of positive roots based on the adjoint decomposition. Here the indices $i$, $j$ run from $1$ to the number of nodes on the left-hand side of the mirror ($BC$). The tilde nodes are the reflected branes and the indices continue running as $\tilde{i} = N-i$ where $N$ is the total number of nodes in the diagram.}
\label{tab:classicalRootTable}
\end{table}
\end{landscape}

\begin{figure}[t!]
\centering
\includegraphics{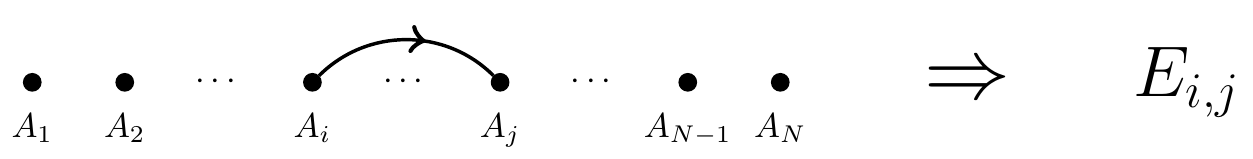}
\caption{Separating a collection of A-type branes leads to a deformation of $\mathfrak{su}(N)$ to the Cartan subalgebra. Open strings
stretched between distinct branes are associated with specific generators in the complexified Lie algebra. In the figure, this is shown for a string stretched from brane $A_i$ to brane $A_j$.}%
\label{fig:hoppadoo}%
\end{figure}

Ordering the branes $A_{1},..,A_{N}$ from left to right in the plane
transverse to the stack of $7$-branes, we see that we can now populate the
strictly upper triangular portion of a matrix in terms of strings
$A_{i}\rightarrow A_{j}$ for $i<j$ (see figure \ref{fig:hoppadoo}). So in other words, we can populate all
possible nilpotent orbits (in this particular basis). Similar considerations
hold for the other algebras, but clearly, this depends on a number of
additional features such as unoriented open strings (in the case of the
classical SO / Sp algebras) and multi-pronged string junctions (in the case of
the exceptional algebras). A related comment is that we are just constructing
a representative nilpotent element in the orbit of the Lie algebra. What we
will show is that for any deformation onto the Cartan, there is a ``minimal
length'' choice, and all the other elements of the orbit are obtained through
the adjoint action of the Lie algebra.

Our plan in the rest of this section will be to establish in detail how to
construct the corresponding nilpotent orbits for each configuration of
strings. Additionally, we show that not only can we generate all orbits, but
that the combinatorial method of \textquotedblleft adding extra
strings\textquotedblright\ automatically generates a partial ordering on the
space of nilpotent orbits, which reproduces the standard partial ordering of the
nilpotent cone. The essential information for the classical Lie algebras, and
in particular the list of simple and positive roots, is illustrated in table
\ref{tab:classicalRootTable}. We elaborate on the content of this table (as
well the exceptional analogs) in the following subsections.

\subsection{$SU(N)$: Partition by Grouping Branes with Strings}

\begin{figure}[t!]
\centering
\includegraphics{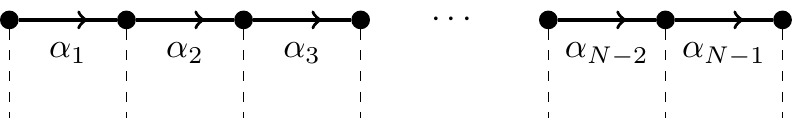}
\caption{Brane diagram of strings/roots stretching between the $A$-branes
yielding an $SU(N)$ flavor symmetry (see \cite{DeWolfe:1998zf}). The dashed
lines represent the position of branch cuts. Since they do not contribute to
our analysis, they are not drawn in subsequent pictures.}%
\label{fig:SUNroots}%
\end{figure}

In the case of an $SU(N)$ flavor we simply have $N$ perturbative $A$-branes
with $[p,q]=[1,0]$ charges. The $N-1$ simple roots of $SU(N)$ can be
represented by strings joining two adjacent $A$-branes as shown in figure
\ref{fig:SUNroots}. We refer to these as ``simple strings'' due to their
correspondence to the simple roots. The remaining (non-simple) roots are then
described by strings connecting any two $A$-branes. The positive roots are
represented by strings stretching from left to right while the negative ones
would go in the opposite direction (as indicated by the arrows). That is we
choose a basis for the generators of the $\mathfrak{su}_{N}$ algebra to be
given by:

\begin{itemize}
\item $N(N-1)/2$ nilpositive elements $E_{i,j}$ with $1 \leq i < j
\leq N$ corresponding to strings stretching from the $i^{th}$ to the $j^{th}$
$A$-brane (with the arrow pointing from left to right).

\item $N(N-1)/2$ nilnegative elements $E_{j,i} = X_{k}^{T}$ with $1
\leq i < j \leq N$ corresponding to strings stretching from the $j^{th}$ to
the $i^{th}$ $A$-brane (with the arrow now pointing from right to left).

\item $(N-1)$ Cartans $[E_{i,i+1}, E_{i+1,i}]$ for $1 \leq i \leq N-1$.
\end{itemize}

Through out this paper we denote $E_{i,j}$ to be matrix with value $+1$ in the
entry $(i,j)$ but zeros everywhere else. The positive simple roots are given
by $\alpha_{i}$ $(1 \leq i \leq rank(G))$, with the corresponding matrix
representation labelled $E_{\alpha_{i}}$. Any non-simple root can then be labelled
explicitly in terms of its simple roots constituents: $\alpha_{i,j,k,\dots
,p,q} = \alpha_{i} + \alpha_{j} + \alpha_{k} + \dots+ \alpha_{p} + \alpha_{q}$
and the corresponding matrix representation is obtained from nested commutators.

In this basis, the simple positive roots are $E_{i,i+1}$ for $1
\leq i \leq N-1$, as illustrated by their corresponding directed strings in
figure \ref{fig:SUNroots}. Furthermore, we use the convention of
\cite{DeWolfe:1998zf} to keep track of the different monodromies. Namely, we only display the directions
transverse to the $7$-brane, thus representing each $7$-brane as a point. In this picture
the associated branch cut for $SL(2,\mathbb{Z})$ monodromy stretches vertically downward to infinity.
This will not enter our analysis in any material way so in order not to overcrowd the
figures, we will mostly not draw the branch cuts.

\begin{figure}[t!]
\centering
  \includegraphics{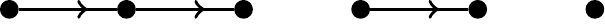}
  \par
  \includegraphics{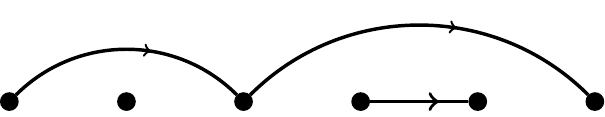}
  \par
  \includegraphics{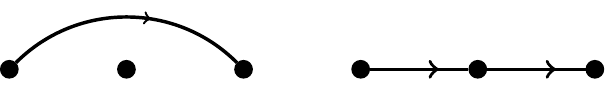}
\caption{Three equivalent ways of describing the partition $[3,2,1]$ in the
set of nilpotent orbits of $SU(6)$. To each picture is associated a different
matrix, but they all have the same Jordan block decomposition and thus belong
to the same equivalence class.}%
\label{fig:equivSU}%
\end{figure}

We have already seen that nilpotent orbits of $SU(N)$ are parametrized by
partitions of $N$ (with no restriction whatsoever). Thus it becomes natural to
classify nilpotent orbits by how branes are grouped together. Namely, we can
group any set of $A$-branes by stretching strings between them, giving rise to
a particular partition of the $N$ branes. This partition is then in one-to-one
correspondence with its corresponding nilpotent orbit. As an equivalence
class, we have many different string configurations belonging to the same
orbit (just like many different matrices have the same Jordan block
decomposition). For instance, the three string junctions of figure
\ref{fig:equivSU} all represent the same $[3,2,1]$ partition:

\begin{itemize}
\item The first string junction picture has a matrix representation $M_{1} =
E_{1,2}+E_{2,3}+E_{4,5}$.

\item The second configuration has matrix representation $M_{2} = E_{1,3}+E_{3,6}+E_{4,5}$.

\item And finally, the third one has matrix representation $M_{3} =E_{1,3}+E_{4,5}+E_{5,6}$.
\end{itemize}

To each nilpotent orbit of $SU(N)$ we can then associate one of many possible string junction pictures. To keep the picture as simple as
possible, we choose to use only ``simple'' positive strings, that is strings stretching
from left to right between two adjacent $A$-branes. This ensures that we
only make use of simple roots. This typically does not completely fix a string
junction representative, so we are free to make a convenient choice of the remaining possibilities.

\begin{figure}[t!]
  \centering
  \includegraphics{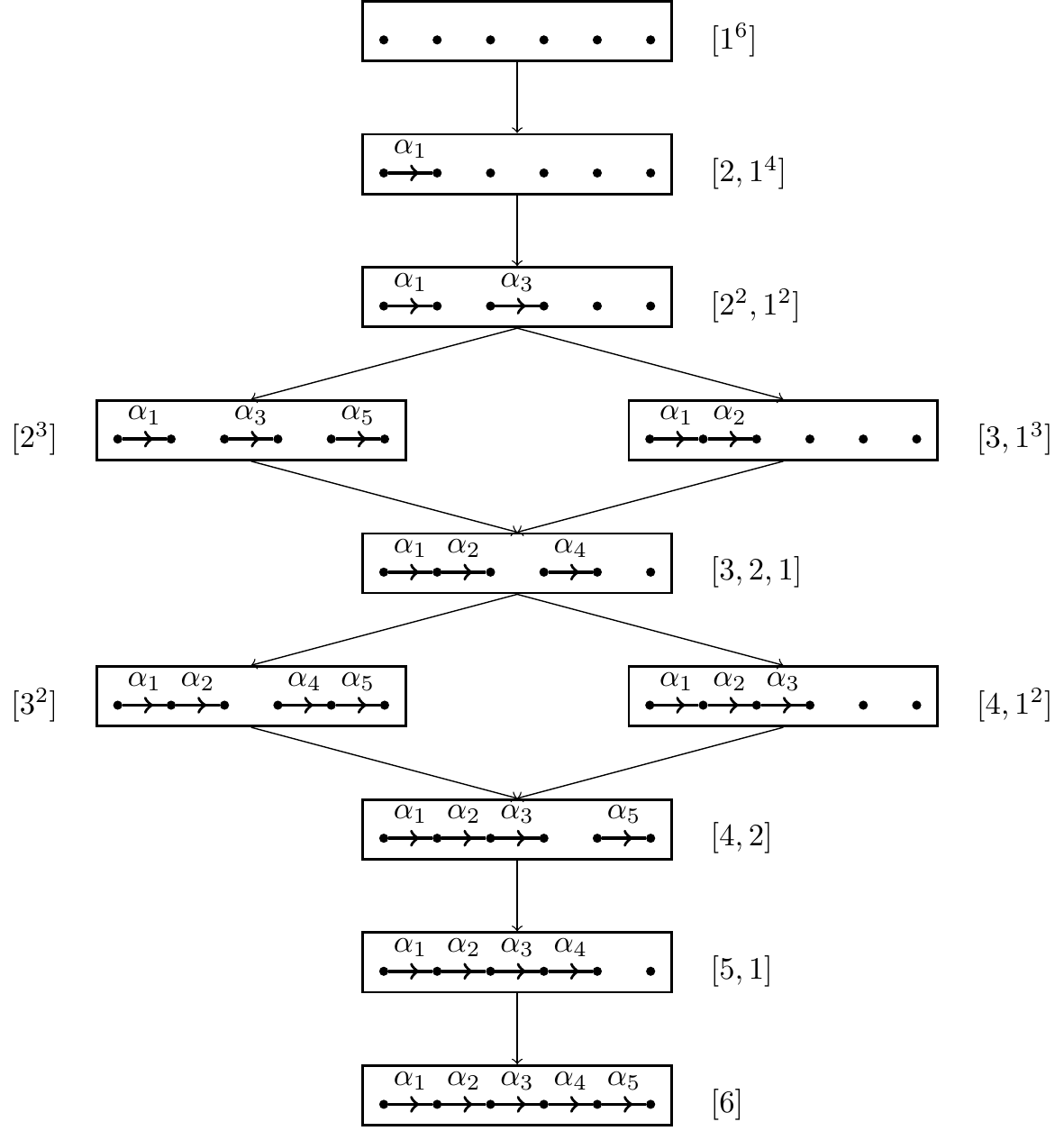}
  \caption{Hasse diagram of $SU(6)$ nilpotent deformations going from top (UV) to bottom (IR) where all simple roots are turned on and all corresponding ``simple strings'' connect the $A$-branes.}
  \label{fig:SU6hasse}
\end{figure}

By starting with a configuration with no string attached (a $[1^{N}]$
partition) we can add more and more strings to go from the $[2,1^{N-2}]$ orbit
all the way to the $[N]$ partition. This generates a whole Hasse diagram of
nilpotent orbits which exactly matches that which is mathematically predicted.
Figure \ref{fig:SU6hasse} illustrates this diagram for the case of $SU(6)$
where we associate a ``standard'' string junction picture to each
nilpotent orbit according to how the branes are partitioned as we add more and
more strings.

\newcommand{\rsep}{0.5cm}
\newcommand{\vsep}{-1.5}

More precisely, in order to flow from one point of the Hasse diagram to the
next, one simply needs to add a small perturbation, that is, an oriented
string (moving from left to right) corresponding to a positive root. By the definition
of the partial ordering of nilpotent orbits, this guarantees that the RG flow
indeed always takes us deeper into the IR. Weyl transformations / brane permutations can then be used to reduce
the obtained diagram back to one of the standard ones which only relies on the
simple roots.

\begin{figure}[t!]
  \centering
  \includegraphics{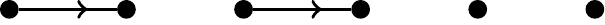}
  \par
{\Huge \rotatebox{-45}{$\Downarrow$} }
\par
{\Huge \vspace{-0.5cm}
  \includegraphics{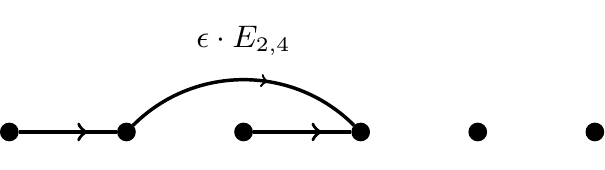}
  \hspace{1cm} $\Leftrightarrow$ \hspace{1cm}
  \includegraphics{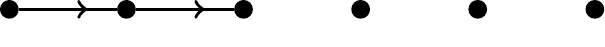}
}\caption{One way of flowing from the $[2^{2},1^{2}]$ nilpotent orbit (top) to
the $[3,1^{3}]$ orbit (bottom). In the top figure we have the matrix
representation $M_{1}=E_{1,2}+E_{3,4}$. The flow is then induced by adding an
extra string stretching between the $2^{nd}$ and $3^{rd}$ branes, as
illustrated in the bottom left figure. This corresponds to the matrix
$M_{2}=E_{1,2}+E_{3,4}+\epsilon\cdot E_{2,4}$. This matrix is similar to
$M_{2}^{\prime}=E_{1,2}+E_{2,3}$ corresponding to the bottom right diagram.
Thus, both bottom string junctions belong to the same nilpotent orbit
$[3,1^{3}]$.}%
\label{fig:SUspecialFlow}%
\end{figure}

The flows involving only the addition of a simple root (corresponding to
linking two more branes together) are fairly clear. The only cases where that
is not so obvious are the ones corresponding to flows that are similar to the
one described in figure \ref{fig:SU6hasse} by going from $[2^{2},1^{2}]$ to
$[3,1^{3}]$. For this we can add the string $\alpha_{2}+\alpha_{3}=a_{2}%
-a_{4}$, corresponding to a small deformation $\epsilon\cdot E_{2,4}$. This
particular flow is illustrated in figure \ref{fig:SUspecialFlow}. Generalizing
this procedure to arbitrary $SU(N)$ shows that the intermediate RG flows are
guaranteed to be physically realizable in the same fashion.

\subsection{$SO(2N)$ and $SO(2N-1)$ \label{subsubsec:SO2Nflav}}

In F-theory, the $SO(2N)$ and $SO(2N-1)$ geometries
are realized by the presence of $A^{N}BC$-branes. In type IIB however, the
$BC$-branes turn into an $O7^{-}$ orientifold plane (as discussed in
\cite{Sen:1996vd}) which we refer here as the ``BC-mirror''. This mirror
reflects the $N$ $A$-branes across, yielding a total of $2N$ branes (half of
which are physical, half of which are ``image'' branes). We thus
represent $SO(2N)$ by $2N$ dots separated by a vertical line representing the
$BC$-mirror, and $SO(2N-1)$ by merging one $A$-brane with its mirror image
onto the orientifold so that we have $N-1$ $A$-branes on the left, $N-1$
mirror $A$-branes on the right, and a single $A$-brane squeezed onto the
vertical line representing the mirror.

\begin{figure}[t!]
\centering
\includegraphics{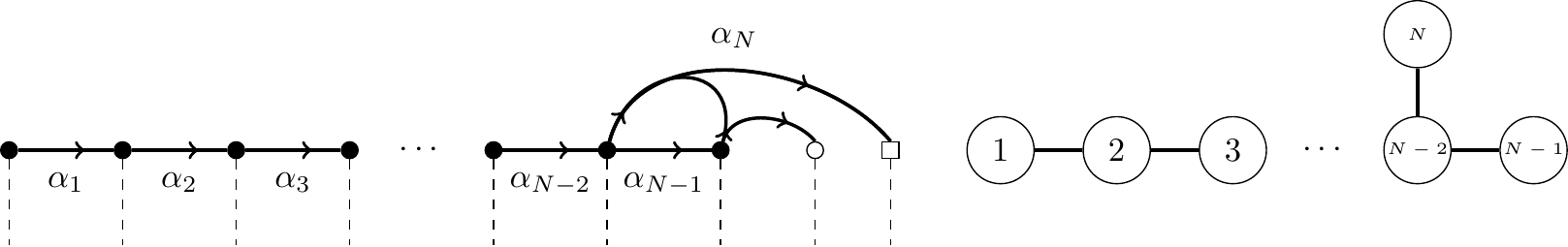}
\caption{Brane diagram of strings/roots stretching between the $A$, $B$, and
$C$-branes, making up the $SO(2N)$ symmetry (see \cite{DeWolfe:1998zf}). The
$A$-branes are denoted by black circles, the $B$-brane by an empty circle, and
the $C$-brane by an empty square. The dashed lines represent the position of
branch cuts, which (once again) are not drawn in subsequent pictures. To the right
we give the corresponding Dynkin diagram with simple roots numbered.}%
\label{fig:SO2NrootsFtheory}%
\end{figure}

Furthermore, \cite{DeWolfe:1998zf} provides us with a set of string junctions
to represent the simple roots of $SO(2N)$, as illustrated in figure
\ref{fig:SO2NrootsFtheory}. We can then obtain the corresponding roots for
$SO(2N-1)$ via the standard projection (or branching) of $SO(2N) \rightarrow
SO(2N-1)$. We see that much like $SU(N)$, we can have strings stretching
between any pair of $A$-branes, and the simple strings correspond to those
stretching between adjacent pairs. However, the presence of the $B$ and $C$
branes allows for a new kind of string: a two-pronged string which takes two
$A$-branes and connects them to the $B$ and $C$-branes. All these
configurations are regulated by charge conservation: the $A$-branes
all have charges $[1,0]$ so that a fundamental string can stretch between any pair
of them, but the $B$-brane has charge $[1,-1]$, and the $C$-brane has charge
$[1,1]$. Thus, no string can stretch directly between a $B$ and a $C$-brane.
However, these two branes together have an overall charge of $[2,0]$, which is exactly
twice that of an $A$-brane. Therefore, by combining two $A$-branes with the
$B$ and $C$-branes, charge can be conserved. This combination is achieved
through the introduction of a two-pronged string denoted $\alpha_{N}$ in
figure \ref{fig:SO2NrootsFtheory}.

\begin{figure}[t!]
\centering
\includegraphics{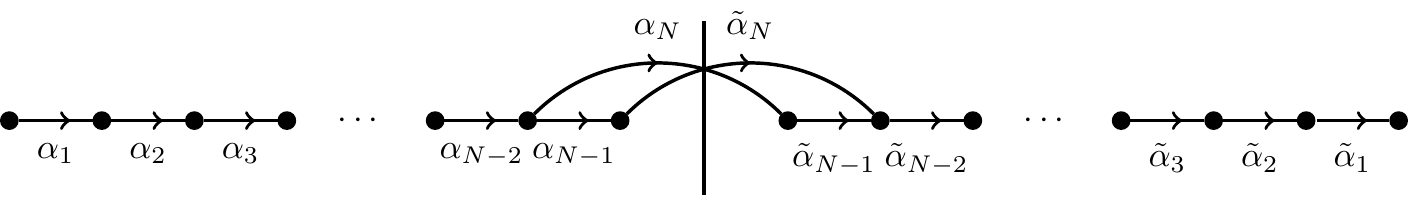}
\caption{Brane diagram of strings/roots stretching, for $SO(2N)$. The $B$ and
$C$-branes are turned into an orientifold, which is denoted by a mirror (vertical line).
The strings corresponding to simple roots are illustrated by arrows stretching
between the branes and reflected across the mirror. We note that the
distinguished root $\alpha_{N}$ corresponds to the two-pronged string and
indeed it is made of two legs moving across the $BC$-mirror in order to
respect the difference in charges between the $A$, $B$, and $C$ branes.}%
\label{fig:SO2Nroots}%
\end{figure}

\begin{figure}[t!]
\centering
\includegraphics{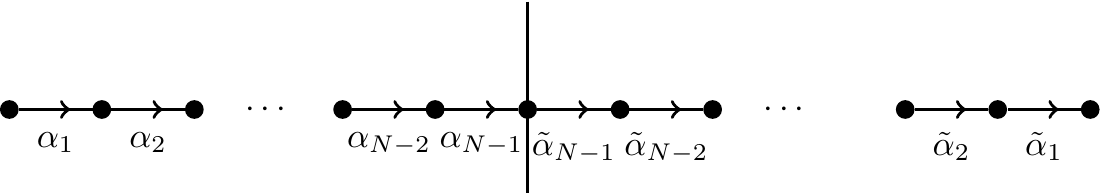}
\caption{Brane diagram of strings/roots stretching, for $SO(2N-1)$. The $B$
and $C$-branes are turned into an orientifold denoted by a mirror (vertical
line) and one of the $A$-branes is squeezed onto it. The strings corresponding
to simple roots are illustrated by arrows stretching between the branes and
reflected across the mirror.}%
\label{fig:SO2Nm1roots}%
\end{figure}

We then visualize this $SO(N)$ geometry by introducing the orientifold, which reflects the strings as well as the $A$-branes. This is
illustrated in figures \ref{fig:SO2Nroots} and \ref{fig:SO2Nm1roots} for
$SO(2N)$ and $SO(2N-1)$ respectively.

As we can see, the presence of the mirror guarantees that even parts (in the
partition of $2N$ or $2N-1$) appear an even number of times whenever we use
any of the regular one-pronged simple strings. Thus, using the same rules as
with $SU(N)$, we can generate most allowed partitions corresponding to $SO$
groups. We note that unlike $SU(N)$, we also have the presence of a two-pronged
string coming as a result of the distinguished root $\alpha_{N}$ of $SO(2N)$.
This can result in configurations where the partitions are not so obvious from
the string junction picture. We can thus turn to the equivalent matrix
representation and read off the corresponding partition from the equivalence
class it belongs to. To do that, we once again need to specify what basis we
are using. Generalizing the rules from $\mathfrak{su}_{N}$ listed in the
previous section to $\mathfrak{so}_{2N}$, we have the following $N(N-1)$
nilpositive elements:

\begin{itemize}
\item Half of them are: $E_{\mathrm{1-pronged}} = E_{i,j}-(-1)^{j-i}
E_{2N-j+1,2N-i+1}$ with $1 \leq i < j \leq N$ corresponding to one-pronged
strings stretching from the $i^{th}$ to the $j^{th}$ $A$-brane, as well as
their reflections--namely, the strings stretching between the $(2N-j+1)^{th}$
and the $(2N-i+1)^{th}$ nodes, which are on the right-hand side of the mirror.
These correspond to the $\mathfrak{su}_{N} \subset\mathfrak{so}_{2N}$
nilpositive generators.

\item The other half are: $E_{\mathrm{2-pronged}} = E_{i,2N-j+1}- (-1)^{j-i}
E_{j,2N-i+1}$ with $1 \leq i < j \leq N$ corresponding to two-pronged strings
stretching between the $i^{th}$ and $(2N-j+1)^{th}$ nodes as well as the
$j^{th}$ and $(2N-i+1)^{th}$ nodes.
\end{itemize}

The associated $N(N-1)$ nilnegative elements are simply $E^{T}_{\mathrm{1-pronged}}$
and $E^{T}_{\mathrm{2-pronged}}$. These correspond to the same one- and two-pronged
strings but with their directions reversed. Finally, we have $N$ Cartans:
The first $(N-1)$ come from one-pronged strings: $H_{i} = [E_{i,i+1}%
+E_{2N-i,2N-i+1},E_{i+1,i}+E_{2N-i+1,2N-i}]$ for $1 \leq i \leq N-1$. These
correspond to the $\mathfrak{su}_{N} \subset\mathfrak{so}_{2N}$ Cartan
generators. The last generator is then given by $H_{N} = [E_{N-1,N+1}%
+E_{N,N+2},E_{N+1,N-1}+E_{N+2,N}]$

Note the presence of negative values introduced by the reflection across
the $BC$-mirror. We choose our convention such that simple roots only contain
positive entries. The minus signs are then imposed to some non-simple roots
simply because they are given by commutators of simple root. For instance the
non-simple string $\alpha_{1}+\alpha_{2}$ inside $SO(8)$ is represented by the
matrix $[E_{1,2}+E_{7,8}, E_{2,3}+E_{6,7}] = E_{1,2} \cdot E_{2,3} - E_{6,7} \cdot E_{7,8} = E_{1,3}-E_{6,8}$.

As a result of the above equations, the simple positive roots (corresponding
to the simple strings of figure \ref{fig:SO2Nroots}) are then given by the
matrices $E_{i,i+1}+E_{2N-i,2N-i+1}$ for $1 \leq i \leq N-1$
and $X^{SO(2N)}_{N} = E_{N-1,N+1}+E_{N,N+2}$. The positive simple roots for
$SO(2N-1)$ are identical, except for the last one. Indeed, we have:
$E_{i,i+1}+E_{2N-i,2N-i+1}$ for $1 \leq i \leq N-2$ (as
before) but the shorter simple root is $\sqrt{2} \left(E_{N-1,N} + E_{N,N+1}\right)  $.
The remaining non-simple roots are simply
obtained by taking the appropriate commutators.

As an example of a partition which is not immediately obvious from the string
junction picture, we can stretch the two strings $\alpha_{N}$ and $\alpha
_{N-1}$ from figure \ref{fig:SO2Nroots}. The associated matrix makes it
obvious what orbit such configuration belongs to: in particular, it corresponds to
the $2N \times2N$ matrix $M=E_{N-1,N}+E_{N+1,N+2}+E_{N-1,N+1}+E_{N,N+2}$ which
belongs to the nilpotent orbit of $[3,1^{2N-3}]$.

\begin{figure}[t!]
\centering
\includegraphics{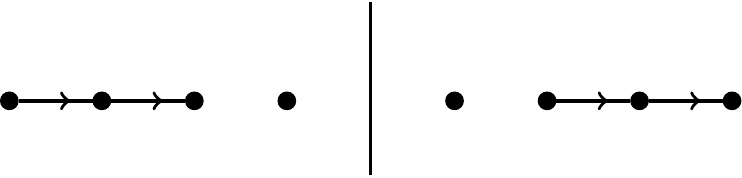}
\par
\includegraphics{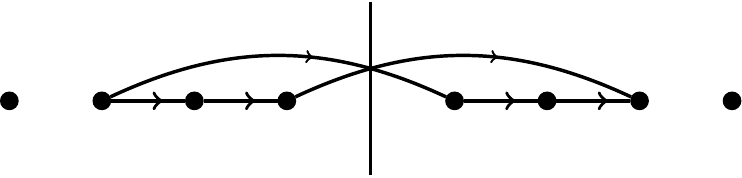}
\par
\includegraphics{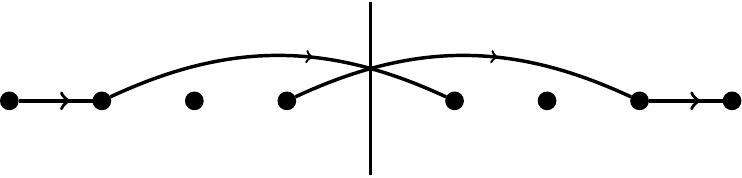}
\caption{Three equivalent ways of describing the partition $[3^{2},1^{2}]$ in
the set of nilpotent orbits of $SO(8)$. To each picture is associated a
different matrix, but there exists an inner automorphism that can bring them
all to the same Jordan block decomposition. Therefore, they belong to the same
equivalence class.}%
\label{fig:equivSO}%
\end{figure}

With this set of strings and corresponding matrices we can now associate to
each partition a string junction picture. Just like for $SU(N)$ we have many
choices. For instance, the three diagrams of figure \ref{fig:equivSO} all
represent the same $[3^{2},1^{2}]$ partition:

\begin{itemize}
\item The first string junction picture has a matrix representation $M_{1} =
E_{1,2}+E_{7,8}+E_{2,3}+E_{6,7}$.

\item The second configuration has matrix representation $M_{2} =
E_{2,3}+E_{6,7}+E_{3,4}+E_{5,6}+E_{2,5}-E_{4,7}$.

\item The third has matrix representation $M_{3}%
=E_{1,2}+E_{7,8}+E_{2,5}-E_{4,7}$.
\end{itemize}

\begin{figure}[!htb]
  \centering
  \includegraphics{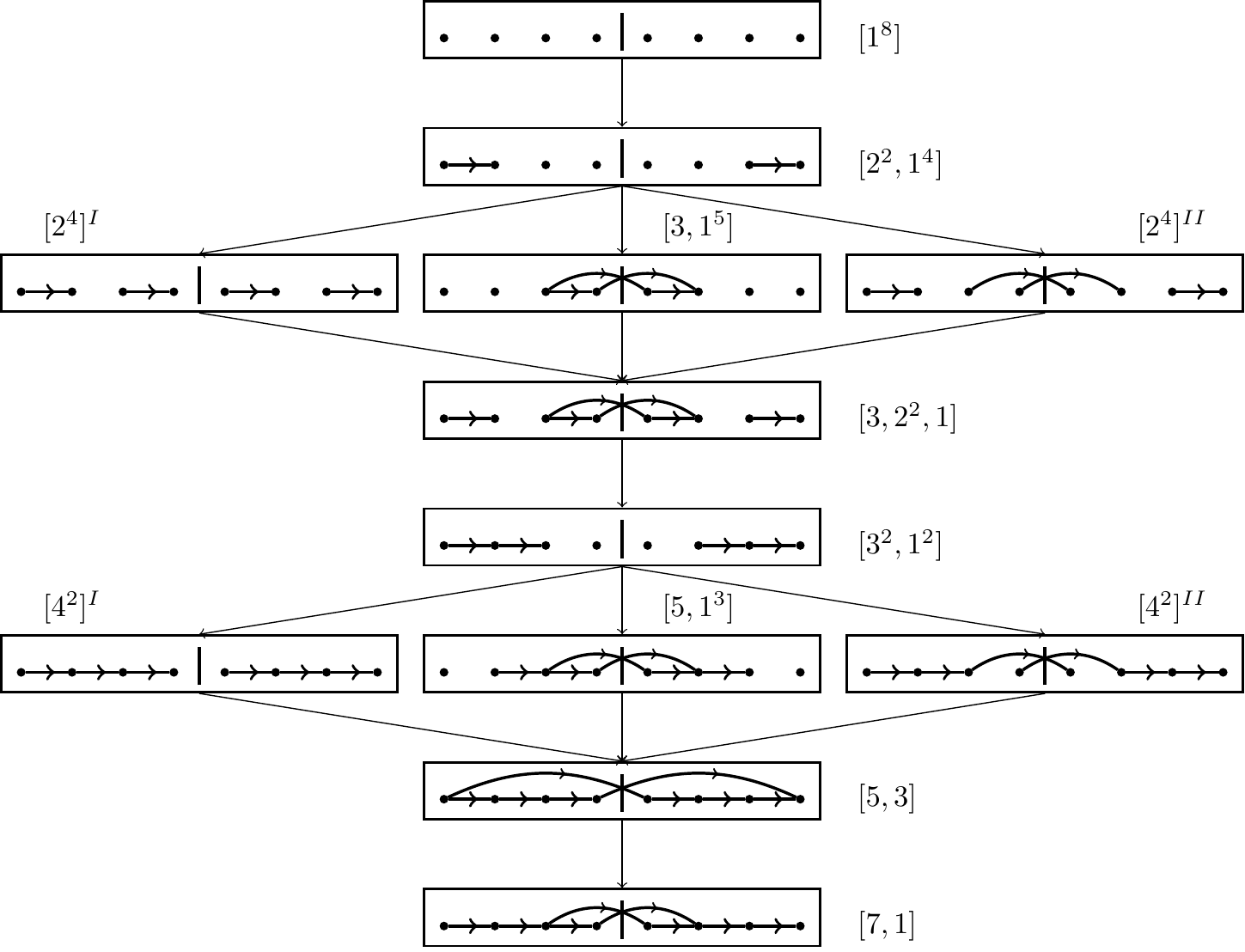}
  \caption{Hasse diagram of $SO(8)$ nilpotent deformations going from top (UV) to bottom (IR) where all simple roots are present and every corresponding simple string is connecting adjacent $A$-branes or in the case of the last simple root, two $A$-branes are connected to the $BC$-mirror.}
    \label{fig:SO8hasse}
\end{figure}

In order to keep our diagrams as simple as possible, we chose representatives
which only make use of the simple strings from figure \ref{fig:SO2Nroots},
whenever possible. However, unlike $SU(N)$, the $SO(2N)$ and $SO(2N-1)$ algebras
also contain distinguished orbits. These orbits cannot be described
with only simple roots and must therefore involve one or more non-simple
strings. We observe such a special case in the distinguished orbit $[5,3]$ of
$SO(8)$ (see figure \ref{fig:SO8hasse}). Our string junction diagrams then
allow us to recognize distinguished orbits as those requiring the presence of
one or more non-simple strings.

\begin{figure}[t!]
\centering
\includegraphics{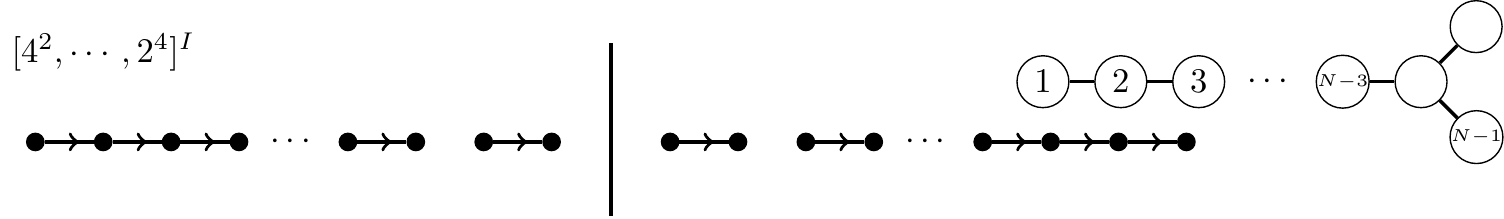}
\par
\includegraphics{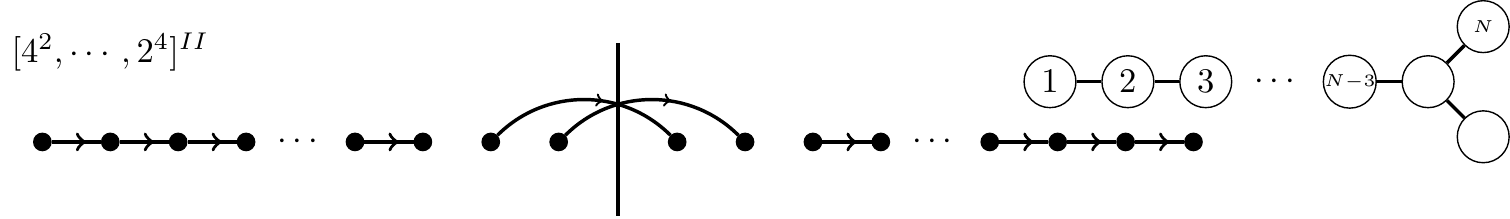}
\caption{Two very even partitions that yield the same partition but do not
belong to the same nilpotent orbit. The first one only involves one-pronged
strings and is labeled $[4^{2},\cdots,2^{4}]^{I}$ while the second one
replaces $\alpha_{N-1}$ with the two-pronged string $\alpha_{N}$ and is
labeled $[4^{2},\cdots,2^{4}]^{II}$. To the right we give the Dynkin diagrams
with the corresponding strings turned on.}%
\label{fig:veryEven}%
\end{figure}

The groups $SO(4N)$ contain ``very even'' orbits. These are orbits
with corresponding partition given by only even parts. Such partitions split into two
separate orbits, such as $[2^{4}]^{I}$ and $[2^{4}]^{II}$ or $[4^{2}]^{I}$ and
$[4^{2}]^{II}$ in $SO(8)$. That is, the matrix representation of a
$[\lambda^{\mu}]^{I}$ and a $[\lambda^{\mu}]^{II}$ configuration have the same
Jordan block decomposition and are therefore related by an \textit{outer}
automorphism. However, they are not related by any \textit{inner} automorphism
and thus do not actually belong to the same nilpotent orbit. This splitting to
two orbits for the very even partitions simply comes from the symmetry of the
Dynkin diagram for $D_{n}$: namely, the exchange of the last two roots
$\alpha_{N-1}$ and $\alpha_{N}$. This means that a very even partition
involving $\alpha_{N-1}$ (a one-pronged string) will be labeled $[\lambda
^{\mu}]^{I}$ while its companion very even partition involving $\alpha_{N}$
instead (a two-pronged string) will be labeled $[\lambda^{\mu}]^{II}$. This is
illustrated in figure \ref{fig:veryEven}.

\begin{figure}[t!]
\centering
\includegraphics{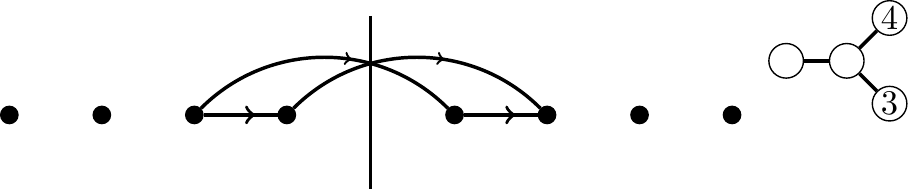}
\par
\includegraphics{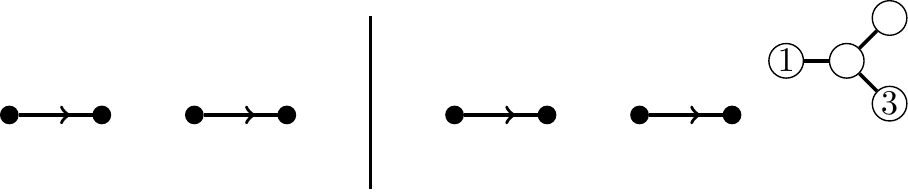}
\par
\includegraphics{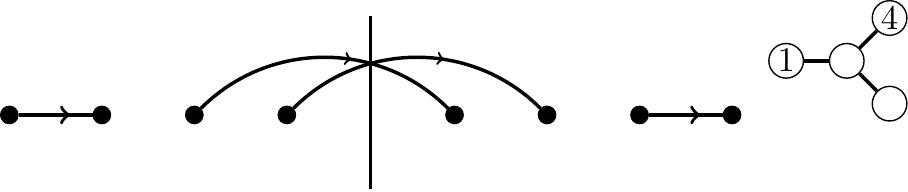}
\caption{Triality of $SO(8)$ illustrated by the three different
representations corresponding to partitions $[3,1^{5}]$ (top), $[2^{4}]^{I}$
(middle), and $[2^{4}]^{II}$ (bottom). The corresponding simple roots used are
illustrated in the adjacent Dynkin diagrams. The first has a matrix
representation $M_{1} = E_{3,4}+E_{5,6}+E_{3,5}+E_{4,6}$, the second is given
by $M_{2} = E_{1,2}+E_{7,8} + E_{3,4}+E_{5,6}$, and the last by $M_{3} =
E_{1,2}+E_{7,8}+E_{3,5}+E_{4,6}$. These all correspond to different nilpotent
orbits because there exists no inner automorphism between these three
matrices.}%
\label{fig:triality}%
\end{figure}

We briefly mention the triality automorphism of $SO(8)$ in figure
\ref{fig:triality}. Namely, we know that the nilpotent orbits with partitions
$[3,1^{5}]$, $[2^{4}]^{I}$, and $[2^{4}]^{II}$ are all related by the triality
outer automorphism. Indeed, they are represented by the following set of
roots: $\{\alpha_{3},\alpha_{4}\}$, $\{\alpha_{1},\alpha_{3}\}$, and
$\{\alpha_{1},\alpha_{4}\}$ respectively. Similarly the partitions $[5,1^{3}%
]$, $[4^{2}]^{I}$, and $[4^{2}]^{II}$ also form a trio. There is no
inner automorphism that exists between these representations, which implies that
they do indeed belong to different nilpotent orbits.

\begin{figure}[t!]
\centering
\includegraphics{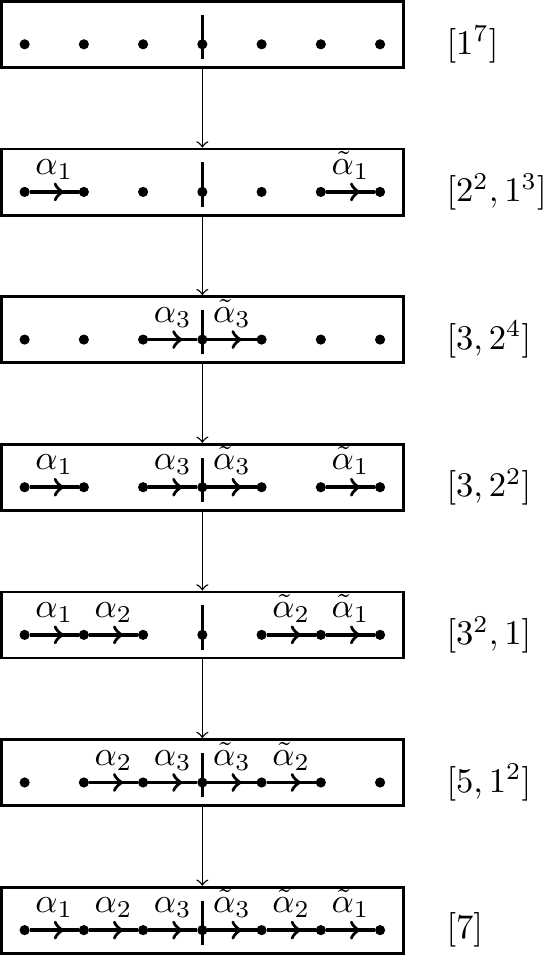}
\caption{Hasse diagram of $SO(7)$ nilpotent deformations going from the
smallest orbits at the top to largest orbits at the bottom. All simple roots
are present and every corresponding simple string is connecting the
$A$-branes. In the case of the last simple root, one $A$-brane is connecting
to the middle $A$-brane located on the $BC$-mirror.}%
\label{fig:SO7hasse}%
\end{figure}

By starting with a configuration with no string attached
($[1^{2N-1}]$ partition for $SO(2N-1)$ or $[1^{2N}]$ for $SO(2N)$) we can add
more and more strings to go from the $[2^2,1^{2N-5}]$ or $[2^{2},1^{2N-4}]$
orbit all the way to the $[2N-1]$ or $[2N]$ partitions. We summarize all of
the nilpotent orbits of $SO(7)$ and $SO(8)$ in figures \ref{fig:SO7hasse} and
\ref{fig:SO8hasse} respectively.

\newcommand{\rsepSOe}{0.5cm}
\newcommand{\vsepSOe}{-1.5}

Finally, much like what we have seen in $SU(N)$, most flows include the simple
addition of a root/string and therefore are obvious. However, there are a few
cases that are not so immediately clear. We work them out here in the case of
$SO(8)$ and note that the methods below extend to the higher rank $SO$ groups.

\begin{itemize}
\item {$[2^{2},1^{4}]\rightarrow[3,1^{5}]$:} We can add to $\alpha_{1}$ the
highest positive root $\alpha_{2,1,3,2,4}=\alpha_{1}+2 \alpha_{2}+\alpha
_{3}+\alpha_{4}$ (identified with the matrix $E_{1,7}+E_{2,8}$). This setup is
represented by the matrix $E_{1,2}+E_{7,8}+\epsilon\left(  E_{1,7}+E_{2,8}
\right)  $, which belongs to the same orbit as $E_{3,4}+E_{5,6}+E_{3,5}%
+E_{4,6}$ and corresponds to the diagram involving the set of simple strings
$\{\alpha_{3}, \alpha_{4}\}$.

\item {$[3,2^{2},1]\rightarrow[3^{2},1^{2}]$:} We can add the non-simple
string $\alpha_{2}+\alpha_{3}+\alpha_{4}$ to the initial set $\{\alpha
_{1},\alpha_{3},\alpha_{4}\}$. This gives the matrix $E_{1,2}+E_{7,8}%
+E_{3,4}+E_{5,6}+E_{3,5}+E_{4,6}+\epsilon\left(  E_{2,6}+E_{3,7} \right)  $
which is similar to the matrix $E_{1,2}+E_{7,8}+E_{2,3}+E_{6,7}$.

\item {$[3^{2},1^{2}] \rightarrow[5,1^{3}]$:} We can add the non-simple string
$\alpha_{2}+\alpha_{3}+\alpha_{4}$ to the set of simple roots $\{\alpha_{1},
\alpha_{2}\}$ to obtain the matrix $E_{1,2}+E_{7,8}+E_{2,3}+E_{6,7}%
+\epsilon\left(  E_{2,6}+E_{3,7} \right)  $. This matrix is similar to the one
corresponding to the set of strings $\{\alpha_{2}, \alpha_{3}, \alpha_{4}\}$.

\item {$[5,1^{3}], [4^{2}]^{II} \rightarrow[5,3]$} Starting from the set of simple
roots $\{\alpha_{2}, \alpha_{3}, \alpha_{4}\}$ of $[5,1^{3}]$ we can add the
positive root $\alpha_{1}+\alpha_{2}+\alpha_{3}$ to obtain the equivalent set
$\{\alpha_{1}, \alpha_{2}, \alpha_{3}, \alpha_{2}+\alpha_{3}+\alpha_{4}\}$.

Similarly, starting from the set of simple roots $\{\alpha_{1}, \alpha_{2},
\alpha_{4}\}$ of $[4^{2}]^{II}$ we can add the positive non-simple root
$\{\alpha_{2}, \alpha_{3}, \alpha_{4}\}$ again to obtain the same Weyl
equivalent set $\{\alpha_{1}, \alpha_{2}, \alpha_{3}, \alpha_{2}+\alpha
_{3}+\alpha_{4}\}$.
\end{itemize}

\subsection{$Sp(N)$}

Recall that in F-theory, we realize the $Sp(N)$-type gauge theories by a non-split $I_N$ fiber.
In terms of 7-branes, this involves the transverse intersection of a stack of D7-branes with an $O7^{-}$-plane along a common
6D subspace. In the IIA realization of this algebra, we can also consider a stack of D6-branes on top of an $O6^{+}$-plane.

\begin{figure}[t!]
\centering
\includegraphics{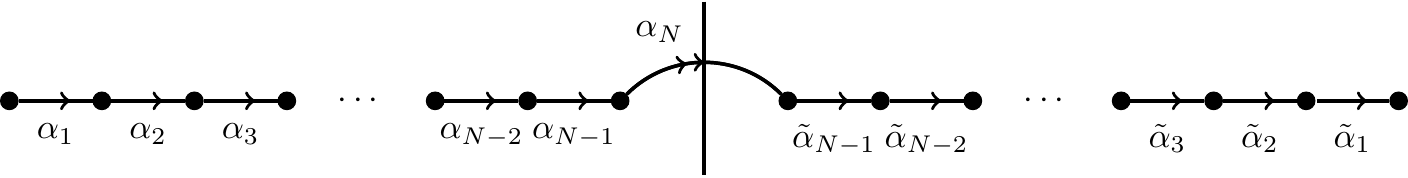}
\caption{Brane diagram of strings/roots stretching, for $Sp(N)$. The
orientifold is once again represented by a mirror (vertical line). The strings
corresponding to simple roots are illustrated by arrows stretching between the
branes and reflected across the mirror. We note that the longer root
$\alpha_{N}$ corresponds to the two-pronged string being squeezed into a
single double arrow crossing the mirror, ensuring that the charge differences
are still respected.}%
\label{fig:SpNroots}%
\end{figure}

For our present purposes, we can merge the $A$-branes pairwise on each side of the mirror. This then yields $N$ nodes on each side of the mirror but with the particularity that a two-pronged string can stretch from a single composite node, as seen in table \ref{tab:classicalRootTable}. Zooming out, the two-pronged string -- which corresponds to the long simple root of $Sp(N)$ -- gets squished into a double arrow coming out of the same node and connecting to its mirror-image across the $BC$-branes. This means that, unlike with $SO(2N)$ algebras, we can now draw a double string stretching from the same node and crossing the $BC$-mirror. The simple root $\alpha_N$ of figure  \ref{fig:SpNroots} is one example of the $N$ double string connections that can be stretched that way. In terms of the IIA description, the change in orientation
of the mirror means we can now draw all of the same string junctions as for
$SO(2N)$, but we also have an additional $2N$ possible roots which correspond to
double connections coming out of the same node (something that was not allowed
in $SO(2N)$). The set of simple roots/strings for $Sp(N)$ is given in figure
\ref{fig:SpNroots}.

The set of simple strings (as illustrated in figure \ref{fig:SpNroots}) along
with the reflecting mirror ensures that odd parts in the partition of
$2N$ must appear with even multiplicity. This exactly matches the constraint
that, in the partitions used to parametrize the nilpotent orbits of $Sp(N)$,
the multiplicity of odd parts must be even. Furthermore, $Sp(N)$ also contains
distinguished orbits, which involve the presence of one or more non-simple root.

Following the same conventions as before, we use the following matrices as the
nilpositive part of the basis for $\mathfrak{sp}_{N}$:

\begin{itemize}
\item $N(N-1)/2$ one-pronged strings $E_{\mathrm{1-pronged}} = E_{i,j}%
-(-1)^{j-i} E_{2N-j+1,2N-i+1}$ with $1 \leq i < j \leq N$ corresponding to
one-pronged strings stretching from the $i^{th}$ to the $j^{th}$ $A$-brane as
well as their reflections. That is the strings stretching between the
$(2N-j+1)^{th}$ and the $(2N-i+1)^{th}$ nodes which are on the right-hand side
of the mirror. These correspond to the $\mathfrak{su}_{N} \subset
\mathfrak{sp}_{N}$ nilpositive generators.

\item $N(N-1)/2$ two-pronged strings $E_{\mathrm{2-pronged}} = E_{i,2N-j+1} +
(-1)^{j-i} E_{j,2N-i+1}$ with $1 \leq i < j \leq N$ corresponding to
two-pronged strings stretching between the $i^{th}$ and $(2N-j+1)^{th}$ nodes
as well as the $j^{th}$ and $(2N-i+1)^{th}$ nodes.

\item $N$ double strings $X_{\mathrm{doubled}} =2 E_{i,2N-i+1}$ with $1 \leq i
\leq N-1$ and the long simple string $X_{N} = E_{N,N+1}$. These correspond to
double-pronged strings merged together into single double connections. They
stretched from the $i^{th}$ to the $(2N-i+1)^{th}$ node.
\end{itemize}

The $N$ doubled strings coming out of the same node are the only new roots
which were not present in $\mathfrak{so}_{2N}$.

\begin{figure}[t!]
\centering
\includegraphics{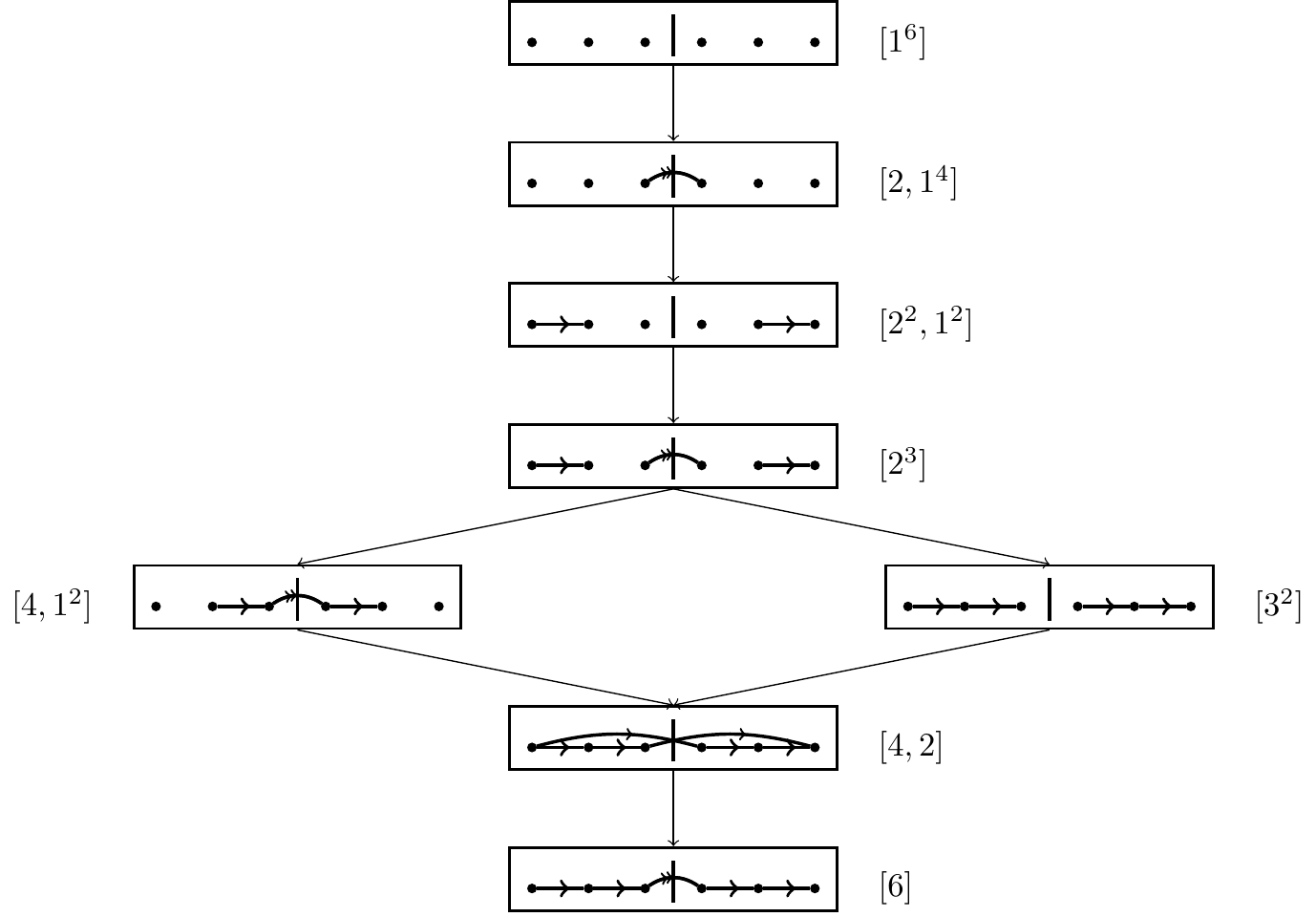}
\caption{Hasse diagram of $Sp(3)$ nilpotent deformations going from top (UV)
to bottom (IR) where all simple roots are turned on and every corresponding
simple strings are connecting the $A$-branes. In the case of the last simple
root, a double connection stretches from the last node and connects across the
mirror, ensuring charge conservation.}%
\label{fig:Sp3hasse}%
\end{figure}

We give the Hasse diagram of nilpotent orbits for $Sp(3)$ in figure
\ref{fig:Sp3hasse} to illustrate the possible string junctions. Flows between
each level in the Hasse diagrams follow the same rules as for the $SO$ groups.

\subsection{An Almost Classical Algebra: $G_{2}$}

\label{subsubsec:AlmostClassical} We next consider the exceptional Lie
group $G_{2}$. Even though the Lie algebra of $G_{2}$ is technically an exceptional
Lie group, the fact that it can easily be embedded inside the Lie algebra of $SO(7)$
makes it behave almost identically. Furthermore, as we are going to encounter
this algebra even when dealing only with classical quivers it is useful to
have a closer look at exactly how one might want to describe it.

\begin{figure}[t!]
\centering
\includegraphics{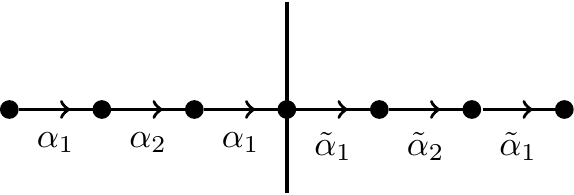}
\caption{Brane diagram of strings/roots stretching, for $G_{2}$. The $B$ and
$C$-branes are turned into an orientifold denoted by a mirror (vertical line)
and one of the $A$-branes is squeezed onto it. Furthermore, the first
$A$-brane is ``linked" to the middle one (as if it were also merged onto the
mirror), so that the first and third root of $SO(7)$ join together as the
first root of $G_{2}$ (as dictated by the quotient which takes $SO(7) \rightarrow
G_{2}$). The strings corresponding to simple roots are illustrated by arrows
stretching between the branes and reflected across the mirror.}%
\label{fig:G2roots}%
\end{figure}

First, we note that the monodromy of $G_{2}$ is the
same as for $SO(7)$ and $SO(8)$ that is, there are a
total of four $A$-branes and a $B$ with a $C$ brane. Thus, we can start from
the $SO(7)$ configuration which has four $A$-branes with one stuck on the
$BC$-mirror (see figure \ref{fig:SO7hasse}). Then, we note
that for $G_{2}$, the roots $\alpha_{1}$ and $\alpha_{3}$ are identified while
$\alpha_{2}$ is left untouched. Namely, the branching $SO(7) \rightarrow G_{2}$
takes $\alpha_{1} + \alpha_{3} \rightarrow\alpha_{1}$ and $\alpha_{2}
\rightarrow\alpha_{2}$. Therefore, we obtain the positive roots listed in
figure \ref{fig:G2roots}.

The matrix representation is taken directly from $SO(7)$. For the
positive simple roots we have:
\begin{align}
X_{1}  &  \equiv E_{1,2}+E_{6,7}+\sqrt{2} \left(  E_{3,4}+E_{4,5}\right)  ,\\
X_{2}  &  \equiv E_{2,3}+E_{5,6}.
\end{align}
The other four positive roots are given by:
\begin{align}
\left[  X_{1},X_{2}\right]   &  = E_{1,3} - E_{5,7} - \sqrt{2} \left(
E_{2,4}-E_{4,6}\right)  ,\\
\left[  \left[  X_{1},X_{2}\right]  ,X_{1}\right]   &  = 2 \sqrt{2} \left(
E_{1,4}+E_{4,7}\right)  -2\left(  E_{2,5}+E_{3,6}\right)  ,\\
\left[  \left[  \left[  X_{1},X_{2}\right]  ,X_{1}\right]  ,X_{1}\right]   &
= 6 \left(  E_{1,5}-E_{3,7}\right)  ,\\
\left[  \left[  \left[  \left[  X_{1},X_{2}\right]  ,X_{1}\right]
,X_{1}\right]  ,X_{2}\right]   &  = 6 \left(  E_{1,6} + E_{2,7}\right)  .
\end{align}

\begin{figure}[t!]
\centering
\includegraphics{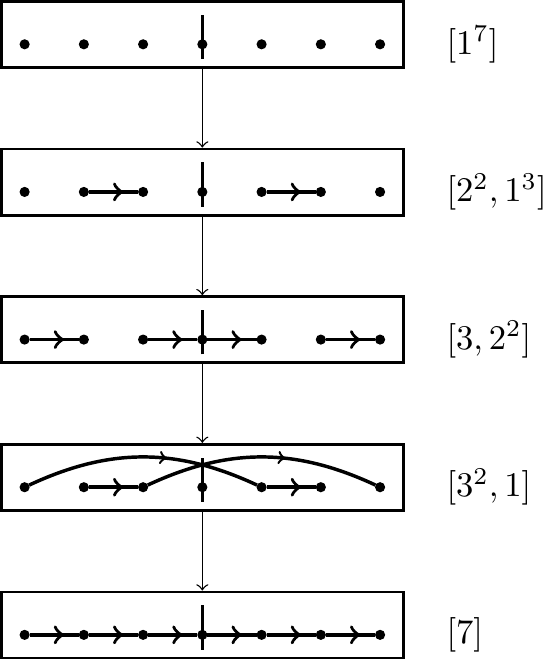}
\caption{Hasse diagram of $G_{2}$ nilpotent deformations going from top (UV)
to bottom (IR) where both simple roots are present so that both corresponding
simple strings are there to connect all $7$-branes and mirror image branes}%
\label{fig:G2hasse}%
\end{figure}

As a result, we can now give the four non-trivial nilpotent orbits of $G_{2}$
in terms of strings (see figure \ref{fig:G2hasse}). We note that, once again, we
have a simple correspondence with partitions of $7$, illustrated by the
groupings allowed from the associated string junctions. The ordering is a total ordering rather than a mere partial ordering (unlike
for most larger groups), and the flows from one orbit to the other follow from the fact that they are projections of the previously studied $SO(7)$ symmetry.

\subsection{Nilpotent Orbits for Exceptional Algebras}

\label{sec:NilpotentExceptionals}
We now turn our attention to the exceptional Lie algebras $E_{6,7,8}$. These
distinguish themselves from the classical algebras in several ways. First,
their nilpotent orbits are not simply described by partitions but rather by
Bala-Carter labels. These labels are in one-to-one correspondence with a
weighted Dynkin diagram and a set of roots. Interestingly, when the matrix
representations of these roots are added together, their Jordan block
decomposition still yields a unique partition. Thus, we can still parametrize
the nilpotent orbits of $E_{6,7,8}$ by partitions of $27$, $56$, and $248$
(corresponding to the dimension of their respective fundamental
representations). These partitions arise from the branching of the
fundamental of $E_{N}$ to the $SU(2)$ associated to the nilpotent orbit. However,
there does not exist a simple set of rules or restriction on these partitions like we
have seen for the classical Lie algebras. Thus this classification is very limited.

\begin{figure}[t!]
\centering
\includegraphics{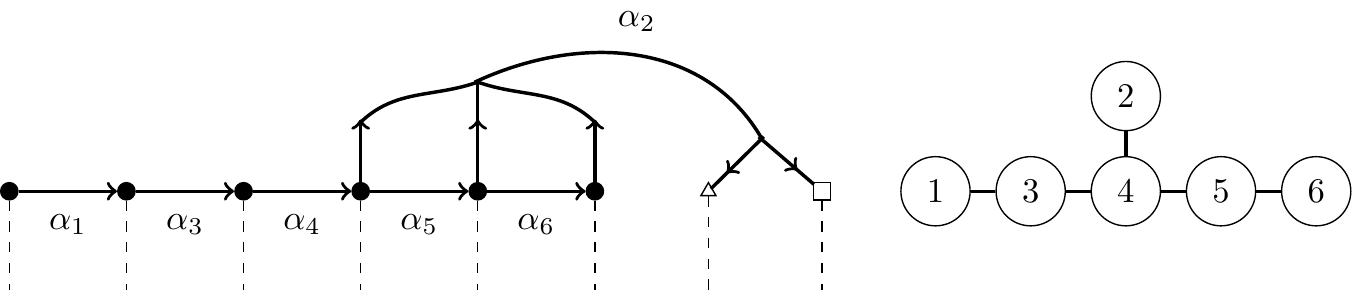}
\vspace{.5cm}
\par
\includegraphics{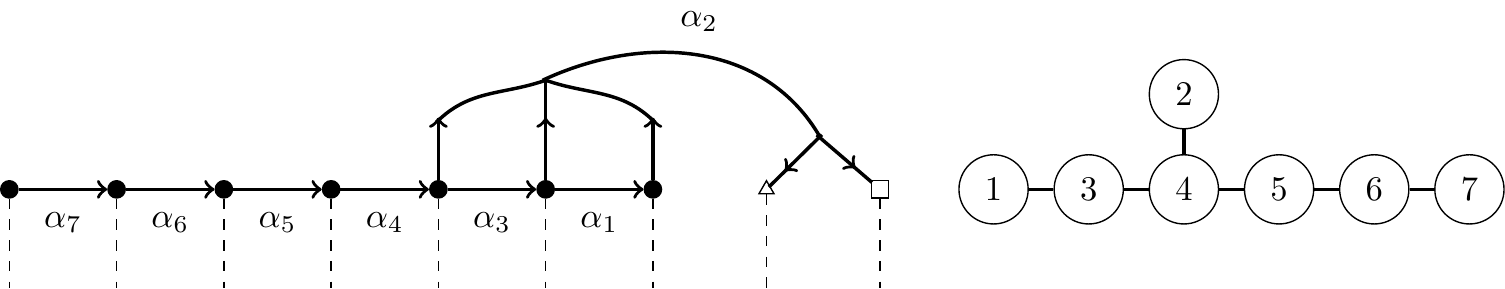}
\vspace{.5cm}
\par
\includegraphics{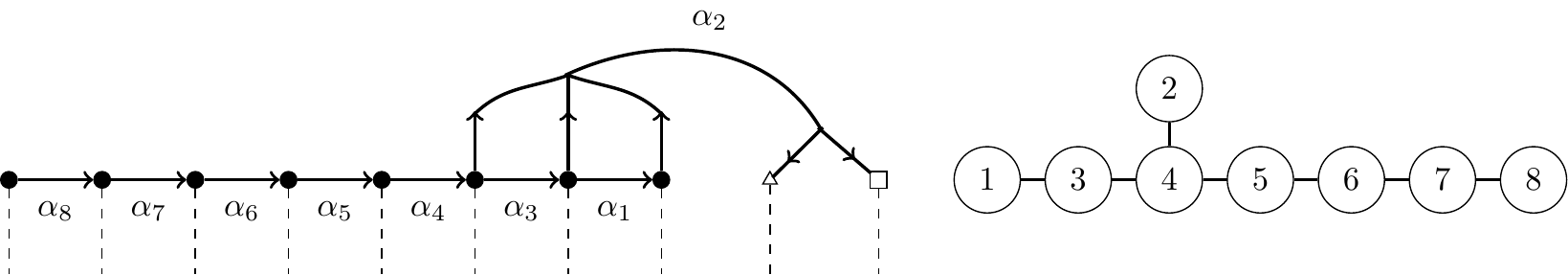}
\caption{Brane diagram of strings/roots stretching between the $A$, $X$, and
$C$-branes making up the $E_{6,7,8}$ symmetry (see \cite{DeWolfe:1998eu}). The
$A$-branes are denoted by black circles, the $X$-brane by an empty triangle
and the $C$-brane by an empty square. The dashed lines represent the position
of branch cuts. Again, these branch cuts are not drawn in subsequent pictures. To the
right we give the corresponding Dynkin diagram with simple roots numbered.}%
\label{fig:ENrootsFtheory}%
\end{figure}

By making use of string junctions and the brane configuration describing these
algebras, it is however possible to gain a little more insight into the
structure of nilpotent orbits for these exceptional groups. Physically, we know that the $E_{N}$ symmetries are given by $A^{N-1}BC^{2}$ or
equivalently $A^{N}XC$ brane configurations. The advantage of using the description with
an $X$-brane is that we can now branch $E_{N}$ to
$SU(N) \times U(1)$, where the $SU(N)$ piece is represented by $N$ $A$-branes
and $N-1$ ordinary open strings (i.e. one beginning and one end) stretching between them.
States charged under the $U(1)$ factor necessarily involve multi-prong strings which attach to
this stack of $A$-branes and also involve the $XC$ stack. This procedure matches identically the
initial setup used for describing $SO(2N)$ symmetries. The only difference is
that we now have a generalized mirror made out of an $X$ and a $C$ brane
instead of simply a $B$ and $C$ branes. This means that it now takes a
three-pronged string stretching from three $A$-branes to attach to the
$XC$-mirror in order to conserve the charges. Indeed, the charges from an $X$
and a $C$ brane now sum to $[3,0]$ which is exactly three times that of an $A$
brane. As a result we obtain the brane and string configurations given in
figure \ref{fig:ENrootsFtheory}.

\begin{figure}[t!]
\centering
\includegraphics{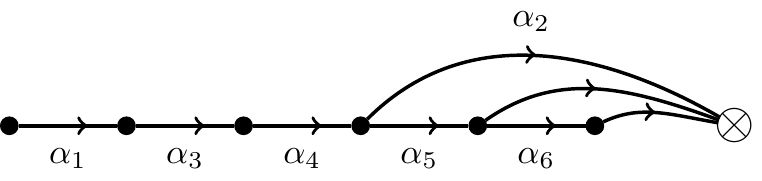}
\vspace{.5cm}
\par
\includegraphics{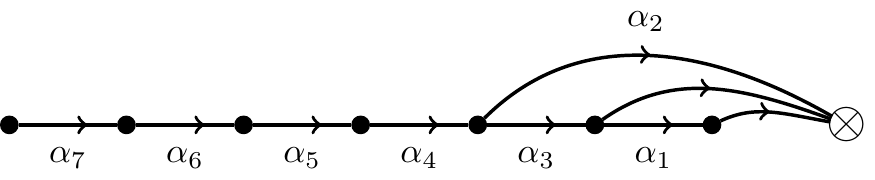}
\vspace{.5cm}
\par
\includegraphics{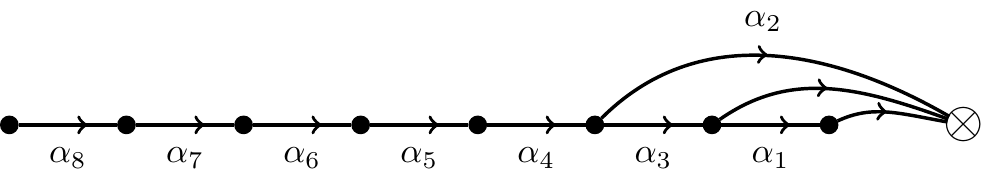}
\caption{Brane diagram of strings/roots stretching, for $E_{6,7,8}$. The $X$
and $C$-branes are turned into a generalized mirror denoted by a crossed
circle. The strings corresponding to simple roots are illustrated by arrows
stretching between the branes. We note that the distinguished root $\alpha
_{2}$ corresponds to the three-pronged string and indeed is made of three-legs
attaching to the $XC$-mirror in order to respect the difference in charges
between the $A$, $X$, and $C$ branes.}%
\label{fig:ENroots}%
\end{figure}

We then treat the $X$ and $C$ branes together as a generalized mirror and use
the short-hand picture of figure \ref{fig:ENroots} where the $XC$-mirror is
represented by an $\times$ inside a circle to differentiate it from the
vertical line that represented the $BC$-mirror for the orientifold seen in
the $SO(N)$ symmetries.

\begin{figure}[t!]
\centering
\includegraphics{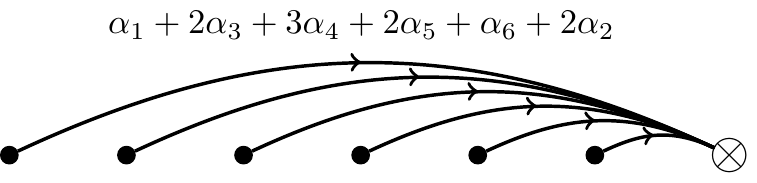}
\caption{Highest roots of $E_{6}$ represented by its corresponding six-pronged
string. It stretches from all six $A$-branes and attaches to the $X$ and $C$
branes represented by the crossed circle.}%
\label{fig:highestE6}%
\end{figure}

\begin{figure}[t!]
\centering
\includegraphics{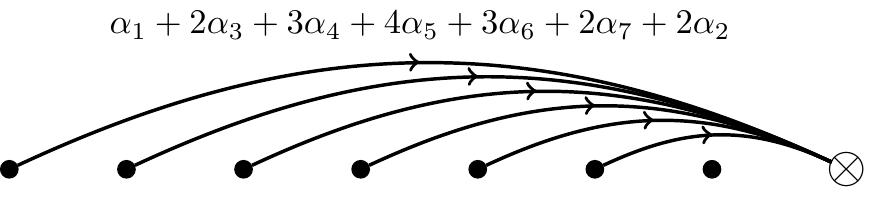}
\caption{Highest roots of $E_{7}$ represented by its corresponding six-pronged
string. It stretches from the six left-most $A$-branes and attaches to the $X$
and $C$ branes represented by the crossed circle.}%
\label{fig:highestE7}%
\end{figure}

\begin{figure}[t!]
\centering
\includegraphics{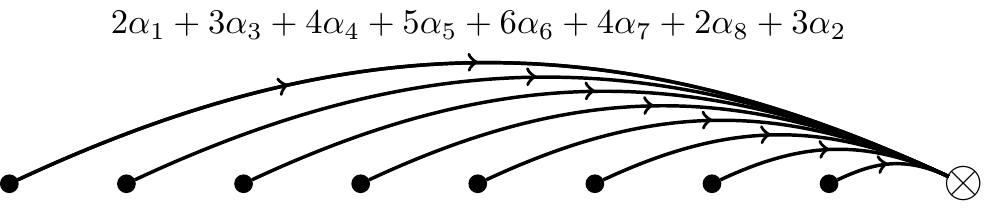}
\caption{Highest roots of $E_{8}$ represented by its corresponding
nine-pronged string. It stretches from all eight $A$-branes (attaching twice
onto the first one) to the $X$ and $C$ branes represented by the crossed
circle.}%
\label{fig:highestE8}%
\end{figure}

This $XC$-mirror is more complicated than the simply reflecting mirror for
the classical algebras. Indeed, we can think of this mirror as fragmenting the
partitions of $27$, $56$, and $248$ according to their branching rules.
The fundamental representation of $E_{N}$ branches to irreducible representations of
$SU(N) \times U(1)$ as:
\begin{align}
&  \mathbf{27} \rightarrow\mathbf{\overline{15}}_{0} + \mathbf{6}_{1} +
\mathbf{6}_{\scalebox{0.75}[1.0]{-}1}, & \text{for } E_{6} \rightarrow SU(6)
\times U(1),\\
&  \mathbf{56} \rightarrow\mathbf{\overline{21}}_{\scalebox{0.75}[1.0]{-}2} +
\mathbf{21}_{2} +\mathbf{\overline{7}}_{6} + \mathbf{7}%
_{\scalebox{0.75}[1.0]{-}6}, & \text{for } E_{7} \rightarrow SU(7) \times
U(1),\\
&  \mathbf{248} \rightarrow\mathbf{63}_{0} + \mathbf{56}_{3} +
\mathbf{\overline{56}}_{\scalebox{0.75}[1.0]{-}3} + \mathbf{28}%
_{\scalebox{0.75}[1.0]{-}6} + \mathbf{\overline{28}}_{6} + \mathbf{\overline
{8}}_{\scalebox{0.75}[1.0]{-}9} +\mathbf{8}_{9} + \mathbf{1}_{0}, & \text{for
} E_{8} \rightarrow SU(8) \times U(1).
\end{align}
Here, $\mathbf{15}$ is the two-index anti-symmetric representation of $SU(6)$
and $\mathbf{21}$ is the two-index anti-symmetric representation of $SU(7)$.
For the $E_{8}$ case, $\mathbf{63}$ is the adjoint, $\mathbf{28}$ is the
two-index anti-symmetric, $\mathbf{56}$ is the three-index anti-symmetric and
$\mathbf{8}$ is the fundamental representation of $SU(8)$. For the adjoint
representations of $E_{6}$ and $E_{7}$ we also have:
\begin{align}
&  \mathbf{78} \rightarrow + \mathbf{35}_{0} + \mathbf{20}_{1}
+ \mathbf{20}_{\scalebox{0.75}[1.0]{-}1}
+  \mathbf{1}_{2}+ \mathbf{1}_{\scalebox{0.75}[1.0]{-}2}
+  \mathbf{1}_{0}, & \text{for } E_{6} \rightarrow SU(6) \times U(1),\\
&  \mathbf{133} \rightarrow \mathbf{45}_{0} + \mathbf{35}_{\scalebox{0.75}[1.0]{-}4} + \mathbf{\overline{35}}_{4} +\mathbf{7}_{8} +\mathbf{\overline{7}}_{\scalebox{0.75}[1.0]{-}8} +   \mathbf{1}_{0} , & \text{for } E_{7}
\rightarrow SU(7) \times U(1).
\end{align}
By embedding $SU(N)$ inside $E_N$ in this manner, we see that positive strings can be described by any
set of one-pronged strings between the $N$ $A$-branes or any three-pronged string
attaching to three $A$-branes and stretching to the $XC$-mirror.
Furthermore, $E_{6}$ also allows a six-pronged string attaching all of its
$A$-branes to the $XC$-mirror, as illustrated by the trivial representation
$\mathbf{1}_{2}$ in its branching. This string corresponds to the highest root
of $E_{6}$. $E_{7}$ also allows six-pronged strings, as seen by the presence of
$\mathbf{\overline{7}}_{\scalebox{0.75}[1.0]{-}8}$ in its branching (this is
indeed the six index anti-symmetric representation of $SU(7)$). Finally,
$E_{8}$ not only allows six-pronged strings (as seen by the six index
anti-symmetric $\mathbf{\overline{28}}_{6}$ representation), but it also
allows for eight different nine-pronged strings, which connect all eight
$A$-branes to the $XC$-mirror with a double connection stretching from one of
the eight $A$-branes. These rules follow directly from the structure of the
exceptional algebras, as shown in \cite{DeWolfe:1998zf,DeWolfe:1998eu}.
To illustrate these situations, we depict the highest roots of $E_{6}$, $E_{7}$
and $E_{8}$ in figures \ref{fig:highestE6}, \ref{fig:highestE7}, and
\ref{fig:highestE8}.

In order to describe each nilpotent orbit, we now need to rely more heavily on
the matrix representation. As a result, we associate to each simple string of
figure \ref{fig:ENrootsFtheory} a matrix in the fundamental representation of
$E_{N}$. Any choice of basis will yield the same results, but for reference we
give the simple roots in Appendix \ref{app:ExpMat} and use the method of
\cite{howlett2001matrix} to obtain the remaining non-simple roots.

Next, we proceed just as with the classical algebras. Namely, we start with $N$
$A$-branes next to an $XC$-mirror and start attaching more and more small
string deformations until we reach the deepest nilpotent orbit. To every
string junction diagram we associate a matrix representation which belongs to
some nilpotent orbit. We can differentiate between nilpotent orbits based on
the Bala-Carter label or the partition associated to the matrix (by Jordan
block decomposition). For instance, the diagram involving the first two simple
roots of $E_{6}$ is represented by the matrix $X_{1}+X_{3}$ where
\begin{align*}
X_{1}  &  = E_{1,2} + E_{12,13} + E_{15,16} + E_{17,18} + E_{19,20} +
E_{21,22},\\
X_{3}  &  = E_{2,3} + E_{10,12} + E_{11,15} + E_{14,17} + E_{20,23} +
E_{22,24}.
\end{align*}
This matrix $X_{1}+X_{3}$ has Jordan block decomposition $[3^{6},1^{9}]$ and
is associated to the Bala-Carter label $A_{2}$.

\begin{figure}[ptb]
\centering
\includegraphics{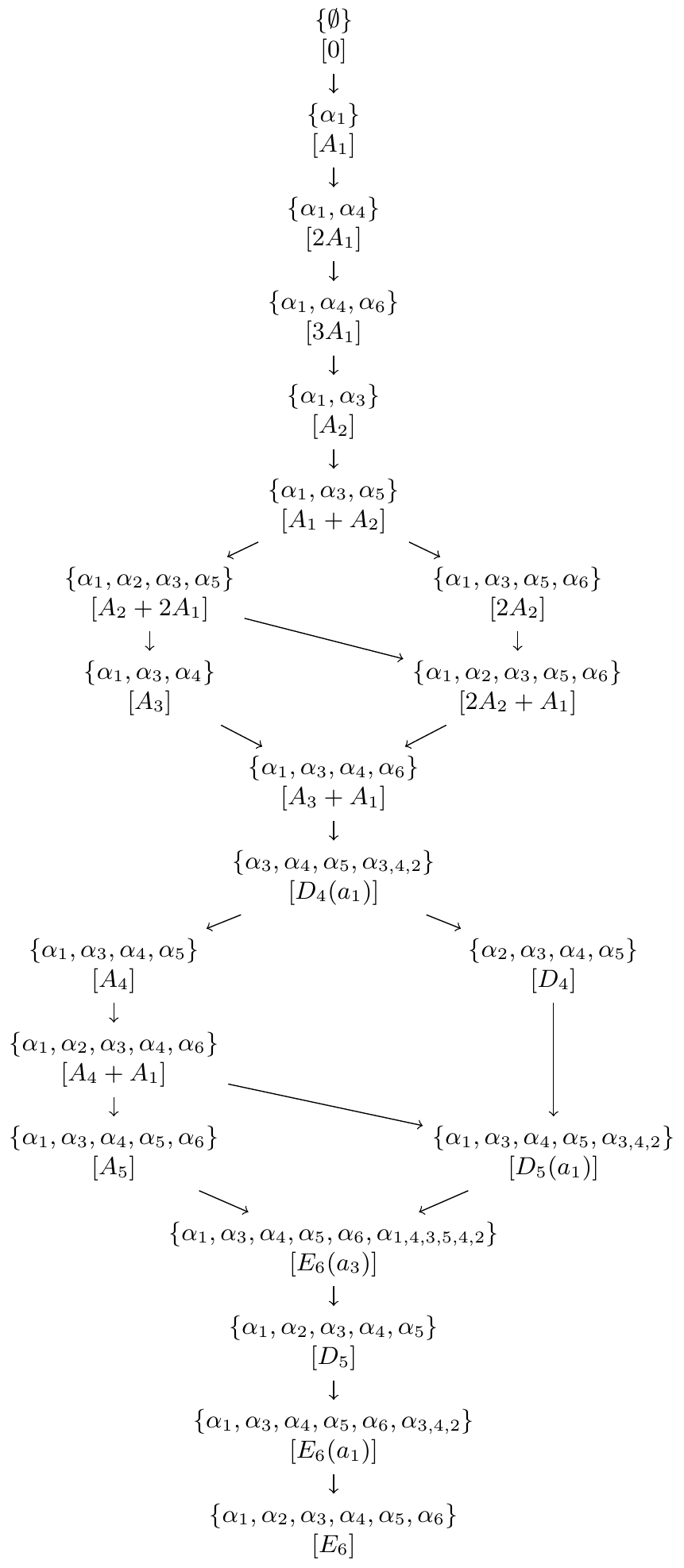}
\caption{Hasse diagram of $E_{6}$ nilpotent deformations going from top (UV)
to bottom (IR) where all simple roots are present, and every corresponding
simple string connects adjacent $A$-branes, or in the case of the second
simple root, three $A$-branes are connected to the $XC$-mirror. For ease of
exposition we only list the set of strings rather than the complete string
junction diagrams for each case.}%
\label{fig:E6hasse}%
\end{figure}

\begin{figure}[ptb]
\centering
\vspace{-1cm}
\includegraphics{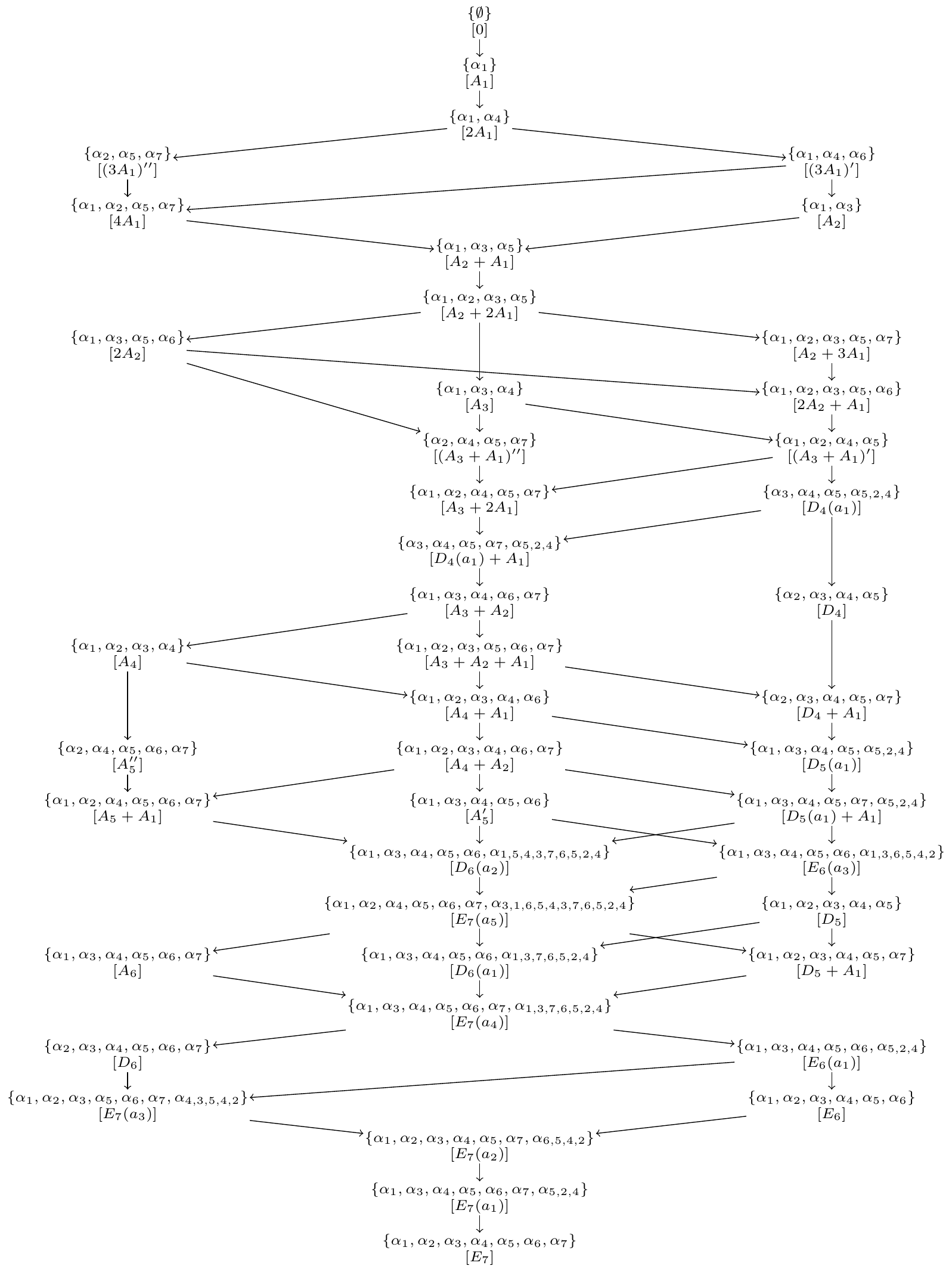}
\caption{Hasse diagram of $E_{7}$ nilpotent deformations going from top (UV)
to bottom (IR) where all simple roots are present, and every corresponding
simple string connects adjacent $A$-branes, or in the case of the second
simple root, three $A$-branes connect to the $XC$-mirror.}%
\label{fig:E7hasse}%
\end{figure}

\begin{figure}[ptb]
  \includegraphics{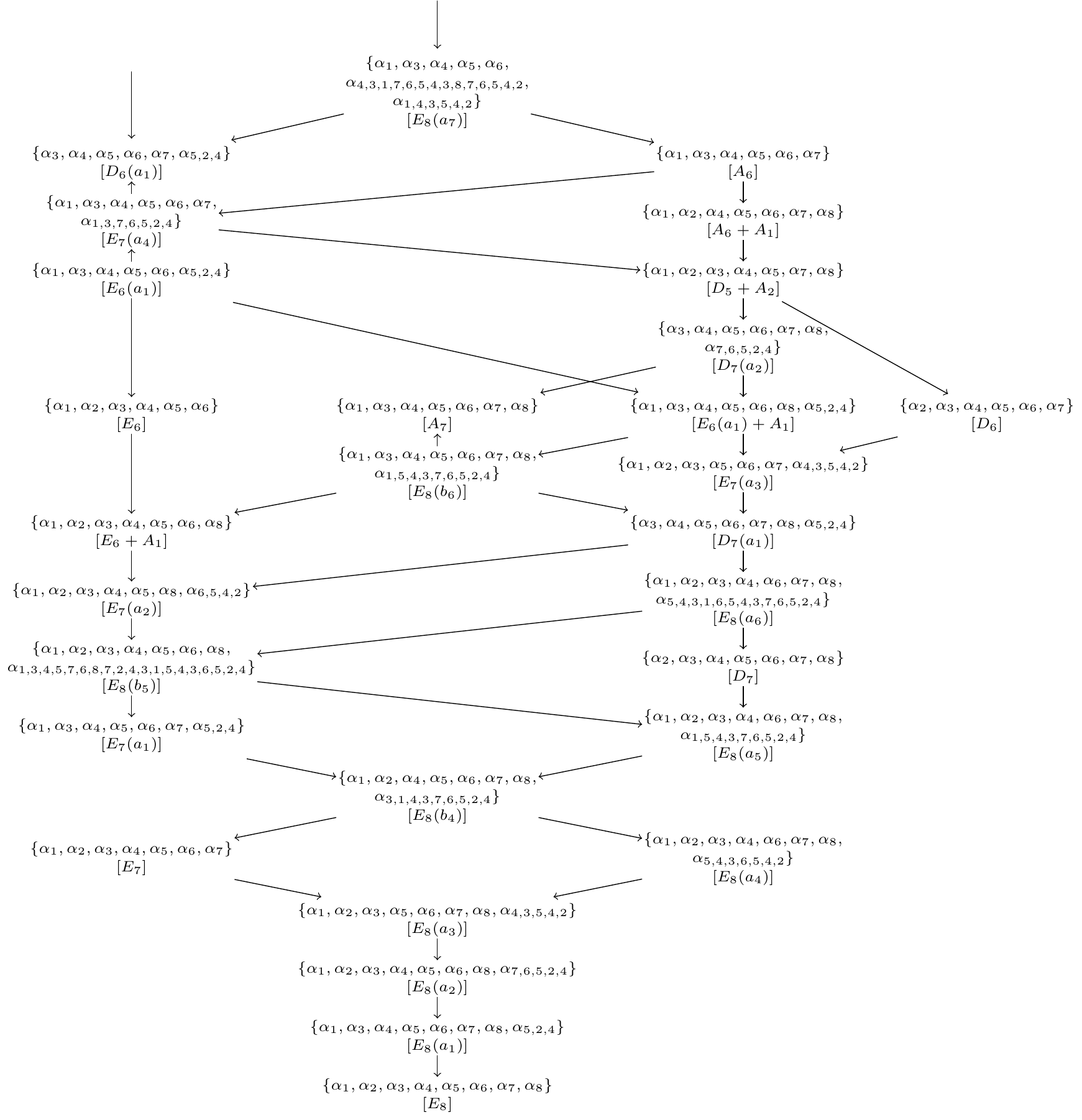}
\caption{Hasse diagram of $E_8$ nilpotent deformations going from top (UV) to bottom (IR) where all simple roots are present, and every corresponding simple string connects adjacent $A$-branes, or in the case of the second simple root, three $A$-branes connect to the $XC$-mirror.}%
\label{fig:E8hasse}%
\end{figure}

Much as in the case of the classical algebras, multiple diagrams belong to the same
equivalence class. Thus, in order to keep our diagrams as simple as possible,
we choose representative string junction diagrams that only make use of the
simple strings from figure \ref{fig:ENrootsFtheory} whenever possible.
Indeed, once again we identify some distinguished orbits as those which cannot
be described solely by a set of simple roots and necessarily involve non-simple
roots. Furthermore, while any string junction yielding the proper partition
is valid, for simplicity we select configurations with the minimum number of strings required (with as
few non-simple strings as possible) so that the addition of only a single
positive root $\epsilon\cdot X_{k}$ is required to flow to the nearest
nilpotent orbit. We illustrate the nilpotent orbits of $E_{6}$, $E_{7}$, and
$E_{8}$ in figures \ref{fig:E6hasse}, \ref{fig:E7hasse}, \ref{fig:E8hasse}.
The Hasse diagrams labeled by just their Bala-Carter labels can be found in e.g. the Appendix of \cite{Chacaltana:2012zy}, which summarizes several aspects regarding nilpotent orbits of exceptional algebras.

We see that we can move from one nilpotent orbit to the next by small deformations, just like we did for the classical groups.
Furthermore, we can describe every orbit using only simple strings except for
the distinguished ones. These distinguished orbits once again require the
presence of one (or two, for $E_{8}(a_{7})$) non-simple roots.

\subsubsection{The Non-Simply Laced $F_{4} \subset E_{6}$}

\label{subsubsect:F4} Finally, we note that $F_{4} \subset E_{6}$ is obtained
from $E_{6}$ by a very simple identification of simple roots:
\begin{align}
\alpha_{2}^{E_{6}}  &  = \alpha_{1}^{F_{4}},\nonumber\\
\alpha_{4}^{E_{6}}  &  = \alpha_{2}^{F_{4}},\nonumber\\
\alpha_{3}^{E_{6}}+\alpha_{5}^{E_{6}}  &  = \alpha_{3}^{F_{4}},\nonumber\\
\alpha_{1}^{E_{6}}+\alpha_{6}^{E_{6}}  &  = \alpha_{4}^{F_{4}},
\end{align}
where $\alpha_{1}^{F_{4}}$ and $\alpha_{2}^{F_{4}}$ denote the first two short
roots of $F_{4}$ while $\alpha_{3}^{F_{4}}$ and $\alpha_{4}^{F_{4}}$ denote
the longer ones. As a result, we can also simply give the Hasse diagram of
$F_{4}$ as a subset of the one from $E_{6}$.

\newcommand{\Eeigh}{1.5mm}
\begin{figure}[ptb]
\centering
  \vspace{-1cm}
  \includegraphics{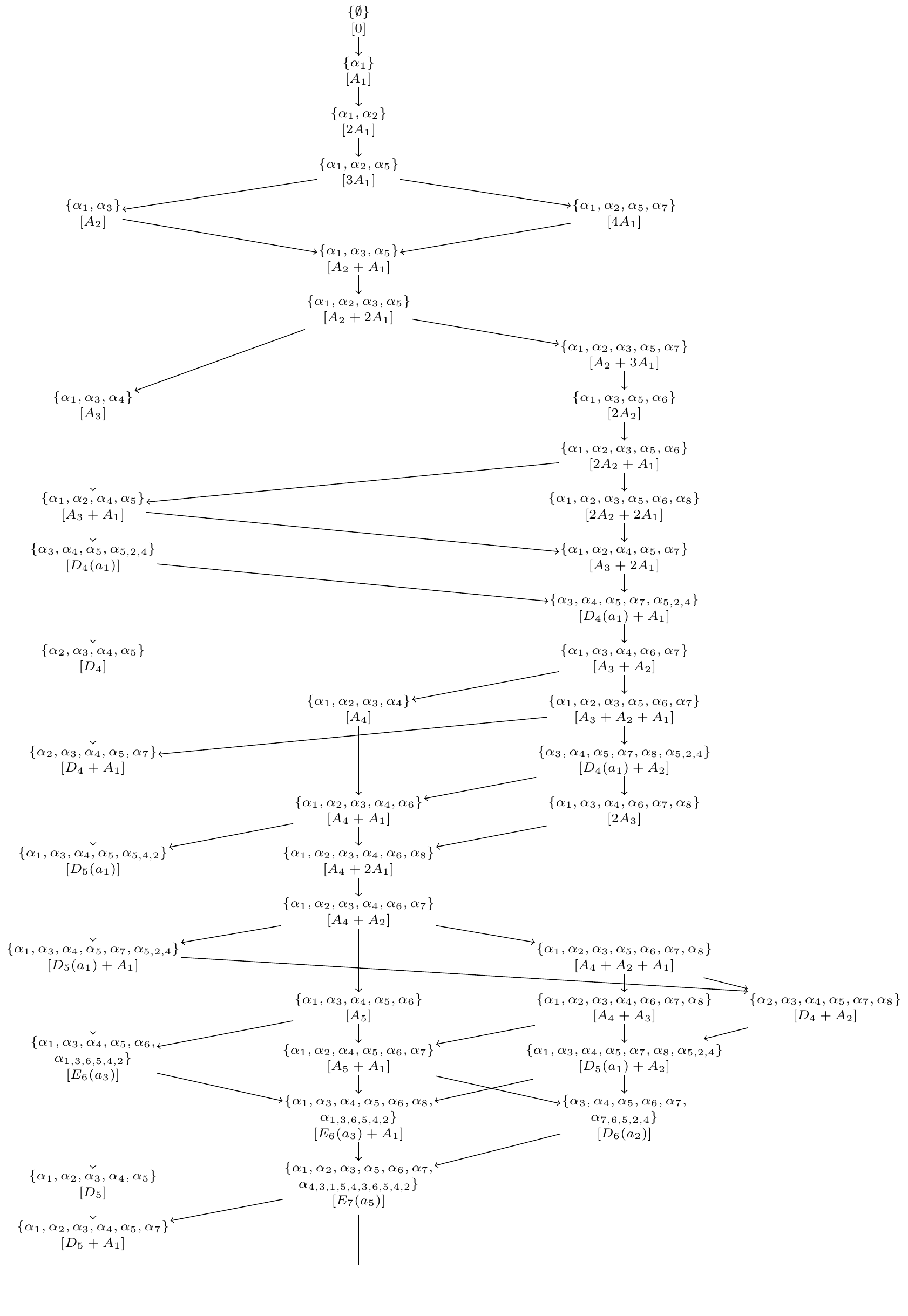}
\end{figure}

\section{Higgsing and Brane Recombination \label{sec:RECOMBO}}

In the previous section, we showed how to generate the entire nilpotent cone of
a semi-simple algebra using the combinatorics of string junctions. In
particular, the operation of \textquotedblleft adding a
string\textquotedblright\ reproduces the expected partial ordering based on
orbit inclusion. We now use this analysis to study Higgs branch flows for
6D\ SCFTs. Our main task here will be to study the effects of brane
recombination triggered by vevs for 6D conformal matter.

We first remark that the picture in terms of string junctions leads to a simple
description of Higgsing with semi-simple deformations. Recall that a semi-simple element is one that is
diagonalizable (in particular, not nilpotent). Since all the quiver-like gauge theories consist of stacks of $A^N$ branes
with either a $BC$ or $XC$ plane, we may join an open string from one stack of $A$-branes to the next, continuing
from left to right across the entire quiver. This leads to a ``peeling off'' of the corresponding $7$-brane, and has the effect of reducing the rank of each of the gauge algebras by one in both the classical case and the exceptional case.

\begin{figure}[t!]
\centering
\includegraphics{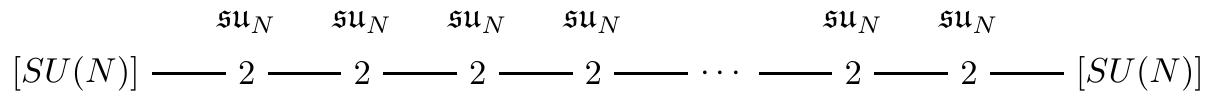}
\caption{Tensor branch of the UV quiver-like theory with just $SU(N)$ gauge algebras.}%
\label{fig:SUNUV}%
\end{figure}

\begin{figure}[t!]
\centering
\includegraphics{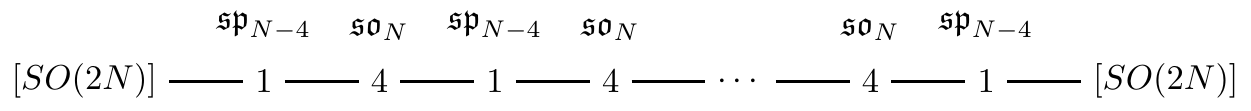}
\caption{Tensor branch of the UV quiver-like theory with just $SO(2N)$ gauge algebras. The full tensor branch also
includes additional $Sp(N-4)$ gauge algebras coming from blowing up the conformal matter between D-type collisions.}%
\label{fig:SO2NUV}%
\end{figure}

\begin{figure}[t!]
\centering
\includegraphics{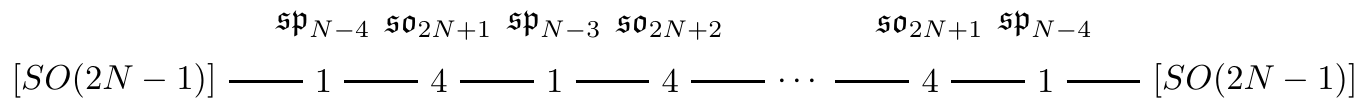}
\caption{Tensor branch of the UV theory with just $SO(2N-1)$ gauge algebras. The full tensor branch also includes
additional $Sp$ gauge algebras coming from blowing up the conformal matter between D-type collisions.
Any deformation with partition $\mu = [\{\mu_i\}]$ in $SO(2N-1)$ is equivalent to the partition $\nu = [\{\mu_i\},3]$ in $SO(2N+2)$.}%
\label{fig:SO2Nm1UV}%
\end{figure}

\begin{figure}[t!]
\centering
\includegraphics{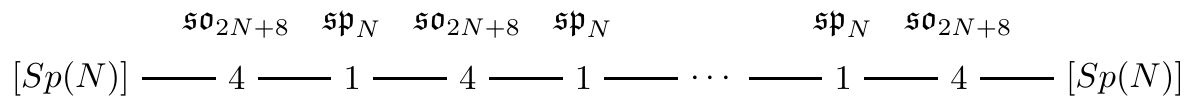}
\caption{UV theory for $Sp(N)$.}%
\label{fig:SpNUV}%
\end{figure}

Much more subtle is the case of T-brane deformations.
For the most part, we confine our analysis to the case of quiver-like theories in which all the gauge groups are classical (see figures \ref{fig:SUNUV}, \ref{fig:SO2NUV}, \ref{fig:SO2Nm1UV}, \ref{fig:SpNUV}).
Even in these cases, the matter content of the partial tensor branch can still be strongly coupled, as evidenced by
$SO-SO$ 6D conformal matter. Nonetheless, we will still be able to develop systematic sets of rules to extract the IR fixed point obtained from a given T-brane deformation in such cases.

To some extent, the necessary data is encoded by judiciously applying Hanany-Witten moves involving suspended D6-branes. Such moves were used in \cite{Gaiotto:2014lca}, for instance, to extract different presentations of a given 6D SCFT. To apply the Hanany-Witten analysis of that work to the case at hand, we will need to extend it in two respects. First of all, to cover the case of quiver-like theories with $SO$ gauge algebras, such brane maneuvers sometimes
result in a formally negative number of D6-branes \cite{Heckman:2016ssk, Mekareeya:2016yal}. Additionally, in the case of short quivers, the data specified by
pairs of nilpotent orbits will produce correlated effects in the resulting IR fixed points. To address both points, we will need to extend the available results in the literature.

As we have already mentioned, our main focus will be on tracking brane recombinations as
triggered by the condensation of open strings. In the
context of 6D\ SCFTs, all of this occurs in a small localized region of the base of the non-compact elliptic threefold.
Macroscopic data such as the surviving flavor symmetries corresponds to the asymptotic behavior of non-compact $7$-branes that pass
through this singular region, but which also extend out to the boundary of the
non-compact base. This also means that, provided we hold fixed the total
asymptotic $7$-brane charge present in the configuration, we can consider
any number of \textquotedblleft microscopic processes\textquotedblright\ which
could appear in the physics of brane recombination.

One such process which we shall often use is the
creation of brane / anti-brane pairs localized in the region
near the 6D\ SCFT. We denote such an anti-brane
by $\overline{A}$ and use it in annihilation processes such as:
\begin{equation}\label{AAbar}
A+\overline{A}\rightarrow\text{no branes.}%
\end{equation}
Strictly speaking, such a physical process would generate radiation. The only sense in which
we are really using these objects is to count the overall Ramond Ramond charge asymptotically far away from
the configuration. In this sense, there will be little distinction between an anti-brane and a ``negative / ghost-brane.''
Since we are primarily interested in determining the end outcome of Higgsing, we use these $\overline{A}$-branes as a
combinatorial tool which must disappear at the final stages of our analysis through processes such as line (\ref{AAbar}).
We refer to this as having a \textquotedblleft Dirac sea\textquotedblright\ of $A/\overline{A}$ pairs of
$7$-branes.

Much as in the case of a general configuration of plus and minus charges in electrodynamics, a lowest energy configuration is obtained by allowing charges to freely move through a material. In much the same way, we shall allow the branes and anti-branes to redistribute. Our main physical condition is that the net $7$-brane charge is unchanged by such processes, and also, that no anti-brane charge remains uncanceled in any
final configuration obtained after Higgsing.

We also remark that from the standpoint of renormalization group flow, these sorts of microscopic details are expected to be irrelevant at long 
distances. Said differently, while there could, a priori, be different UV completions in the full framework of quantum gravity, 
such details will not matter in determining possible fixed points obtained after a Higgs branch deformation. The brane maneuvers indicated 
here are of this sort and are used as a tool to analyze possible fixed points.

Including these formal structures is useful in that it allows us to make sense of the resulting 6D\ SCFT, even when the ranks of the intermediate gauge groups are negative numbers of small magnitude. This procedure has been used
in \cite{Heckman:2016ssk, Mekareeya:2016yal, Apruzzi:2017iqe, Heckman:2018pqx, Frey:2018vpw}
as a way to track the effects of Higgs branch flows
in certain 6D\ SCFTs. We will return to this point in section \ref{sec:GETSHORTY}.

Our main focus in this section will be on determining the Higgs branch flows
associated with the classical algebras, since in these cases there is also a
gauge theory description available for some Higgs branch flows in terms of
vevs of conventional hypermultiplets. Any nilpotent orbit is then described by stretching the appropriate strings as
described in section \ref{sec:NILPJUNC}. We then need to propagate
the deformation by removing some strings as we move deeper into the quiver, which allows us
to read off the resulting gauge symmetries that are left over in the IR. We explain these propagation rules in the following
section.

Before that, however, we need to introduce the possibility of
anti-branes. Indeed, while the nodes in the $SU(N)$ quivers all have the same
number of branes on each level (namely $N$ $A$-branes), the other
classical algebras do not. For instance, a quiver with $SO(2N)$ flavor in the
UV will alternate between $N$ and $N-4$ $A$-branes on the $\mathfrak{so}_{2N}$
and $\mathfrak{sp}_{N}$ levels respectively. This introduces an additional complication in that we may end up with configurations that have
more strings stretching between branes (as dictated by the nilpotent orbit
configuration of section \ref{sec:NILPJUNC}) than are
available according to the gauge group on the quiver node. We remedy this
situation by extracting as many extra $A$ branes as necessary out of the brane
/ anti-brane \textquotedblleft Dirac sea\textquotedblright\ to draw the proper
number of string junctions. These extra branes are then immediately
canceled with the same number of anti-branes.

For example, the theory with $SO(8)$ flavor symmetry
has gauge symmetries alternating between $\mathfrak{sp}_{0}$ (i.e. a trivial gauge group associated with an ``unpaired tensor'' \cite{Morrison:2016djb}) and
$\mathfrak{so}_{8}$, and the nilpotent orbit $[4^{2}]^I$ uses strings
stretching between every brane (i.e. all four $A$-branes and their images have at least one
string attached). However, $\mathfrak{sp}_{0}$ only has the $BC$-mirror
and no $A$-brane. So, in order to describe the $[4^{2}]^I$ nilpotent orbit, we
must introduce four $A$-branes through which we can stretch strings (on each
side of the mirror) and then add them with four anti-branes. This also applies to
the non-simply laced classical algebras, since they can be
obtained from Higgs branch flows of $SO($even$)$ quiver-like theories \cite{Heckman:2015bfa}.

Notably, there are a few cases, even for SO- and Sp-type quivers,
which require non-perturbative ingredients such as E-string / small instanton
deformations. In these cases, the number of tensor multiplets in the
6D\ SCFT\ also decreases. Our method using brane / anti-brane pairs
carries over to these situations and allows us to obtain a complete
picture of Higgs branch flows in these cases as well. We use this
feature in section \ref{sec:GETSHORTY} to determine IR fixed points in the case of short quivers.

Our plan in the rest of this section is as follows: first, we discuss a IIA realization of quiver-like theories
with classical gauge groups, and especially the treatment of Hanany-Witten moves in such setups. After this, we
state our rules for how a T-brane propagates into the interior of a quiver with classical gauge algebras. We then illustrate with several examples the general procedure for Higgsing such theories. This provides a uniform account of brane recombination and also
agrees in all cases with the result expected from related F-theory methods (when available). We also
comment on some of the subtleties associated with extending this to the case of quiver-like theories with exceptional
algebras.

\subsection{IIA Realizations of Quivers with Classical Gauge Groups}\label{subsec:HananyWitten}

To aid in our investigation of Higgs branch flows for 6D SCFTs, it will also prove convenient to
use the type IIA realizations of the quiver-like theories with classical algebras, as used previously in references
\cite{Hanany:1997gh, Brunner:1997gk, Brunner:1997gf, Gaiotto:2014lca}. In the case of quivers with $SU$ gauge group factors,
each classical gauge group factor is obtained from a collection of D6-branes suspended between spacetime filling NS5-branes, with non-compact ``flavor'' D8-branes emanating ``out to infinity.'' The case of $SO$ algebras on the partial tensor branch is obtained by also including $O6^{-}$-planes coincident with each stack of D6-branes. In this case, the NS5-branes can fractionate to $\frac{1}{2}$ NS5-branes. Working in terms of these fractional branes, there is an alternating sequence of $O6^+$ and $O6^-$ planes, and correspondingly an alternating sequence of $SO$ and $Sp$ gauge group factors. This all matches up with the F-theory realization of these theories, where each
$SO$ factor originates from an $I_n^{\ast}$ fiber and each $Sp$ factor from a non-split $I_m$ fiber.

The utility of this suspended brane description is that we can write several equivalent brane configurations which realize the same IR fixed point via ``Hanany-Witten moves,'' much as in the original reference \cite{Hanany:1996ie} and its application to 6D SCFTs in reference \cite{Gaiotto:2014lca}. This provides a convenient way to uniformly organize the data of Higgs branch deformations generated by nilpotent orbits. In fact, we will shortly demonstrate that using these brane moves along with some additional data (such as the appearance of brane / anti-brane pairs) provides an intuitive method for determining the resulting fixed points in both long and short quivers.

Since we will be making heavy use of the IIA realization in our analysis of Higgs branch flows, we now discuss such constructions in greater detail. In our analysis, we will also consider formal versions of Hanany-Witten moves which would seem to involve a negative number of branes. These cases are closely connected with strong coupling phenomena (such as the appearance of small instanton transitions and spinor representations) and can be fully justified in the corresponding F-theory realization of the same SCFT. Indeed, the description in terms of Hanany-Witten moves extends to the F-theory description, so we will interchangeably use the two conventions when the context is clear.

\subsubsection{SU($N$)}
We begin with a quiver-like theory with $L-1$ tensor multiplets and for each one a paired $SU(N)$ gauge group factor. The UV theory has a tensor branch given by the quiver
\begin{center}
\includegraphics{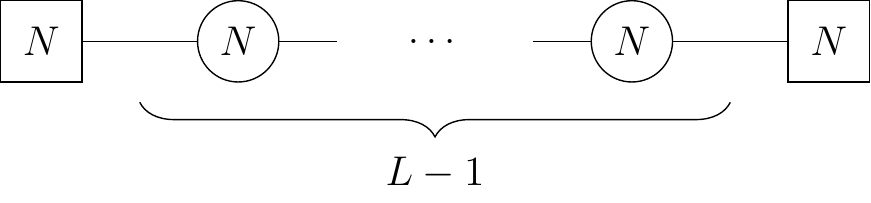},
\end{center}
which is realized in terms of the IIA brane setup:
\begin{center}
\includegraphics{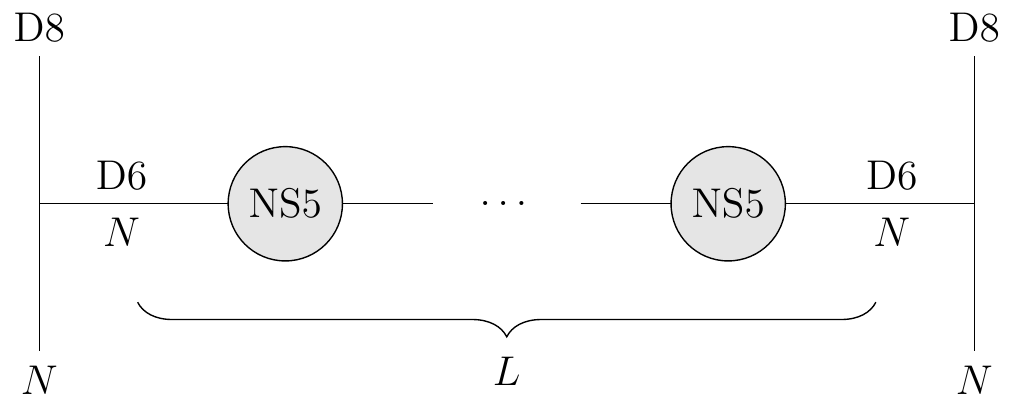}\,.
\end{center}
From the point of view of the D6-branes, the D8-branes specify boundary conditions, which are controlled by the Nahm equations \cite{Diaconescu:1996rk}. These pick three ($X^i$, $i=1,2,3$) out of the $N^2-1$ scalars controlling the Higgs branch and relates them to the distance $t$ of the intersection point by
\begin{equation}
  X^i \sim \frac{T^i}{t}.
\end{equation}
The generators $T^i$ describe an $SU(2)$ subgroup of the flavor symmetry SU($N$), whose embedding is captured by a partition of $N$. This happens on both sides of the quiver. Thus all the data we need in order to study Higgs branch flows of the UV theory are two partitions $\mu_L$ and $\mu_R$ of $N$ and the length $L$.

\begin{figure}[t!]
\centering
\includegraphics{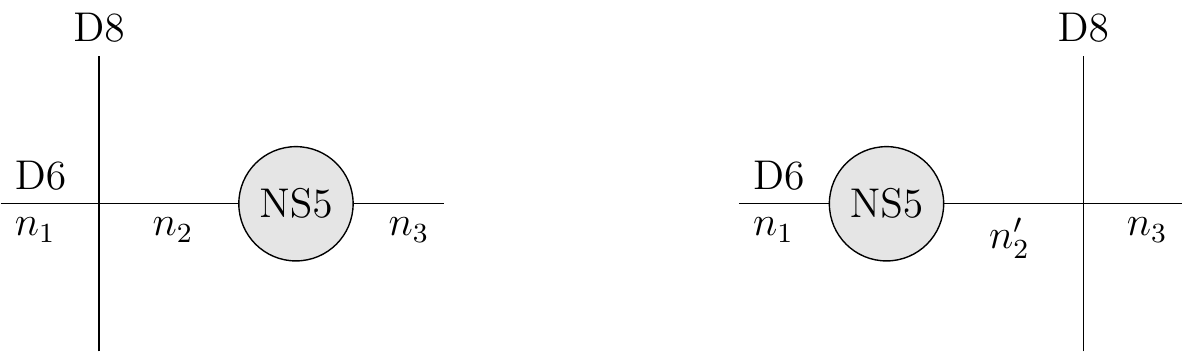}
\caption{The basic operation of swapping a D8- and NS5-branes. The relation between the number of D6-branes stretching between the D8-brane and the NS5-brane before ($n_2$) and after ($n_2'$) the swapping is given by $n_2' = n_1 + n_3 - n_2 + 1$.}\label{fig:hwmove}
\end{figure}

A partition $\mu$ of $N$ is given in terms of $l\le N$ integers $\mu^i$ with $\mu^1 \ge \mu^2 \ge \dots \ge \mu^l$ and $\mu^1 + \mu^2 + \dots \mu^l = N$. In the corresponding brane realization, the two partitions describe the separation of the stack of $N$ D8-branes on each side into smaller stacks
\begin{center}
\includegraphics{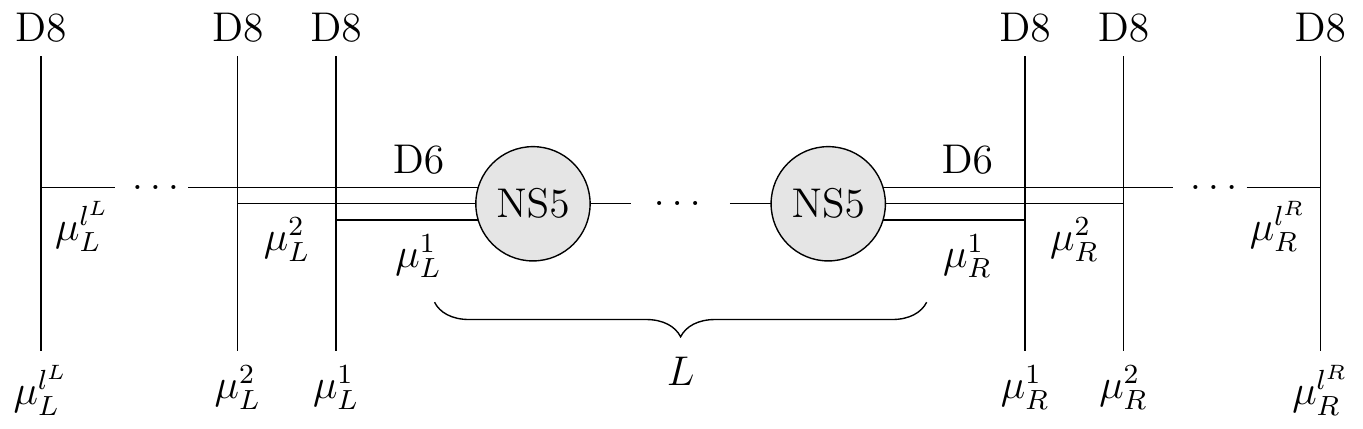}\,.
\end{center}
The brane picture is particularly useful because we can easily read off the IR theory from it. This works by applying Hanany-Witten moves, which swap a D8-brane and an NS5-brane, until all of the D8-branes are balanced. Looking at the stack of $\mu_L^1$  D8-branes left of the first NS5-branes, we can measure its imbalance by the difference $\Delta n$ of D6-branes departing from the right and arriving on the left. A balanced stack would have $\Delta n=0$, but for the setup depicted above we find $\Delta n = \mu_L^1$ instead. After performing the Hanany-Witten move described in figure~\ref{fig:hwmove}, $\Delta n$ becomes
\begin{equation}
  \Delta n' = \Delta n - 1 \qquad \text{with} \qquad
    \Delta = n_2 - n_1 \quad \text{and} \quad
    \Delta' = n_3 - n_2' \,.
\end{equation}
Hence, we have to perform exactly $\Delta n = \mu_L^1$ Hanany-Witten moves to balance this stack.
One can always balance all D8-branes provided that the length of the quiver $L$ is large enough. This constraint will become important when we discuss short quivers in section~\ref{sec:GETSHORTY}. Once all D8-branes are balanced, the resulting IR quiver gauge theory can be read off by using the building blocks
\begin{center}
\includegraphics{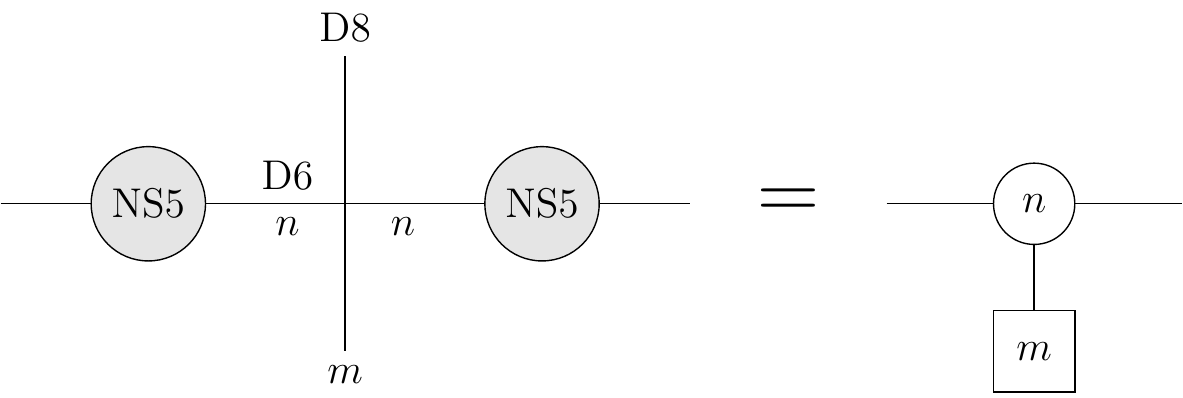}\,.
\end{center}

Applying subsequent Hanany-Witten moves results in a simple, algebraic description of the Higgs branch flows. Let us, for simplicity, consider very long quivers. In this case it is sufficient to just focus on one partition, i.e. $\mu_L$, since the analysis on the right-hand side is perfectly analogous. Using the fact that a stack of $\mu^i_L$ D8-branes moves $\mu^i_L$ NS5-branes to the right until it is balanced, we can read off the flavor symmetries of the IR theory directly from the partition. However, obtaining the number of D6-branes stretched between each pair of adjacent NS5s is slightly more complicated. If we denote this number as $n_i$ between the $i$'s and $i+1$'s NS5s we find the following recursion relation
\begin{equation}
  (n_i)_j = \begin{cases} (n_i)_{j-1} - \mu_L^j + i & \text{for } i < \mu_L^j \\
    (n_i)_{j-1} & \text{otherwise}\,.
  \end{cases}
\end{equation}
Here $(n_i)_j$ denotes the $n_i$ after the $j$'th stack of NS5-branes has been balanced. Hence, the initial condition is $(n_i)_0 = N$, and we are interested in $(n_i)_{l_L}$, which describes the number of D6-branes once all D8-branes have been balanced. An example for $N=6$ is $\mu=[3\,2\,1]$, for which we find
\begin{equation}
  \begin{aligned}
    (n_i)_1 &= \begin{pmatrix} 4 & 5 & 6 & 6 & \dots \end{pmatrix} \\
    (n_i)_2 &= \begin{pmatrix} 3 & 5 & 6 & 6 & \dots \end{pmatrix} \\
    (n_i)_3 &= \begin{pmatrix} 3 & 5 & 6 & 6 & \dots \end{pmatrix}
  \end{aligned}
\end{equation}
with the resulting IR quiver
\begin{center}
  \includegraphics{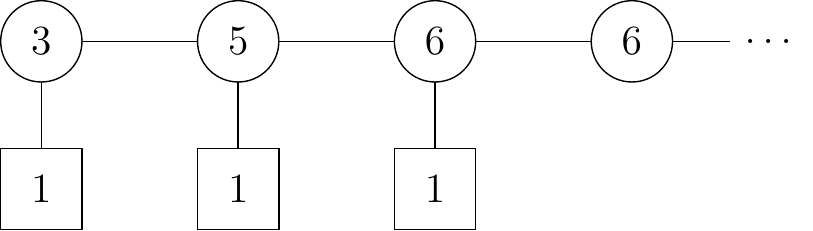}\,.
\end{center}

\subsection{SO($2N$), SO($2N + 1$) and Sp($N$)}
Gauge groups SO($2N$), SO($2N + 1$) and Sp($N$) arise if the setup from the last subsection is extended to include O6 orientifold planes placed on top of the D6-branes. In particular, assume we have $N$ physical D6-branes. Each of these has a mirror image under the $\mathbb{Z}_2$ orientifold action $\Omega$, and thus we have in total $2N$ 1/2 D6-branes. Their Chan-Paton factors transform under $\Omega$ as $\Omega \lambda = M \lambda^T M^{-1}$. Since $\Omega^2 = 1$, we therefore find two different solutions for $M$, which are denoted as $M_\pm = \pm M_\pm^T$. Each of these solutions gives rise to a distinguished orientifold action $\Omega_\pm$. Only massless open string excitations satisfying $\Omega_\pm \lambda = - \lambda^T$ survive the orientifold projection. Depending on whether $\Omega_-$ (O6$^-$) or $\Omega_+$ (O6$^+$) is used, the resulting gauge group is either SO($2N$) or Sp($N$). If a single 1/2 D6-branes is exactly on top of the O6$^-$ plane, it becomes its own mirror and we obtain the gauge group SO($2N+1$). Similar to the D6-branes, a single NS5-branes on the orientifold plane splits into two half NS5-branes:
\begin{center}
\includegraphics[scale=.9]{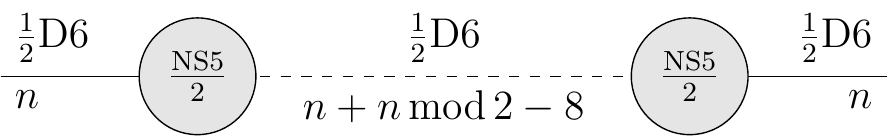}
\end{center}
Here, we depict a stack of 1/2 D6-branes on O6$^-$ with a solid line and a stack of 1/2 D6-branes on O6$^+$ with a dashed line. Because the D6-charge of the O6$^+$ differs by 4 from the one of the O6$^-$ the number of 1/2 D6-branes changes from $n$ to $n + n\,\mathrm{mod}\,2 - 8$ and back.

\begin{figure}[htb]
  \centering
  \begin{tabularx}{.9\textwidth}{l l}
  SO($2N-1$) & \multicolumn{1}{m{2cm}}{\includegraphics[scale=.9]{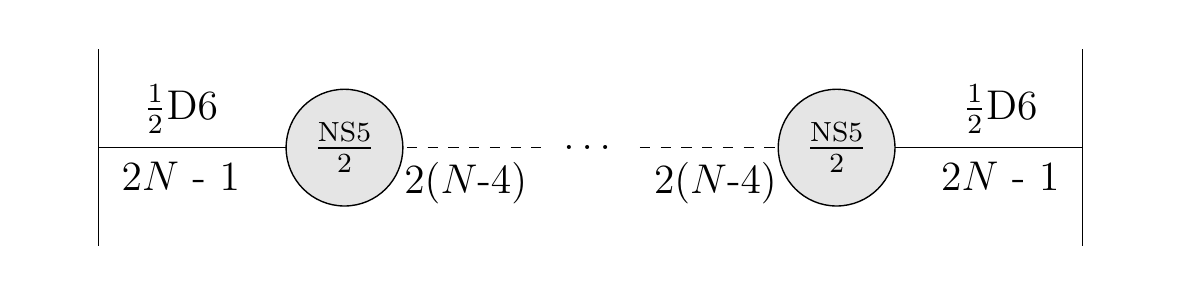}} \\
  SO($2N$)   & \multicolumn{1}{m{2cm}}{\includegraphics[scale=.9]{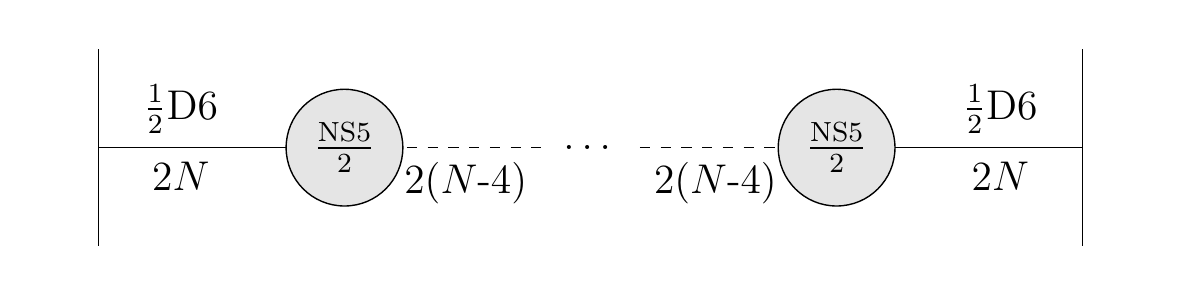}} \\
  Sp($N$)    & \multicolumn{1}{m{2cm}}{\includegraphics[scale=.9]{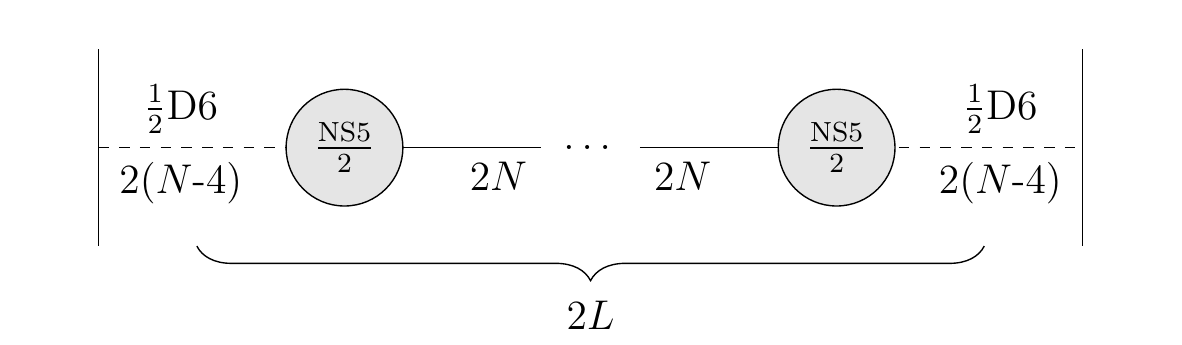}}
  \end{tabularx}
  \caption{Suspended brane realization of UV quivers with SO($2N$-1), SO($2N$), and Sp($N$) flavor symmetries.}
  \label{fig:sospbranes}
\end{figure}
There are three different classes UV SCFTs which we can now realize in terms of suspended branes depicted in figure~\ref{fig:sospbranes}.
To study their Higgs branch flow, we follow the same approach as in the SU($N$) case: first, we choose two partitions, which each describe an embedding of $\mathfrak{su}_2$ into the corresponding flavor symmetry algebra. These control how the stacks of 1/2 D8-branes on the left and right side of the quiver are split into smaller stacks. Finally, we apply Hanany-Witten moves to these stack until they are balanced.

It is convenient to combine the D6-brane charge of the orientifold planes with the contribution from the 1/2 D6-branes. In this case, rules for the Hanany-Witten shown in figure~\ref{fig:hwmove} still apply and we can use the results from the last subsection. The only thing we have to keep in mind is that we are now counting 1/2 D6-branes.

\subsection{Propagation Rules}

In this section, we present a set of rules for working out Higgs branch
deformations in the case of quivers with classical gauge algebras. The main idea is
to consider each stack of $7$-branes wrapped over a curve and
strings that stretch from one stack to the next. To visualize the possible
locations where such strings can begin and end, we will use the same
diagrammatic analysis developed in section \ref{sec:NILPJUNC} to track these breaking
patterns. When such a string is present, it signals the presence of a brane
recombination move, and the corresponding brane becomes non-dynamical (having
become attached to a non-compact $7$-brane on the boundary of the quiver).
On each layer, we introduce a directed graph, as dictated by a choice of nilpotent orbit. This
tells us how to connect the branes into ``blobs'' after recombination. We want to see how these
blobs recombine, both with the non-compact branes at the end of quiver and the compact branes further in the interior.

On each consecutive level of the quiver (i.e. for each gauge algebra in the quiver), we draw the same string configuration with a few modifications according to the following rules for propagating
Higgs branch flows into the interior of a quiver:

\begin{itemize}
\item First, we consider blobs made only of $A$-branes. That is, only
one-pronged strings are involved and there is no crossing or touching the
mirror. These configurations cover all possible orbits of $SU(N)$. In such
cases, the one-pronged strings get removed one at a time (per blob) so that
one $A$-brane is added back (to each blob) at each step. These steps can be
visualized in the example of $SU(6)$ nilpotent orbits given in figure
\ref{fig:SU6quiver}.

\item Next, we consider cases with a two-pronged string, but in which both legs are
disjoint (unlike $\alpha_{N}$ for $Sp(N)$) so that no loop is formed. In this
case, the propagation follows the same rule as for one-pronged strings. Indeed
in such configurations each leg becomes independent and they individually
behave like one-pronged strings. This is the case for $SO(2N)$ whenever the
two-pronged string $\alpha_{N}$ is present but not the string $\alpha_{N-1}$
below it. (See for instance the partition $[2^{4}]^{II}$ for $SO(8)$ in figure
\ref{fig:SO8quiver}).

\item Now suppose (without loss of generality) that branes $A_{1}, A_{2},
\cdots, A_{n}$ are connected via simple one-pronged strings and a two-pronged
string attaches the $i^{th}$ and $n^{th}$ brane to the mirror ($1 \leq i <
n$). Then, for the next $n-i$ levels, the right-most leg moves one step to the
left (attaching to the brane $A_{n-1}, A_{n-2}, \cdots,A_{n-i}$) and the
right-most one-pronged string below it is removed, namely $\alpha_{n}$
followed by $\alpha_{n-1}, \cdots, \alpha_{n-i}$. After these $n-i$ steps, both
legs overlap and the right-most leg cannot move any further. Instead, we then
move the second leg one step to the left so that one leg stretches from
$\alpha_{n-i-1}$ and the other stretches from $\alpha_{n-i}$. We can now
repeat the previous steps once by moving the right-most leg one brane to the
left (and removing $\alpha_{n-i-1}$) so that it overlaps with the left-most
leg. This process ends whenever the two-pronged string with both legs
overlapping is the last one of the group and it is then simply removed for the
next node in the quiver. (See for instance the partitions $[5,3]$ or $[7,1]$
for $SO(8)$ in figure \ref{fig:SO8quiver}).

\item Finally we can have groups of $K$ branes involving the short root
$\alpha_{N-1}$ of $SO(2N-1)$, which connects the $N^{th}$ $A$-brane to the one
merged onto the mirror. In this case, the first step consists of lifting the
short string above the middle brane so that it becomes a doubled-arrow string
crossing the mirror and connecting $K-1$ branes. The next steps in the
propagation are then identical to the ones described in the previous point.
(See for instance the partitions $[7,1^{2}]$ or $[9]$ for $SO(9)$ in figure
\ref{fig:SO9quiver}).
\end{itemize}

We note that in terms of partitions, these steps simply translate into every
part being reduced by $1$, so that the partition $[\mu_{1}, \mu_{2}, \cdots,
\mu_{i}, 1^{k}]$ goes to $[\mu_{1}-1, \mu_{2}-1, \cdots, \mu_{i}-1, 1^{k+i}]$
after each step until there are no more parts with $\mu_{i} > 1$, and we are left with
the trivial partition (corresponding to the total absence of strings).

\subsection{Higgsing and Brane Recombination}\label{ssec:Higgsing}

Once we have propagated the strings according to the above rules, we are ready
to read off the residual gauge symmetry on each node. To do so, we note that
the strings force connected branes on each side of the mirror to coalesce so that a blob of $K$ $A$-branes behaves like a single
$A$-brane. We can then directly read off the gauge symmetry that is described
by the resulting collapsed brane configuration.

For $SU(N)$ quivers, which do not involve a mirror, strings group $A$-branes
without any ambiguity, as no $B$ or $C$ brane is present. Thus, the residual
gauge symmetry is given by the number of groups formed at each level. For
instance, if only one simple string stretches between two $A$-branes, these branes
coalesce, and we are left with $N-1$ separate groups of strings on
this level. This yields the residual gauge symmetry $\mathfrak{su}_{N-1}$ as
illustrated in the first orbit of $SU(6)$ (see figure \ref{fig:SU6quiver}).

\begin{figure}[H]
  \vspace{-1.0cm} \includegraphics{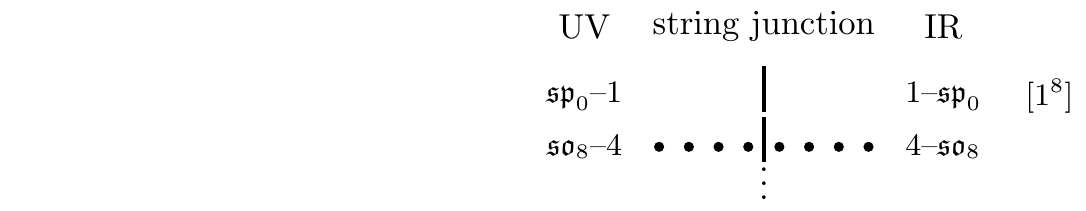}
  \par
  \includegraphics{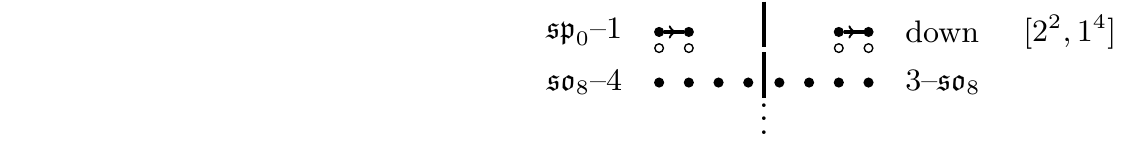}
  \par
  \begin{tabular}{ccc}
    \hspace{-2.26cm}
    \includegraphics{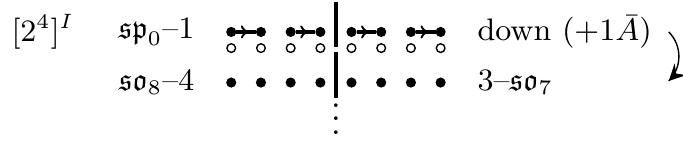} & \includegraphics{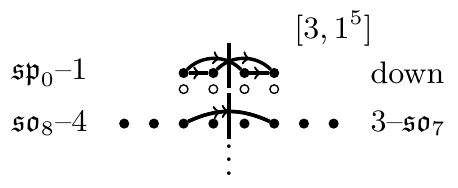} & \includegraphics{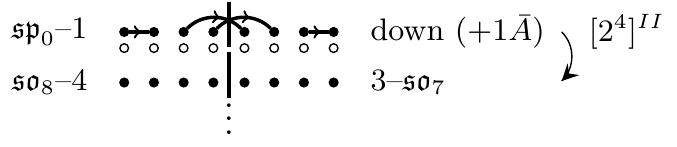}
  \end{tabular}
  \par
  \includegraphics{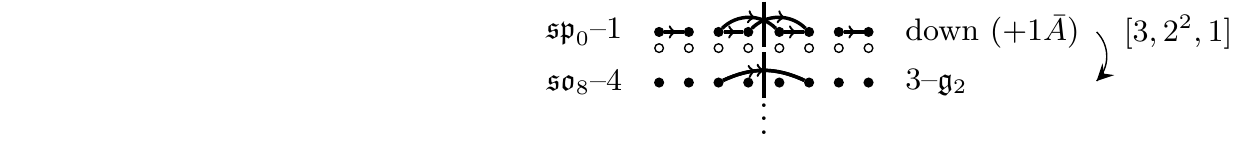}
  \par
  \includegraphics{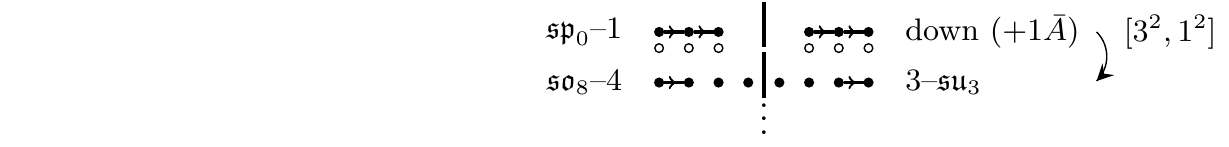}
  \par
  \begin{tabular}{ccc}
    \hspace{-2.38cm}
    \includegraphics{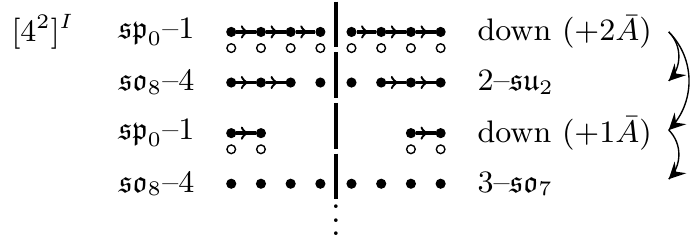} & \includegraphics{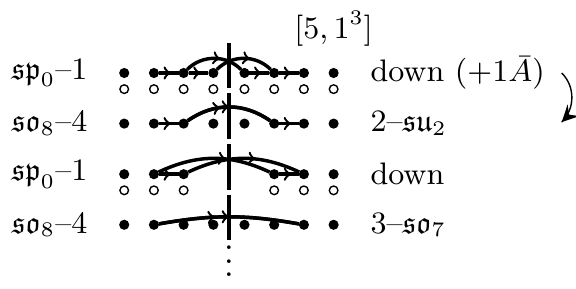} & \includegraphics{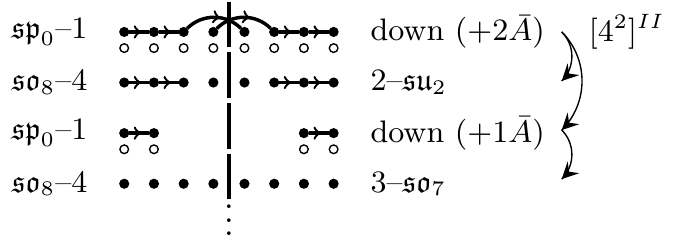} \vspace{-0.1cm}
  \end{tabular}
  \par
  \includegraphics{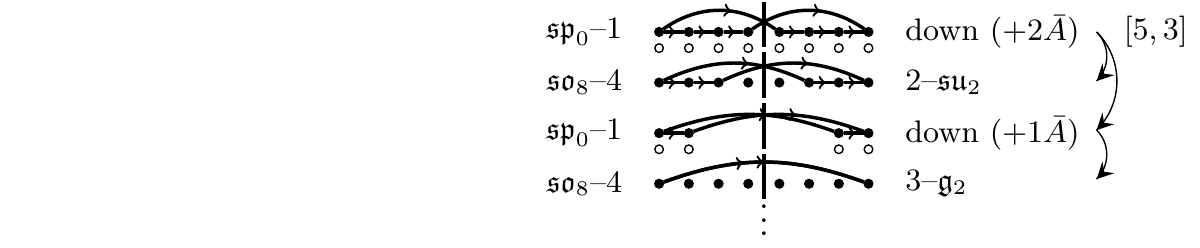}
  \par
  \includegraphics{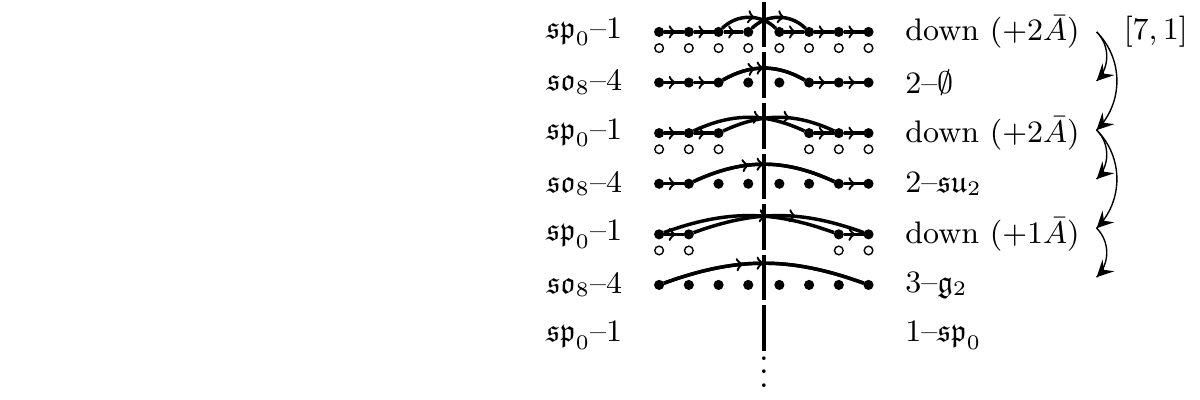}
  \caption{Nilpotent deformations of the $SO(8)$ quiver from the UV configuration of figure \ref{fig:SO2NUV}. Each subfigure corresponds to the quiver diagram of a nilpotent obit with strings propagating into the interior of the quiver. The quivers have been rotated to go from top to bottom (rather than left to right) to fit on the page. On the left-hand side of each subfigure, we have the initial UV theory with alternating $-1$ and $-4$ curves containing $\mathfrak{sp}_0$ and $\mathfrak{so}_8$ respectively. On the right-hand side, we give the IR theory induced by the strings stretched in the middle diagram. The vertical line denotes the $BC$-mirror. Whenever anti-branes are required, they are denoted by white circle below their $A$-brane counterparts. In some cases, there are extra anti-branes indicated in parentheses on the right (which occur when there are more groups of $A$-branes than anti-branes). The first one is used to blow-down the $-1$ curve it is on (indicated by the word ``down''), while the others get distributed on the following quiver nodes as indicated by the side arrows on the right. The theories are ordered from top to bottom according to their partial ordering of RG flows. The corresponding partitions are given on the side.}
  \label{fig:SO8quiver}
\end{figure}

Similarly, a blob with $K$ branes connected by strings on each side of
a mirror turns an $\mathfrak{so}_{2N}$ algebra into $\mathfrak{so}_{2(N-K+1)}$,
$\mathfrak{so}_{2N-1}$ into $\mathfrak{so}_{2(N-K+1)-1}$, and $\mathfrak{sp}_{N}$
into $\mathfrak{sp}_{N-K+1}$. The same is true if the blob consists of branes on both sides of the mirror connected by double-pronged strings. However, if the blob consists of branes connected by a double-arrowed string, then the blob of connected branes gets merged onto the mirror.
As a result, an $\mathfrak{so}_{2K}$ algebra will turn into $\mathfrak{so}_{2K-1}$, and $\mathfrak{so}_{2K-1}$ into $\mathfrak{so}_{2K-2}$. (See for instance the [7,1] diagrams at the bottom of figure \ref{fig:SO8quiver}.) We note that
the propagation rules listed above prevent such a configuration from ever
appearing on a level with $\mathfrak{sp}_{N}$ gauge symmetry.

In some cases, the $\mathfrak{so}$ quivers require the
introduction of ``anti-branes.'' In our figures, we denote a brane by a filled in circle (black dot)
and an anti-brane by an open circle. At the final step, all such anti-branes must disappear by
pairing up with other coalesced branes, deleting such blobs from the resulting configuration.
This further reduces the number of leftover blobs which generate the residual gauge symmetry.

Note that there are also situations where the number of anti-branes is larger than the number of available blobs of branes on a given layer. This occurs whenever the number of D6-branes in the type IIA suspended brane realization formally becomes negative, signaling that the perturbative type IIA description has broken down, and F-theory is required to construct the theory in question. Nevertheless, it is still useful to write down a ``formal IIA quiver,'' which includes negative numbers of D6-branes and hence negative gauge group ranks. Additionally, as we will now show with examples, our picture of brane / anti-brane nucleation can be adapted to these situations if we allow extra anti-branes at a given layer to move to other layers and annihilate other blobs of branes.

Consider, for instance, the partition $[5,3]$ of $SO(8)$ requires the presence of four $A$-branes on the first
quiver node, which only has $\mathfrak{sp}_{0}$ symmetry. Thus, we also need to
introduce four anti-branes to compensate. Only one blob of branes is formed
on each side of the mirror, so only one of the four anti-branes is used to
cancel it, and we are left with three anti-branes. The first anti-brane is used
to collapse the $-1$ curve it is on. The second anti-brane is distributed to
the next $\mathfrak{so}$ quiver node and the third anti-brane is distributed
to the next $\mathfrak{sp}$ quiver node, where it is used to either reduce the
gauge symmetry from $\mathfrak{sp}_{K}$ to $\mathfrak{sp}_{K-1}$ or, if $K=0$, to blow down
this next $-1$ curve. The
anti-brane that lands on a quiver node with an $\mathfrak{so}$ algebra also reduces the residual symmetry according to the following rules:
\begin{align}
  \mathfrak{so}_{N} &\overset{\overline{A}}{\rightarrow} \mathfrak{so}_{N-1} \text{ for } N \geq 8, \nonumber\\
  \mathfrak{so}_{7} &\overset{\overline{A}}{\rightarrow} \mathfrak{g}_{2},\nonumber\\
  \mathfrak{g}_{2}  &\overset{\overline{A}}{\rightarrow} \mathfrak{su}_{3},\nonumber\\
  \mathfrak{so}_{6} \simeq \mathfrak{su}_{4} &\overset{\overline{A}}{\rightarrow} \mathfrak{su}_{3},\nonumber\\
  \mathfrak{su}_{3} &\overset{\overline{A}}{\rightarrow} \mathfrak{su}_{2},\nonumber\\
  \mathfrak{so}_{5} \simeq \mathfrak{sp}_{2} &\overset{\overline{A}}{\rightarrow} \mathfrak{sp}_{1} \simeq \mathfrak{su}_{2},\nonumber\\
  \mathfrak{so}_{4} &\overset{\overline{A}}{\rightarrow} \mathfrak{so}_{3} \simeq \mathfrak{su}_{2},\nonumber\\
  \mathfrak{so}_{3} \simeq \mathfrak{su}_{2} &\overset{\overline{A}}{\rightarrow} \mathfrak{su}_{1} \simeq \emptyset.
  \label{eq:rules}
\end{align}
Note that for classical quiver theories, there can never be more than four anti-branes, since the
quiver nodes with $\mathfrak{sp}$ gauge symmetry only have four fewer branes
than their neighboring $\mathfrak{so}$ nodes.

\begin{figure}[H]
  \includegraphics{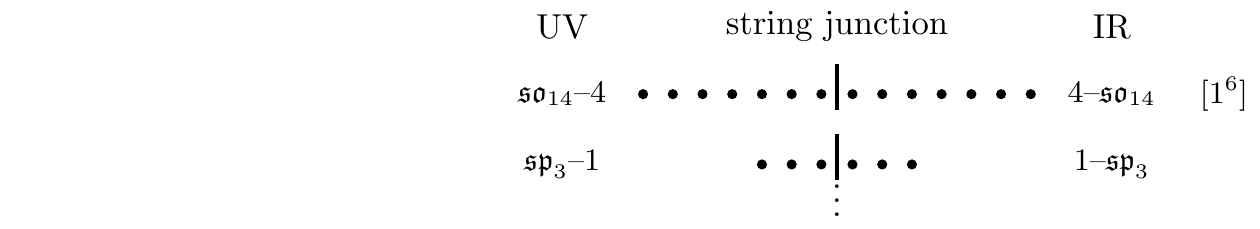}
  \par
  \includegraphics{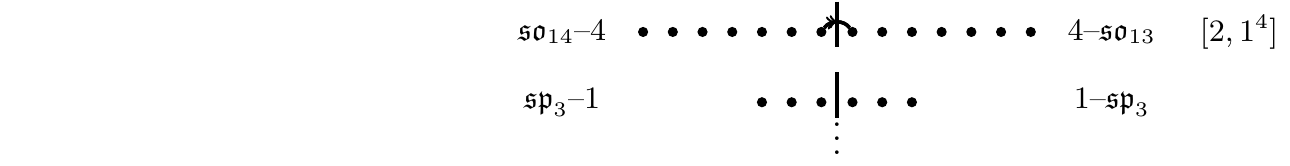}
  \par
  \includegraphics{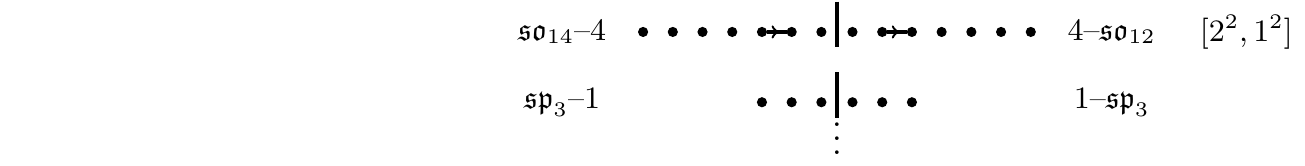}
  \par
  \includegraphics{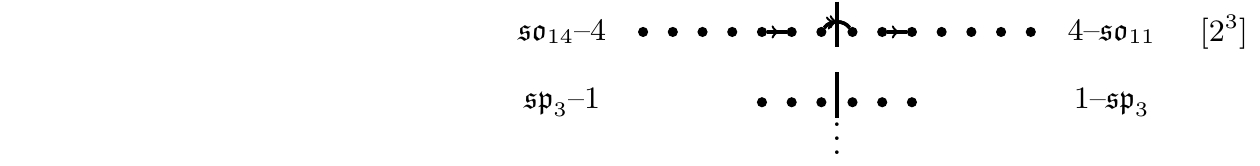}
  \par
  \begin{center}
    \hspace{0cm} \includegraphics{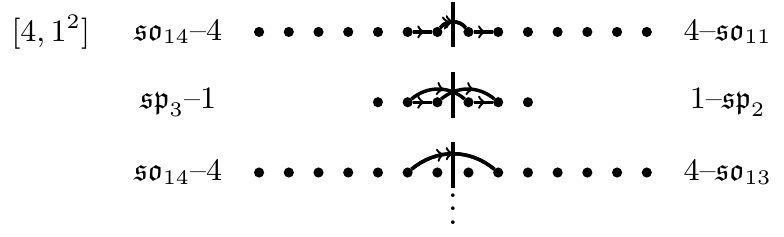} \hspace{0.5cm} \includegraphics{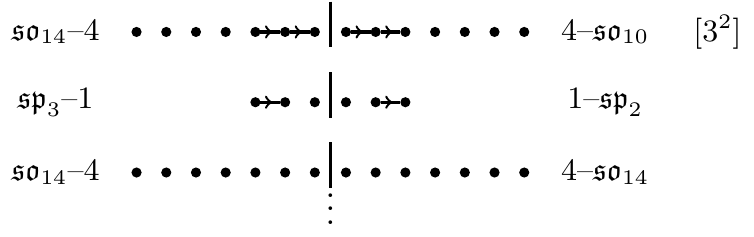} \hspace{0cm}
  \end{center}
  \par
  \includegraphics{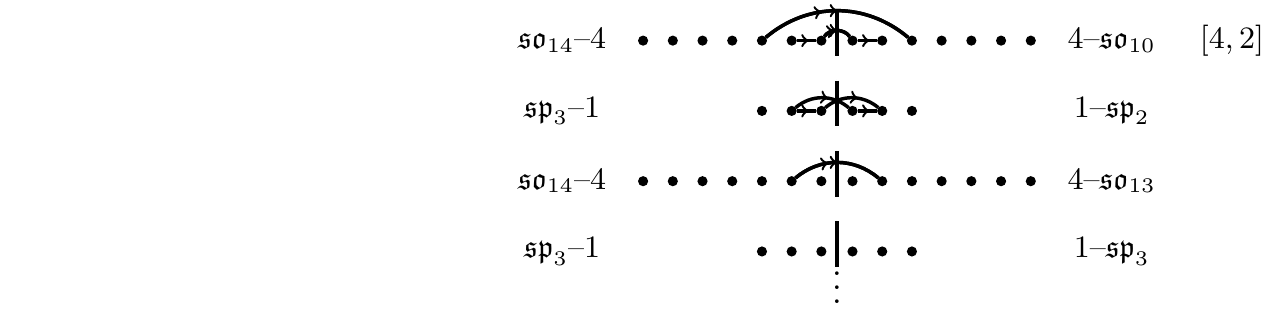}
  \par
  \includegraphics{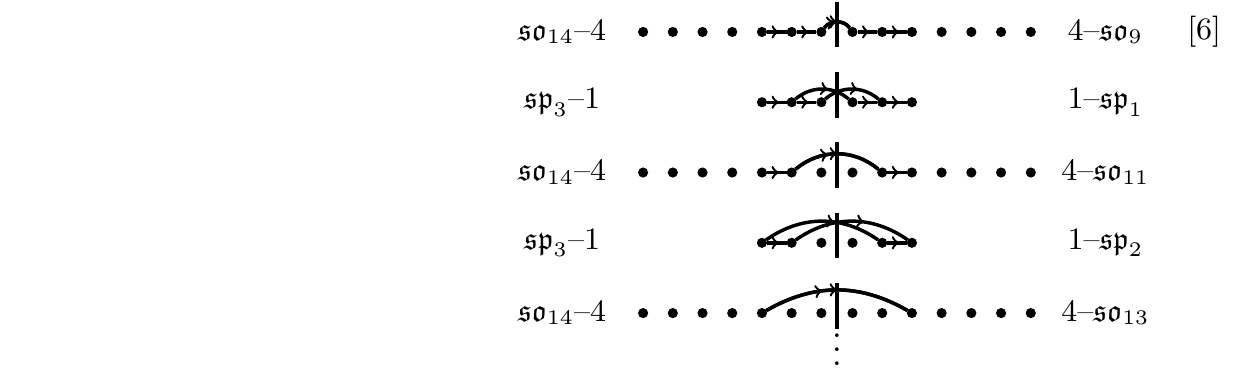}
  \par
  \caption{Nilpotent deformations of the $Sp(3)$ quiver from the UV configuration of figure \ref{fig:SpNUV}. See figure \ref{fig:SO8quiver} for additional details on the notation and conventions.}
  \label{fig:Sp3quiver}
\end{figure}

We illustrate all of these steps through the examples of $SU(6)$, $SO(8)$,
$SO(10)$, $SO(9)$, and $Sp(3)$ in figures \ref{fig:SU6quiver},
\ref{fig:SO8quiver}, \ref{fig:SO10quiver}, \ref{fig:SO9quiver}, and
\ref{fig:Sp3quiver} respectively. Explicit examples of $\mathfrak{g}_{2} \overset{\overline{A}}{\rightarrow} \mathfrak{su}_{3}$ and $\mathfrak{su}_{3} \overset{\overline{A}}{\rightarrow} \mathfrak{su}_{2}$ can only be found when dealing with ``short quivers,'' which we discuss in section \ref{sec:GETSHORTY}.
\newcommand{\rightsep}{0.5cm}
\newcommand{\vertsep}{0.25cm}
\newcommand{\SUscale}{0.8}
\newcommand{\SUvspace}{0.5cm}
\newcommand{\nodesSUsixa}{\node[paramsAbranes] (a1) {}; \node[paramsAbranes] (a2) [right=\rightsep of a1] {}; \node[paramsAbranes] (a3) [right=\rightsep of a2] {}; \node[paramsAbranes] (a4) [right=\rightsep of a3] {}; \node[paramsAbranes] (a5) [right=\rightsep of a4] {}; \node[paramsAbranes] (a6) [right=\rightsep of a5] {};}
\newcommand{\nodesSUsixb}{\node[paramsAbranes] (b1) [below=\vertsep of a1] {}; \node[paramsAbranes] (b2) [right=\rightsep of b1] {}; \node[paramsAbranes] (b3) [right=\rightsep of b2] {}; \node[paramsAbranes] (b4) [right=\rightsep of b3] {}; \node[paramsAbranes] (b5) [right=\rightsep of b4] {}; \node[paramsAbranes] (b6) [right=\rightsep of b5] {}}
\newcommand{\nodesSUsixc}{\node[paramsAbranes] (c1) [below=\vertsep of b1] {}; \node[paramsAbranes] (c2) [right=\rightsep of c1] {}; \node[paramsAbranes] (c3) [right=\rightsep of c2] {}; \node[paramsAbranes] (c4) [right=\rightsep of c3] {}; \node[paramsAbranes] (c5) [right=\rightsep of c4] {}; \node[paramsAbranes] (c6) [right=\rightsep of c5] {};}
\newcommand{\nodesSUsixd}{\node[paramsAbranes] (d1) [below=\vertsep of c1] {}; \node[paramsAbranes] (d2) [right=\rightsep of d1] {}; \node[paramsAbranes] (d3) [right=\rightsep of d2] {}; \node[paramsAbranes] (d4) [right=\rightsep of d3] {}; \node[paramsAbranes] (d5) [right=\rightsep of d4] {}; \node[paramsAbranes] (d6) [right=\rightsep of d5] {};}
\newcommand{\nodesSUsixe}{\node[paramsAbranes] (e1) [below=\vertsep of d1] {}; \node[paramsAbranes] (e2) [right=\rightsep of e1] {}; \node[paramsAbranes] (e3) [right=\rightsep of e2] {}; \node[paramsAbranes] (e4) [right=\rightsep of e3] {}; \node[paramsAbranes] (e5) [right=\rightsep of e4] {}; \node[paramsAbranes] (e6) [right=\rightsep of e5] {};}
\newcommand{\nodesSUsixf}{\node[paramsAbranes] (f1) [below=\vertsep of e1] {}; \node[paramsAbranes] (f2) [right=\rightsep of f1] {}; \node[paramsAbranes] (f3) [right=\rightsep of f2] {}; \node[paramsAbranes] (f4) [right=\rightsep of f3] {}; \node[paramsAbranes] (f5) [right=\rightsep of f4] {}; \node[paramsAbranes] (f6) [right=\rightsep of f5] {};}

\begin{figure}[H]
  \centering
  \scalebox{\SUscale}{\begin{tabular}{ccc}
     &UV \hspace{1cm} string junction \hspace{1cm} IR& \\
    {\color{white}$[1^6]$} & \includegraphics{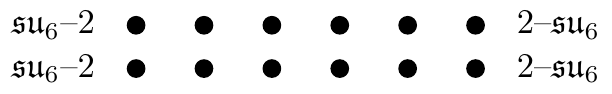} &$[1^6]$ \vspace{-0.3cm} \\
     &$\vdots$ &
  \end{tabular}}\\
  \vspace{\SUvspace}
  \scalebox{\SUscale}{\begin{tabular}{ccc}
   {\color{white}$[2,1^4]$} & \includegraphics{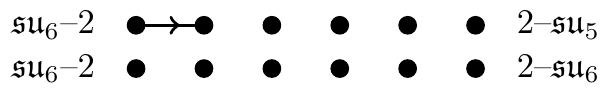} &$[2,1^4]$ \vspace{-0.3cm} \\
     &$\vdots$ &
  \end{tabular}}\\
  \vspace{\SUvspace}
  \scalebox{\SUscale}{\begin{tabular}{ccc}
    {\color{white}$[2^2,1^2]$} & \includegraphics{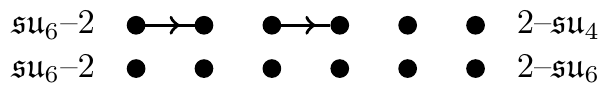} &$[2^2,1^2]$ \vspace{-0.3cm} \\
     &$\vdots$ &
  \end{tabular}}\\
  \vspace{\SUvspace}
  \scalebox{\SUscale}{\begin{tabular}{ccccccccc}
      $[2^3]${\color{white}$,1$} & \includegraphics{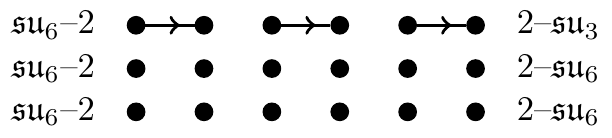}
      &&&&&&
      \includegraphics{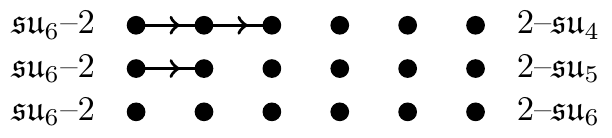}
    & $[3,1^3]$ \vspace{-0.3cm} \\
    &$\vdots$ &&&&&& $\vdots$&\\
  \end{tabular}}\\
  \vspace{\SUvspace}
  \scalebox{\SUscale}{\begin{tabular}{ccc}
    {\color{white}$[3,2,1]$} & \includegraphics{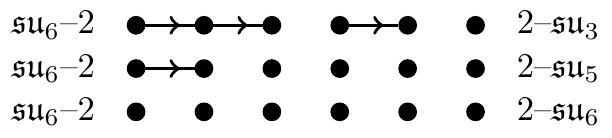} & $[3,2,1]$ \vspace{-0.3cm} \\
     &$\vdots$ &
  \end{tabular}}\\
  \vspace{\SUvspace}
  \scalebox{\SUscale}{\begin{tabular}{ccccccccc}
      $[3^2]${\color{white}$,1$}& \includegraphics{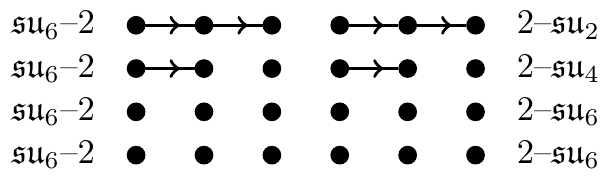}
    &&&&&&
    \includegraphics{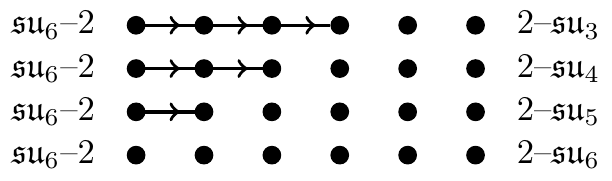} &$[4,1^2]$ \vspace{-0.3cm} \\
    &$\vdots$ &&&&&& $\vdots$&\\
  \end{tabular}} \\
  \vspace{\SUvspace}
  \scalebox{\SUscale}{\begin{tabular}{ccc}
    {\color{white}$[4,2]$} & \includegraphics{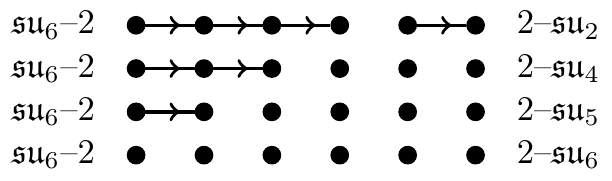} & $[4,2]$ \vspace{-0.3cm} \\
     &$\vdots$&
  \end{tabular}}\\
  \vspace{\SUvspace}
 \scalebox{\SUscale}{ \begin{tabular}{ccc}
    {\color{white}$[5,1]$} & \includegraphics{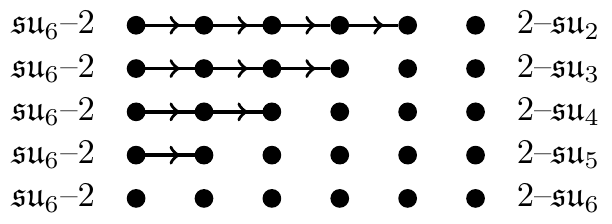} & $[5,1]$ \vspace{-0.3cm} \\
     &$\vdots$ &
  \end{tabular}}\\
  \vspace{\SUvspace}
  \scalebox{\SUscale}{\begin{tabular}{ccc}
    {\color{white}$[6]$} & \includegraphics{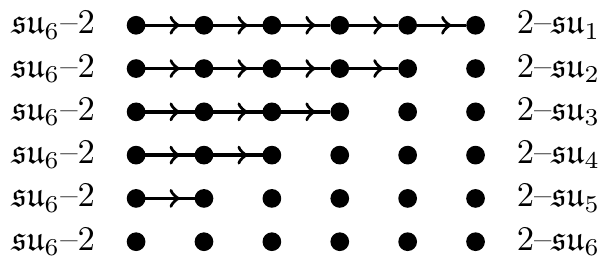} & $[6]$\vspace{-0.3cm} \\
     &$\vdots$ &
  \end{tabular}}
\caption{Nilpotent deformations of the $SU(6)$ quiver from the UV configuration of figure \ref{fig:SUNUV}. Each subfigure corresponds to the quiver diagram of a nilpotent orbit with strings propagating through. The quivers have been rotated to go from top to bottom (rather than left to right) to fit on the page. On the left-hand side of each subfigure we have the setting in the UV with each $-2$ curve containing an $\mathfrak{su}_6$ gauge algebra, while on the right-hand side we give the IR theory induced by the strings stretched in the middle diagram. The theories are ordered from top to bottom according to their partial ordering of RG flows, which matches their mathematical ordering. The corresponding partitions are given on the side.}
  \label{fig:SU6quiver}
\end{figure}

\newcommand{\SOeightrightsep}{0.2cm}
\newcommand{\SOeightvertsep}{0.5cm}
\newcommand{\SOeightscale}{0.8}
\newcommand{\SOeightvspace}{-0.2cm}
\newcommand{\SOeightmirup}{0.3cm}
\newcommand{\SOeightmirdown}{0.15cm}
\newcommand{\nodesSOeighta}{
\node[circle,draw=black, fill=black, inner sep=0pt,minimum size=0pt]   (amir) {};
\node[circle,draw=black, fill=black, inner sep=0pt,minimum size=0pt]   (amirup)   [above=\SOeightmirup of amir] {};
\node[circle,draw=black, fill=black, inner sep=0pt,minimum size=0pt]   (amirdown) [below=\SOeightmirdown of amir] {};
\draw[-, line width=1pt, shorten <=0, shorten >=0] (amirup) -- (amirdown);
}
\newcommand{\nodesSOeightb}{
\node[circle,draw=black, fill=black, inner sep=0pt,minimum size=0pt]     (bmir) [below=\SOeightvertsep of amir] {};
\node[circle,draw=black, fill=black, inner sep=0pt,minimum size=0pt]   (bmirup) [above=\SOeightmirup of bmir] {};
\node[circle,draw=black, fill=black, inner sep=0pt,minimum size=0pt] (bmirdown) [below=\SOeightmirdown of bmir] {};
\node[paramsbranes]   (b4) [left=\SOeightrightsep/2 of bmir] {};
\node[paramsbranes]   (b3) [left=\SOeightrightsep of b4] {};
\node[paramsbranes]   (b2) [left=\SOeightrightsep of b3] {};
\node[paramsbranes]   (b1) [left=\SOeightrightsep of b2] {};
\node[paramsbranes]   (bt4)   [right=\SOeightrightsep/2 of bmir] {};
\node[paramsbranes]   (bt3) [right=\SOeightrightsep of bt4] {};
\node[paramsbranes]   (bt2) [right=\SOeightrightsep of bt3] {};
\node[paramsbranes]   (bt1) [right=\SOeightrightsep of bt2] {};
\draw[-, line width=1pt, shorten <=0, shorten >=0] (bmirup) -- (bmirdown);
}
\newcommand{\nodesSOeightc}{
\node[circle,draw=black, fill=black, inner sep=0pt,minimum size=0pt]     (cmir) [below=\SOeightvertsep of bmir] {};
\node[circle,draw=black, fill=black, inner sep=0pt,minimum size=0pt]   (cmirup) [above=\SOeightmirup of cmir] {};
\node[circle,draw=black, fill=black, inner sep=0pt,minimum size=0pt] (cmirdown) [below=\SOeightmirdown of cmir] {};
\draw[-, line width=1pt, shorten <=0, shorten >=0] (cmirup) -- (cmirdown);
}
\newcommand{\nodesSOeightd}{
\node[circle,draw=black, fill=black, inner sep=0pt,minimum size=0pt]     (dmir) [below=\SOeightvertsep of cmir] {};
\node[circle,draw=black, fill=black, inner sep=0pt,minimum size=0pt]   (dmirup) [above=\SOeightmirup of dmir] {};
\node[circle,draw=black, fill=black, inner sep=0pt,minimum size=0pt] (dmirdown) [below=\SOeightmirdown of dmir] {};
\node[paramsbranes]   (d4) [left=\SOeightrightsep/2 of dmir] {};
\node[paramsbranes]   (d3) [left=\SOeightrightsep of d4] {};
\node[paramsbranes]   (d2) [left=\SOeightrightsep of d3] {};
\node[paramsbranes]   (d1) [left=\SOeightrightsep of d2] {};
\node[paramsbranes]   (dt4) [right=\SOeightrightsep/2 of dmir] {};
\node[paramsbranes]   (dt3) [right=\SOeightrightsep of dt4] {};
\node[paramsbranes]   (dt2) [right=\SOeightrightsep of dt3] {};
\node[paramsbranes]   (dt1) [right=\SOeightrightsep of dt2] {};
\draw[-, line width=1pt, shorten <=0, shorten >=0] (dmirup) -- (dmirdown);
}
\newcommand{\nodesSOeighte}{
\node[circle,draw=black, fill=black, inner sep=0pt,minimum size=0pt]     (emir) [below=\SOeightvertsep of dmir] {};
\node[circle,draw=black, fill=black, inner sep=0pt,minimum size=0pt]   (emirup) [above=\SOeightmirup of emir] {};
\node[circle,draw=black, fill=black, inner sep=0pt,minimum size=0pt] (emirdown) [below=\SOeightmirdown of emir] {};
\draw[-, line width=1pt, shorten <=0, shorten >=0] (emirup) -- (emirdown);
}
\newcommand{\nodesSOeightf}{
\node[circle,draw=black, fill=black, inner sep=0pt,minimum size=0pt]     (fmir) [below=\SOeightvertsep of emir] {};
\node[circle,draw=black, fill=black, inner sep=0pt,minimum size=0pt]   (fmirup) [above=\SOeightmirup of fmir] {};
\node[circle,draw=black, fill=black, inner sep=0pt,minimum size=0pt] (fmirdown) [below=\SOeightmirdown of fmir] {};
\node[paramsbranes]   (f4) [left=\SOeightrightsep/2 of fmir] {};
\node[paramsbranes]   (f3) [left=\SOeightrightsep of f4] {};
\node[paramsbranes]   (f2) [left=\SOeightrightsep of f3] {};
\node[paramsbranes]   (f1) [left=\SOeightrightsep of f2] {};
\node[paramsbranes]   (ft4) [right=\SOeightrightsep/2 of fmir] {};
\node[paramsbranes]   (ft3) [right=\SOeightrightsep of ft4] {};
\node[paramsbranes]   (ft2) [right=\SOeightrightsep of ft3] {};
\node[paramsbranes]   (ft1) [right=\SOeightrightsep of ft2] {};
\draw[-, line width=1pt, shorten <=0, shorten >=0] (fmirup) -- (fmirdown);
}
\newcommand{\nodesSOeightg}{
\node[circle,draw=black, fill=black, inner sep=0pt,minimum size=0pt]     (gmir) [below=\SOeightvertsep of fmir] {};
\node[circle,draw=black, fill=black, inner sep=0pt,minimum size=0pt]   (gmirup) [above=\SOeightmirup of gmir] {};
\node[circle,draw=black, fill=black, inner sep=0pt,minimum size=0pt] (gmirdown) [below=\SOeightmirdown of gmir] {};
\draw[-, line width=1pt, shorten <=0, shorten >=0] (gmirup) -- (gmirdown);
}

\newcommand{\SOtenrightsep}{0.2cm}
\newcommand{\SOtenvertsep}{0.5cm}
\newcommand{\SOtenscale}{0.8}
\newcommand{\SOtenvspace}{0.3cm}
\newcommand{\SOtenmirup}{0.3cm}
\newcommand{\SOtenmirdown}{0.15cm}
\newcommand{\nodesSOtena}{
\node[circle,draw=black, fill=black, inner sep=0pt,minimum size=0pt]   (amir) {};
\node[circle,draw=black, fill=black, inner sep=0pt,minimum size=0pt]   (amirup)   [above=\SOtenmirup of amir] {};
\node[circle,draw=black, fill=black, inner sep=0pt,minimum size=0pt]   (amirdown) [below=\SOtenmirdown of amir] {};
\draw[-, line width=1pt, shorten <=0, shorten >=0] (amirup) -- (amirdown);
}
\newcommand{\nodesSOtenb}{
\node[circle,draw=black, fill=black, inner sep=0pt,minimum size=0pt]     (bmir) [below=\SOtenvertsep of amir] {};
\node[circle,draw=black, fill=black, inner sep=0pt,minimum size=0pt]   (bmirup) [above=\SOtenmirup of bmir] {};
\node[circle,draw=black, fill=black, inner sep=0pt,minimum size=0pt] (bmirdown) [below=\SOtenmirdown of bmir] {};
\node[paramsbranes]   (b5) [left=\SOtenrightsep/2 of bmir] {};
\node[paramsbranes]   (b4) [left=\SOtenrightsep of b5] {};
\node[paramsbranes]   (b3) [left=\SOtenrightsep of b4] {};
\node[paramsbranes]   (b2) [left=\SOtenrightsep of b3] {};
\node[paramsbranes]   (b1) [left=\SOtenrightsep of b2] {};
\node[paramsbranes]   (bt5) [right=\SOtenrightsep/2 of bmir] {};
\node[paramsbranes]   (bt4) [right=\SOtenrightsep of bt5] {};
\node[paramsbranes]   (bt3) [right=\SOtenrightsep of bt4] {};
\node[paramsbranes]   (bt2) [right=\SOtenrightsep of bt3] {};
\node[paramsbranes]   (bt1) [right=\SOtenrightsep of bt2] {};
\draw[-, line width=1pt, shorten <=0, shorten >=0] (bmirup) -- (bmirdown);
}
\newcommand{\nodesSOtenc}{
\node[circle,draw=black, fill=black, inner sep=0pt,minimum size=0pt]     (cmir) [below=\SOtenvertsep of bmir] {};
\node[circle,draw=black, fill=black, inner sep=0pt,minimum size=0pt]   (cmirup) [above=\SOtenmirup of cmir] {};
\node[circle,draw=black, fill=black, inner sep=0pt,minimum size=0pt] (cmirdown) [below=\SOtenmirdown of cmir] {};
\draw[-, line width=1pt, shorten <=0, shorten >=0] (cmirup) -- (cmirdown);
}
\newcommand{\nodesSOtend}{
\node[circle,draw=black, fill=black, inner sep=0pt,minimum size=0pt]     (dmir) [below=\SOtenvertsep of cmir] {};
\node[circle,draw=black, fill=black, inner sep=0pt,minimum size=0pt]   (dmirup) [above=\SOtenmirup of dmir] {};
\node[circle,draw=black, fill=black, inner sep=0pt,minimum size=0pt] (dmirdown) [below=\SOtenmirdown of dmir] {};
\node[paramsbranes]   (d5) [left=\SOtenrightsep/2 of dmir] {};
\node[paramsbranes]   (d4) [left=\SOtenrightsep of d5] {};
\node[paramsbranes]   (d3) [left=\SOtenrightsep of d4] {};
\node[paramsbranes]   (d2) [left=\SOtenrightsep of d3] {};
\node[paramsbranes]   (d1) [left=\SOtenrightsep of d2] {};
\node[paramsbranes]   (dt5) [right=\SOtenrightsep/2 of dmir] {};
\node[paramsbranes]   (dt4) [right=\SOtenrightsep of dt5] {};
\node[paramsbranes]   (dt3) [right=\SOtenrightsep of dt4] {};
\node[paramsbranes]   (dt2) [right=\SOtenrightsep of dt3] {};
\node[paramsbranes]   (dt1) [right=\SOtenrightsep of dt2] {};
\draw[-, line width=1pt, shorten <=0, shorten >=0] (dmirup) -- (dmirdown);
}
\newcommand{\nodesSOtene}{
\node[circle,draw=black, fill=black, inner sep=0pt,minimum size=0pt]     (emir) [below=\SOtenvertsep of dmir] {};
\node[circle,draw=black, fill=black, inner sep=0pt,minimum size=0pt]   (emirup) [above=\SOtenmirup of emir] {};
\node[circle,draw=black, fill=black, inner sep=0pt,minimum size=0pt] (emirdown) [below=\SOtenmirdown of emir] {};
\draw[-, line width=1pt, shorten <=0, shorten >=0] (emirup) -- (emirdown);
}
\newcommand{\nodesSOtenf}{
\node[circle,draw=black, fill=black, inner sep=0pt,minimum size=0pt]     (fmir) [below=\SOtenvertsep of emir] {};
\node[circle,draw=black, fill=black, inner sep=0pt,minimum size=0pt]   (fmirup) [above=\SOtenmirup of fmir] {};
\node[circle,draw=black, fill=black, inner sep=0pt,minimum size=0pt] (fmirdown) [below=\SOtenmirdown of fmir] {};
\node[paramsbranes]   (f5) [left=\SOtenrightsep/2 of fmir] {};
\node[paramsbranes]   (f4) [left=\SOtenrightsep of f5] {};
\node[paramsbranes]   (f3) [left=\SOtenrightsep of f4] {};
\node[paramsbranes]   (f2) [left=\SOtenrightsep of f3] {};
\node[paramsbranes]   (f1) [left=\SOtenrightsep of f2] {};
\node[paramsbranes]   (ft5) [right=\SOtenrightsep/2 of fmir] {};
\node[paramsbranes]   (ft4) [right=\SOtenrightsep of ft5] {};
\node[paramsbranes]   (ft3) [right=\SOtenrightsep of ft4] {};
\node[paramsbranes]   (ft2) [right=\SOtenrightsep of ft3] {};
\node[paramsbranes]   (ft1) [right=\SOtenrightsep of ft2] {};
\draw[-, line width=1pt, shorten <=0, shorten >=0] (fmirup) -- (fmirdown);
}
\newcommand{\nodesSOteng}{
\node[circle,draw=black, fill=black, inner sep=0pt,minimum size=0pt]     (gmir) [below=\SOtenvertsep of fmir] {};
\node[circle,draw=black, fill=black, inner sep=0pt,minimum size=0pt]   (gmirup) [above=\SOtenmirup of gmir] {};
\node[circle,draw=black, fill=black, inner sep=0pt,minimum size=0pt] (gmirdown) [below=\SOtenmirdown of gmir] {};
\draw[-, line width=1pt, shorten <=0, shorten >=0] (gmirup) -- (gmirdown);
}
\newcommand{\nodesSOtenh}{
\node[circle,draw=black, fill=black, inner sep=0pt,minimum size=0pt]     (hmir) [below=\SOtenvertsep of gmir] {};
\node[circle,draw=black, fill=black, inner sep=0pt,minimum size=0pt]   (hmirup) [above=\SOtenmirup of hmir] {};
\node[circle,draw=black, fill=black, inner sep=0pt,minimum size=0pt] (hmirdown) [below=\SOtenmirdown of hmir] {};
\node[paramsbranes]   (h5) [left=\SOtenrightsep/2 of hmir] {};
\node[paramsbranes]   (h4) [left=\SOtenrightsep of h5] {};
\node[paramsbranes]   (h3) [left=\SOtenrightsep of h4] {};
\node[paramsbranes]   (h2) [left=\SOtenrightsep of h3] {};
\node[paramsbranes]   (h1) [left=\SOtenrightsep of h2] {};
\node[paramsbranes]   (ht5) [right=\SOtenrightsep/2 of hmir] {};
\node[paramsbranes]   (ht4) [right=\SOtenrightsep of ht5] {};
\node[paramsbranes]   (ht3) [right=\SOtenrightsep of ht4] {};
\node[paramsbranes]   (ht2) [right=\SOtenrightsep of ht3] {};
\node[paramsbranes]   (ht1) [right=\SOtenrightsep of ht2] {};
\draw[-, line width=1pt, shorten <=0, shorten >=0] (hmirup) -- (hmirdown);
}
\newcommand{\nodesSOteni}{
\node[circle,draw=black, fill=black, inner sep=0pt,minimum size=0pt]     (imir) [below=\SOtenvertsep of hmir] {};
\node[circle,draw=black, fill=black, inner sep=0pt,minimum size=0pt]   (imirup) [above=\SOtenmirup of imir] {};
\node[circle,draw=black, fill=black, inner sep=0pt,minimum size=0pt] (imirdown) [below=\SOtenmirdown of imir] {};
\draw[-, line width=1pt, shorten <=0, shorten >=0] (imirup) -- (imirdown);
}
\begin{figure}[H]
  \vspace{-0.5cm}
  \includegraphics{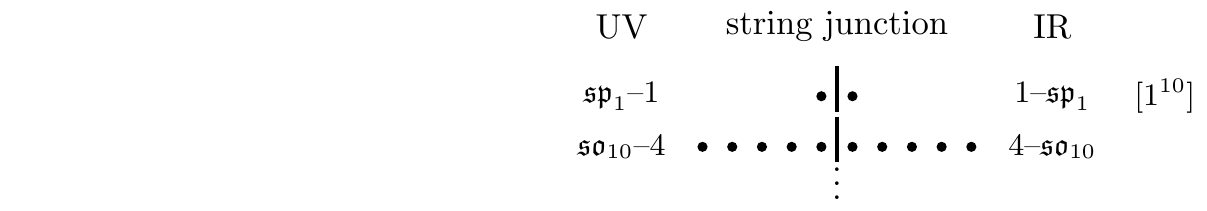}
  \par
  \includegraphics{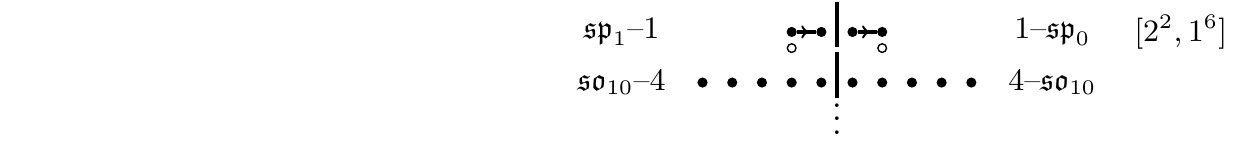}
  \par
  \begin{center}
    \hspace{-2cm} \includegraphics{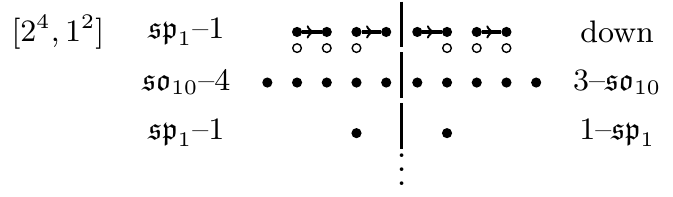} \hspace{0.5cm} \includegraphics{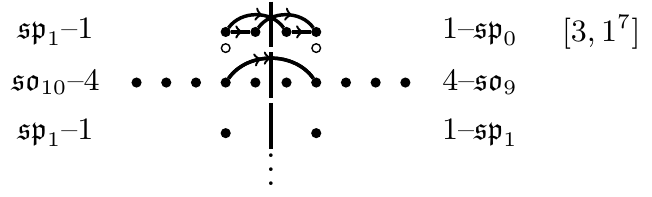} \hspace*{-2cm}
  \end{center}
  \par
  \includegraphics{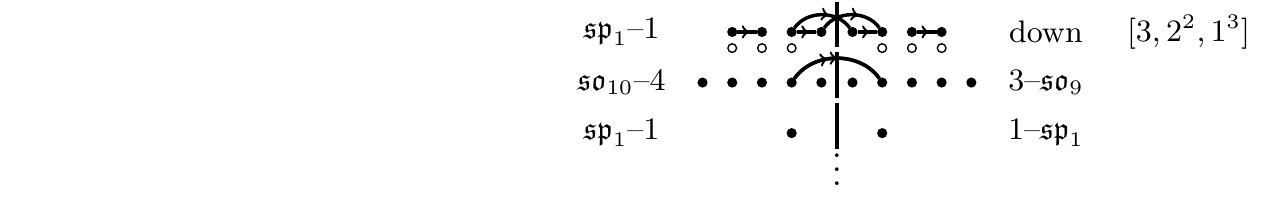}
  \par
  \includegraphics{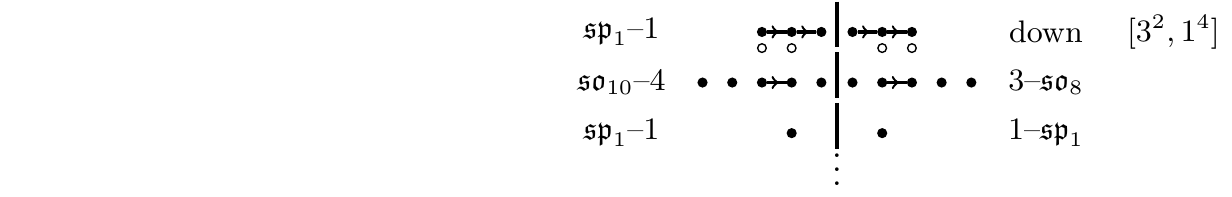}
  \par
  \begin{center}
    \hspace{-2cm} \includegraphics{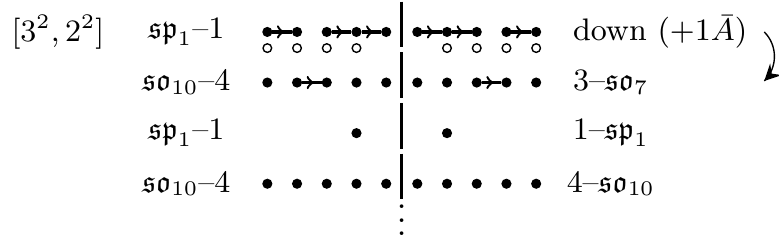} \hspace{1cm} \includegraphics{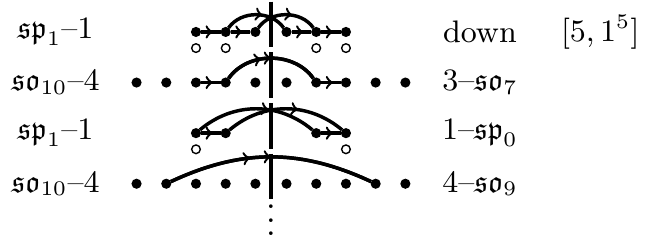} \hspace*{-2cm}
  \end{center}
  \par
  \hspace{-4.25cm} \includegraphics{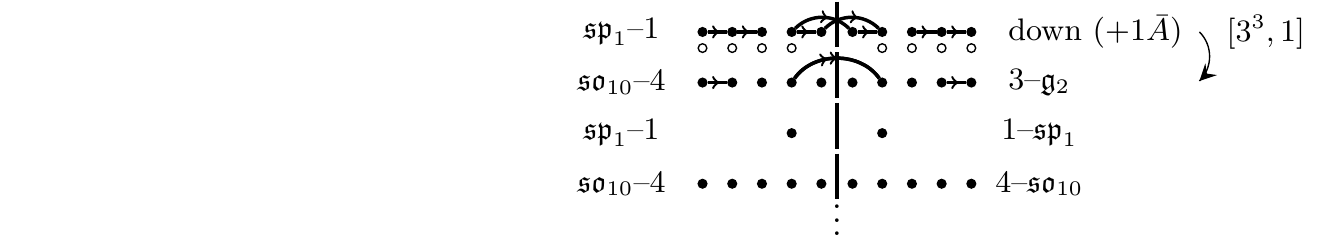}
  \par
  \begin{center}
    \hspace{-2cm} \includegraphics{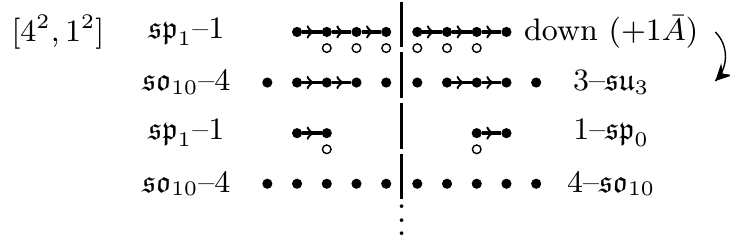} \hspace{1cm} \includegraphics{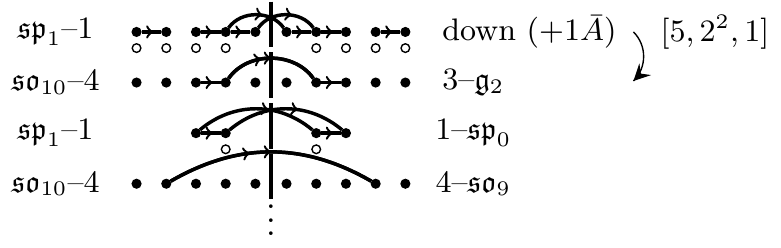} \hspace*{-2cm}
  \end{center}
 \caption{Nilpotent deformations of the $SO(10)$ quiver from the UV configuration of figure \ref{fig:SO2NUV}. See figure \ref{fig:SO8quiver} for additional details on the notation and conventions.}
\end{figure}
\begin{figure}\ContinuedFloat
  \includegraphics{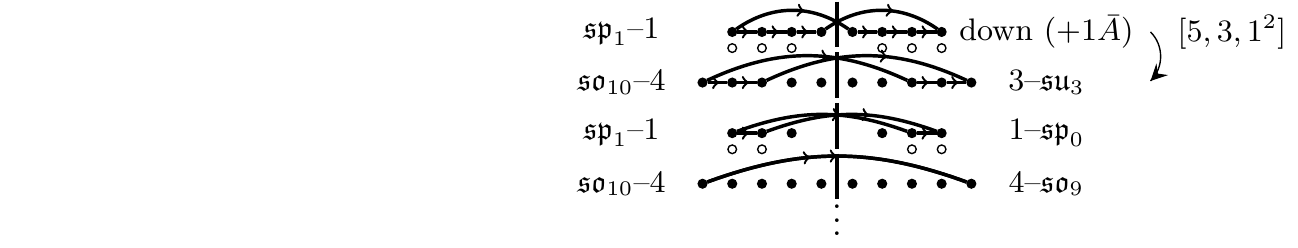}
  \par
  \begin{center}
    \hspace{-2cm} \includegraphics{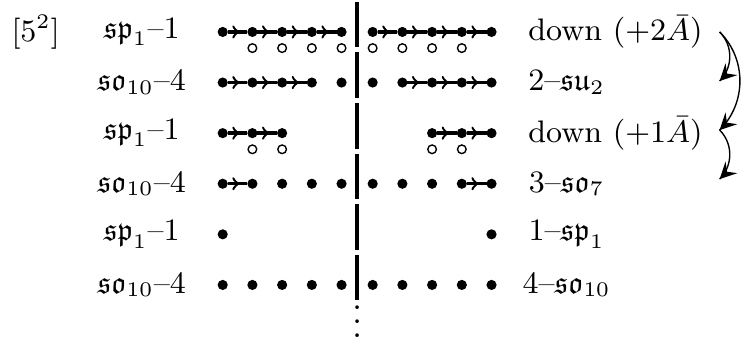} \hspace{0.5cm} \includegraphics{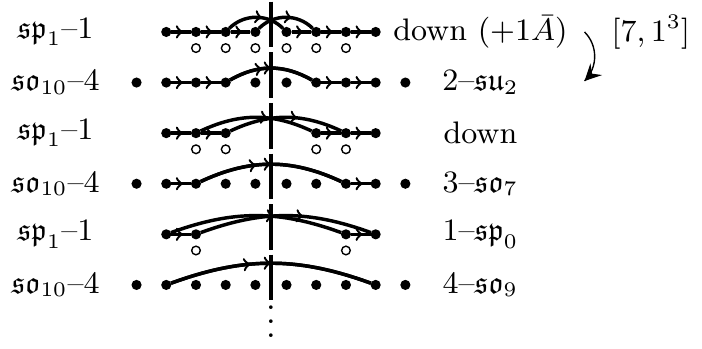} \hspace*{-2cm}
  \end{center}
  \par
  \includegraphics{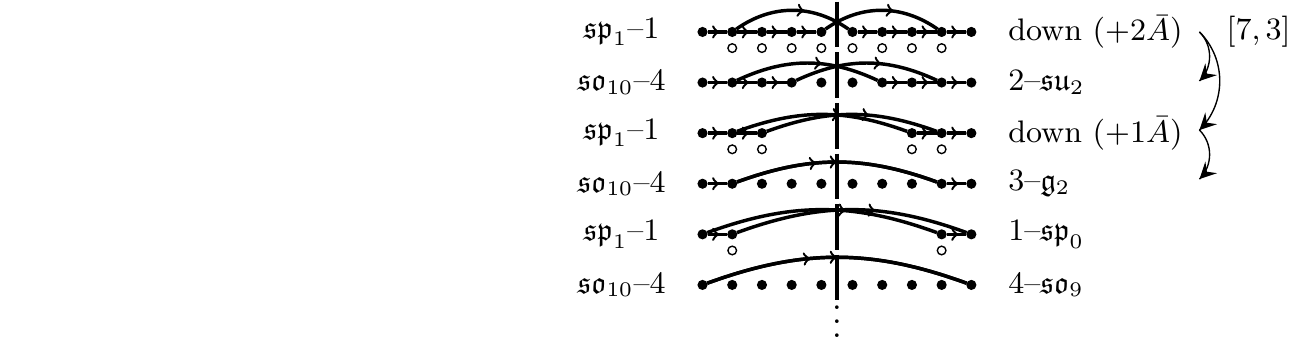}
  \par
  \includegraphics{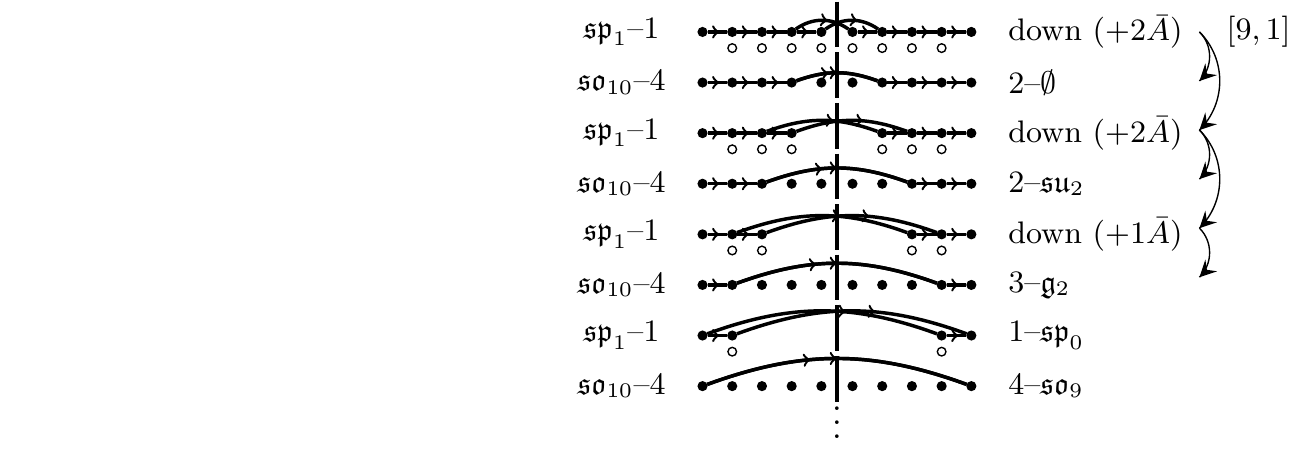}
\caption{(continued) Nilpotent deformations of the $SO(10)$ quiver from the UV configuration of figure \ref{fig:SO2NUV}. See figure \ref{fig:SO8quiver} for additional details on the notation and conventions.}
  \label{fig:SO10quiver}
\end{figure}

\newcommand{\SOninerightsep}{0.2cm}
\newcommand{\SOninevertsep}{0.6cm}
\newcommand{\SOninescale}{0.8}
\newcommand{\SOninevspace}{0.3cm}
\newcommand{\SOninemirup}{0.3cm}
\newcommand{\SOninemirdown}{0.15cm}
\newcommand{\nodesSOninea}{
\node[circle,draw=black, fill=black, inner sep=0pt,minimum size=0pt]   (amir) {};
\node[circle,draw=black, fill=black, inner sep=0pt,minimum size=0pt]   (amirup)   [above=\SOninemirup of amir] {};
\node[circle,draw=black, fill=black, inner sep=0pt,minimum size=0pt]   (amirdown) [below=\SOninemirdown of amir] {};
\draw[-, line width=1pt, shorten <=0, shorten >=0] (amirup) -- (amirdown);
}
\newcommand{\nodesSOnineb}{
\node[paramsbranes]     (bmir) [below=\SOninevertsep of amir] {};
\node[circle,draw=black, fill=black, inner sep=0pt,minimum size=0pt]   (bmirup) [above=\SOninemirup of bmir] {};
\node[circle,draw=black, fill=black, inner sep=0pt,minimum size=0pt] (bmirdown) [below=\SOninemirdown of bmir] {};
\node[paramsbranes]   (b5) [left=\SOninerightsep/2 of bmir] {};
\node[paramsbranes]   (b4) [left=\SOninerightsep of b5] {};
\node[paramsbranes]   (b3) [left=\SOninerightsep of b4] {};
\node[paramsbranes]   (b2) [left=\SOninerightsep of b3] {};
\node[paramsbranes]   (b1) [left=\SOninerightsep of b2] {};
\node[paramsbranes]   (bt5)   [right=\SOninerightsep/2 of bmir] {};
\node[paramsbranes]   (bt4)   [right=\SOninerightsep of bt5] {};
\node[paramsbranes]   (bt3) [right=\SOninerightsep of bt4] {};
\node[paramsbranes]   (bt2) [right=\SOninerightsep of bt3] {};
\node[paramsbranes]   (bt1) [right=\SOninerightsep of bt2] {};
\draw[-, line width=1pt, shorten <=0, shorten >=0] (bmirup) -- (bmirdown);
}
\newcommand{\nodesSOninec}{
\node[circle,draw=black, fill=black, inner sep=0pt,minimum size=0pt]   (cmir) [below=\SOninevertsep of bmir] {};
\node[circle,draw=black, fill=black, inner sep=0pt,minimum size=0pt]   (cmirup)   [above=\SOninemirup of cmir] {};
\node[circle,draw=black, fill=black, inner sep=0pt,minimum size=0pt]   (cmirdown) [below=\SOninemirdown of cmir] {};
\draw[-, line width=1pt, shorten <=0, shorten >=0] (cmirup) -- (cmirdown);
}
\newcommand{\nodesSOnined}{
\node[circle,draw=black, fill=black, inner sep=0pt,minimum size=0pt]     (dmir) [below=\SOninevertsep of cmir] {};
\node[circle,draw=black, fill=black, inner sep=0pt,minimum size=0pt]   (dmirup) [above=\SOninemirup of dmir] {};
\node[circle,draw=black, fill=black, inner sep=0pt,minimum size=0pt] (dmirdown) [below=\SOninemirdown of dmir] {};
\node[paramsbranes]  (d6)  [left=\SOninerightsep/2 of dmir] {};
\node[paramsbranes]  (d5)  [left=\SOninerightsep of d6] {};
\node[paramsbranes]  (d4)  [left=\SOninerightsep of d5] {};
\node[paramsbranes]  (d3)  [left=\SOninerightsep of   d4] {};
\node[paramsbranes]  (d2)  [left=\SOninerightsep of   d3] {};
\node[paramsbranes]  (d1)  [left=\SOninerightsep of   d2] {};
\node[paramsbranes]  (dt6)[right=\SOninerightsep/2 of dmir] {};
\node[paramsbranes]  (dt5)[right=\SOninerightsep of dt6] {};
\node[paramsbranes]  (dt4)[right=\SOninerightsep of dt5] {};
\node[paramsbranes]  (dt3)  [right=\SOninerightsep of dt4] {};
\node[paramsbranes]  (dt2)  [right=\SOninerightsep of dt3] {};
\node[paramsbranes]  (dt1)  [right=\SOninerightsep of dt2] {};
\draw[-, line width=1pt, shorten <=0, shorten >=0] (dmirup) -- (dmirdown);
}
\newcommand{\nodesSOninee}{
\node[circle,draw=black, fill=black, inner sep=0pt,minimum size=0pt]   (emir) [below=\SOninevertsep of dmir] {};
\node[circle,draw=black, fill=black, inner sep=0pt,minimum size=0pt]   (emirup)   [above=\SOninemirup of emir] {};
\node[circle,draw=black, fill=black, inner sep=0pt,minimum size=0pt]   (emirdown) [below=\SOninemirdown of emir] {};
\draw[-, line width=1pt, shorten <=0, shorten >=0] (emirup) -- (emirdown);
}
\newcommand{\nodesSOninef}{
\node[paramsbranes]     (fmir) [below=\SOninevertsep of emir] {};
\node[circle,draw=black, fill=black, inner sep=0pt,minimum size=0pt]   (fmirup) [above=\SOninemirup of fmir] {};
\node[circle,draw=black, fill=black, inner sep=0pt,minimum size=0pt] (fmirdown) [below=\SOninemirdown of fmir] {};
\node[paramsbranes]  (f6)  [left=\SOninerightsep/2 of fmir] {};
\node[paramsbranes]  (f5)  [left=\SOninerightsep of f6] {};
\node[paramsbranes]  (f4)  [left=\SOninerightsep of f5] {};
\node[paramsbranes]  (f3)  [left=\SOninerightsep of   f4] {};
\node[paramsbranes]  (f2)  [left=\SOninerightsep of   f3] {};
\node[paramsbranes]  (f1)  [left=\SOninerightsep of   f2] {};
\node[paramsbranes]  (ft6)[right=\SOninerightsep/2 of fmir] {};
\node[paramsbranes]  (ft5)[right=\SOninerightsep of ft6] {};
\node[paramsbranes]  (ft4)[right=\SOninerightsep of ft5] {};
\node[paramsbranes]  (ft3)  [right=\SOninerightsep of ft4] {};
\node[paramsbranes]  (ft2)  [right=\SOninerightsep of ft3] {};
\node[paramsbranes]  (ft1)  [right=\SOninerightsep of ft2] {};
\draw[-, line width=1pt, shorten <=0, shorten >=0] (fmirup) -- (fmirdown);
}
\newcommand{\nodesSOnineg}{
\node[circle,draw=black, fill=black, inner sep=0pt,minimum size=0pt]   (gmir) [below=\SOninevertsep of fmir] {};
\node[circle,draw=black, fill=black, inner sep=0pt,minimum size=0pt]   (gmirup)   [above=\SOninemirup of gmir] {};
\node[circle,draw=black, fill=black, inner sep=0pt,minimum size=0pt]   (gmirdown) [below=\SOninemirdown of gmir] {};
\draw[-, line width=1pt, shorten <=0, shorten >=0] (gmirup) -- (gmirdown);
}
\newcommand{\nodesSOnineh}{
\node[paramsbranes]     (hmir) [below=\SOninevertsep of gmir] {};
\node[circle,draw=black, fill=black, inner sep=0pt,minimum size=0pt]   (hmirup) [above=\SOninemirup of hmir] {};
\node[circle,draw=black, fill=black, inner sep=0pt,minimum size=0pt] (hmirdown) [below=\SOninemirdown of hmir] {};
\node[paramsbranes]  (h6)  [left=\SOninerightsep/2 of hmir] {};
\node[paramsbranes]  (h5)  [left=\SOninerightsep of h6] {};
\node[paramsbranes]  (h4)  [left=\SOninerightsep of h5] {};
\node[paramsbranes]  (h3)  [left=\SOninerightsep of   h4] {};
\node[paramsbranes]  (h2)  [left=\SOninerightsep of   h3] {};
\node[paramsbranes]  (h1)  [left=\SOninerightsep of   h2] {};
\node[paramsbranes]  (ht6)[right=\SOninerightsep/2 of hmir] {};
\node[paramsbranes]  (ht5)[right=\SOninerightsep of ht6] {};
\node[paramsbranes]  (ht4)[right=\SOninerightsep of ht5] {};
\node[paramsbranes]  (ht3)  [right=\SOninerightsep of ht4] {};
\node[paramsbranes]  (ht2)  [right=\SOninerightsep of ht3] {};
\node[paramsbranes]  (ht1)  [right=\SOninerightsep of ht2] {};
\draw[-, line width=1pt, shorten <=0, shorten >=0] (hmirup) -- (hmirdown);
}
\begin{figure}[H]
  \vspace{0.5cm}
  \includegraphics{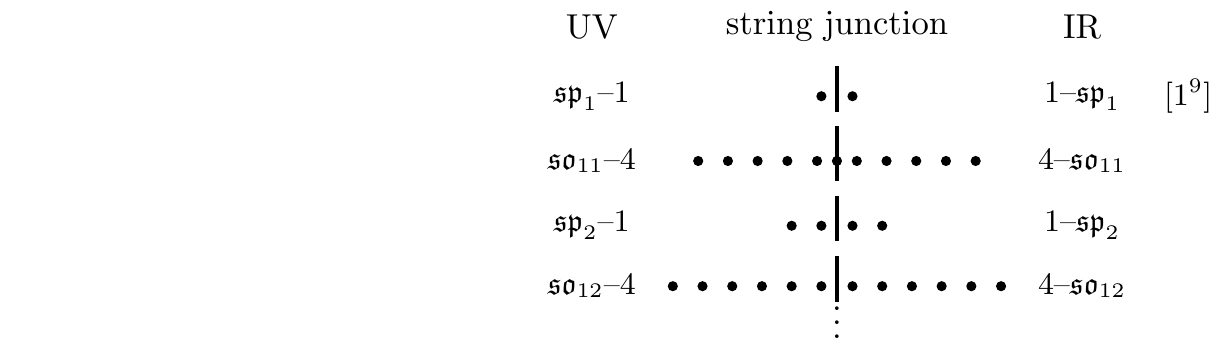}
  \par
  \includegraphics{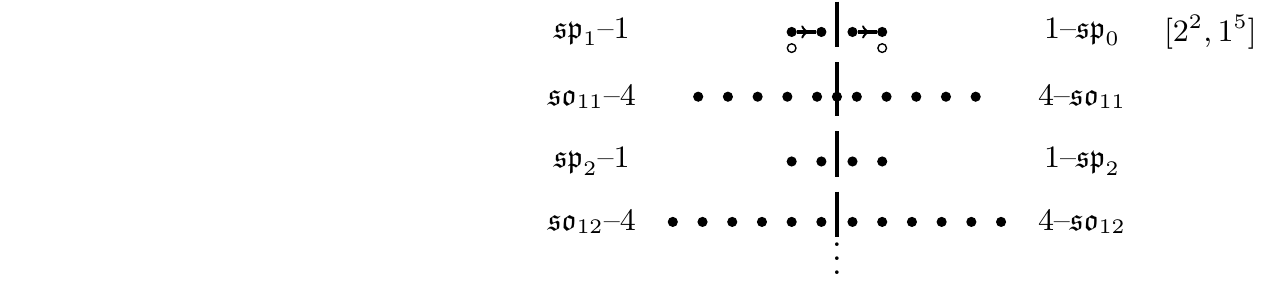}
  \par
  \begin{center}
    \hspace{-2cm} \includegraphics{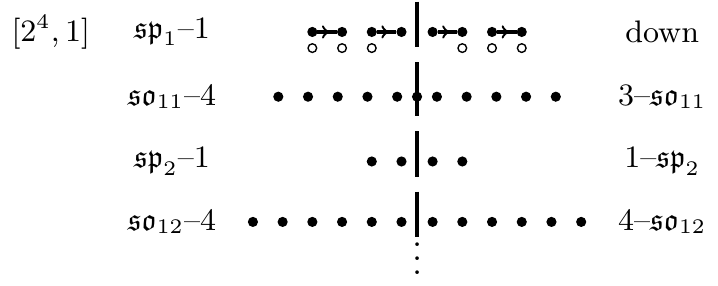} \hspace{1.5cm} \includegraphics{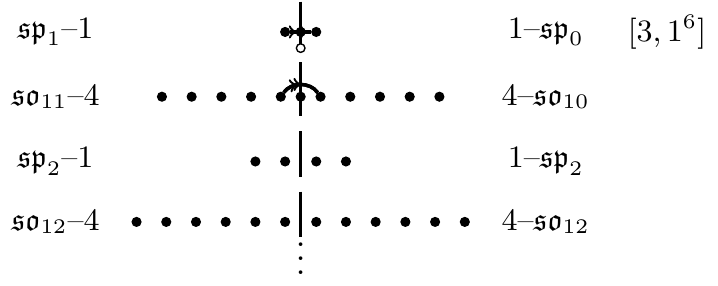} \hspace*{-2cm}
  \end{center}
  \par
  \includegraphics{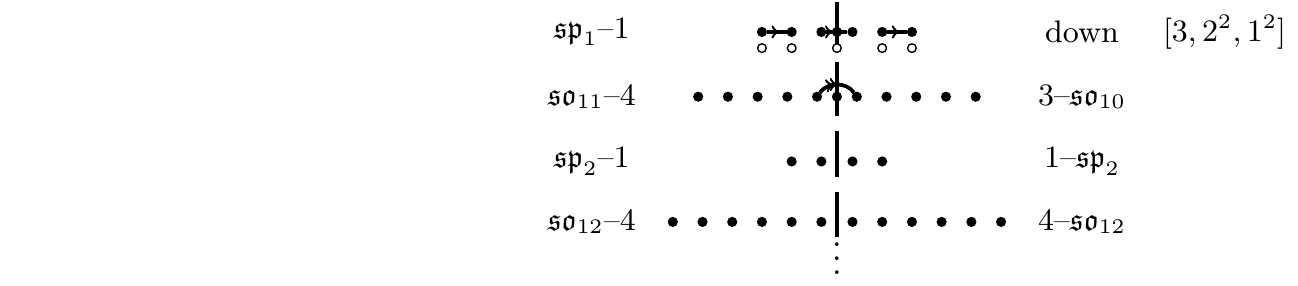}
  \par
  \includegraphics{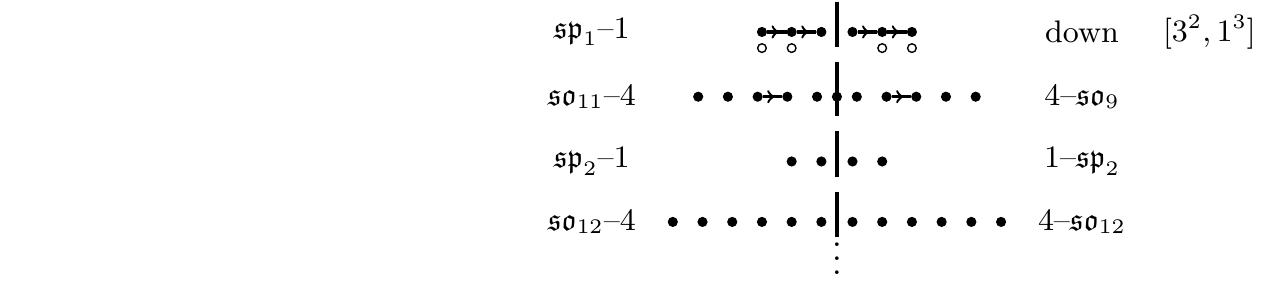}
  \par
  \begin{center}
    \hspace{-2cm} \includegraphics{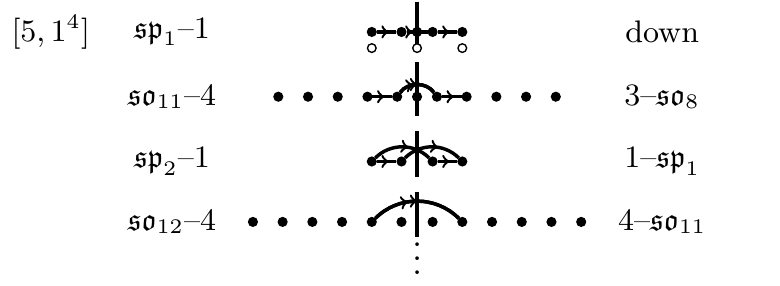} \hspace{0.5cm} \includegraphics{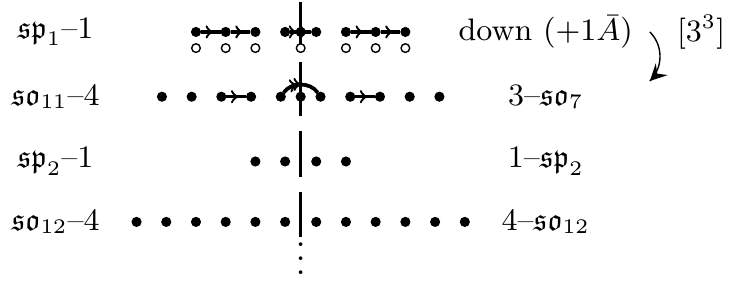} \hspace*{-2cm}
  \end{center}
  \caption{Nilpotent deformations of the $SO(9)$ quiver from the UV configuration of figure \ref{fig:SO2Nm1UV}. See figure \ref{fig:SO8quiver} for additional details on the notation and conventions.}
\end{figure}
\begin{figure}\ContinuedFloat
  \begin{center}
    \hspace{-2cm} \includegraphics{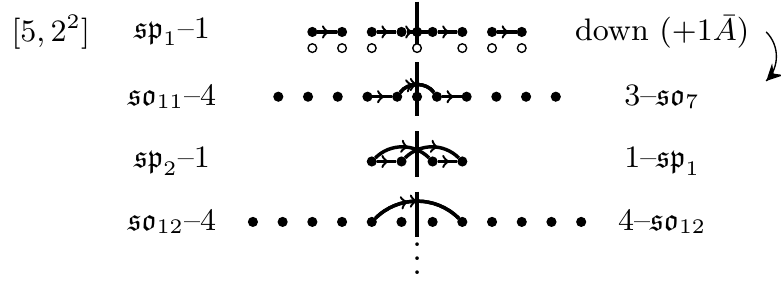} \hspace{0.5cm} \includegraphics{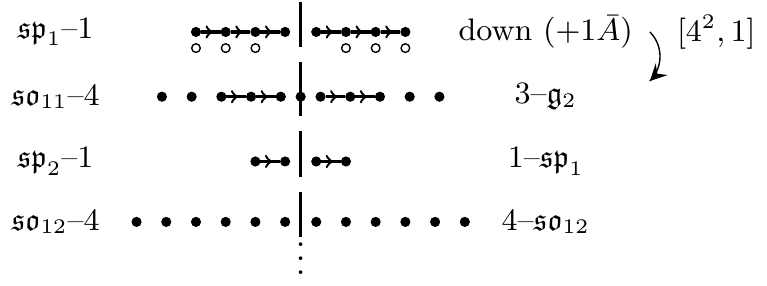} \hspace*{-2cm}
  \end{center}
  \par
  \includegraphics{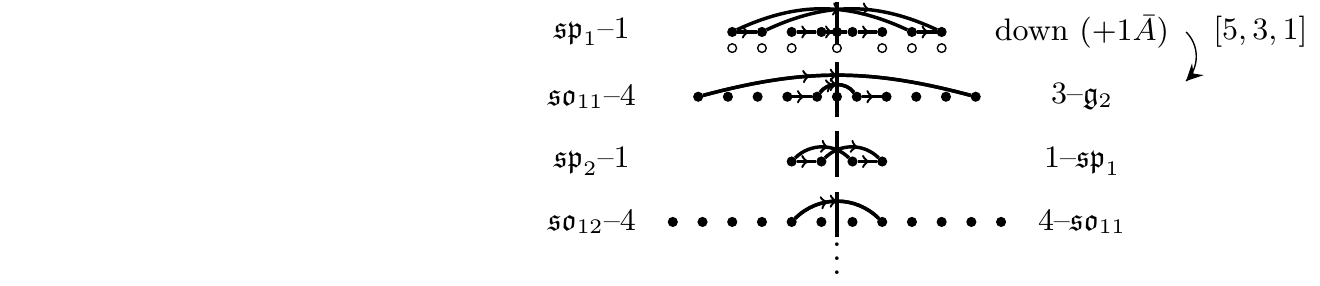}
  \par
  \includegraphics{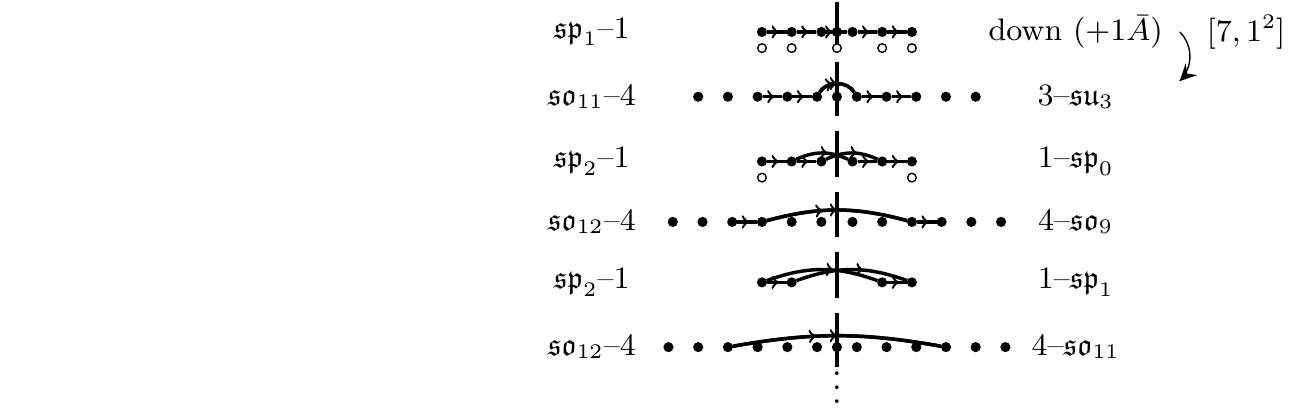}
  \par
  \includegraphics{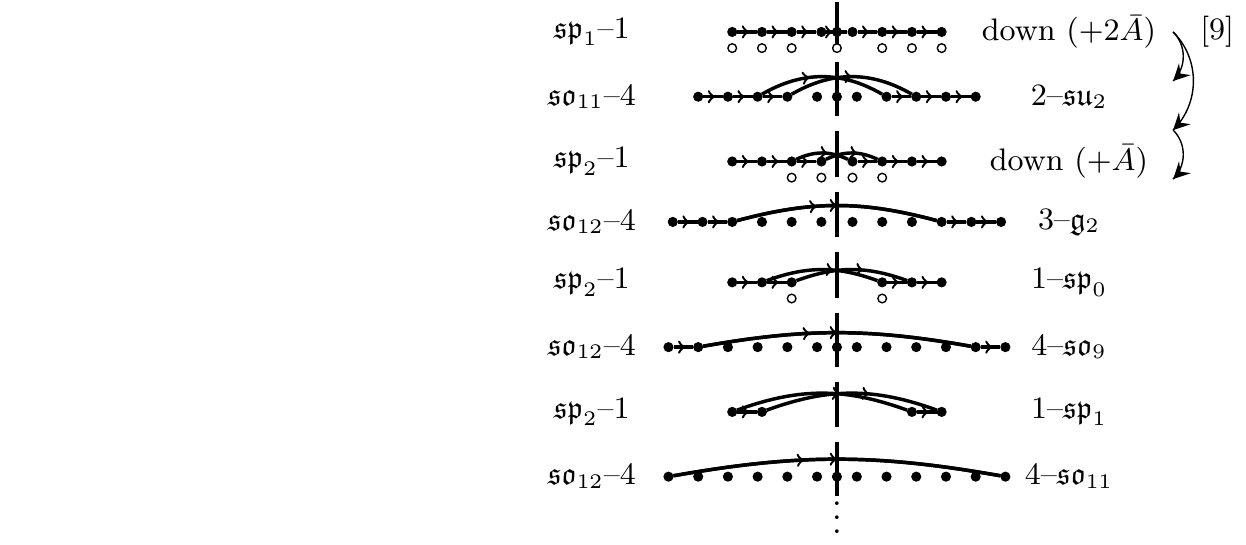}
\caption{(continued) Nilpotent deformations of the $SO(9)$ quiver from the UV configuration of figure \ref{fig:SO2Nm1UV}. See figure \ref{fig:SO8quiver} for additional details on the notation and conventions.}
  \label{fig:SO9quiver}
\end{figure}

\newcommand{\Spthreerightsep}{0.2cm}
\newcommand{\Spthreevertsep}{0.7cm}
\newcommand{\Spthreescale}{0.8}
\newcommand{\Spthreevspace}{0.3cm}
\newcommand{\Spthreemirup}{0.3cm}
\newcommand{\Spthreemirdown}{0.15cm}
\newcommand{\nodesSpthreea}{
\node[circle,draw=black, fill=black, inner sep=0pt,minimum size=0pt]   (amir) {};
\node[circle,draw=black, fill=black, inner sep=0pt,minimum size=0pt]   (amirup)   [above=\Spthreemirup of amir] {};
\node[circle,draw=black, fill=black, inner sep=0pt,minimum size=0pt]   (amirdown) [below=\Spthreemirdown of amir] {};
\draw[-, line width=1pt, shorten <=0, shorten >=0] (amirup) -- (amirdown);
\node[paramsbranes]  (a7)   [left=\Spthreerightsep/2 of  amir] {};
\node[paramsbranes]  (at7)  [right=\Spthreerightsep/2 of amir] {};
\node[paramsbranes]  (a6)   [left=\SOtenrightsep of      a7] {};
\node[paramsbranes]  (at6)  [right=\SOtenrightsep of     at7] {};
\node[paramsbranes]  (a5)   [left=\SOtenrightsep of      a6] {};
\node[paramsbranes]  (at5)  [right=\SOtenrightsep of     at6] {};
\node[paramsbranes]  (a4)   [left=\SOtenrightsep of      a5] {};
\node[paramsbranes]  (at4)  [right=\SOtenrightsep of     at5] {};
\node[paramsbranes]  (a3)   [left=\SOtenrightsep of      a4] {};
\node[paramsbranes]  (at3)  [right=\SOtenrightsep of     at4] {};
\node[paramsbranes]  (a2)   [left=\SOtenrightsep of      a3] {};
\node[paramsbranes]  (at2)  [right=\SOtenrightsep of     at3] {};
\node[paramsbranes]  (a1)   [left=\SOtenrightsep of      a2] {};
\node[paramsbranes]  (at1)  [right=\SOtenrightsep of     at2] {};
}
\newcommand{\nodesSpthreeb}{
\node[circle,draw=black, fill=black, inner sep=0pt,minimum size=0pt]     (bmir) [below=\Spthreevertsep of amir] {};
\node[circle,draw=black, fill=black, inner sep=0pt,minimum size=0pt]   (bmirup) [above=\Spthreemirup of bmir] {};
\node[circle,draw=black, fill=black, inner sep=0pt,minimum size=0pt] (bmirdown) [below=\Spthreemirdown of bmir] {};
\draw[-, line width=1pt, shorten <=0, shorten >=0] (bmirup) -- (bmirdown);
\node[paramsbranes]  (b3)   [left=\Spthreerightsep/2 of  bmir] {};
\node[paramsbranes]  (bt3)  [right=\Spthreerightsep/2 of bmir] {};
\node[paramsbranes]  (b2)   [left=\SOtenrightsep of      b3] {};
\node[paramsbranes]  (bt2)  [right=\SOtenrightsep of     bt3] {};
\node[paramsbranes]  (b1)   [left=\SOtenrightsep of      b2] {};
\node[paramsbranes]  (bt1)  [right=\SOtenrightsep of     bt2] {};
}
\newcommand{\nodesSpthreec}{
\node[circle,draw=black, fill=black, inner sep=0pt,minimum size=0pt]     (cmir) [below=\Spthreevertsep of bmir] {};
\node[circle,draw=black, fill=black, inner sep=0pt,minimum size=0pt]   (cmirup) [above=\Spthreemirup of cmir] {};
\node[circle,draw=black, fill=black, inner sep=0pt,minimum size=0pt] (cmirdown) [below=\Spthreemirdown of cmir] {};
\draw[-, line width=1pt, shorten <=0, shorten >=0] (cmirup) -- (cmirdown);
\node[paramsbranes]  (c7)   [left=\Spthreerightsep/2 of  cmir] {};
\node[paramsbranes]  (ct7)  [right=\Spthreerightsep/2 of cmir] {};
\node[paramsbranes]  (c6)   [left=\SOtenrightsep of      c7] {};
\node[paramsbranes]  (ct6)  [right=\SOtenrightsep of     ct7] {};
\node[paramsbranes]  (c5)   [left=\SOtenrightsep of      c6] {};
\node[paramsbranes]  (ct5)  [right=\SOtenrightsep of     ct6] {};
\node[paramsbranes]  (c4)   [left=\SOtenrightsep of      c5] {};
\node[paramsbranes]  (ct4)  [right=\SOtenrightsep of     ct5] {};
\node[paramsbranes]  (c3)   [left=\SOtenrightsep of      c4] {};
\node[paramsbranes]  (ct3)  [right=\SOtenrightsep of     ct4] {};
\node[paramsbranes]  (c2)   [left=\SOtenrightsep of      c3] {};
\node[paramsbranes]  (ct2)  [right=\SOtenrightsep of     ct3] {};
\node[paramsbranes]  (c1)   [left=\SOtenrightsep of      c2] {};
\node[paramsbranes]  (ct1)  [right=\SOtenrightsep of     ct2] {};
}
\newcommand{\nodesSpthreed}{
\node[circle,draw=black, fill=black, inner sep=0pt,minimum size=0pt]     (dmir) [below=\Spthreevertsep of cmir] {};
\node[circle,draw=black, fill=black, inner sep=0pt,minimum size=0pt]   (dmirup) [above=\Spthreemirup of dmir] {};
\node[circle,draw=black, fill=black, inner sep=0pt,minimum size=0pt] (dmirdown) [below=\Spthreemirdown of dmir] {};
\draw[-, line width=1pt, shorten <=0, shorten >=0] (dmirup) -- (dmirdown);
\node[paramsbranes]  (d3)   [left=\Spthreerightsep/2 of  dmir] {};
\node[paramsbranes]  (dt3)  [right=\Spthreerightsep/2 of dmir] {};
\node[paramsbranes]  (d2)   [left=\SOtenrightsep of      d3] {};
\node[paramsbranes]  (dt2)  [right=\SOtenrightsep of     dt3] {};
\node[paramsbranes]  (d1)   [left=\SOtenrightsep of      d2] {};
\node[paramsbranes]  (dt1)  [right=\SOtenrightsep of     dt2] {};
}
\newcommand{\nodesSpthreee}{
\node[circle,draw=black, fill=black, inner sep=0pt,minimum size=0pt]     (emir) [below=\Spthreevertsep of dmir] {};
\node[circle,draw=black, fill=black, inner sep=0pt,minimum size=0pt]   (emirup) [above=\Spthreemirup of emir] {};
\node[circle,draw=black, fill=black, inner sep=0pt,minimum size=0pt] (emirdown) [below=\Spthreemirdown of emir] {};
\draw[-, line width=1pt, shorten <=0, shorten >=0] (emirup) -- (emirdown);
\node[paramsbranes]  (e7)   [left=\Spthreerightsep/2 of  emir] {};
\node[paramsbranes]  (et7)  [right=\Spthreerightsep/2 of emir] {};
\node[paramsbranes]  (e6)   [left=\SOtenrightsep of      e7] {};
\node[paramsbranes]  (et6)  [right=\SOtenrightsep of     et7] {};
\node[paramsbranes]  (e5)   [left=\SOtenrightsep of      e6] {};
\node[paramsbranes]  (et5)  [right=\SOtenrightsep of     et6] {};
\node[paramsbranes]  (e4)   [left=\SOtenrightsep of      e5] {};
\node[paramsbranes]  (et4)  [right=\SOtenrightsep of     et5] {};
\node[paramsbranes]  (e3)   [left=\SOtenrightsep of      e4] {};
\node[paramsbranes]  (et3)  [right=\SOtenrightsep of     et4] {};
\node[paramsbranes]  (e2)   [left=\SOtenrightsep of      e3] {};
\node[paramsbranes]  (et2)  [right=\SOtenrightsep of     et3] {};
\node[paramsbranes]  (e1)   [left=\SOtenrightsep of      e2] {};
\node[paramsbranes]  (et1)  [right=\SOtenrightsep of     et2] {};
}

\subsection{Comments on Quiver-like Theories with Exceptional Algebras}

It is natural to ask whether the propagation
rules given for quivers with classical algebras also
extend to theories with exceptional algebras. In principle, we expect this to follow from our description of
the nilpotent cone in terms of multi-pronged string junctions. Indeed, we have already explained that at least for
semi-simple deformations, there is no material distinction between the quivers of classical and exceptional type.

That being said, we expect our analysis of nilpotent deformations to be more subtle in this case.
Part of the issue is that even in the case of the D-type algebras, to really describe the physics of brane recombination, we had to go onto the
full tensor branch so that both $SO$ and $Sp$ gauge algebras could be manipulated (via brane recombination). From this perspective, we need to
understand brane recombination in 6D conformal matter for the following configurations
of $(E_{N},E_{N})$ conformal matter:
\begin{align}
& \lbrack E_{6}],1,3,1,[E_{6}]\\
& \lbrack E_{7}],1,2,3,2,1,[E_{7}]\\
& \lbrack E_{8}],1,2,2,3,1,5,1,3,2,2,1,[E_{8}].
\end{align}
Said differently, a breaking pattern which connects two E-type algebras will necessarily
involve a number of tensor multiplets. For the most part, one can work out a set of ``phenomenological''
rules which cover nearly all cases involving quivers with $E_6$ gauge algebras, but its generalization to $E_7$
and $E_8$ appears to involve some new ingredients beyond the ones introduced already in this paper. For all these
reasons, we defer a full analysis of these cases to future work.


\section{Short Quivers \label{sec:GETSHORTY}}

In the previous section, we demonstrated that the physics of brane
recombination accurately recovers the expected Higgs branch flows for
6D\ SCFTs. It is reassuring to see that these methods reproduce -- but also
extend -- the structure of Higgs branch flows obtained through other methods.
The main picture we have elaborated on is the
propagation of T-brane data into the interior of a quiver-like gauge theory.

The main assumption made in these earlier sections is the presence of a
sufficient number of gauge group factors in the interior of the quiver so that this
propagation is independent of other T-brane data associated with other
flavor symmetry factors. In this section
we relax this assumption by considering \textquotedblleft short quivers\textquotedblright\ in which
the number of gauge group factors is too low
to prevent such an overlap. There has been very little analysis in the 6D SCFT
literature on this class of RG flows.

Using the brane recombination picture developed in the previous section, we
show how to determine the corresponding 6D\ SCFTs generated by such
deformations. We mainly focus on quivers with classical algebras, since this
is the case we presently understand most clearly. Even here, there is a rather
rich structure of possible RG\ flows.

There are two crucial combinatorial aspects to our analysis. First of all,
we use open strings to collect recombined branes into ``blobs.'' Additionally, to determine the scope of possible
deformations, we introduce brane / anti-brane pairs, as prescribed by the rules of section \ref{sec:RECOMBO}. To track the effects of
having a short quiver, we gradually reduce the number of gauge group factors until the brane moves on either side of the quiver become correlated. As a result, we sometimes reach configurations in which the anti-branes cannot be eliminated. We take this to mean
that we have not actually satisfied the D-term constraints in the quiver-like gauge theory.

The procedure we outline also has some overlap with the formal proposal of reference \cite{Mekareeya:2016yal} (see also \cite{Apruzzi:2017iqe}), which analyzed Higgs branch flows by analytically continuing the rank of gauge groups to negative values. Using our description in terms of anti-branes, we show that in many cases, the theory we obtain has an anomaly polynomial which matches to these proposed theories. We also find,
however, that in short quivers (which were not analyzed in \cite{Mekareeya:2016yal})
this analytic continuation method sometimes does not produce a sensible IR fixed point.
This illustrates the utility of the methods developed in this paper.

In the case of sufficiently long quiver-like theories, there is a natural partial ordering set by the nilpotent
orbits in the two flavor symmetry algebras. In the case of shorter quivers, the partial ordering becomes more complicated because
there is (by definition) some overlap in the symmetry breaking patterns on the two sides of a quiver. In many cases, different pairs of nilpotent orbit wind up generating the same IR fixed point simply because most or all of the gauge symmetry in the quiver has already been Higgsed.
We show in explicit examples how to obtain the corresponding partially ordered set of theories labeled by pairs of overlapping nilpotent orbits. We refer to these as ``double Hasse diagrams'' since they merge two Hasse diagrams of a given flavor symmetry algebra.

To illustrate the main points of this analysis, we primarily focus on illustrative examples in which the number of gauge group factors in the interior of a quiver is sufficiently small and / or in which the size of the nilpotent orbits is sufficiently large so that there is non-trivial overlap between the breaking patterns on the left and right. For this reason, we often work with low rank gauge algebras such as $\mathfrak{su}(4)$ and $\mathfrak{so}(8)$ and a small number of interior gauge group factors, though we stress that our analysis works in the same way for all short quivers.

The rest of this section is organized as follows. First, we show how to obtain short quivers as a limiting case in which we gradually reduce the number of gauge group factors in a long quiver. We then turn to a study of nilpotent hierarchies in these models, and we conclude this section with a brief discussion of the residual global symmetries after Higgsing in a short quiver.

\subsection{From Long to Short Quivers \label{subsec:introShortQuiver}}

In this subsection, we determine how T-brane data propagating from the two sides of a quiver becomes intertwined as we decrease the number of gauge groups / tensor multiplets. It is helpful to split up this analysis according to the choice of gauge group appearing, so we present examples for each different choices of gauge algebras.

\subsubsection{$SU(N)$ Short Quivers}

We begin with quiver-like theories with $\mathfrak{su}$ gauge algebras.
Applying the Hanany-Witten rules from section \ref{subsec:HananyWitten} to
the type IIA realization of the $SU(N)$ theories, we have that:
\begin{align}
  &k_{\textrm{NS5}} \geq \textrm{Max}\{\mu_{L}^{1}, \mu_{R}^{1}\} + 1
\end{align}
for left and right partitions $\mu_{L} = [\mu^{i}]$, $\mu_{R}=[\mu^{j}]$
respectively. Here, $k_{\textrm{NS5}}$ denotes the number of NS5-branes in the
corresponding type IIA picture. When this condition is violated, it is impossible to balance the D8-branes. Note that $k_{\textrm{NS5}}$ is also equal to one plus the number of $-2$ curves $N_{-2} = N_{T}$ the number of tensor multiplets in the UV quiver, so we may equivalently write this condition as
\begin{equation}
  \mathrm{Max}\{\mu_L^1,\mu_R^1\} \leq N_{-2},
  \label{eq:HWconstraintSU}
\end{equation}
where $N_{-2}$ denotes the number of $-2$ curves in the UV quiver.
This is equivalent to saying that, when only one nilpotent deformation (either $\mu_L$ or $\mu_R$) is implemented over the UV quiver (either the left or right partition), there has to be at least one $-2$ curve whose fiber remains untouched by the deformation.

Assuming this restriction is obeyed, we can straightforwardly produce any short $SU(N)$ quiver given a UV quiver and a pair of nilpotent orbits. Before giving the general formula, however, let us look at a concrete example: consider a UV theory of $SU(5)$ over five $-2$ curves, and apply the
nilpotent deformations of $[3, 2]$ -- $[2^{2}, 1]$, where no interaction
between the orbits take place. This theory can be written as:
\begin{equation}
[3, 2]:\,\, \overset{\mathfrak{su}(2)}{2} \,\,
\underset{[N_f = 1]}{\overset{\mathfrak{su}(4)}{2}} \,\,
\underset{[N_f = 1]}{\overset{\mathfrak{su}(5)}{2}}
\,\,\underset{[SU(2)]}{\overset{\mathfrak{su}(5)}{2}} \,\,
\underset{[N_f = 1]}{\overset{\mathfrak{su}(3)}{2}}\,\,: [2^{2}, 1]
\end{equation}
where the notation $[N_f = 1]$ refers to having one additional flavor on each corresponding gauge algebra.

We now decrease the length of the quiver and gradually turn it into a short quiver. We decrease the number of $-2$ curves one at a time, and when the nilpotent deformation from the left and right overlaps, we simply add the rank reduction
effect together linearly. After each step we get:
\begin{equation}
[3, 2]:\,\, \overset{\mathfrak{su}(2)}{2} \,\,
\underset{[N_f = 1]}{\overset{\mathfrak{su}(4)}{2}} \,\, \underset{[
SU(3)]}{\overset{\mathfrak{su}(5)}{2}} \,\,
\underset{[N_f = 1]}{\overset{\mathfrak{su}(3)}{2}}\,\,: [2^{2}, 1]
\end{equation}
\begin{equation}
[3, 2]:\,\, \overset{\mathfrak{su}(2)}{2} \,\,
\underset{[SU(3)]}{\overset{\mathfrak{su}(4)}{2}} \,\,
\underset{[SU(2)]}{\overset{\mathfrak{su}(3)}{2}}\,\,: [2^{2}, 1]
\end{equation}
At this stage we are unable to decrease the length of the quiver any further without violating the constraint of (\ref{eq:HWconstraintSU}).

We note that each step changes the global symmetry, the
gauge symmetry, or both. In particular, after the second step we no longer
see a node with the UV gauge group $SU(5)$. The global symmetries also change at each step, which will be discussed further in \ref{subsec:globalSymmetryShortQuiver}.

Let us consider another example of a short quiver with $SU(N)$ gauge groups. If we take the UV quiver theory to be:
\begin{equation}
[SU(6)] \,\, \overset{\mathfrak{su}(6)}{2} \,\, \overset{\mathfrak{su}(6)}{2}
\,\, \overset{\mathfrak{su}(6)}{2}\,\, \overset{\mathfrak{su}(6)}{2}\,\,
\overset{\mathfrak{su}(6)}{2} \,\, [SU(6)]\label{eqn:SUshortexample1:1}%
\end{equation}
and apply the following pair of nilpotent deformations denoted by partitions
$\mu_{L,R}$:
\begin{equation}
\mu_{L} = [5, 1],\ \ \mu_{R} = [2^{3}]
\end{equation}
we obtain the resulting IR theory:
\begin{equation}
\underset{[N_f = 1]}{\overset{\mathfrak{su}(2)}{2}} \,\, \overset{\mathfrak{su}%
(3)}{2} \,\, \overset{\mathfrak{su}(4)}{2}\,\,
\underset{[SU(3)]}{\overset{\mathfrak{su}(5)}{2}}\,\,
\underset{[N_f = 1]}{\overset{\mathfrak{su}(3)}{2}} \,.\label{eqn:SUshortexample1:3}%
\end{equation}

\begin{figure}[ptb]
  \centering
  \includegraphics{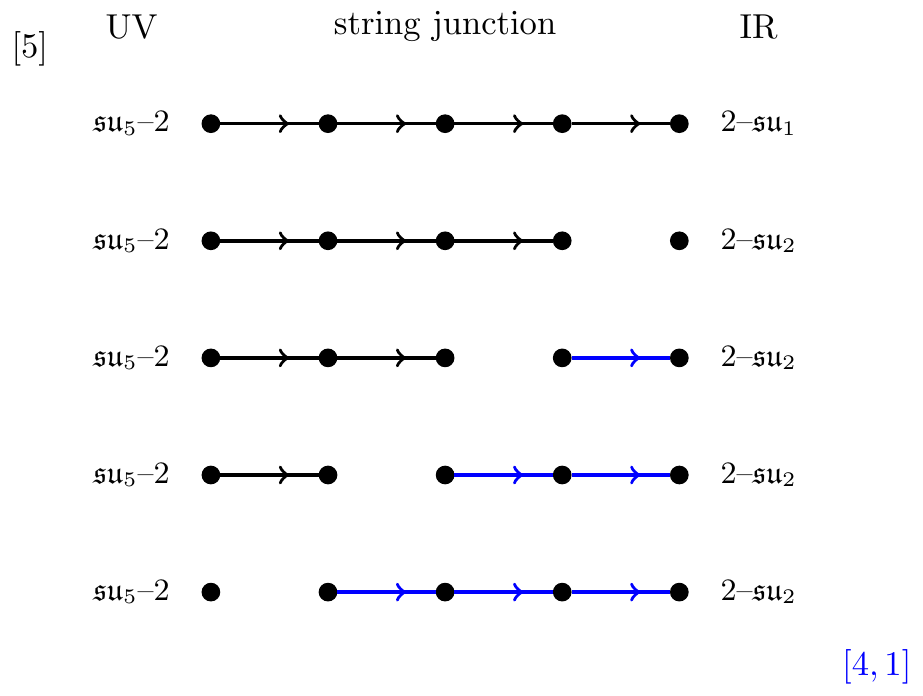}
\caption{An $SU(N)$ short quiver brane picture, the pair of nilpotent
deformation being $\mu_{L}=[5]$, $\mu_{R}=[4, 1]$ on $SU(5)$ UV theory and
four $-2$ curves. The figure is arranged so that the left deformation starts
from the top and propagates downwards (in black) while the right deformation
starts on the bottom and propagates upwards (blue).}%
\label{fig:SUshortBrane}%
\end{figure}

We illustrate another example with $SU(5)$ UV gauge group and partitions
$\mu_L=[5]$, $\mu_R = [4,1]$ in figure \ref{fig:SUshortBrane}, making the
brane recombination explicit.

In general, let us define the conjugate partitions of the left and right nilpotent
orbits to be $\rho_{L}:= \mu_L^T$ and $\rho_R := \mu_R^T$ and denote their number of elements as
$N_{L}^{\prime}$ and $N_{R}^{\prime}$, with the index counting from each of their
starting point, respectively. Then, the gauge group rank at the $m^{\text{th}}$ node is
given by %
\begin{equation}
r_{m}=N-\sum_{i=m+1}^{N_{L}^{\prime}} \rho_{i}^{L}-\sum_{j=(N_{-2})-m+1}%
^{N_{R}^{\prime}} \rho_{j}^{R},
\label{eq:SUshort}
\end{equation}
with the UV gauge group equal to $SU(N)$.

\subsubsection{Interlude: $SO$ and $Sp$ Short Quivers}

In the case of quivers with $SU$ gauge groups, the Higgsing of
the corresponding quiver-like gauge theories is controlled by vevs for
weakly coupled hypermultiplets. In this case, the physics of brane recombination primarily serves to simplify the combinatorics associated with
correlated breaking patterns in the quiver. Now, an important feature of the other quiver-like theories with flavor groups $SO$ or
$Sp$ is the more general class of possible Higgs branch flows as generated by 6D conformal matter. Recall
that on the full tensor branch of such a theory, we have a gauge group consisting of alternating classical gauge groups.
These gauge groups typically have bifundamental matter (in half-hypermultiplets of $SO \times Sp$ representations), which in turn
leads to Higgs flows generated by ``classical matter,'' much as in the case of the $SU$ quivers. There are, however,
more general Higgs branch flows connected with vevs for conformal matter. Recall that these are associated with a smoothing deformation for a collapsed $-1$ curve, namely the analog of a small instanton transition as in the case of the E-string theory. The combinatorics associated with this class of Higgs branch flows is more subtle, but as we have already remarked, the brane / anti-brane description correctly computes the resulting IR fixed points in this case as well.

By definition, in the case of a short quiver, the effects of Higgsing on the two sides of the quiver become correlated. It is therefore helpful to distinguish a few specific cases of interest as the size of the nilpotent orbit / breaking pattern continues to grow. As the size of the nilpotent orbit grows, the appearance of a small instanton deformation becomes inevitable. The distinguishing feature is the extent to which small instanton transitions become necessary to realize the corresponding Higgs branch flow. When there is at least one $-1$ curve remaining in the tensor branch description of the Higgsed theory, we refer to this as a case where the nilpotent orbits are ``touching.'' The end result is that so many small instanton deformations are generated that the tensor branch of the resulting IR theory has no $-1$ curves at all. We refer to this as a ``kissing case'' since the partitions are now more closely overlapping. Increasing the size of a nilpotent orbit beyond a kissing case leads to a problematic configuration: There are no more small instanton transitions available (as the $-1$ curves have all been used up). We refer to these as ``crumpled cases.'' In terms of our brane / anti-brane analysis, this leads to configurations with $\overline{A}$ branes which cannot be canceled off. Such crumpled configurations are inconsistent, and must be discarded. Summarizing, we refer to the different sorts of overlapping nilpotent orbit configurations as:
\begin{itemize}

\item A ``touching'' configuration is one in which all gauge groups of the quiver-like theory are at least partially broken, but at least one $-1$ curve remains in the tensor branch of the Higgsed theory.

\item A ``kissing'' configuration is defined as one in which all groups of the quiver-like theory are at least partially broken, and there are no $-1$ curves remaining in the Higgsed theory.

\item A ``crumpled'' configuration is defined as one in which the orbits have become so large that there are left over $\overline{A}$ branes which cannot be canceled off, and therefore such configurations are to be discarded.

\end{itemize}

\begin{figure}[ptb]
  \centering
  \vspace{-1.5cm}
  \hspace{-1cm}
  \begin{subfigure}{0.5\textwidth}
    \includegraphics[width=0.82\textwidth]{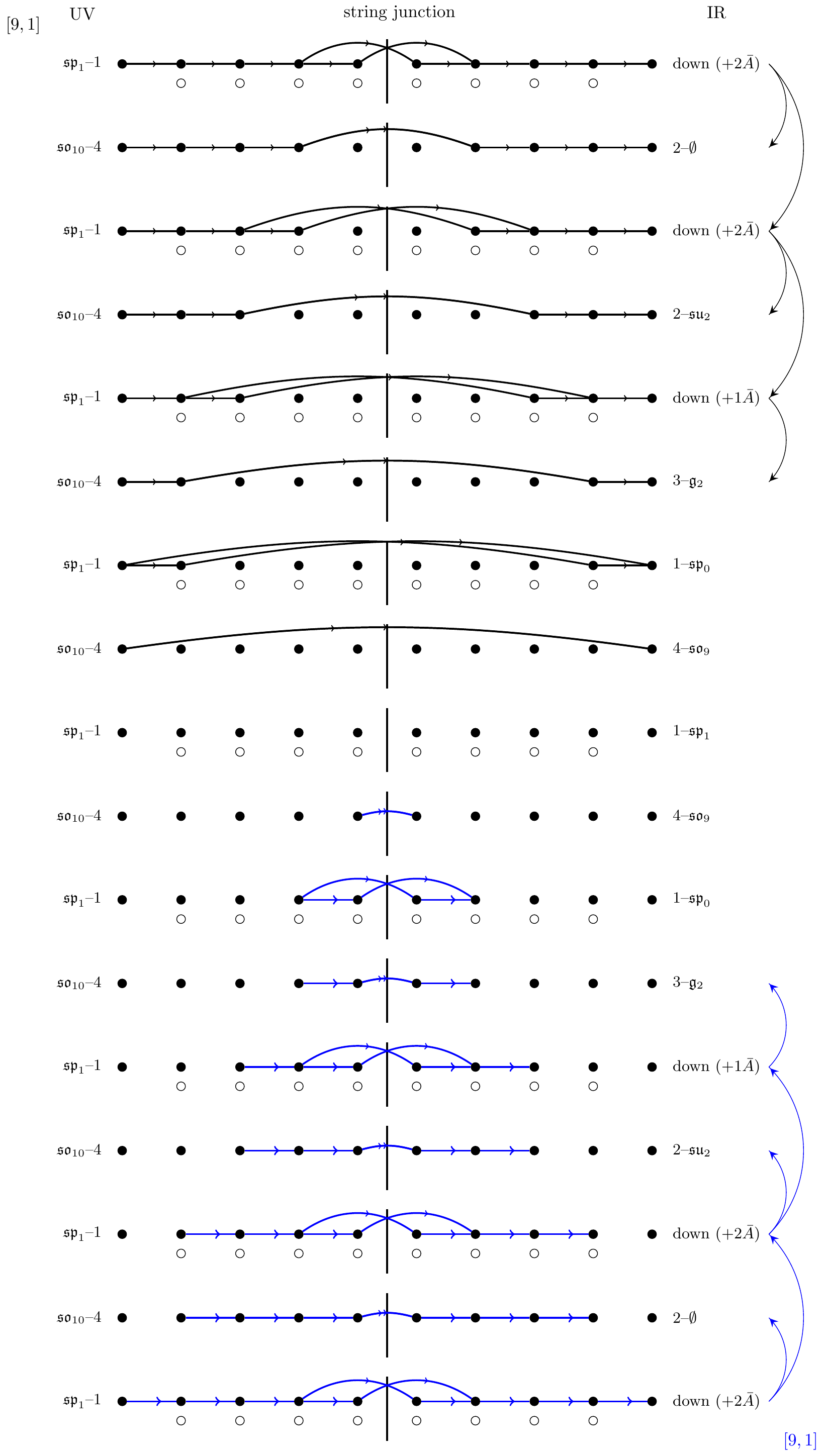}
    \captionsetup{font=footnotesize,labelfont=footnotesize}
    \caption{Independent example: Partitions $\mu_L=\mu_R=[9,1]$ on $17$ curves.} \label{subfig:independent}
  \end{subfigure} \hspace{0.25cm}
  \begin{subfigure}{0.5\textwidth}
    \includegraphics[width=0.82\textwidth]{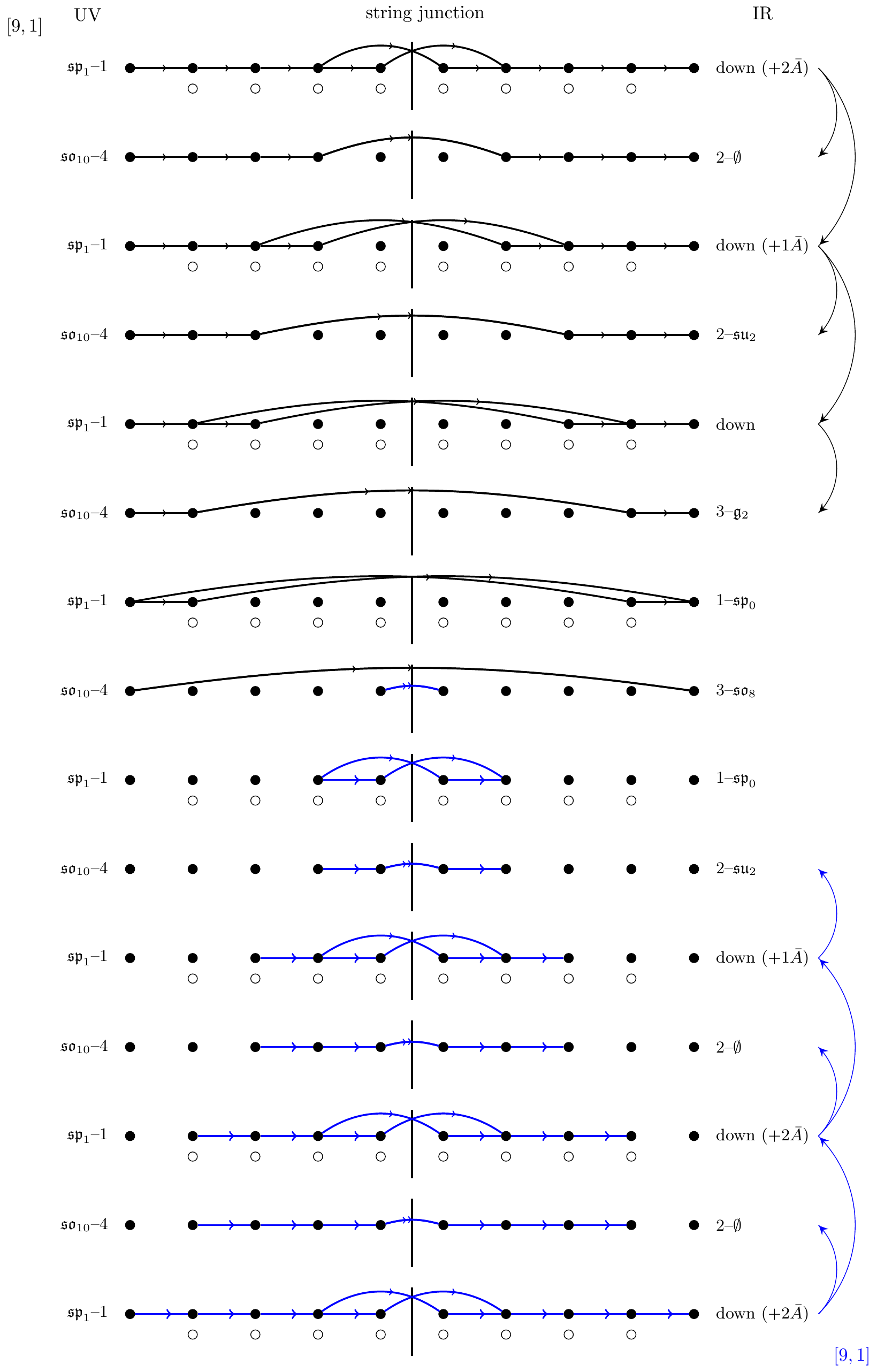}
    \captionsetup{font=footnotesize,labelfont=footnotesize}
    \caption{Touching example: Partitions $\mu_L=\mu_R=[9,1]$ on $15$ curves. Some but not all $-1$ curves participate in small instanton deformations.} \label{subfig:touch}
  \end{subfigure} \hspace{-1cm}

  \vspace{0.5cm} \hspace{-1cm}
  \begin{subfigure}{0.5\textwidth}
    \includegraphics[width=0.82\textwidth]{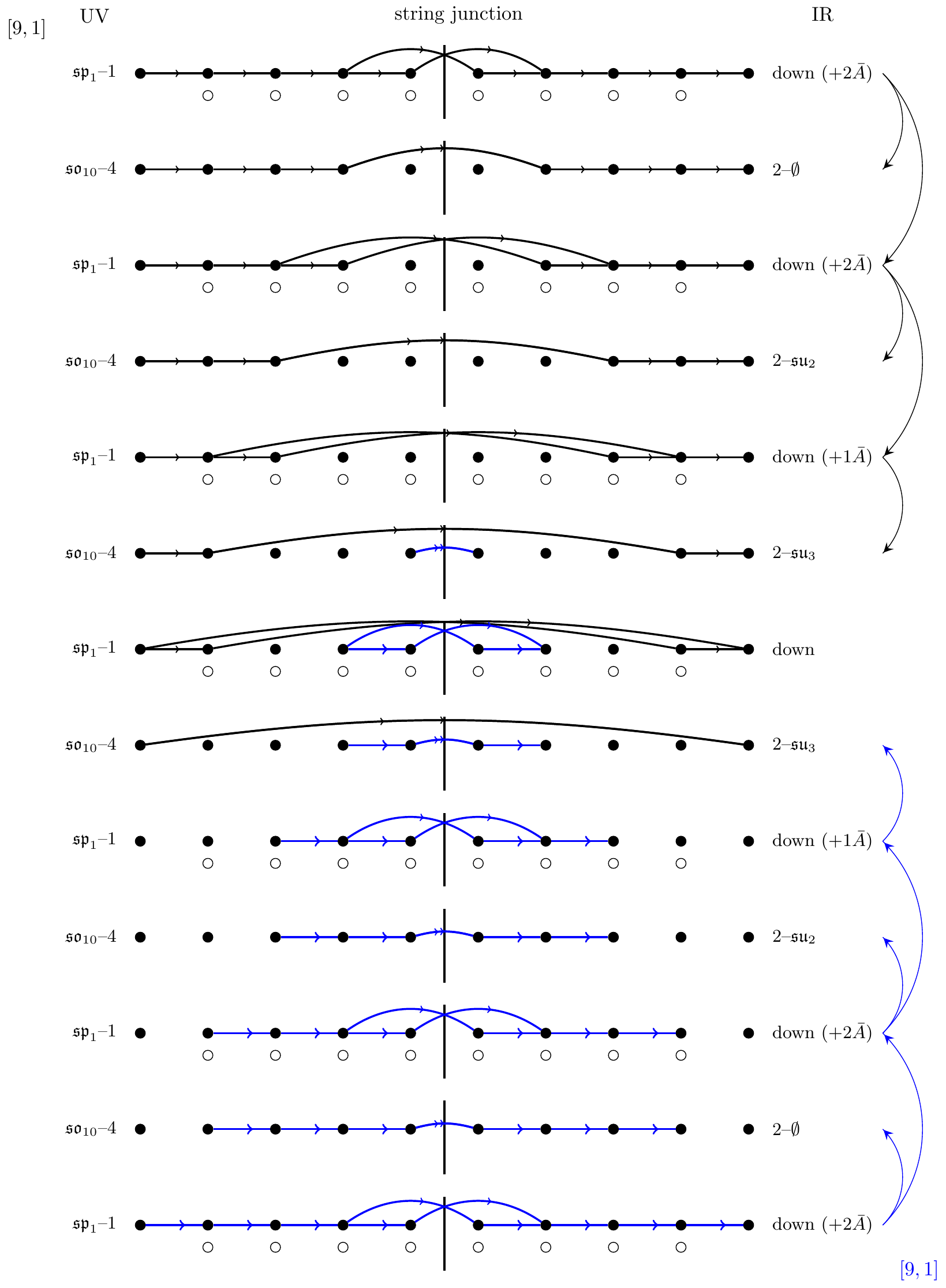}
    \captionsetup{font=footnotesize,labelfont=footnotesize}
    \caption{Kissing configuration: Partitions $\mu_L=\mu_R=[9,1]$ on $13$ curves. Every $-1$ curve participates in a small instanton / smoothing deformation.} \label{subfig:kiss}
  \end{subfigure}  \hspace{0.25cm}
  \begin{subfigure}{0.5\textwidth}
    \includegraphics[width=0.82\textwidth]{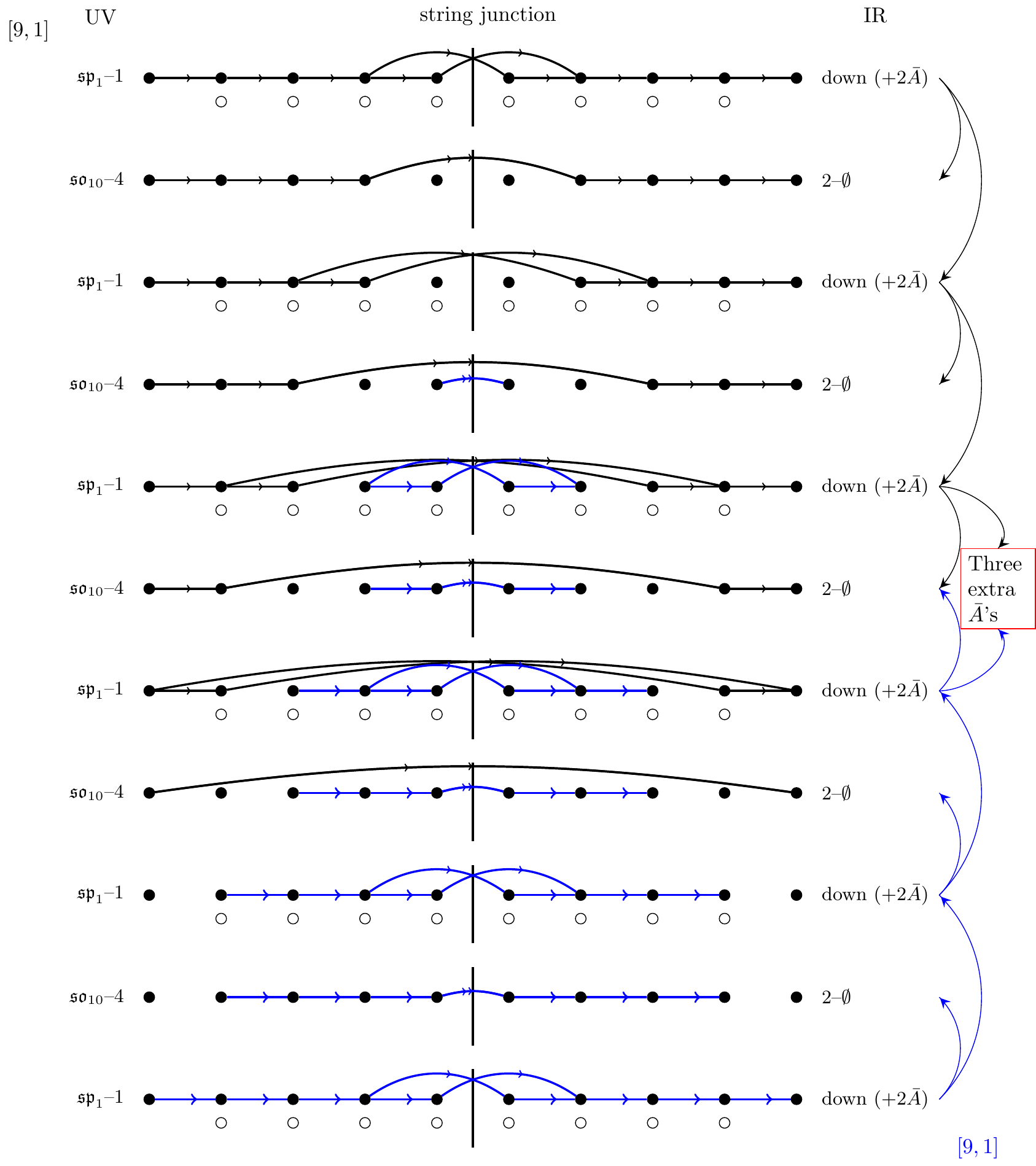}
    \captionsetup{font=footnotesize,labelfont=footnotesize}
    \caption{Crumpled configuration: Partitions $\mu_L=\mu_R=[9,1]$ on only $11$ curves. Too many $\overline{A}$'s are generated.} \label{subfig:crumple}
  \end{subfigure} \hspace{-1cm}
  \caption{Holding fixed the partitions $\mu_L=\mu_R=[9,1]$ we can decrease the number of curves to go from a long quiver (where the deformations are independent) all the way to a forbidden crumpled configuration.}\label{fig:ToShort}
\end{figure}

Of course, there are also nilpotent orbits which are uncorrelated, as will occur whenever the quiver is sufficiently long or the nilpotent orbits are sufficiently small, which we can view as ``independent cases.''
Such ``independent / touching cases'' fall within the scope of the long quiver analysis that we have discussed previously -- the latter just marginally so. We illustrate all four configurations in figure \ref{fig:ToShort} for $SO(10)$ with partitions $\mu_L=\mu_R=[9,1]$ going from an ``independent'' (long) quiver configuration all the way down to a forbidden ``crumpled'' configuration.

Following the IIA realization from section \ref{subsec:HananyWitten}, we
can formally perform Hanany-Witten moves even when small instanton transitions
occur by allowing for a negative number of D6-branes, or in the
string-junction picture by allowing brane / anti-brane
pairs as intermediate steps in our analysis.
 The formula (\ref{eq:HWconstraintSU}) generalizes to the
other quiver-like theories with classical algebras:
\begin{align}
  & k_{\frac{1}{2} \textrm{NS5}} \geq \textrm{Max}\{\mu_{L}^{1}, \mu_{R}^{1}\} + 1, \text{ rounded up to the nearest even number.} \\
  \iff & N_{T} \geq \textrm{Max}\{\mu_{L}^{1}, \mu_{R}^{1}\}. \label{eq:HWconstraint}
\end{align}
Here $k_{\frac{1}{2} \textrm{NS5}}$ is the number of half NS5-branes in the
corresponding type IIA picture, and equals one plus the number of tensor multiplets in the UV quiver ($N_{T}=2N_{-4}+1$) in the UV.
One might worry that this becomes meaningless whenever small instanton transitions occur. Indeed, the quivers described after such transitions all have matter with spinor representations and therefore no perturbative type IIA representation. While we can formally draw suspended brane diagrams with gauge groups of negative ranks, physically there is no corresponding suspended brane diagram. However, by analytically continuing the anomaly polynomials of these quivers to the case of negative ranks, we find perfect agreement with the anomaly polynomials of the actual, physical theory constructed via F-theory. This gives us strong reason to believe that the rules for Hanany-Witten moves should likewise carry over to the formal IIA brane diagrams, which implies that the formal quiver must be of length at least $\textrm{Max}\{\mu_{L}^{1}, \mu_{R}^{1}\}$.

Finally, from the brane / anti-brane analysis, we note that there should not be any residual $\overline{A}$'s in the IR theories. Any configuration yielding extra $\overline{A}$'s that cannot be canceled are said to ``crumple'' and are therefore forbidden. This further restricts the above constraints from Hanany-Witten moves.

As an example, an $SO(2N)$ quivers with partitions
\begin{equation}
  \mu_L=\mu_R=[2 N - 1,1]
\end{equation}
requires that
\begin{equation}
  k_{\frac{1}{2} \textrm{NS5}} \geq 2N + 4,
\end{equation}
which is a strictly stronger lower bound than the one imposed by equation \ref{eq:HWconstraint}. This particular example is illustrated for $SO(10)$ with partitions $\mu_L=\mu_R=[9,1]$ in the ``crumpling'' example of subfigure \ref{subfig:crumple}.

\subsubsection{$SO(2N)$ Short Quivers} \label{subsec:braneShortQuiver}

As we did in the $SU(N)$ case, we now show how to produce short $SO(2N)$ quivers beginning from long ones. For our first example, we consider the following formal $SO(8)$ quiver:
\begin{equation}
[5, 3]: \,\, \overset{\mathfrak{sp}(-3)}{1} \,\, \overset{\mathfrak{so}(4)}{4}
\,\, \overset{\mathfrak{sp}(-1)}{1} \,\, \overset{\mathfrak{so}(7)}{4} \,\, 1
\,\, \overset{\mathfrak{so}(8)}{4} \,\, \overset{\mathfrak{sp}(-1)}{1} \,\,
\overset{\mathfrak{so}(4)}{4}\,\,\overset{\mathfrak{sp}(-3)}{1}\,\, :[4^{2}],
\end{equation}
which is converted into the following F-theory quiver:
\begin{equation}
[5, 3]: \,\, \overset{\mathfrak{su}(2)}{2} \,\, \overset{\mathfrak{g}_{2}}{3}
\,\, 1 \,\, \underset{[SU(2)]}{\overset{\mathfrak{so}(7)}{3}} \,\, {\overset{\mathfrak{su}(2)}{2}}
\,\, [4^{2}].
\end{equation}
If we reduce the length by one, we would get a kissing theory (that is, every $-1$ curve has been blown-down):
\begin{equation}
[5, 3]: \,\, \underset{[N_f = 1]}{\overset{\mathfrak{su}(2)}{2}} \,\, \underset{[SU(2)]}{\overset{\mathfrak{su}%
(3)}{2}}\,\, \underset{[N_f = 1]}{\overset{\mathfrak{su}(2)}{2}} \,\, [4^{2}] \,.
\end{equation}
However, if we try to further reduce the length, we will reach a case that ``crumples'' due to an excess of $\overline{A}$'s that cannot be canceled, and therefore is invalid.

We can also keep the length of the quiver fixed and follow the RG flows along the nilpotent orbits (we will discuss this part in more detail in section
\ref{subsec:DoubleNilpotentHierarchy}). Consider the same example, but now increase the right nilpotent orbit from $[4^2]$ to $[5, 3]$. We still get an ``independent'' theory:
\begin{equation}
[5, 3]: \,\, \overset{\mathfrak{su}(2)}{2} \,\, \overset{\mathfrak{g}_{2}}{3}
\,\, \underset{[SU(2)]}{1} \,\, \overset{\mathfrak{g}_{2}}{3} \,\, \overset{\mathfrak{su}(2)}{2}
\,\, [5, 3] \,.
\end{equation}
If we further increase the right nilpotent orbit to $[7, 1]$, we will instead get
a kissing theory:
\begin{equation}
[5, 3]: [SU(2) \times SU(2)] \,\, {\overset{\mathfrak{su}(2)}{2}} \,\,
\overset{\mathfrak{su}(2)}{2} \,\, \underset{[N_f = 3/2]}{\overset{\mathfrak{su}(2)}{2}} \,\, 2 \,\,
[7, 1] \,.
\end{equation}
At this step, increasing the left orbit also up to $[7, 1]$ would give a crumpled configuration, which is not allowed.

\begin{figure}[ptb]
  \centering
  \includegraphics{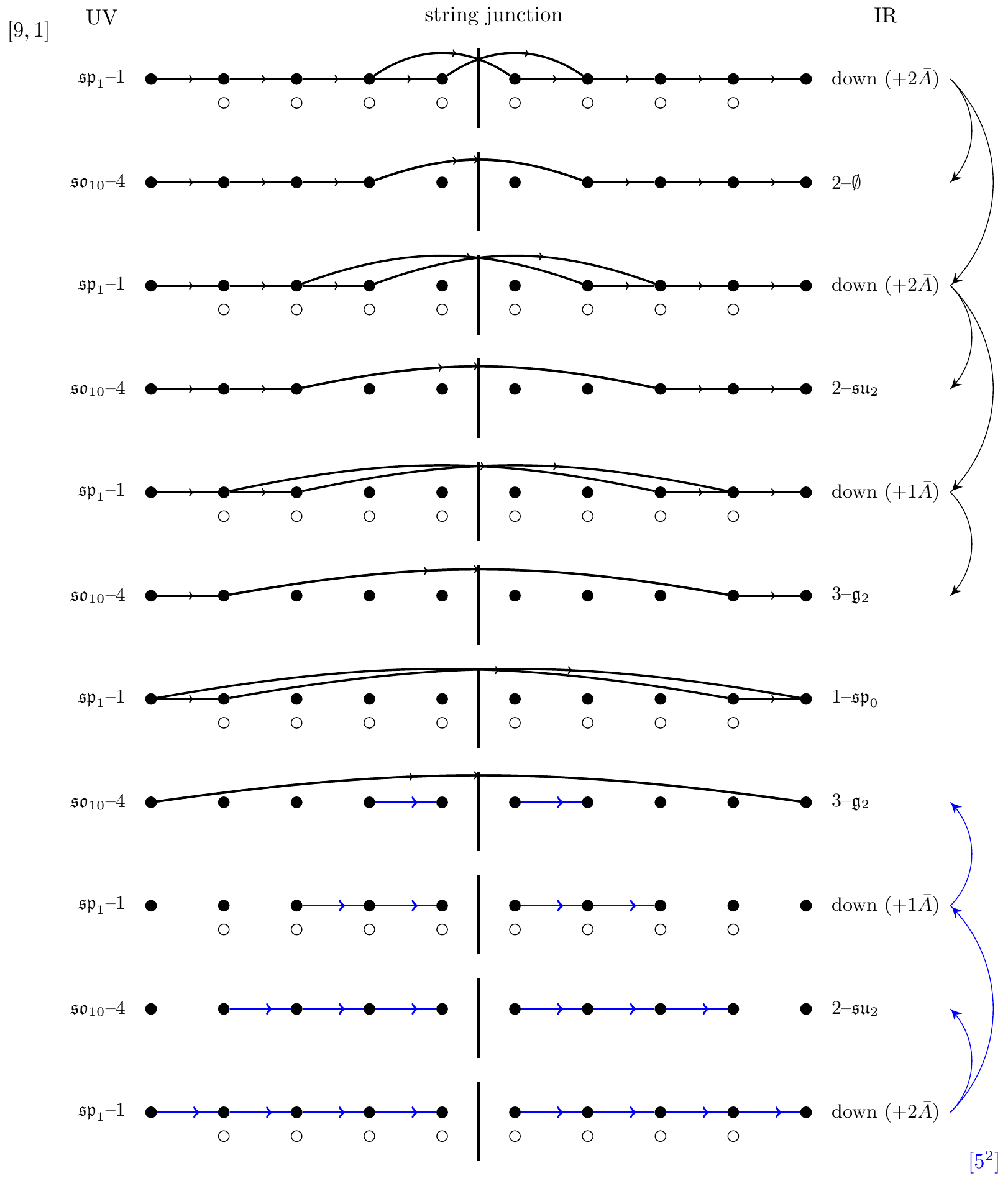}
\caption{An $SO(10)$ short quiver brane picture for nilpotent deformations $\mu_{L} = [9, 1]$, $\mu_{R}=[5^{2}]$. Additional branes are needed in order to construct the associated string diagrams, which in turn introduces anti-branes (depicted by white circles). The figure is arranged so that the left deformation
starts from the top and propagates downwards (in black) while the right
deformation starts on the bottom and propagates upwards (in blue). After the
blowdown and  Higgsing procedures, all
but one of the $-1$ curves are blown down, and the remaining curves now have self-intersection $-2$ or $-3$.}%
\label{fig:SOshortBrane}%
\end{figure}

We can describe all of this in general using the string junction picture previously developed.
Following our previous proposal for long quiver brane pictures, we start
from the outermost curves of the quiver, where we initialize our nilpotent deformation in terms of the string junction picture. Then, following the $SO/Sp$
propagation rule, we propagate the clusters from both sides towards the middle
simultaneously. In the case of short quivers, strings from both sides might
end up touching, sharing different intermediate layers, in which case the
gauge group reduction effects from both sides add together. For example,
figure \ref{fig:SOshortBrane} illustrates the action of
$\mu_L=[9,1]$, $\mu_R=[5^2]$ for $SO(10)$ in a theory with $11$ curves.

\begin{figure}[ptb]
  \centering
  \hspace{-1cm}
  \begin{subfigure}{0.5\textwidth}
    \includegraphics[width=\textwidth]{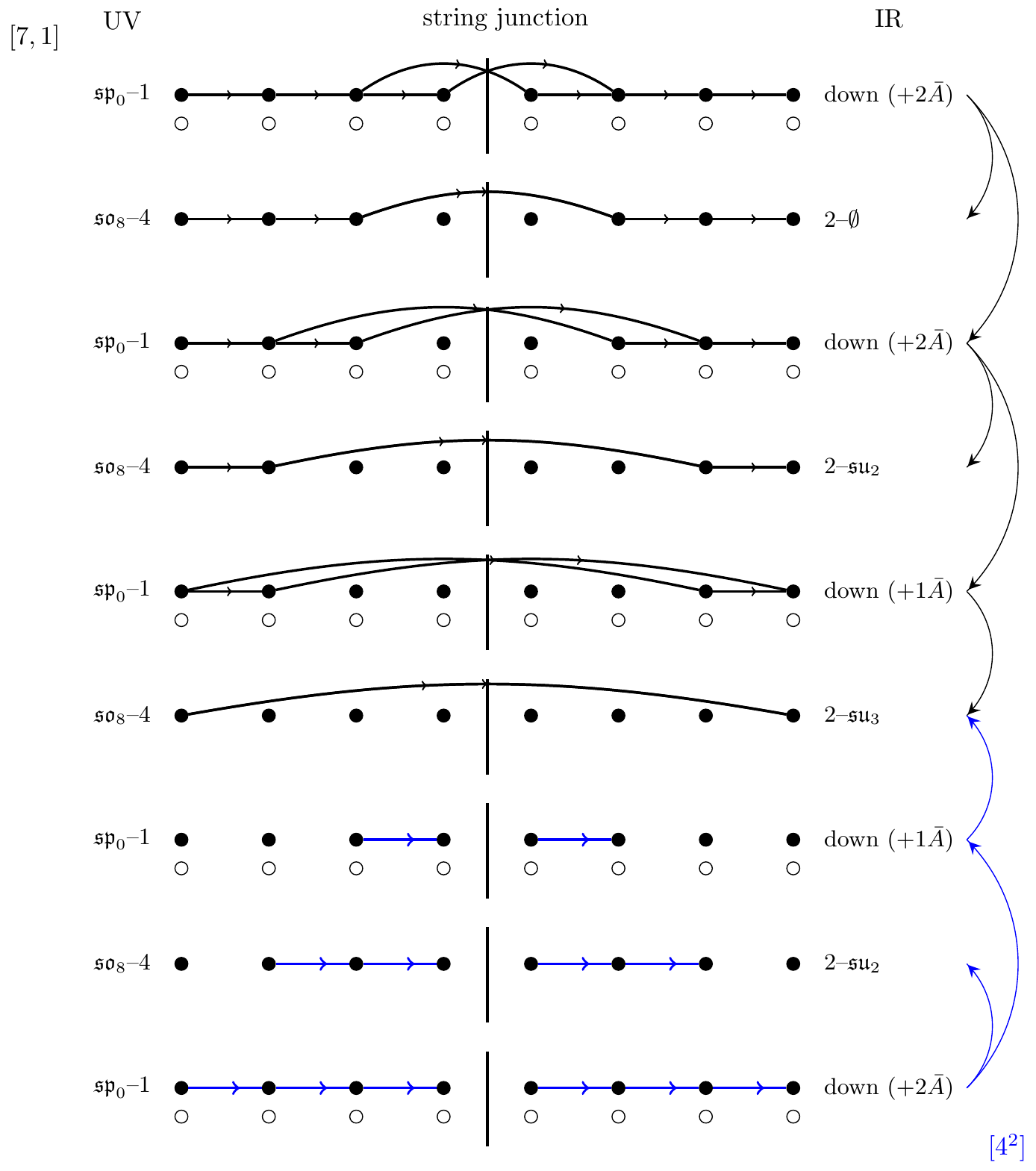}
    \captionsetup{font=footnotesize,labelfont=footnotesize}
    \caption{An example of a configuration that was not found for long quivers: partitions $\mu_L=[7,1]$, $\mu_R=[4^2]$ for a short quiver with $9$ curves. Note that two $\overline{A}$'s land on the third $-4$ curve, one from the top (left partition) and one from the bottom (right partition). There, the gauge group is reduced according to $\mathfrak{so}_7 \overset{\overline{A}}{\rightarrow} \mathfrak{g}_2 \overset{\overline{A}}{\rightarrow} \mathfrak{su}_3$.} \label{subfig:newShort1}
  \end{subfigure} \hspace{0.25cm}
  \begin{subfigure}{0.5\textwidth}
    \includegraphics[width=\textwidth]{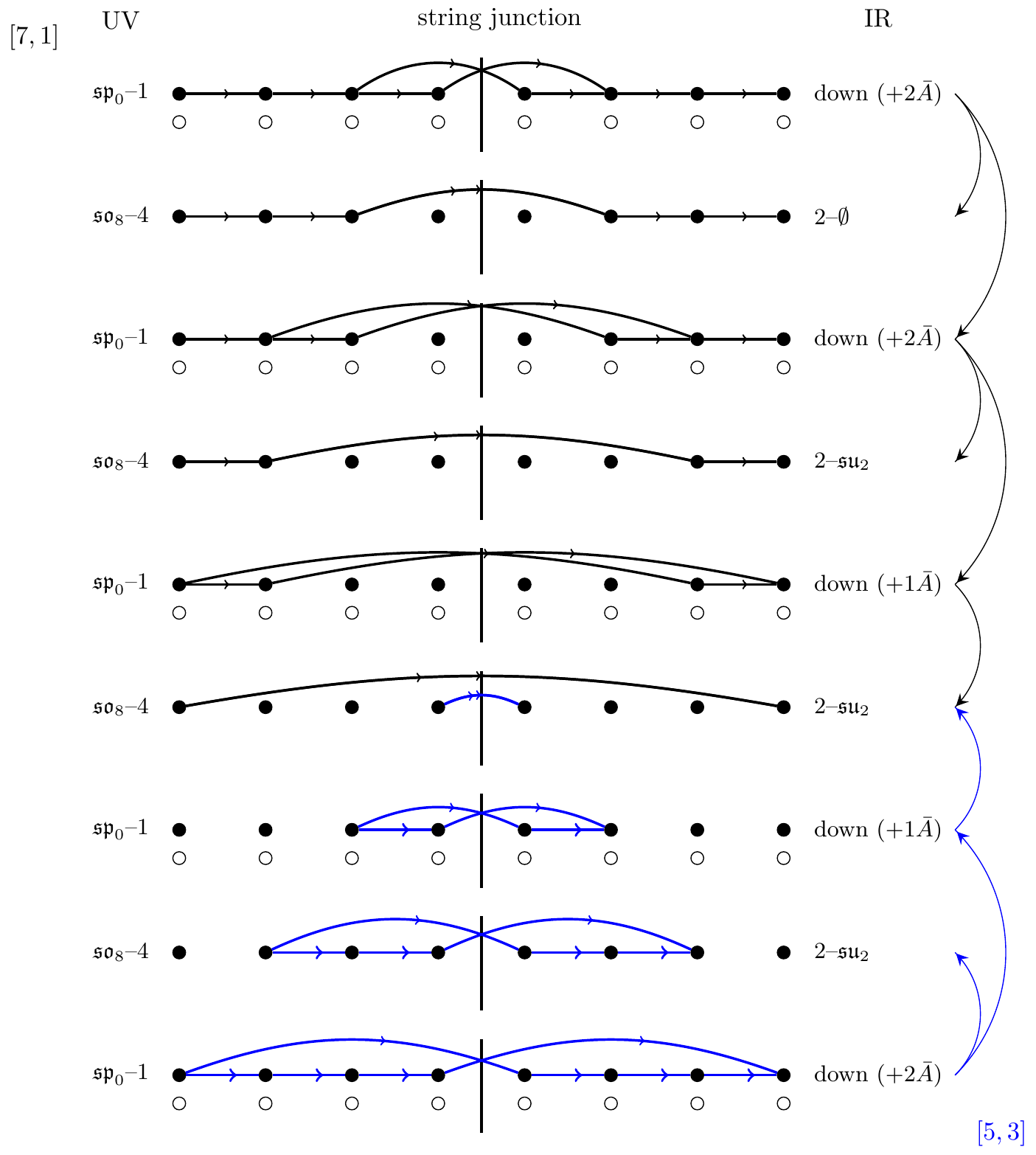}
    \captionsetup{font=footnotesize,labelfont=footnotesize}
    \caption{A second example of a configuration that was not found for long quivers: partitions $\mu_L=[7,1]$, $\mu_R=[5,3]$ for a short quiver with $9$ curves. Note that two $\overline{A}$'s land on the third $-4$ curve, one from the top (left partition) and one from the bottom (right partition). There, the gauge group is reduced according to $\mathfrak{so}_6 \simeq \mathfrak{su}_4 \overset{\overline{A}}{\rightarrow} \mathfrak{su}_3 \overset{\overline{A}}{\rightarrow} \mathfrak{su}_2$.} \label{subfig:newShort2}
  \end{subfigure} \hspace{-1cm}
  \caption{Two interesting examples where two $\overline{A}$'s land on the same $-4$ curve resulting in a chain of Higgsings that was not previously observed for long quivers.}\label{fig:newShort}
\end{figure}

We note that we can have new situations that could not previously occur in long quivers. The first novelty comes from the fact that levels with $\mathfrak{so}$ gauge algebra can now be Higgsed by two $\overline{A}$'s: one from the left nilpotent deformation and one from the right. As a result, we get configurations where two anti-branes accumulate on the same $-4$ curve and reduce it to a $-2$ curve. The resulting gauge algebra is then given by two applications of the rules for anti-brane reductions given in section \ref{ssec:Higgsing}. Figure \ref{fig:newShort} illustrates this phenomenon for a pair of theories, which respectively involve the reductions:
\begin{align}
  \mathfrak{so}_7 \overset{\overline{A}}{\rightarrow} \mathfrak{g}_2 \overset{\overline{A}}{\rightarrow} \mathfrak{su}_3 \\
  \mathfrak{so}_6 \simeq \mathfrak{su}_4 \overset{\overline{A}}{\rightarrow} \mathfrak{su}_3 \overset{\overline{A}}{\rightarrow} \mathfrak{su}_2.
\end{align}

The second novelty is that, in the $SO(8)$ case, partitions related by the triality outer automorphism do not necessarily yield the same IR theory! We saw previously that the long quivers for $\mu=[2^4]^{I,II}$ and $\mu=[3,1^5]$ are identical, as well as long quivers with deformations $\mu=[4^{2}]^{I,II}$ and $\mu=[5,1^{3}]$.
In the case of a long quiver, both of the $[4^{2}]$ and $[5, 1^{3}]$ deformations reduces the UV theory to the following IR theory \cite{Heckman:2016ssk}:
\begin{equation}
\overset{\mathfrak{su}(2)}{2} \,\, \underset{[SU(2)]}{\overset{\mathfrak{so}%
(7)}{3}} \,\, 1 \,\, \overset{\mathfrak{so}(8)}{4} \dots[SO(8)]\,.
\end{equation}
However, if we go to the short quiver cases from a UV theory of three $-4$
curves, we see that the pairs of $[4^{2}]$ -- $[4^{2}]$ and $[4^{2}]$ -- $[5,1^{3}]$ both yield the following quiver theory:
\begin{equation}
\underset{[N_f = 1/2]}{\overset{\mathfrak{su}(2)}{2}} \,\,
\underset{[Sp(2)]}{{\overset{\mathfrak{g_{2}}}{2}}} \,\,
\underset{[N_f = 1/2]}{\overset{\mathfrak{su}(2)}{2}}\,.%
\end{equation}
However, the pair of deformation $[5, 1^{3}]$ -- $[5, 1^{3}]$ gives a different short quiver theory:
\begin{equation}
{\overset{\mathfrak{su}(2)}{2}} \,\,
\underset{[SU(4)]}{{\overset{\mathfrak{su}(4)}{2}}} \,\,
{\overset{\mathfrak{su}(2)}{2}}\,.%
\end{equation}
This is a new effect regarding the outer automorphism of $SO(8)$, which is specific to having a short quiver.
The main point is that is that both $[4^{2}]$ -- $[4^{2}]$ and $[4^{2}]$ -- $[5,1^{3}]$ have one or two $\overline{A}$ branes involved, making it possible to reduce the gauge symmetry to $\mathfrak{g}_{2}$, while the $[5, 1^{3}]$ -- $[5, 1^{3}]$ does not involve $\overline{A}$ branes. Instead, the strings break the UV gauge group down to $\mathfrak{so}(6) \simeq \mathfrak{su}(4)$.

\begin{figure}[ptb]
  \centering
  \hspace{-1cm}
  \begin{subfigure}{0.5\textwidth}\vspace{0.78cm}
    \includegraphics[width=\textwidth]{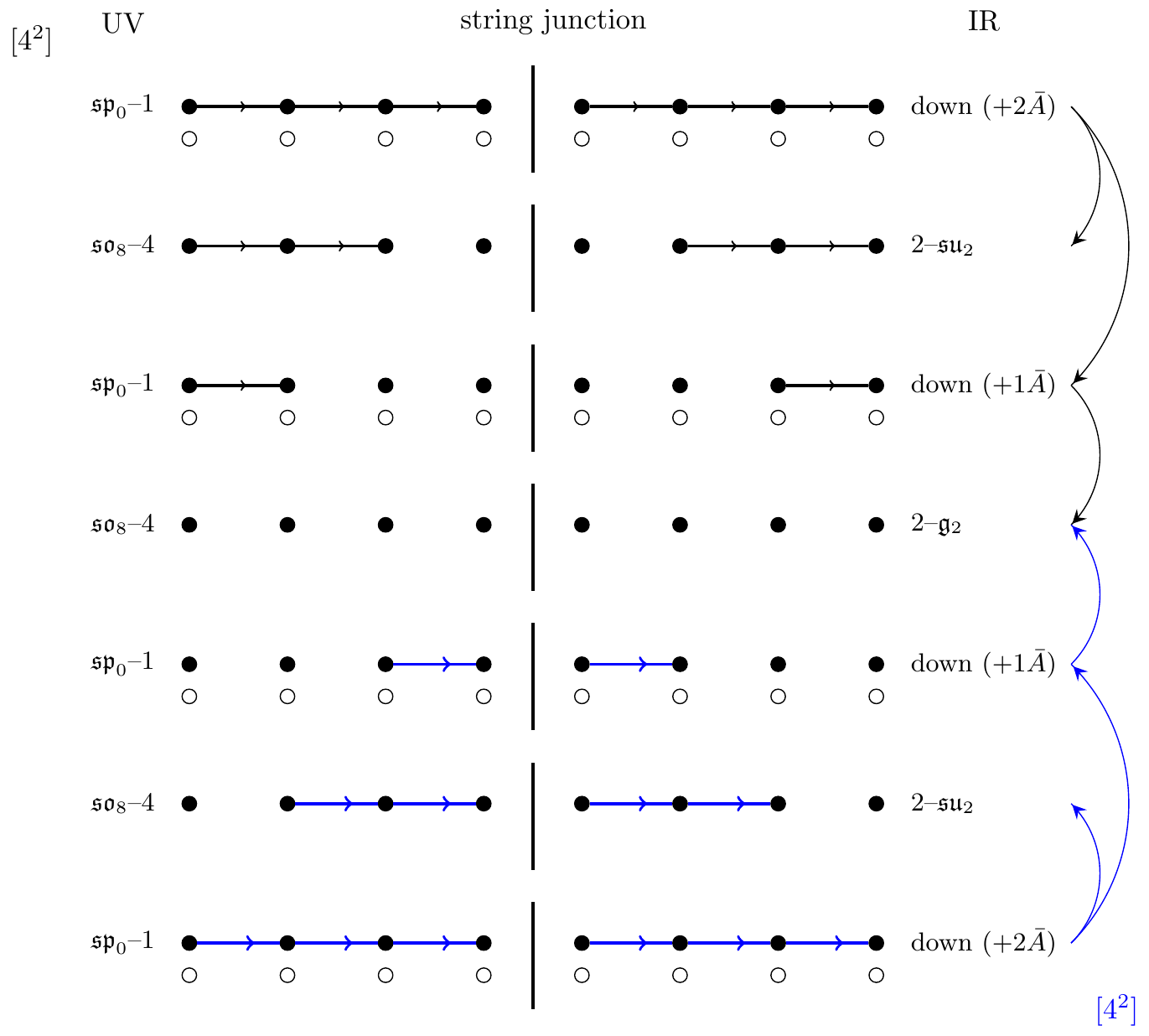}
    \captionsetup{font=footnotesize,labelfont=footnotesize}
    \caption{Partitions $\mu_L=\mu_R=[4^2]$ for a short quiver with $7$ curves. We note that in contrast to long quivers, we obtain a different IR theory than for the partitions $\mu_L=\mu_R=[5,1^3]$. Two $\overline{A}$'s land on the middle $-4$ curve, one from the top (left partition) and one from the bottom (right partition). There, the gauge group is reduced according to $\mathfrak{so}_8 \overset{\overline{A}}{\rightarrow} \mathfrak{so}_7 \overset{\overline{A}}{\rightarrow} \mathfrak{g}_2$.} \label{subfig:compare1}
  \end{subfigure} \hspace{0.25cm}
  \begin{subfigure}{0.5\textwidth}
    \includegraphics[width=\textwidth]{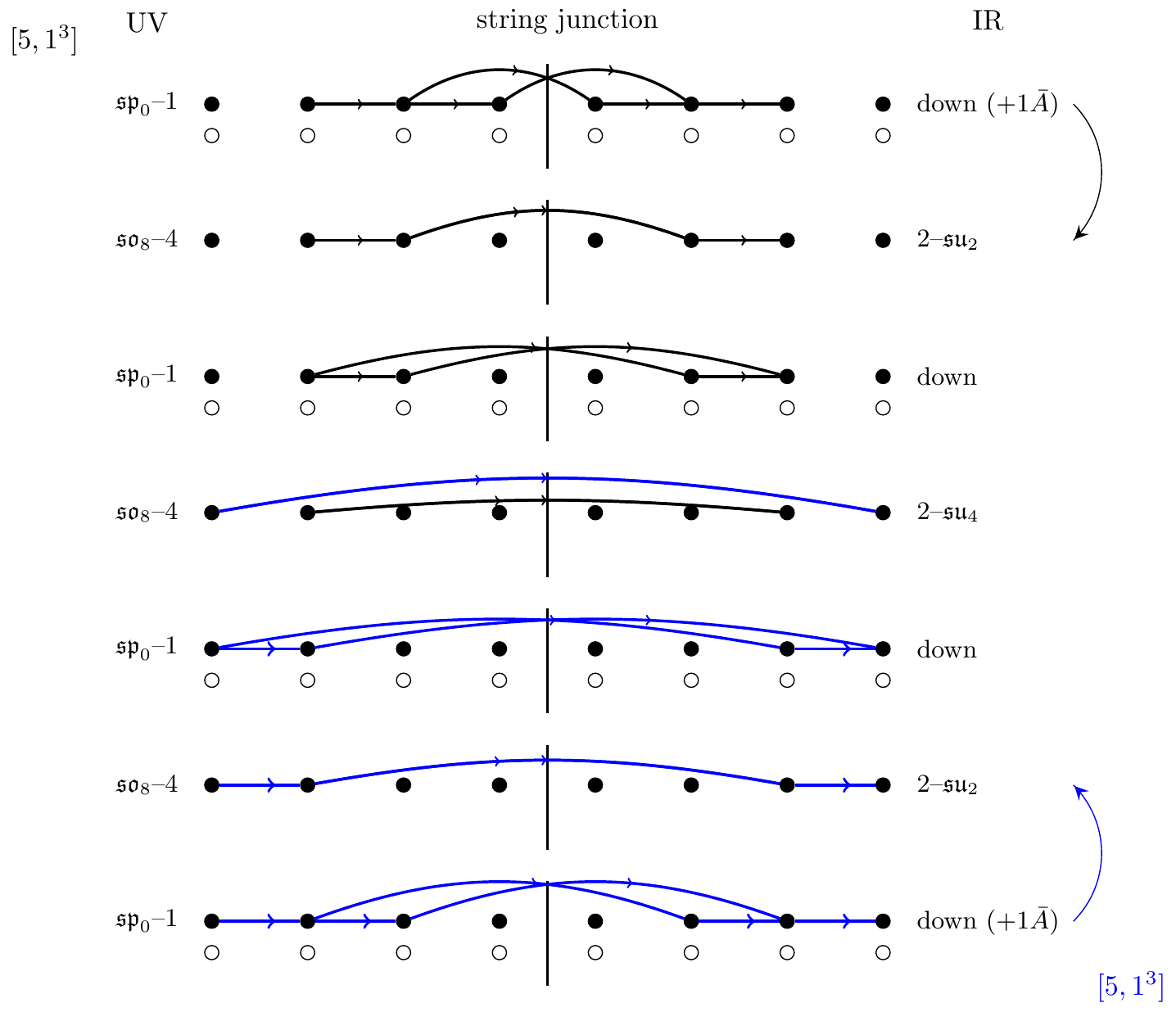}
    \captionsetup{font=footnotesize,labelfont=footnotesize}
    \caption{Partitions $\mu_L=\mu_R=[5,1^3]$ for a short quiver with $7$ curves. We note that in contrast to long quivers we obtain a different IR theory than for the partitions $\mu_L=\mu_R=[4^2]$. On the middle $-4$ curve we now have $\mathfrak{so}_6 \simeq \mathfrak{su}_4$ gauge algebra.} \label{subfig:compare2}
  \end{subfigure} \hspace{-1cm}
  \caption{Nilpotent orbits with $\mu=[5,1^3]$ or $\mu=[4^2]$ yield the same IR theories for long quivers (see figure \ref{fig:SO8quiver} for instance). However, here we see a clear difference for short quivers.}\label{fig:compare}
\end{figure}

These phenomena are recorded in figures \ref{fig:so8ShortFlowsOne}, \ref{fig:so8ShortFlowsTwo}, and \ref{fig:so8ShortFlowsThree}, but we show explicitly the string junction pictures in figure \ref{fig:compare} for the partitions $\mu_L=\mu_R=[4^{2}]$ vs. the partitions $\mu_L=\mu_R=[5,1^{3}]$. In section \ref{subsubsec:ShortAnomalyRules}, we will justify this surprising conclusion by an analysis of the anomaly polynomials for these respective theories.

\subsubsection{$SO($odd$)$ Case} \label{subsubsec:SOodd}

In general, $SO(2N-1)$ short quivers can be reinterpreted as $SO(2N+2)$ short quivers deformed by a pair of nilpotent orbits.
For example, suppose we start from an $SO(7)$ short quiver UV theory, written as:%
\begin{equation}
[SO(7)] \ \ 1 \ \ \overset{\mathfrak{so}(9)}{4}
\ \ \underset{[N_f = 1]}{\overset{\mathfrak{sp}(1)}{1}}
\ \ \overset{\mathfrak{so}(9)}{4} \ \ 1 \ \ [SO(7)].
\end{equation}
This can be reinterpreted as starting from the following $SO(10)$ UV theory:
\begin{equation}
[SO(10)] \ \ \overset{\mathfrak{sp}(1)}{1}
\ \ \overset{\mathfrak{so}(10)}{4} \ \ \overset{\mathfrak{sp}(1)}{1} \ \ \overset{\mathfrak{so}(10)}{4}
\ \ \overset{\mathfrak{sp}(1)}{1} \ \ [SO(10)],
\end{equation}
and applying the pair of nilpotent deformations $[3, 1^{7}]$ -- $[3,1^{7}]$.

In general, any $SO(2N-p)$ quiver with deformations parametrized by the partitions $\mu_L^{\mathrm{odd}}$, $\mu_R^{\mathrm{odd}}$ of $2N-p$ can be reinterpreted as an $SO(2N)$ quiver with associated partitions $\mu_L^{\mathrm{even}}$, $\mu_R^{\mathrm{even}}$ obtained by simply adding a ``$p$'' to the partitions $\mu_L^{\mathrm{odd}}$ and $\mu_R^{\mathrm{odd}}$, respectively. For instance, for the minimal choice $p=3$ with $\mu_L^{\mathrm{odd}} = [1^9]$, $\mu_R^{\mathrm{odd}} = [7, 1^2]$, we can equivalently express the theory as an $SO(12)$ quiver with $\mu_L^{\mathrm{even}} = [3,1^9]$, $\mu_R^{\mathrm{even}}=[7,3,1^2]$. In this way, the rules we developed for $SO(2N)$ quivers above carry over straightforwardly to $SO(2n-p)$ quivers for $p$ odd.

\subsubsection{$Sp$ Case}
We now turn to quiver-like theories in which the flavor symmetries are a pair of $Sp$-type. The first thing we should note is that no blow-downs can happen.
As a result, there are no ``kissing'' or ``crumpled'' configurations. The only constraint that needs to be imposed comes from the Hanany-Witten moves:
\begin{equation}
  N_{T} \geq \textrm{Max}\{\mu_L^1,\mu_R^1\},
\end{equation}
with $N_{T}$ the number of tensor multiplets in the UV theory.

The behavior of the $Sp$ short quivers is then the same as for $SO(2N)$, where the contributions from each side can overlap, but without any of the complications found due to small instanton transitions or anti-branes. Indeed, no anti-branes are necessary for $Sp$ -- $Sp$ quivers.

\subsubsection{Mixed $[G]$--$[G^\prime]$ Case}
It is interesting to consider mixed quivers where the left and right flavors are not equal. The advantage of our analysis is that it straightforwardly generalizes to these cases. Indeed, without loss of generality let $M \leq N$, then
\begin{itemize}
  \item Quivers with $SU(M)$ -- $SU(N)$, $M<N$, flavor symmetries are obtained from partitions of $N$ with $\mu_L=[\nu_L^i,N-M]$ and $\mu_R=[\mu_R^i]$, where $[\nu_L^i]$ is a partition of $M$.
  \item Quivers with $SO(2M)$ -- $SO(2N)$, $M<N$, flavor symmetries are similarly obtained from partitions of $2N$ with $\mu_L=[\nu_L^i,(N-M)^2]$ and $\mu_R=[\mu_R^i]$, where $[\nu_L^i]$ is a partition of $2M$.
  \item Quivers with $SO(\mathrm{even})$ -- $Sp$ flavors can be viewed as two $SO(\mathrm{even})$ flavor symmetries with the right most $-1$ curve decompactified. Small instanton transitions of the interior $-1$ curves on the right-hand side of this quiver are allowed only if the resulting base is given by 223 or 23.
  \item Any quiver involving $SO($odd$)$ flavor symmetries can be embedded inside an $SO($even$)$ quiver, as seen in subsection \ref{subsubsec:SOodd}. Thus, these reduce to the cases above.
\end{itemize}

\subsection{Anomaly Matching for Short Quivers}\label{subsec:matching}

In this subsection, we propose a method for computing the anomalies of short quivers with classical algebras. We begin by introducing the notion of a ``formal $SO$ quiver.'' We then show how these can be useful in determining the true F-theory quiver of a 6D SCFT via anomaly polynomial matching. In some cases of short quivers, there is a mismatch between the anomaly polynomial computed via the formal $SO$ quiver and the quiver obtained through the string junction picture described previously. However, this mismatch seems to take a universal form, indicating that the string junction approach may nonetheless give the correct answer, even when there is a disagreement with the formal quiver approach. We conclude the subsection with illustrative examples.

\subsubsection{Formal $SO$ theories}

``Formal'' $SO$ quivers involve
analytically continuing the gauge algebra $SO(8+m)$ or $Sp(n)$ so that $m, n
\leq0$. This is only an intermediate step, and the motivation for introducing
such formal quiver is to help determine the actual F-theory quiver via anomaly
polynomial matching (see \cite{Mekareeya:2016yal} for a detailed construction
of such formal quivers). Here, we present a brief review of how this is done.

We start from the long quiver case, where we make a comparison between a long
$SO(8)$ quiver theory and its formal quiver theory and show that the the anomaly polynomials between the two agree. The actual F-theory quiver is obtained by a $[5, 3]$
deformation to the left:
\begin{equation}
[5, 3]: \,\, \overset{\mathfrak{su}(2)}{2} \,\, \overset{\mathfrak{g}_{2}}{3}
\,\, 1 \,\, \overset{\mathfrak{so}(8)}{4} \,\, \cdots\,\, 1 \,\, [SO(8)] \,\,
:[1^{8}]\,.
\label{eq:53}
\end{equation}
On the other hand, we can also express this in terms of a formal quiver by allowing for gauge groups with negative rank:
\begin{equation}
[5, 3]: \,\, \overset{\mathfrak{sp}(-3)}{1} \,\, \overset{\mathfrak{so}(4)}{4}
\,\, \overset{\mathfrak{sp}(-1)}{1} \,\, \overset{\mathfrak{so}(7)}{4} \,\, 1
\,\, \overset{\mathfrak{so}(8)}{4}\,\, \cdots\,\, 1 \,\, [SO(8)] \,\,
:[1^{8}]\,.
\end{equation}
If we truncate both of these theories, keeping only the part of the quiver to the left of the ``$\cdots
$", then their anomaly polynomials are both given by
\begin{equation}
I_8=\frac{6337}{168}c_{2}(R)^{2} + \frac{25}{336}c_{2}(R)p_{1}(T) + \frac
{631}{40320}p_{1}(T)^{2} -  \frac{79}{1440} p_{2}(T).
\label{eq:53I}
\end{equation}
In the case of the formal quiver, this anomaly polynomial computation is performed by analytically continuing the formula for an $Sp-SO$ quiver to negative gauge group rank (see \cite{Mekareeya:2016yal}).

This example illustrates the utility of the formal quiver for anomaly matching. In
our short quiver theories, the actual F-theory quivers can be difficult to read off, whereas these formal $SO$ quivers are easy to determine. As a result, we can use them together with their associated anomaly polynomials relation to check our proposal for the F-theory quiver, as described below.

The general formula for formal quivers--both long and short--is similar to the formula (\ref{eq:SUshort}) for the $SU$ case.
Define the partition of the left and right nilpotent orbits of $SO(2N)$ to be $\mu_{L}^{j},
\mu_{R}^{j}$ and define their conjugate partitions $\rho_{L}^{j}, \rho_{R}^{j}$. We have an alternating
sequence of $SO$ and $Sp$ gauge algebras on the full tensor branch. Indexing the gauge algebras by a parameter $m$
which starts with $Sp(q_1)$ on the left and continues to $SO(p_2)$, ... and terminating with an $Sp$ factor, we have the
assignments:
\begin{equation}
SO(p_m),~~~p_{m}=2N-\sum_{i=m+1}^{N_{L}^{\prime}} \rho_{i}^{L}-\sum_{j=N_T-m+2}%
^{N_{R}^{\prime}} \rho_{j}^{R} \quad(m\,\, \mathrm{even})
\end{equation}
\begin{equation}
Sp(q_m),~~~q_{m}=\frac{1}{2}(2N-\sum_{i=m+1}^{N_{L}^{\prime}} \rho_{i}^{L}-\sum
_{j=N_T-m+2}^{N_{R}^{\prime}} \rho_{j}^{R}) - 4 \quad(m\,\, \mathrm{odd})\,.
\end{equation}
Here, $N_T$ is the number of tensor multiplets in the UV
F-theory description and $N_L^{'}, N_R^{'}$ are the lengths of left and right conjugate partitions, respectively.

Let us illustrate the construction of short quiver formal $SO$ theories by starting with a sufficiently long formal theory and then reducing the length. Consider the $SO(8)$ theory with $[5, 3]$ and $[3^{2}, 1^{2}]$ nilpotent
deformations and four $-4$ curves, so that the
pair of deformations does not overlap:
\begin{equation}
[5, 3]: \,\, \overset{\mathfrak{sp}(-3)}{1} \,\, \overset{\mathfrak{so}(4)}{4}
\,\, \overset{\mathfrak{sp}(-1)}{1} \,\, \overset{\mathfrak{so}(7)}{4} \,\, 1
\,\, \overset{\mathfrak{so}(8)}{4} \,\, 1 \,\, \overset{\mathfrak{so}%
(4)}{4}\,\,\overset{\mathfrak{sp}(-1)}{1}\,\, :[3^{2}, 1^{2}]\,.
\end{equation}
Now we decrease the length of the quiver. In each step, we start from a
shorter UV theory by removing one group of $(-1, -4)$ curves. We get the
following set of theories after each step:%
\begin{equation}
[5, 3]: \,\, \overset{\mathfrak{sp}(-3)}{1} \,\, \overset{\mathfrak{so}(4)}{4}
\,\, \overset{\mathfrak{sp}(-1)}{1} \,\, \overset{\mathfrak{so}(7)}{4} \,\, 1
\,\, \overset{\mathfrak{so}(4)}{4}\,\,\overset{\mathfrak{sp}(-1)}{1}\,\,
:[3^{2}, 1^{2}]
\end{equation}
\begin{equation}
[5, 3]: \,\, \overset{\mathfrak{sp}(-3)}{1} \,\, \overset{\mathfrak{so}(4)}{4}
\,\, {\overset{\mathfrak{sp}(-1)}{1}} \,\,
\overset{\mathfrak{so}(5)}{4}\,\,{\overset{\mathfrak{sp}%
(-2)}{1}}\,\, :[3^{2}, 1^{2}]\,.
\end{equation}

We stop at this point, following the constraints from the Hanany-Witten moves.
We see that the formal gauge algebra goes down to the unphysical values of $\mathfrak{sp}%
(-3)$ and $\mathfrak{so}(2)$.

However, from such a quiver we may still extract its anomaly polynomial by analytically continuing the formulae developed in the physical regime, $\mathfrak{sp}(m), m > 0$ and $\mathfrak{so}(n),
n \geq8$. In the long quiver case, the anomaly polynomial of the formal quiver exactly matches that of the actual quiver \cite{Mekareeya:2016yal}, as in the example in (\ref{eq:53})-(\ref{eq:53I}). This serves as a strong motivation for us to test the relationship between $SO$ short quivers and their formal counterparts via anomaly matching.

\subsubsection{Anomaly Polynomial Matching and Correction Terms}
    \label{subsubsec:ShortAnomalyRules}

    For theories with long quivers, there is a well-defined prescription in the literature for producing the F-theory quiver of a given formal type IIA quiver (see \cite{Mekareeya:2016yal}). For short quiver theories, however, the situation becomes much more complicated, and there is at present no well-defined proposal in the literature. Nonetheless, the rules we have introduced in section \ref{sec:RECOMBO} carry over to the case of short quivers, so we may check that these rules give the correct answer by comparing the anomaly polynomials of the proposed short quiver theories to those obtained from the formal quiver. This check has been done explicitly for all cases in the catalogs \ref{table:SO8tangentialShortQuiver} and \ref{table:SO10TangentshortQuiver} in Appendix \ref{apdx:shortQuiverCatalogs}.

In general, we find that there is frequently a mismatch in the $p_1(T)^2$ and $p_2(T)$ coefficients of the anomaly polynomials computed via the formal quiver vs. the actual F-theory quiver. However, this is not very concerning, as the mismatch can always be canceled by adding an appropriate number of neutral hypermultiplets, each of which contributes $(4p_1(T)^2-7p_2(T))/5760$ to the anomaly polynomial. Indeed, such a mismatch in short quiver theories was previously noted in \cite{Heckman:2018pqx}.

More concerning are the mismatches in the coefficients of the $c_2(R)^2 $ coefficient and the $c_2(R) p_1(T)$ coefficient (denoted $\alpha$ and $\beta$, respectively). These mismatches are relatively rare, arising only in a smaller number of kissing cases (see tables \ref{table:SO8tangentialShortQuiver} and \ref{table:SO10TangentshortQuiver} in Appendix \ref{apdx:shortQuiverCatalogs}). This could be an indication that these theories are sick and should be discarded. However, we note that these mismatches seem to follow a universal set of rules, which indicates that our proposed F-theory quiver may nonetheless represent an accurate translation of the formal quiver.

Theories with mismatches always involve two anti-branes acting on a curve carrying an $\mf{so}$ gauge algebra according to the rules in (\ref{eq:rules}), and it depends on the size of the gauge group. In particular, denoting the mismatch in the anomaly polynomial coefficients $\alpha$ and $\beta$ by $\Delta \alpha$, $\Delta \beta$, respectively, we have:
\begin{enumerate}[(1)]
    \item
    \begin{equation}
    \mathfrak{so}(8) \overset{2\overline{A}}{\rightarrow} \mathfrak{g}_{2}: (\Delta \alpha, \Delta \beta) = (0,0)
    \end{equation}
    (see figure \ref{subfig:compare1} for an example)

    \item
    \begin{equation}
    \mathfrak{so}(7) \overset{2\overline{A}}{\rightarrow} \mathfrak{su}(3):  (\Delta \alpha, \Delta \beta) = (\frac{1}{24},\frac{1}{48})
    \label{eq:rule2}
    \end{equation}
    (see figure \ref{subfig:newShort1} for an example)

    \item
    \begin{equation}
    \mathfrak{so}(6) \simeq \mathfrak{su}(4) \overset{2\overline{A}}{\rightarrow} \mathfrak{su}(2):  (\Delta \alpha, \Delta \beta) = (\frac{1}{12},\frac{1}{24})
    \label{eq:rule3}
    \end{equation}
    (see figure \ref{subfig:newShort2} for an example)

    \item
    \begin{equation}
    \mathfrak{so}(5)  \overset{2\overline{A}}{\rightarrow} \mathfrak{su}(1):  (\Delta \alpha, \Delta \beta) = (\frac{1}{6},\frac{1}{12})
    \end{equation}

    \item
    \begin{equation}
    \textrm{All remaining cases}:  (\Delta \alpha, \Delta \beta) = (0,0).
    \end{equation}
    \end{enumerate}
Note that the kissing condition and Hanany-Witten constraints only allow one $-4$ curve to have 2 $\overline{A}$'s simultaneously attach to the curve. There is one borderline case involving $\mathfrak{so}(4)$ gauge symmetry and a pair of $\overline{A}$'s. In both long and short quivers, we have a consistent rule $\mathfrak{so}(4) \overset{\overline{A}}{\rightarrow} \mathfrak{su}(2)$, but adding an additional $\overline{A}$ brane appears to be problematic in general. Including this case would generate a curve without any gauge symmetry, which in many examples leads to a quiver where the ``convexity condition'' required of gauge group ranks is violated. This is best illustrated with an example.
Consider the UV quiver:
\begin{equation*}
  [1^{16}] \,\,
  \overset{\mathfrak{sp}(4)}{1}\,\,\overset{\mathfrak{so}(16)}{4}\,\,
  \overset{\mathfrak{sp}(4)}{1} \,\,\overset{\mathfrak{so}(16)}{4}\,\,
  \overset{\mathfrak{sp}(4)}{1} \,\,\overset{\mathfrak{so}(16)}{4}\,\,
  \overset{\mathfrak{sp}(4)}{1}\,\,
          [1^{16}]
\end{equation*}
If we were to na\"ively assume that $\mathfrak{so}(4)  \overset{2\overline{A}}{\rightarrow} \emptyset$ without crumpling, then the deformation $\mu_L=\mu_R = [7^2,1^2]$ would yield the following sick IR theory:
\begin{equation*}
  [7^2,1^2] \,\,
  \overset{\mathfrak{su}(2)}{2}\,\,
  \overset{\emptyset}{2}\,\,
  \overset{\mathfrak{su}(2)}{2}\,\,
          [7^2,1^2]
\end{equation*}
From this, we conclude that whenever $\mathfrak{so}(4)$ is hit by two $\overline{A}$'s simultaneously, it must crumple, so we forbid these configurations.

In summary, in cases without a double $\overline{A}$ Higgsing chain (``All remaining cases") we never have such a mismatch, and in many cases with a double $\overline{A}$ Higgsing chain, there is also no mismatch. There are a few cases where there is a mismatch, which always involve two $\overline{A}$'s in the Higgsing chain. The above proposal has been explicitly verified in the $SO(8)$ and $SO(10)$ catalogs of
Appendix \ref{apdx:shortQuiverCatalogs}.

What is the physical interpretation of these mismatches? We note that in case (1), where there is no mismatch, the gauge group is reduced from $\mathfrak{so}(8)\overset{2\overline{A}}{\rightarrow} \mathfrak{g}_2$, and the brane picture and the string junction root system make perfect sense. However, when there is a mismatch (as in cases (2)-(5)), we always start from an SO brane picture with an orientifold and somehow end up with a SU brane without an orientifold. We leave further explanation of this issue for future work.

\subsubsection{Examples}\label{ssec:examples}

In this section, we present a number of examples to demonstrate our procedure of anomaly matching explicitly and to reveal some of the subtleties of our procedure regarding different quiver lengths, different UV gauge groups, and different types of Higgsing.

\begin{itemize}

\item{\textbf{Example 1}}

We start with the pair of orbits $[5, 1^{3}], [5, 1^{3}]$ on an $SO(8)$ UV
theory with tensor branch given by three $-4$ curves. The resulting description in F-theory
is:
\begin{equation}
\overset{\mathfrak{su}(2)}{2}
\,\,\underset{[SU(4)]}{\overset{\mathfrak{su}(4)}{2}}
\,\,\overset{\mathfrak{su}(2)}{2}
\end{equation}
This theory gives the same anomaly polynomial as the corresponding formal $SO$ quiver:
\begin{equation}
[5, 1^3]: \,\, \overset{\mathfrak{sp}(-2)}{1} \,\, \overset{\mathfrak{so}(5)}{4}
\,\, \overset{\mathfrak{sp}(-1)}{1} \,\, \overset{\mathfrak{so}(6)}{4}\,\, \overset{\mathfrak{sp}(-1)}{1} \,\,
\overset{\mathfrak{so}(5)}{4}\,\,\overset{\mathfrak{sp}(-2)}{1}\,\, :[5,1^3]\,.
\label{eq:SOexample}
\end{equation}
The anomaly polynomial reads:%
\begin{equation}
I_8=\frac{77}{4}c_{2}(R)^{2} - \frac{3}{8}c_{2}(R)p_{1}(T) + \frac{73}{2880}%
p_{1}(T)^{2} - \frac{49}{720}p_{2}(T).
\end{equation}

\item{\textbf{Example 2}}

For a second example, we deform the UV theory of three $-4$ curves by the pair
of orbits of $[4^{2}], [4^{2}]$ (our analysis does not distinguish between the two nilpotent orbits
associated with this partition). The formal theory:
\begin{equation}
[4^{2}]: \,\, \overset{\mathfrak{sp}(-3)}{1} \,\, \overset{\mathfrak{so}%
(4)}{4} \,\, \overset{\mathfrak{sp}(-1)}{1} \,\,
{\overset{\mathfrak{so}(8)}{4}} \,\, \overset{\mathfrak{sp}%
(-1)}{1} \,\, \overset{\mathfrak{so}(4)}{4}\,\,\overset{\mathfrak{sp}%
(-3)}{1}\,\, :[4^{2}]
\end{equation}
gives the following anomaly polynomial:%
\begin{equation}
\frac{463}{24}c_{2}(R)^{2} - \frac{17}{48}c_{2}(R)p_{1}(T) + \frac{73}%
{2880}p_{1}(T)^{2} - \frac{101}{1440}p_{2}(T).
\end{equation}
If we subtract off the contribution of one neutral hypermultiplet $I_{\textrm{neutral}} =
\frac{7p_{1}(T)^{2} - 4p_{2}(T)}{5760}$, we get the F-theory quiver anomaly polynomial:
\begin{equation}
I_{F} = I_{\textrm{formal}} - I_{\textrm{neutral}} = \frac{463}{24}c_{2}(R)^{2} - \frac{17}{48}%
c_{2}(R)p_{1}(T) + \frac{139}{5760}p_{1}(T)^{2} - \frac{97}{1440}p_{2}(T)
\end{equation}
which can be obtained from the F-theory quiver:
\begin{equation}
[4^{2}]: \,\, \underset{[N_f = 1/2]}{\overset{\mathfrak{su}(2)}{2}} \,\, \underset{[Sp(2)]}{\overset{\mathfrak{g}_{2}%
}{2}}\,\, \underset{[N_f = 1/2]}{\overset{\mathfrak{su}(2)}{2}} \,\,: [4^{2}].
\end{equation}

This result is actually quite surprising: the nilpotent deformations considered in these past two examples are related by triality of $SO(8)$. Indeed, their long F-theory quivers are identical, and they have identical anomaly polynomials, even though their formal quivers differ. However, we have just seen that their kissing cases actually differ! We have confirmed this surprising result via anomaly polynomial matching.

\item{\textbf{Example 3}}

Next, we consider a pair of cases with an anomaly polynomial mismatch.

\begin{itemize}

\item{3a}

Consider the theory with $\mu_L=[7, 1]$, $\mu_R = [4^2]$ on an SO(8) UV quiver with four $-4$ curves. The brane pictures for this example are depicted in figure \ref{subfig:newShort1}.
The theory has the following IR quiver:

\begin{equation}
  [7, 1]: \,\,2   \,\, \underset{[N_f = 1/2]}{\overset{\mathfrak{su}(2)}{2}} \,\, \underset{[SU(2)]}{\overset{\mathfrak{su}(3)}{2}}\,\, \underset{[N_f = 1]}{\overset{\mathfrak{su}(2)}{2}} \,\, :[4^2] \,.
\end{equation}

The curve carrying $SU(3)$ na\"ively has $\mf{so}(7)$ gauge algebra, but it is hit by two $\overline{A}$'s, one from the right and one from the left. As a result, the gauge algebra is reduced according to $\mathfrak{so}(7)\overset{2\overline{A}}{\rightarrow}\mathfrak{su}(3)$. This puts us in the situation of rule 2, shown in (\ref{eq:rule2}), so we expect an anomaly correction term of the form $(\Delta \alpha,\Delta\beta) =(1/24, 1/48)$.

Indeed, the formal quiver in this case is given by
\begin{equation}
[7,1]: \,\, \overset{\mathfrak{sp}(-3)}{1} \,\, \overset{\mathfrak{so}%
(3)}{4} \,\, \overset{\mathfrak{sp}(-2)}{1}\,\, \overset{\mathfrak{so}%
(5)}{4} \,\, \overset{\mathfrak{sp}(-1)}{1} \,\,
{\overset{\mathfrak{so}(7)}{4}} \,\, \overset{\mathfrak{sp}%
(-1)}{1} \,\, \overset{\mathfrak{so}(4)}{4}\,\,\overset{\mathfrak{sp}%
(-3)}{1}\,\, :[4^{2}] \,.
\end{equation}
The anomaly polynomial of the F-theory quiver is given by
\begin{equation}
I_{F} = \frac{1331}{60}c_{2}(R)^{2} - \frac{5}{24}%
c_{2}(R)p_{1}(T) + \frac{37}{1440}p_{1}(T)^{2} - \frac{31}{360}p_{2}(T),
\end{equation}
which is indeed the same as $I_{\textrm{formal}}-c_2(R)^2/24-c_2(R) p_1(T)/48- 2 I_{\textrm{neutral}}$.

\item{3b}

Consider the $SO(8)$ theory with nilpotent deformations $[3, 2^2, 1]$ and $[2^4]$ on a UV quiver with a single $-4$ curve. The F-theory quiver is given by:
\begin{equation}
  [3, 2^2, 1]: \,\, \underset{[SU(6)]}{\overset{\mathfrak{su}(3)}{2}} \,\,  :[2^4] \,.
\end{equation}
Here, we again have one anti-brane from both the left and the right, which collide on the $-4$ curve and reduce it as $\mathfrak{so}(7)\overset{2\overline{A}}{\rightarrow}\mathfrak{su}(3)$. The formal quiver is given by
\begin{equation}
[3,2^2,1]: \,\, \overset{\mathfrak{sp}(-2)}{1} \,\, \overset{\mathfrak{so}%
(7)}{4} \,\, \overset{\mathfrak{sp}(-2)}{1}\,\, :[2^{4}]\,.
\end{equation}
The anomaly polynomial of the F-theory quiver is given by
\begin{equation}
I_{F} = \frac{47}{24}c_{2}(R)^{2} - \frac{7}{48}%
c_{2}(R)p_{1}(T) + \frac{31}{1920}p_{1}(T)^{2} - \frac{13}{480}p_{2}(T)\,,
\end{equation}
which is equal to $I_{\textrm{formal}}-c_2(R)^2/24-c_2(R) p_1(T)/48- 4 I_{\textrm{neutral}}$, as expected from (\ref{eq:rule2}).

\end{itemize}
Note that the rule from (\ref{eq:rule2}) has worked correctly for both examples, despite the difference in size of their respective quivers.

\item{\textbf{Example 4}}

As a final example, let us consider a pair of theories with a similar mismatch in the anomaly polynomial but  different UV gauge groups.

\begin{itemize}

\item{4a}

 First, we consider the theory with $SO(8)$ UV gauge groups, nilpotent deformations $[7, 1]$ and $[5, 3]$, and a theory with four $-4$ curves, whose brane diagrams are depicted in figure \ref{subfig:newShort2}. The IR quiver takes the form:
\begin{equation}
  [7, 1]: \,\, 2 \,\, \underset{[N_f = 3/2]}{\overset{\mathfrak{su}(2)}{2}} \,\, {\overset{\mathfrak{su}(2)}{2}}\,\, {\overset{\mathfrak{su}(2)}{2}} \,\, [SU(2)\times SU(2)] \,\, :[5, 3]\,.
\end{equation}
Here, the middle $\mf{su}(2)$ gauge algebra comes from two anti-branes acting on an $\mf{so}(6)$. Per rule 3 of (\ref{eq:rule3}), we expect a mismatch of the form $(\Delta\alpha,\Delta\beta) = (1/12,1/24)$. Indeed, the formal quiver is given by
\begin{equation}
[7,1]: \,\, \overset{\mathfrak{sp}(-3)}{1} \,\, \overset{\mathfrak{so}%
(3)}{4} \,\, \overset{\mathfrak{sp}(-2)}{1}\,\, \overset{\mathfrak{so}%
(5)}{4} \,\, \overset{\mathfrak{sp}(-1)}{1} \,\,
{\overset{\mathfrak{so}(6)}{4}} \,\, \overset{\mathfrak{sp}%
(-1)}{1} \,\, \overset{\mathfrak{so}(4)}{4}\,\,\overset{\mathfrak{sp}%
(-3)}{1}\,\, :[5,3]\,.
\end{equation}
The anomaly polynomial of the F-theory quiver is given by
\begin{equation}
I_{F} = \frac{1943}{120}c_{2}(R)^{2} - \frac{5}{48}%
c_{2}(R)p_{1}(T) + \frac{47}{1920}p_{1}(T)^{2} - \frac{41}{480}p_{2}(T),
\end{equation}
which is indeed the same as $I_{\textrm{formal}}-c_2(R)^2/12-c_2(R) p_1(T)/24- 2 I_{\textrm{neutral}}$.

\item{4b}

 Finally, consider the $SO(10)$ theory with nilpotent deformations $[5^2]$, $[3^2, 2^2]$ on a quiver with two $-4$ curves. This gives:
\begin{equation}
  [5^2]: \,\, [SU(2)] \,\, {\overset{\mathfrak{su}(2)}{2}}\,\, {\overset{\mathfrak{su}(2)}{2}} \,\, [SU(2)\times SU(2)] \,\, :[3^2, 2^2] \,.
\end{equation}
The $\mf{su}(2)$ gauge algebra on the right-hand side again comes from two anti-branes acting on $\mf{so}(6)$. The formal quiver is given by
\begin{equation}
[5^2]: \,\, \overset{\mathfrak{sp}(-3)}{1} \,\, \overset{\mathfrak{so}%
(4)}{4} \,\, \overset{\mathfrak{sp}(-1)}{1}\,\, \overset{\mathfrak{so}%
(6)}{4} \,\, \overset{\mathfrak{sp}(-2)}{1}\,\, :[3^2,2^2] \,.
\end{equation}
The anomaly polynomial of the F-theory quiver is given by
\begin{equation}
I_{F} = \frac{23}{6}c_{2}(R)^{2} - \frac{1}{12}%
c_{2}(R)p_{1}(T) + \frac{11}{720}p_{1}(T)^{2} - \frac{2}{45}p_{2}(T),
\end{equation}
which is indeed the same as $I_{\textrm{formal}}-c_2(R)^2/12-c_2(R) p_1(T)/24- 4 I_{\textrm{neutral}}$, as expected from (\ref{eq:rule3}).

\end{itemize}
Note that the rule from (\ref{eq:rule3}) has worked correctly for both examples, despite the difference in size of their respective quivers as well as their UV gauge groups.

\end{itemize}

Further examples of anomaly polynomial matching can be found in the catalogs in Appendix \ref{apdx:shortQuiverCatalogs}.

\subsection{Nilpotent Hierarchy of Short Quivers \label{subsec:DoubleNilpotentHierarchy}}

Using our analysis above, we now determine a partial ordering for 6D SCFTs based on
pairs of nilpotent orbits, which works in both long and short quivers. We refer to this as a ``double Hasse diagram,'' since it generalizes the independent Hasse diagrams realized by nilpotent orbits on each side of a long quiver (see \cite{Heckman:2016ssk, Bourget:2019aer}) to the case of a short quiver, where the nilpotent deformations overlap. We will see that as we reduce the length of the quiver, several nilpotent orbits will end up generating the same IR fixed point. Said another way, different pairs of nilpotent orbits actually give rise to the same IR theory.

Constructing the double Hasse diagrams proceeds in two steps. First we apply the product order to the tuple of left and right partitions $\mu_L$ and $\mu_R$. It is defined by $(\mu_L, \mu_R) \preceq (\nu_L, \nu_R)$ which holds if and only if $\mu_L \preceq \nu_L$ and $\mu_R \preceq \nu_R$. However, because several deformations in the UV can flow to the same IR theory, we refine this partial ordering in the second step by merging all partitions which result in the same IR quiver. We obtain the same result from a microscopic perspective by appropriately adding strings to the left and right sides of the string junction picture, exactly as we did for the long quivers.

\subsubsection{Example: $SU(4)$}

As a first example, we consider an $SU(4)$ double Hasse diagram. We begin with the UV theory:%
\begin{equation}
[1^{4}]: \underset{[SU(4)]}{\overset{\mathfrak{su}(4)}{2}} \,
\overset{\mathfrak{su}(4)}{2 }\, \underset{[SU(4)]}{\overset{\mathfrak{su}%
(4)}{2}} \, \,:[1^{4}]\,.
\end{equation}

\begin{figure}
\includegraphics{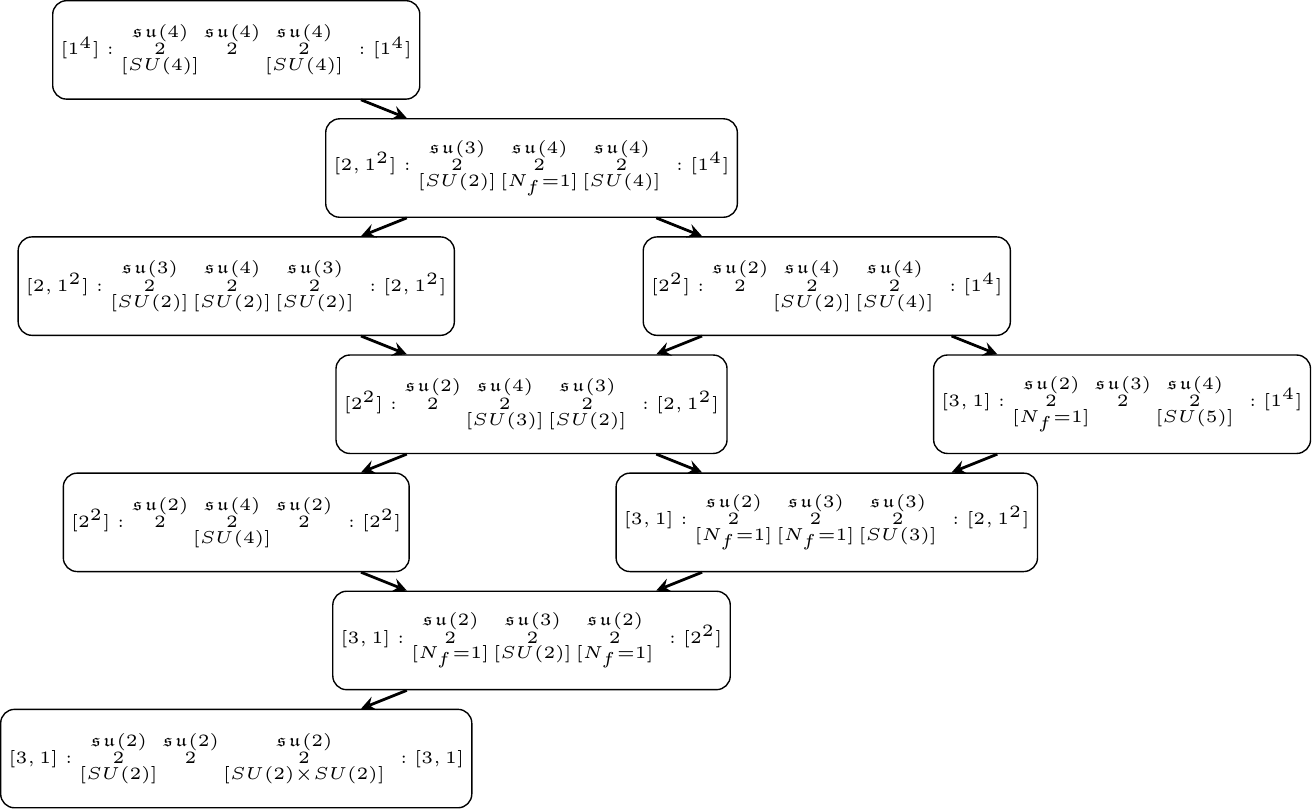}
\caption{Half of the double Hasse diagram of $SU(4)$ short quivers. The full diagram is obtained by reflection across the left-most nodes, as the quivers can always be flipped under the reflection $\mu_L \leftrightarrow \mu_R$.}
\label{fig:su4ShortFlowsThree}%
\end{figure}

Then we turn on nilpotent deformations on both sides, as in the single-sided versions that were plotted in \cite{Heckman:2016ssk}. Note that $SU(4)$ only has
five nilpotent orbits - $[1^{4}], [2, 1^{2}], [2^{2}], [3, 1], [4]$, but the
$[4]$ orbit is prohibited on $N_{-2} = N_{T} = 3$ curves by the Hanany-Witten moves constraint of equation (\ref{eq:HWconstraintSU}).
We are then left with the double Hasse diagram of figure \ref{fig:su4ShortFlowsThree}. This generalizes straightforwardly to all $SU(N)$  quivers.

\subsubsection{Example: $SO(8)$}
\begin{figure}
\includegraphics{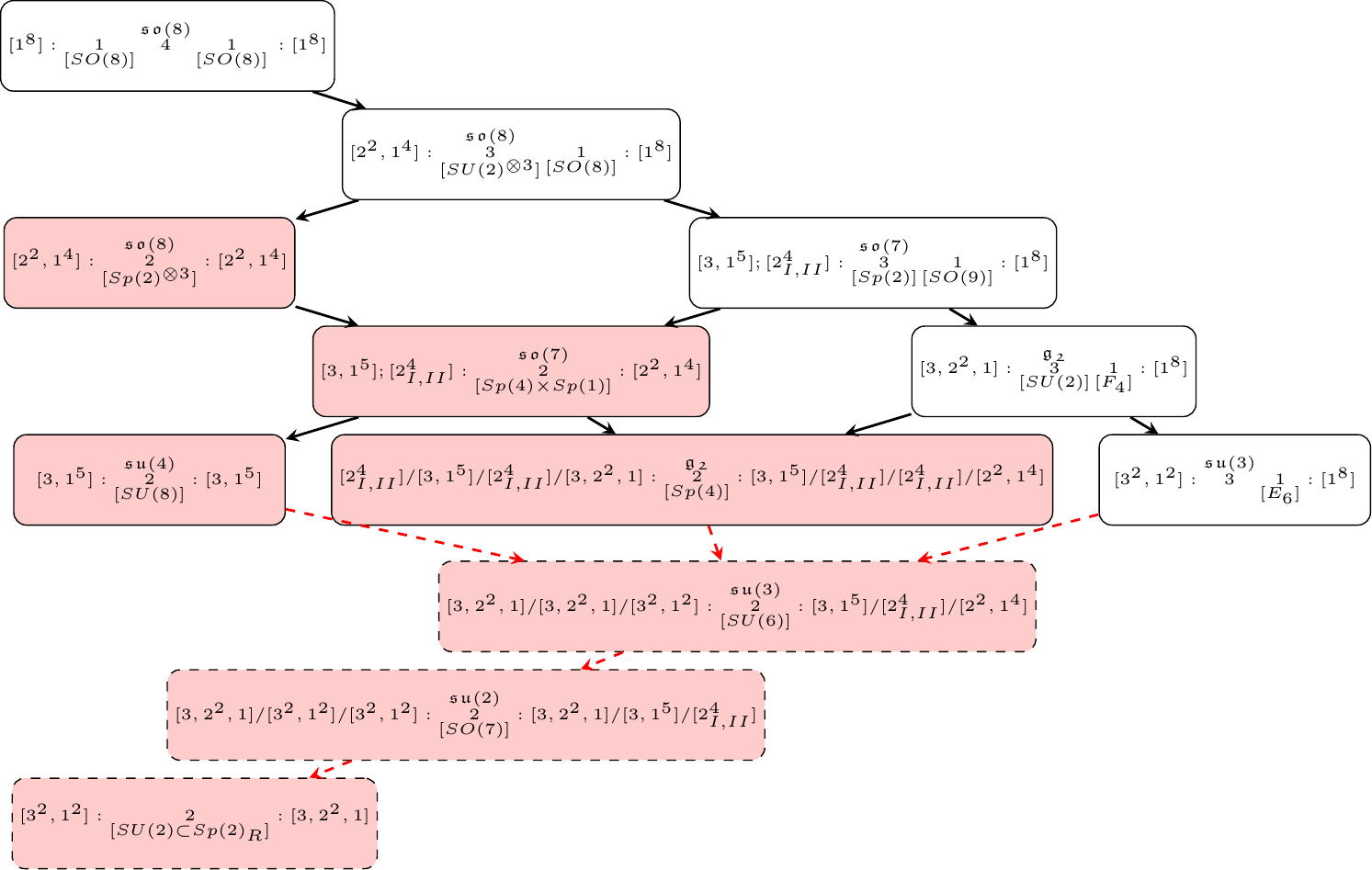}
\caption{Double Hasse diagram for $SO(8)$ short quiver theories with one $-4$ curve in the UV theory. This diagram is again half of a full figure, following the same convention as in figure \ref{fig:su4ShortFlowsThree}. ``Kissing'' configurations are highlighted in red. For concision, several pairs of nilpotent deformations that yield the same IR theory are written in the same box. We separate partitions with semicolons $\mu_L$; $\nu_L$ -- $\mu_R$; $\nu_R$ to denote all possible combinations $\mu_L$ -- $\mu_R$, $\mu_L$ -- $\nu_R$, $\nu_L$ -- $\mu_R$, and $\nu_L$ -- $\nu_R$. On the other hand, slashes denote one-to-one pairings, so $\mu_L/\nu_L$ -- $\mu_R/\nu_R$ means $\mu_L$ -- $\mu_R$ and $\nu_L$ -- $\nu_R$ only. We also mark theories with $(\Delta \alpha, \Delta \beta)$ anomaly mismatches with dashed frames and draw the RG flows towards these cases using red dashed arrows. Note that, whenever there is a dashed frame with more than one possible pair of nilpotent orbits, at least one pair of nilpotent orbits out of them has $(\Delta \alpha, \Delta \beta)$ anomaly mismatch, and in some cases not all of them have such mismatches. See table \ref{table:SO8tangentialShortQuiver} for more details of anomaly mismatches in $SO(8)$ short quiver theories.}%
\label{fig:so8ShortFlowsOne}%
\end{figure}
\begin{figure}
\includegraphics{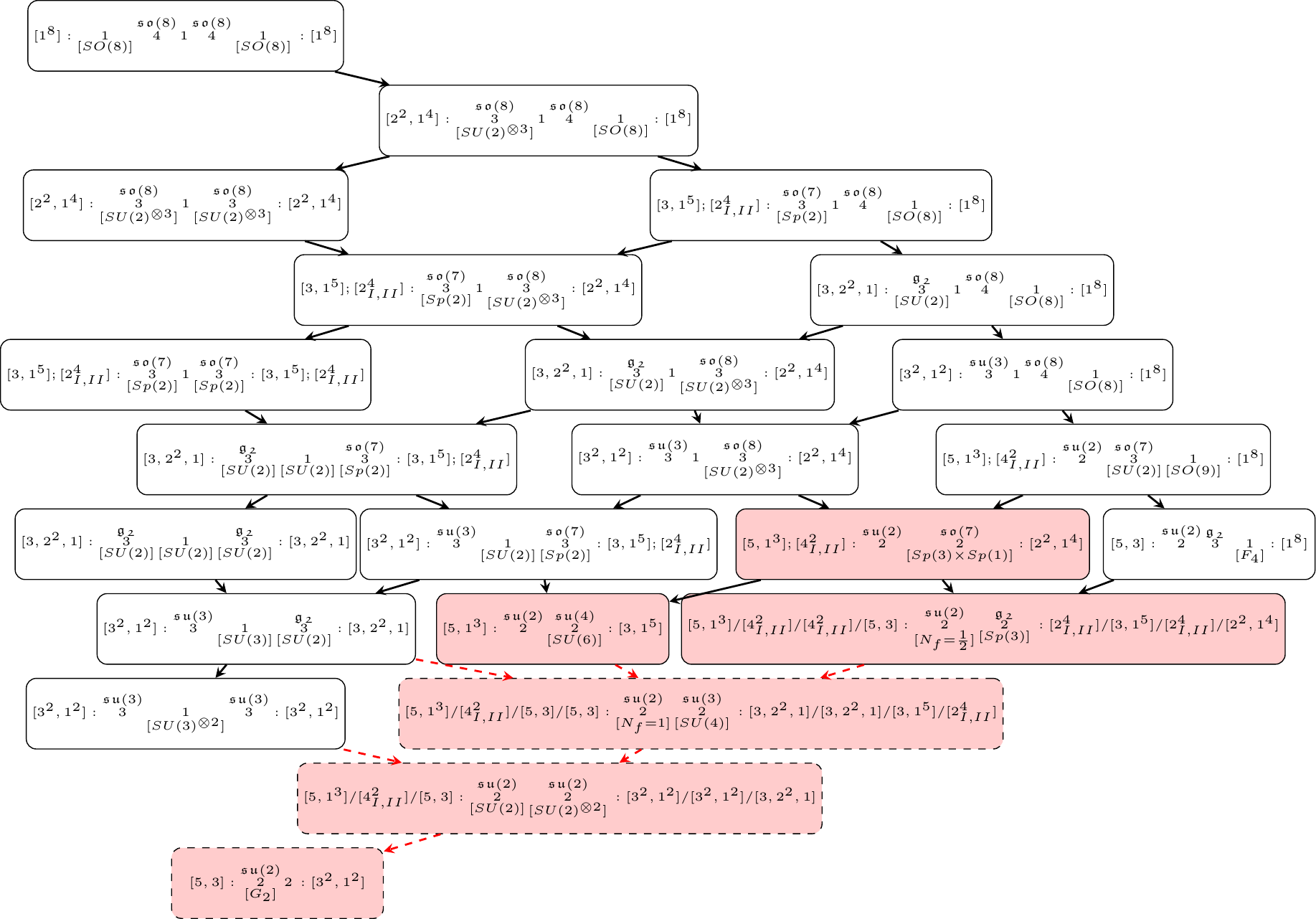}
\caption{Double Hasse diagram of $SO(8)$ short quiver theories over two $-4$ curves in the UV theory. The notation is the same as in figure \ref{fig:so8ShortFlowsOne}.}%
\label{fig:so8ShortFlowsTwo}%
\end{figure}
\begin{figure}[ptb] \hspace{-1cm}
\includegraphics[width = \textwidth + 2cm]{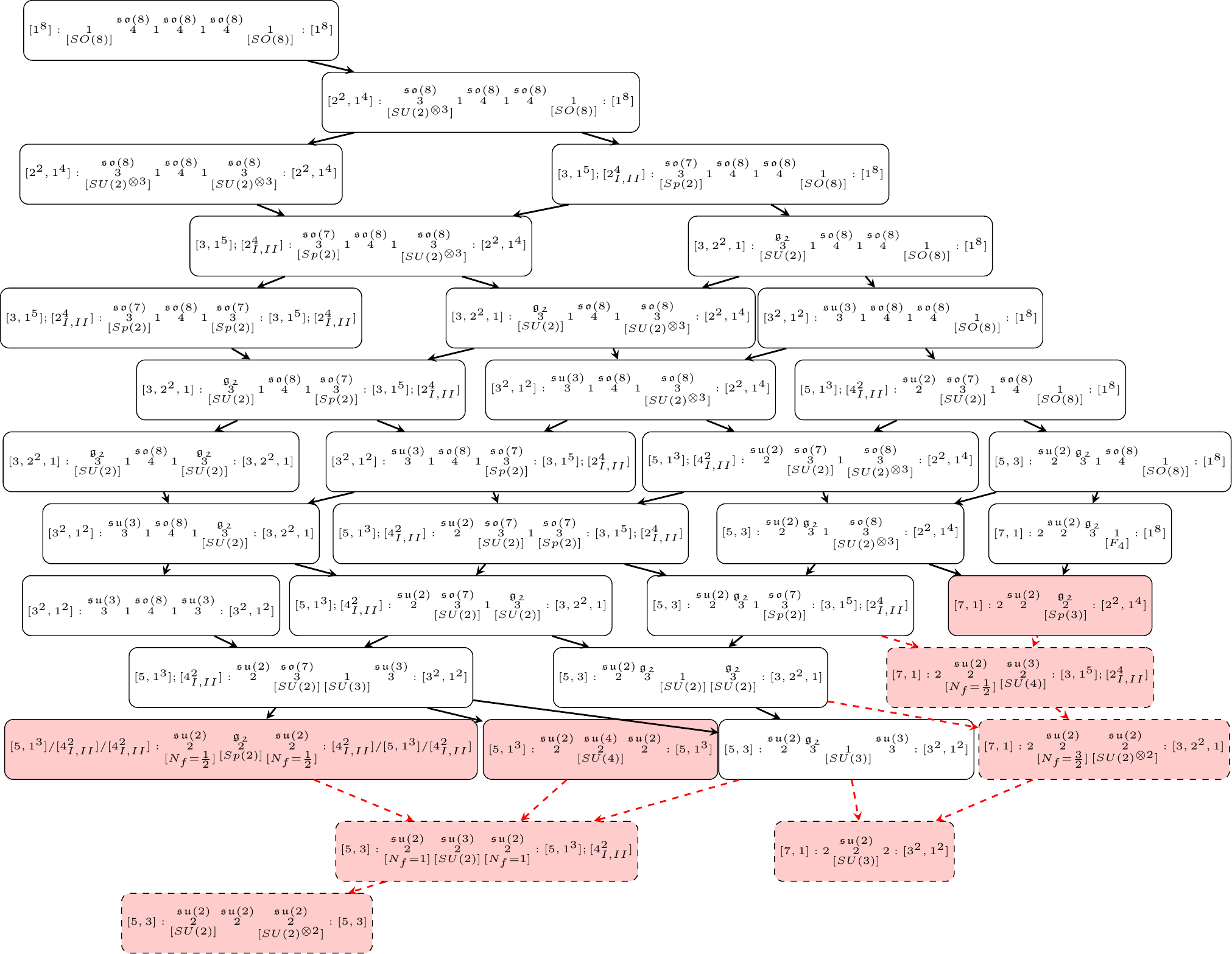}
\caption{Double Hasse diagram of $SO(8)$ short quiver theories over three $-4$ curves in the UV theory. The notation is the same as in figure \ref{fig:so8ShortFlowsOne}.}%
\label{fig:so8ShortFlowsThree}
\end{figure}

Next we look at the double Hasse diagrams for the $SO(8)$ UV theories. For $SO(2N), N>4$ the story is similar, but we choose to illustrate with $SO(8)$ for simplicity. We look at UV quivers with one, two and three $-4$ curves respectively:%
\begin{align}
[1^{8}]: \underset{[SO(8)]}{1} \,  &  {\overset{\mathfrak{so}(8)}{4}} \,
\underset{[SO(8)]}{1} \,:[1^{8}]\\
[1^{8}]: \underset{[SO(8)]}{1} \, {\overset{\mathfrak{so}(8)}{4}} \  &  1 \,
{\overset{\mathfrak{so}(8)}{4}} \, \underset{[SO(8)]}{1} \,:[1^{8}]\\
[1^{8}]: \underset{[SO(8)]}{1} \, {\overset{\mathfrak{so}(8)}{4}} \, 1 \,  &
{\overset{\mathfrak{so}(8)}{4}} \, 1 \, {\overset{\mathfrak{so}(8)}{4}} \,
\underset{[SO(8)]}{1} \,:[1^{8}] \,.
\end{align}
The associated double Hasse diagrams are shown in figures \ref{fig:so8ShortFlowsOne}, \ref{fig:so8ShortFlowsTwo}, and \ref{fig:so8ShortFlowsThree}. We see that as the number of curves decreases, the Hanany-Witten constraints
forbid more and more deformations that were allowed in the long quiver. In each diagram, we highlight in red the ``kissing'' configurations which have all of their $-1$ curves blown-down. We also use dashed lines to indicate theories with an anomaly polynomial mismatch with their associated formal quiver, and we denote flows to these theories with dashed lines.

It is worth pausing here to elaborate on a surprising point noted in example 2 of section \ref{ssec:examples} above: $SO(8)$ nilpotent orbits related by triality always give the same long quiver theory, but they do not not always generate the same short quiver theory. When they do yield the same quiver they are drawn in the same box, but when they give rise to distinct theories, we use separate boxes to denote them.

As an example in which the two disagree, consider the short quivers $[3,1^5]$ -- $[3,1^5]$ and $[2^4]$ -- $[3,1^5]$ on a UV quiver with a single $-4$ curve.
These yield respectively,
\begin{align}
[3, 1^5]:\,\, \overset{\mathfrak{su}_4}{2} \,\, [SU(8)] \,\, : [3, 1^5], \\
[2^4]:\,\, \overset{\mathfrak{g}_2}{2} \,\, [Sp(4)] \,\, : [3, 1^5]\,.
\end{align}
For the first case, with $[3,1^5]$ -- $[3,1^5]$, there are two double strings stretching on the middle curve, so the original $\mathfrak{so}_8$  is Higgsed to $\mathfrak{so}_6 \simeq \mathfrak{su}_4$.
On the other hand the quiver with $[2^4]$ -- $[3,1^5]$ has a single double string stretching on the middle curve (coming from the right deformation) and one extra $\overline{A}$ coming from the left, so the original $\mathfrak{so}_8$ is Higgsed to $\mathfrak{so}_7 \overset{\overline{A}}{\rightarrow} \mathfrak{g}_2$.

The rules that lead us to these quivers can be verified in other examples as well. For instance, consider an $SO(10)$ theory with three $-4$ curves in the UV quiver, deformed by $\mu_L = [7, 1^3]$, $\mu_R = [5, 3, 1^2]$. The resulting theory is given by
\begin{equation}
  [7,1^3]: {\overset{\mathfrak{su}(2)}2}  \,\,   \underset{[SU(4)]}{{\overset{\mathfrak{su}(4)}2}}  \,\,{\overset{\mathfrak{su}(2)}2}:[5,3,1^2] \,.
\end{equation}
In the brane picture, the $\mf{su}(4)$ on the middle $-2$ curve comes from two double strings, one each from the left and right, exactly parallel to the $[3,1^5], [3,1^5]$ case above.

Similarly,
for $\mu_L = [7, 3]$, $\mu_R = [5, 3, 1^2]$, the kissing theory is given by
\begin{equation}
  \underset{[N_f = 1/2]}{\overset{\mathfrak{su}(2)}2}  \,\,   \underset{[Sp(2)]}{{\overset{\mathfrak{g_{2}}}2}}  \,\,\underset{[N_f = 1/2]}{\overset{\mathfrak{su}(2)}2} \,.
\end{equation}
The second $-2$ curve now has a $\mathfrak{g}_2$ gauge algebra, which in the brane picture comes from a single double string coming from one side and an extra $\overline{A}$ coming from the other, just as in the case of the $[2^4]$, $[3,1^5]$ theory above.

 This example nicely illustrates the utility of the string junction approach for determining the nilpotent hierarchy of short quivers, as the short quivers in two cases (which are different) cannot be determined unambiguously from their associated long quivers alone (which are identical).

 Finally, it is also worth noting that additional RG flows have opened up in these short quivers that were not available in the case of long quivers. For instance, in an $SO(8)$ long quiver of fixed size, there is no RG flow from the theory with $\mu_L=[3,2^2,1], \mu_R=[1^8]$ to the theory with $\mu'_L = \mu'_R = [2^4]$, because although $\mu_R \preceq \mu_R'$, we also have $\mu_L \npreceq \mu_L'$.

 However, for a sufficiently-short quiver with these nilpotent orbits, there is a flow from the former to the latter. In particular, there is a flow from
 \begin{equation}
[3,2^2,1]:    \underset{[Sp(1)]}{{\overset{\mathfrak{g}_2}3}}  \,\, \underset{[F_4]}1:[1^8]
\end{equation}
to the theory
 \begin{equation}
[2^4]:    \underset{[Sp(4)]}{{\overset{\mathfrak{g}_2}2}}  :[2^4] .
\label{eq:ex}
\end{equation}
This is related to the fact that short quivers are often degenerate: in particular, the theory of (\ref{eq:ex}) can also be realized by the nilpotent orbits $\mu_L' = [3,2^2,1]$, $\mu_R' = [2^2,1^4]$, which \emph{do} satisfy $\mu_R \preceq \mu_R'$, $\mu_L \preceq \mu_L'$.

\subsection{Flavor Symmetries \label{subsec:globalSymmetryShortQuiver}}

The structure of nilpotent orbits also provides a helpful guide to the analysis of flavor symmetries in 6D SCFTs \cite{Heckman:2016ssk}.
Given a nilpotent orbit, the commutant subalgebra specifies an unbroken symmetry inherited from the UV. For the classical groups, the resulting flavor symmetry algebra associated with a given nilpotent orbit is given simply in terms of the data of partition (see e.g. \cite{Chacaltana:2012zy}):
\begin{equation}
\begin{array}{l@{\quad\text{when}\ }l}
\mathfrak{s}[ \oplus_{i} \mathfrak{u}(r_i)  ] & \mathfrak{g}=\su(N) ,\\
 \op{i\, \text{odd}}\mathfrak{so}(r_i) \oplus \op{i\, \text{even}} \mathfrak{sp}(r_i/2)  & \mathfrak{g}=\mathfrak{so}(2N+1)\, \text{or}\, \mathfrak{so}(2N), \\
\op{i\, \text{odd}}\mathfrak{sp}(r_i/2) \oplus \op{i\, \text{even}} \mathfrak{so}(r_i)  & \mathfrak{g}=\mathfrak{sp}(N). \\
\end{array}\label{classicalflavour}
\end{equation}

In a long quiver, the flavor symmetry inherited from the
parent UV theory is thus given by the products of these flavor symmetries.
For short quivers, on the other hand, we typically observe enhancements of the flavor symmetry whenever flavors coming from the left and from the right end up sharing the same node. As usual, this is easiest to see in theories with $\mf{su}$ gauge symmetries. Here, if flavor symmetries $[SU(m)]_L$ and $[SU(n)]_R$ share the same node, the symmetry enhances from $[SU(m)] \times [SU(n)]$ to $[SU(m+n)]$. For $SO/Sp$ quivers without any small instanton transitions, flavor symmetries of $[SO(m)]_L$ and $[SO(n)]_R$ get enhanced to $[SO(m+n)]$, and similarly for the $Sp$ cases. To illustrate this fact, we start with the theory
\begin{equation}
  [3, 2]:\,\, \overset{\mathfrak{su}(2)}{2} \,\,
\underset{[N_f = 1]}{\overset{\mathfrak{su}(4)}{2}} \,\,
\underset{[N_f = 1]}{\overset{\mathfrak{su}(5)}{2}}
\,\,\underset{[SU(2)]}{\overset{\mathfrak{su}(5)}{2}} \,\,
\underset{[N_f = 1]}{\overset{\mathfrak{su}(3)}{2}}\,\,: [2^{2}, 1] \,.
\end{equation}
We can then shorten the quiver to have only $4$ curves:
\begin{equation}
[3, 2]:\,\, \overset{\mathfrak{su}(2)}{2} \,\,
\underset{[N_f = 1]}{\overset{\mathfrak{su}(4)}{2}} \,\, \underset{[
SU(3)]}{\overset{\mathfrak{su}(5)}{2}} \,\,
\underset{[N_f = 1]}{\overset{\mathfrak{su}(3)}{2}}\,\,: [2^{2}, 1].
\end{equation}
After this first step, we already see an enhancement: the $[SU(3)]$ factor comes from two components: $SU(2)$ from the left and $U(1)$ from the right. Removing yet another curve, we have:
\begin{equation}
[3, 2]:\,\, \overset{\mathfrak{su}(2)}{2} \,\,
\underset{[SU(3)]}{\overset{\mathfrak{su}(4)}{2}} \,\,
\underset{[SU(2)]}{\overset{\mathfrak{su}(3)}{2}}\,\,: [2^{2}, 1].
\end{equation}
Here the enhancement is even greater. Indeed, both of the $[SU(3)]$ and $[SU(2)]$ flavors come from similar enhancements.

Ignoring Abelian factors, enhancements occur in the following two cases:
\begin{itemize}
\item  When flavor symmetries coming from the left and from the right end up sharing the same node.

\item When a $-1$ curve has its surrounding gauge symmetry lowered by short quiver effects (as detailed below). This can happen either for a $-1$ at the edge of the quiver or in the interior.

\end{itemize}
As a first example of the former, consider the theory with nilpotent orbits $[3, 1^5]$ and $[2^4]$ on an $SO(8)$ UV quiver with two $-4$ curves:
  \begin{equation}
    \underset{[Sp(4)]}{\overset{\mathfrak{g_{2}}}2}.
  \end{equation}
  We see that the flavor symmetry $Sp(2) \times Sp(2)$ present in the case of a long quiver has been enhanced to $Sp(4)$.

As another example of the former case, consider the theory with nilpotent orbits $\mu_L=\mu_R=[3,1^{2N-3}]$ on an $SO(2N)$ quiver with one $-4$ curve, which can equivalently be regarded as an $SO(2N-3)$ quiver with $\mu_L=\mu_R=[1^{2N-3}]$:
\begin{equation}
  [SO(2N-2)] \ \ \overset{\mathfrak{sp}(N-5)}{1} \ \ \overset{\mathfrak{so}(2N-2)}{4} \ \ \overset{\mathfrak{sp}(N-5)}{1} \ \ [SO(2N-2)] \,.
\end{equation}
We see that the flavor symmetries of the left and right have been enhanced from $SO(2N-3)$ to $SO(2N-2)$.

Finally, as an example of the latter case, consider the theory of nilpotent orbits $[7, 1]$ and $[1^8]$ on an $SO(8)$ UV quiver with three $-4$ curves:
    \begin{equation}
      2 \, \overset{\ksu(2)}2 \, {\overset{\mathfrak{g_{2}}}3}  \, {1} \,\, [F_4].
    \end{equation}
    The flavor symmetry on the right has been enhanced from $SO(8)$ to $F_4$.

In all cases, we find that the flavor symmetry of a short quiver is enhanced relative to the flavor symmetry of a long quiver associated with the same nilpotent deformations.

\section{Conclusions \label{sec:CONC}}

In this paper we have developed general methods for determining
the structure of Higgs branch RG\ flows in 6D\ SCFTs. In particular, we have analyzed several
aspects of vevs for \textquotedblleft conformal
matter.\textquotedblright\ We have seen that the entire nilpotent cone of a simple Lie algebra,
including its structure as a partially ordered set can be obtained from simple
combinatorial data connected with string junctions stretched between bound
states of $7$-branes. Recombination moves involving
intersecting branes as well as brane / anti-brane pairs fully determine the Higgs branch of
quiver-like 6D\ SCFTs with classical gauge algebras. An added benefit of
this approach is that it also extends to short quiver-like theories where
Higgsing from different nilpotent orbits leads to correlated symmetry breaking constraints.
In the remainder of this section we discuss some other potential areas for future investigation.

In this paper we have primarily focused on Higgsing in quiver-like theories with classical algebras.
We have also seen that we can understand the nilpotent cone of the E-type algebras using multi-pronged
string junctions. This suggests that by including additional $7$-brane
recombination effects, it should be possible to cover
these cases as well. This would provide a nearly complete picture of Higgs
branch flows for 6D\ SCFTs engineered via F-theory.

This work has primarily focused on the case of 6D\ SCFTs in which Higgs
branch deformations can be understood in terms of localized T-brane
deformations of a non-compact $7$-brane. We have already noted how
\textquotedblleft semi-simple\textquotedblright\ deformations fit into
this picture. The other class of Higgs branch deformations which appear quite
frequently involve discrete group homomorphisms from finite subgroups of
$SU(2)$ into $E_{8}$ \cite{Heckman:2015bfa, Mekareeya:2017jgc, Frey:2018vpw}. Obtaining an analogous correspondence in this
case would cover another broad class of Higgs branch deformations in 6D SCFTs.

The main emphasis of this work has centered on combinatorial data connected
with Higgs branch flows and $7$-brane recombination. That being said, it is
also clear that explicit complex structure deformations of the associated
F-theory models should describe some of these deformations as well, a point which
deserves to be clarified.

Lastly, the overarching aim in this work has been to better understand the
structure of all possible 6D\ RG flows obtained from deformations of different
conformal fixed points. The fact that we now have a fairly systematic way to
also understand deformations of short quivers suggests that the time may be
ripe to obtain a full classification of such RG\ flows.

\section*{Acknowledgments}

We thank K. Harigaya and A. Tomasiello for helpful discussions. JJH, TBR\ and
ZYH\ thank the Simons Center for Geometry and Physics program on the Geometry
and Physics of Hitchin Systems as well as the 2019 Summer workshop at the Simons Center for
Geometry and Physics for hospitality during the completion of part
of this work. FH is partially supported by the Spanish Government Research
Grant FPA2015-63667-. The work of JJH is supported by NSF CAREER grant
PHY-1756996. The work of TR is supported by the Carl P. Feinberg Founders
Circle Membership and by NSF grant PHY-1606531.

\appendix

\section{Partial Ordering for Nilpotent Orbits\label{sec:Nilpotents}}

In this Appendix, we review some aspects of nilpotent orbits of simple Lie algebras
and their partial ordering. We refer the interested reader to \cite{collingwood1993nilpotent} for further details.

The general linear group $GL(N,\mathbb{C})$ acts on its Lie algebra
$\mathfrak{gl}_{n}$ of all complex $n \times n$ matrices by conjugation; the
orbits are similarity classes of matrices. The theory of the Jordan
form gives a satisfactory parametrization of these classes and allows us to
regard two kinds of classes as distinguished: those represented by diagonal
matrices, and those represented by strictly upper triangular matrices, i.e., nilpotent matrices. There are only finitely many similarity classes of nilpotent
matrices, which are labeled by partitions of
of $n$. There is a similar parametrization of nilpotent orbits by partitions in any classical semisimple Lie algebra, with some additional restrictions imposed.

Semi-simple orbits are parametrized by points in a fundamental domain for the
action of the Weyl group on a Cartan subalgebra. In particular, there are
infinitely many semi-simple orbits.

\subsection{Weighted Dynkin Diagrams}

Associated to each nilpotent orbit is a unique (completely invariant) weighted
Dynkin diagram \cite{collingwood1993nilpotent}. In general, the Dynkin labels $\alpha_{i}(H)$, $1 \leq i \leq rank(G)$ of a weighted Dynkin diagram
are defined by the commutator relation:
\begin{align}
[H,X_{i}] = \alpha_{i}(H) X_{i},
\end{align}
where the $X_{i}$ are the raising operators corresponding to the positive
simple roots of $\mathfrak{g}$, and $H$ is directly constructed from the
partition $\mathbf{d}=[d_{1},\cdots, d_{n}]$ associated with the nilpotent orbit
as follows:
\begin{align}
H_{[d_{1},\cdots,d_{n}]} =
\begin{pmatrix}
D(d_{1}) & 0 & \cdots & 0\\
0 & D(d_{2}) & \cdots & 0\\
\vdots & \vdots & \ddots & \vdots\\
0 & 0 & \cdots & D(d_{k})
\end{pmatrix}
,
\end{align}
where
\begin{align}
D(d_{i}) =
\begin{pmatrix}
d_{i} -1 & 0 & 0 & \cdots & 0 & 0\\
0 & d_{i}-3 & 0 & \cdots & 0 & 0\\
0 & 0 & d_{i}-5 & \cdots & 0 & 0\\
\vdots & \vdots & \vdots & \ddots & \vdots & \vdots\\
0 & 0 & 0 & \cdots & -d_{i} + 3 & 0\\
0 & 0 & 0 & \cdots & 0 & -d_{i}+1
\end{pmatrix}
\end{align}
The nilpositive element $X$ in the $\{H, X, Y\}$ Jacobson-Morozov
standard triple is then given by:
\begin{align}
X_{[d_{1},\cdots,d_{n}]} =
\begin{pmatrix}
J^{+}(d_{1}) & 0 & \cdots & 0\\
0 & J^{+}(d_{2}) & \cdots & 0\\
\vdots & \vdots & \ddots & \vdots\\
0 & 0 & \cdots & J^{+}(d_{k})
\end{pmatrix}
,
\end{align}
where now
\begin{align}
J^{+}_{i,j}(d_{m})  &  = \delta_{i+1,j} \sqrt{i d_{m}-i^{2}} =
\begin{pmatrix}
0 & \sqrt{d_{m} -1} & 0 & 0 & \cdots & 0 & 0\\
0 & 0 & \sqrt{2 d_{m}-4} & 0 & \cdots & 0 & 0\\
0 & 0 & 0 & \sqrt{3d_{m}-9} & \cdots & 0 & 0\\
\vdots & \vdots & \vdots & \ddots & \vdots & \vdots & \vdots\\
0 & 0 & 0 & \cdots & 0 & \sqrt{ 2 d_{m} -4} & 0\\
0 & 0 & 0 & \cdots & 0 & 0 & \sqrt{d_{m} -1}\\
0 & 0 & 0 & \cdots & 0 & 0 & 0
\end{pmatrix}
\end{align}
and similarly the nilnegative element $Y$ is given by:
\begin{align}
Y_{[d_{1},\cdots,d_{n}]} =
\begin{pmatrix}
J^{-}(d_{1}) & 0 & \cdots & 0\\
0 & J^{-}(d_{2}) & \cdots & 0\\
\vdots & \vdots & \ddots & \vdots\\
0 & 0 & \cdots & J^{-}(d_{k})
\end{pmatrix}
,
\end{align}
where $J^{-} = (J^{+})^{\dagger}$ so that $Y=X^{\dagger}$:
\begin{align}
J^{-}_{i,j}(d_{m}) = \delta_{j+1,i} \sqrt{j d_{m}-j^{2}}.
\end{align}

Direct matrix multiplication then gives the required commutation relations:
\begin{align}
[X,Y]  &  =H,\nonumber\\
[H,X]  &  =2X,\nonumber\\
[H,Y]  &  =-2Y.
\end{align}

This nilpositive matrix is similar to the nilpotent matrix
$X_{\mathcal{O}}$ we used to generate the partition in the first place.
Indeed, any two matrices with the same Jordan block decomposition (and therefore
corresponding to the same partition) are similar matrices and thus belong to
the same nilpotent orbit.

As a summary, the following are equivalent:

\begin{itemize}
\item A nilpotent orbit

\item A given Bala-Carter label

\item A corresponding set of simple roots generating the Levi subalgebra and
one or more positive roots ($X_{\alpha_{i}}$) for the distinguished orbits

\item A corresponding partition

\item An $\{H, X, Y\}$ Jacobson-Morozov standard triple, where $H$ is
explicitly built out of the partitions as described above and $X$ is similar
to the sum of the $X_{\alpha_{i}}$ specified in our brane diagrams.

\item A Weighted Dynkin diagram with weights $\alpha_{i}(H)$ given by the
relation $[H,X_{i}] = \alpha_{i}(H) X_{i}$ for $H$ defined above in the
standard Jacobson-Morozov triple and the $X_{i}$ being the positive simple roots.
\end{itemize}

Finally, we remark that the dimension of the orbit is given by:
\begin{align}
\mathrm{dim} (\mathcal{O}) = \mathrm{dim} (\mathfrak{g}) - \mathrm{dim}
(\mathfrak{g}_{0}) - \mathrm{dim} (\mathfrak{g}_{1}),
\end{align}
where
\begin{align}
\mathfrak{g}_{j} = \{Z \in\mathfrak{g} \, | \, [H,Z] = j Z \}.
\end{align}

\section{Review of Anomaly Polynomial Computations \label{app:ANOMALYPOLY}}

In this Appendix, we briefly review the computation of the anomaly polynomial $I_8$ for any 6D SCFT, as originally developed in \cite{Ohmori:2014kda}. For explicit step-by-step examples of anomaly polynomial computations, we refer
the interested reader to section 7.1 of \cite{Heckman:2018jxk}.

In a theory with a well-defined tensor branch and conventional matter,
the anomaly polynomial can be viewed as a sum of two terms: a 1-loop term and
a Green-Schwarz term,
\begin{equation}
I_{8} = I_{\text{1-loop}}+ I_{\text{GS}} .
\label{eq:Itot}
\end{equation}
The full anomaly polynomial of a 6D SCFT takes the form
\begin{align}
I_{8}  &  = \alpha c_{2}(R)^{2} + \beta c_{2}(R) p_{1}(T) + \gamma
p_{1}(T)^{2} + \delta p_{2}(T)\nonumber\\
&  + \sum_{i} \left[  \mu_{i} \, \mathrm{Tr} F_{i}^{4}
+ \, \mathrm{Tr} F_{i}^{2} \left(  \rho_{i}
p_{1}(T) + \sigma_{i} c_{2}(R) + \sum_{j} \eta_{ij} \, \mathrm{Tr} F_{j}^{2}
\right)  \right]  . \label{eq:anomalypoly}%
\end{align}
Here, $c_{2}(R)$ is the second Chern class of the $SU(2)_{R}$ symmetry,
$p_{1}(T)$ is the first Pontryagin class of the tangent bundle, $p_{2}(T)$ is
the second Pontryagin class of the tangent bundle, and $F_{i}$ is the field
strength of the $i^{th}$ symmetry, where $i$ and $j$ run over the flavor symmetries of the theory.

The 1-loop term receives contributions from
free tensor multiplets, vector multiplets, and hypermultiplets:
\begin{align}
I_{\text{tensor}} = \frac{c_{2}(R)^{2}}{24} + \frac{c_{2}(R)p_{1}(T)}{48}  &
+ \frac{23 p_{1}(T)^{2} -116 p_{2}(T)}{5760},\label{eq:tensor}\\
I_{\text{vector}} = -\frac{\tr_{\text{adj}} F^{4} + 6 c_{2}(R) \tr_{\text{adj}%
} F^{2} + d_{G} c_{2}(R)^{2}}{24}  &  - \frac{\tr_{\text{adj}}F^{2}+d_{G}
c_{2}(R)p_{1}(T)}{48}\nonumber\\
&  - d_{G} \frac{7 p_{1}(T)^{2} - 4 p_{2}(T)}{5760},\label{eq:vector}\\
I_{\text{hyper}} = \frac{\tr_{\rho} F^{4} }{24} + \frac{\tr_{\rho}F^{2}
p_{1}(T)}{48}  &  + d_{\rho}\frac{7 p_{1}(T)^{2} - 4 p_{2}(T)}{5760}.
\label{eq:hyper}%
\end{align}
Here, $\tr_{\rho}$ is the trace in the representation $\rho$, $d_{\rho}$ is
the dimension of the representation $\rho$, and $d_{G}$ is the dimension of
the group $G$. In computing the anomaly polynomial, one should convert the
traces in general representations to the trace in a defining
representation. One may write
\begin{align}
\tr_{\rho}F^{4} = x_{\rho}\, \mathrm{Tr} F^{4} + y_{\rho}(\, \mathrm{Tr}
F^{2})^{2}\\
\tr_{\rho}F^{2} = \text{Ind}_{\rho}\, \mathrm{Tr} F^{2} ,
\end{align}
with $x_{\rho}$, $y_{\rho}$, and $\text{Ind}_{\rho}$ well-known constants in
group theory, which can be found in the Appendix of \cite{Ohmori:2014kda} or \cite{Heckman:2018jxk}. For the adjoint representation, $\text{Ind}_{\rho}$ is also known as
the dual Coxeter number, $h_{G}^{\vee}$. Note that the groups $SU(2)$,
$SU(3)$, $G_{2}$, $F_{4}$, $E_{6}$, $E_{7}$, and $E_{8}$ do not have an independent quartic Casimir $\, \mathrm{Tr} F^{4}$, so $x_\rho = 0$ for all representations of these groups.

The Green-Schwarz term takes the form
\begin{equation}
I_{\text{GS}} = \frac{1}{2} A^{ij} I_{i} I_{j},
\end{equation}
where $A^{ij}$ is a negative-definite matrix given by the inverse of the Dirac pairing on the string charge
lattice.
The term $I_i$ can be written as
\begin{equation}
I_{i} = a_{i} c_{2}(R) + b_{i} p_{1}(T) + \sum_{j} c_{ij} \, \mathrm{Tr}
F_{j}^{2}.
\label{eq:Ii}
\end{equation}
The coefficients $a_i$, $b_i$, and $c_{ij}$ are chosen so that the gauge anomalies $(\mathrm{Tr}
F_{i}^{2})^2$ and mixed gauge-gauge or gauge-global anomalies (e.g. $\mathrm{Tr}
F_{i}^{2} \mathrm{Tr} F_{j}^{2}$, $\mathrm{Tr}F_{i}^{2} c_2(R)$, $\mathrm{Tr} F_{i}^{2} p_1(T)$) vanish. In other words, these anomalies must precisely cancel between the Green-Schwarz term and the 1-loop term. In practice, one need not compute the individual $I_i$: one can simply complete the square with respect to the quadratic Casimir $\mathrm{Tr} F_{i}^{2}$ of each of the gauge groups in turn.
This is guaranteed to cancel out the gauge anomalies and mixed gauge anomalies, and what is left
is simply the total anomaly polynomial $I_8$.


\section{Catalogs of Short Quiver Theories}
\label{apdx:shortQuiverCatalogs}

In this Appendix we present explicit catalogs of ``kissing cases'' for $SO(8)$ and $SO(10)$ short quiver theories, each under a particular UV gauge group but varying UV length. For each case, we give the exact ``kissing case'', together with the ``preceding theory'' obtained from the nilpotent orbit but with a slightly longer quiver to illustrate how such collisions between the nilpotent deformations take place. As in \cite{Heckman:2018pqx}, we may compute the anomaly polynomial of the kissing theory directly, but we can also compute it via analytic continuation from a formal type IIA quiver. In most cases, this procedure gives the same result, but in some cases, there is an additional correction term, which we display in the right-hand columns of the following tables. This additional correction term can also be read off from the brane picture, as explained in section \ref{subsubsec:ShortAnomalyRules}.

\begin{longtable}{|c|c|c|c|c|c|c|}
\hline
$\mathcal{O}_L$ & $\mathcal{O}_R$ & Preceding Theory & Kissing Theory & $\#I_{n}$ & $\Delta \alpha$ & $\Delta \beta$\\ \hline \hline
$[7, 1]$ & $[7, 1]$ & $2  \,\,  {{\overset{\ksu(2)}2}}  \,\,   {\overset{\mathfrak{g_{2}}}3}  \,\, \underset{[SU(2)]}1 \,\, {\overset{\mathfrak{g_{2}}}3} \,\, {{\overset{\ksu(2)}2}}  \,\, {2}$ & ${2}  \,\,   \underset{[N_f = 3/2]}{{\overset{\ksu(2)}2}}  \,\,   \underset{[SU(2)]}{\overset{\ksu(2)}2}  \,\, \underset{[N_f = 3/2]}{{\overset{\ksu(2)}2}}  \,\, {2}$ & 2 & $\frac{1}{12}$ & $\frac{1}{24}$ \\ \hline \hline

$[7, 1]$ & $[4^2]$ &
$2  \,\,   {\overset{\ksu(2)}2}  \,\,   {{\overset{\mathfrak{g_{2}}}3}}  \,\, 1 \,\, \underset{[SU(2)]}{{\overset{\mathfrak{so}(7)}3}}\,\, {\overset{\ksu(2)}2} $
&  ${2}  \,\,   \underset{[N_f = 1/2]}{\overset{\ksu(2)}2}  \,\,   \underset{[SU(2)]}{{\overset{\ksu(3)}2}}  \,\, \underset{[N_f = 1]}{\overset{\ksu(2)}2}$ & 2 & $\frac{1}{24}$ & $\frac{1}{48}$ \\ \hline

$[7, 1]$ & $[5, 1^3]$  &
$2  \,\,   {\overset{\ksu(2)}2}  \,\,   {{\overset{\mathfrak{g_{2}}}3}}  \,\, 1 \,\, \underset{[SU(2)]}{{\overset{\mathfrak{so}(7)}3}}\,\, {\overset{\ksu(2)}2}$
& ${2}  \,\,   \underset{[N_f = 1/2]}{\overset{\ksu(2)}2}  \,\,   \underset{[SU(2)]}{{\overset{\ksu(3)}2}}  \,\, \underset{[N_f = 1]}{\overset{\ksu(2)}2}$ & 2 & 0 & 0 \\ \hline

$[7, 1]$ & $[5, 3]$  &
$2  \,\,   {\overset{\ksu(2)}2}  \,\,   {{\overset{\mathfrak{g_{2}}}3}}  \,\, \underset{[SU(2)]}{1} \,\, {\overset{\mathfrak{g_2}}3}\,\, {\overset{\ksu(2)}2}$
& ${2}  \,\,   \underset{[N_f = 3/2]}{{\overset{\ksu(2)}2}}  \,\,   {\overset{\ksu(2)}2}  \,\, {\overset{\ksu(2)}2}  \,\, [SU(2) \times SU(2)]$ & 2 & $\frac{1}{12}$ & $\frac{1}{24}$ \\ \hline \hline

$[4^2]$ & $[4^2]$&
$ {\overset{\ksu(2)}2}  \,\,   \underset{[SU(2)]}{{\overset{\kso(7)}3}}  \,\, 1  \,\, \underset{[SU(2)]}{{\overset{\kso(7)}3}}  \,\, {\overset{\ksu(2)}2}$
& $ \underset{[N_f = 1/2]}{\overset{\ksu(2)}2}  \,\,   \underset{[Sp(2)]}{{\overset{\mathfrak{g_{2}}}2}}  \,\, \underset{[N_f = 1/2]}{\overset{\ksu(2)}2}$ & 2 & 0 & 0 \\ \hline

$[5, 1^3]$ & $[4^2]$ &
$ {\overset{\ksu(2)}2}  \,\,   \underset{[SU(2)]}{{\overset{\kso(7)}3}}  \,\, 1  \,\, \underset{[SU(2)]}{{\overset{\kso(7)}3}}  \,\, {\overset{\ksu(2)}2}$
& $ \underset{[N_f = 1/2]}{\overset{\ksu(2)}2}  \,\,   \underset{[Sp(2)]}{{\overset{\mathfrak{g_{2}}}2}}  \,\, \underset{[N_f = 1/2]}{\overset{\ksu(2)}2}$ & 1 & 0 & 0 \\ \hline

$[5, 1^3]$ & $[5, 1^3]$  &
$ {\overset{\ksu(2)}2}  \,\,   \underset{[SU(2)]}{{\overset{\kso(7)}3}}  \,\, 1  \,\, \underset{[SU(2)]}{{\overset{\kso(7)}3}}  \,\, {\overset{\ksu(2)}2}$
& $ {\overset{\ksu(2)}2}  \,\,   \underset{[SU(4)]}{{\overset{\ksu(4)}2}}  \,\, {\overset{\ksu(2)}2}$ & 0 & 0 & 0 \\ \hline

$[5, 3]$ & $[4^2]$ &
$ {\overset{\ksu(2)}2}  \,\,  {{\overset{\mathfrak{g_{2}}}3}}  \,\, 1  \,\, \underset{[SU(2)]}{{\overset{\kso(7)}3}}  \,\, {\overset{\ksu(2)}2}$
&  $\underset{[N_f = 1]}{\overset{\ksu(2)}2}  \,\,   \underset{[SU(2)]}{{\overset{\ksu(3)}2}}  \,\, \underset{[N_f = 1]}{\overset{\ksu(2)}2}$ & 2 & $\frac{1}{24}$ & $\frac{1}{48}$ \\ \hline

$[5, 3]$ & $[5, 1^3]$ &
$ {\overset{\ksu(2)}2}  \,\,  {{\overset{\mathfrak{g_{2}}}3}}  \,\, 1  \,\, \underset{[SU(2)]}{{\overset{\kso(7)}3}}  \,\, {\overset{\ksu(2)}2}$
& $\underset{[N_f = 1]}{\overset{\ksu(2)}2}  \,\,   \underset{[SU(2)]}{{\overset{\ksu(3)}2}}  \,\, \underset{[N_f = 1]}{\overset{\ksu(2)}2}$ & 2 & 0 & 0 \\ \hline

$[5, 3]$ & $[5, 3]$ &
$ {\overset{\ksu(2)}2}  \,\,  {\overset{\mathfrak{g_{2}}}3}  \,\, \underset{[SU(2)]}{1}  \,\, {{\overset{\mathfrak{g_{2}}}3}}   \,\, {\overset{\ksu(2)}2}$
& $[SU(2) \times SU(2)] \,\,  {\overset{\ksu(2)}2}  \,\,
{\overset{\ksu(2)}2}  \,\, {\overset{\ksu(2)}2}   \,\,    [SU(2)]$ & 2 & $\frac{1}{12}$ & $\frac{1}{24}$ \\ \hline

$[7, 1]$ & $[2^2, 1^4]$ &
${2}  \,\,   {\overset{\ksu(2)}2}  \,\,   {{\overset{\mathfrak{g_{2}}}3}}  \,\, 1 \,\, \underset{[SU(2)^{\otimes 3}]}{{\overset{\mathfrak{so}(8)}3}}$
&$ 2 \,\,  {\overset{\ksu(2)}2}  \,\,
{\overset{\mathfrak{g_{2}}}2}  \,\,   [Sp(3)]$
& 2 & 0 & 0 \\ \hline

$[7, 1]$ & $[2^4]$ &
${2}  \,\,   {\overset{\ksu(2)}2}  \,\,   {{\overset{\mathfrak{g_{2}}}3}}  \,\, 1 \,\, \underset{[Sp(2)]}{{\overset{\mathfrak{so}(7)}3}}$
&${2}  \,\,  \underset{[N_f = 1/2]}{\overset{\ksu(2)}2}  \,\,   {\overset{\ksu(3)}2}  \,\,   [SU(4)]$
& 3 & $\frac{1}{24}$ & $\frac{1}{48}$ \\ \hline

$[7, 1]$ & $[3, 1^5]$ &
${2}  \,\,   {\overset{\ksu(2)}2}  \,\,   {{\overset{\mathfrak{g_{2}}}3}}  \,\, 1 \,\, \underset{[Sp(2)]}{{\overset{\mathfrak{so}(7)}3}}$
& ${2}  \,\,  \underset{[N_f = 1/2]}{\overset{\ksu(2)}2}  \,\,   {\overset{\ksu(3)}2}  \,\,   [SU(4)]$
& 4 & 0 & 0 \\ \hline

$[7, 1]$ & $[3,2^2,1]$ &
${2}  \,\,   {\overset{\ksu(2)}2}  \,\,   {{\overset{\mathfrak{g_{2}}}3}}  \,\, \underset{[SU(2)]}{1} \,\, \underset{[SU(2)]}{{\overset{\mathfrak{g_2}}3}}$
&  ${2}  \,\,  \underset{[N_f = 3/2]}{\overset{\ksu(2)}2}  \,\,   {\overset{\ksu(2)}2}  \,\,   [SU(2) \times SU(2)]$
& 4 & $\frac{1}{12}$ & $\frac{1}{24}$ \\ \hline

$[7, 1]$ & $[3^2, 1^2]$ &
${2}  \,\,   {\overset{\ksu(2)}2}  \,\,   {{\overset{\mathfrak{g_{2}}}3}}  \,\, \underset{[SU(3)]}{1} \,\, {\overset{\ksu(3)}3}$
&  ${2}  \,\,  \underset{[SU(3)]}{\overset{\ksu(2)}2}  \,\,   2$
& 4 & $\frac{1}{6}$ & $\frac{1}{12}$ \\ \hline \hline

$[4^2]$ & $[2^2, 1^4]$ &
$ {\overset{\ksu(2)}2}  \,\,   \underset{[SU(2)]}{\overset{\kso(7)}3} \,\, 1  \,\,  \underset{[SU(2)^{\otimes 3}]}{\overset{\kso(8)}3}   $
& $ {\overset{\ksu(2)}2}  \,\,   {\overset{\kso(7)}2}  \,\,  [Sp(3)\times Sp(1)]$
& 1 & 0 & 0 \\ \hline

$[4^2]$ & $[2^4]$ &
$ {\overset{\ksu(2)}2}  \,\,   \underset{[SU(2)]}{\overset{\kso(7)}3} \,\, 1  \,\,  \underset{[Sp(2)]}{\overset{\kso(7)}3}  $
&  $ \underset{[N_f = 1/2]}{\overset{\ksu(2)}2}  \,\,   {\overset{\mathfrak{g_{2}}}2}  \,\,  [Sp(3)]$
& 3 & 0 & 0 \\ \hline

$[4^2]$ & $[3, 1^5]$ &
$ {\overset{\ksu(2)}2}  \,\,   \underset{[SU(2)]}{\overset{\kso(7)}3}\,\, 1  \,\,  \underset{[Sp(2)]}{\overset{\kso(7)}3}   $
& $ \underset{[N_f = 1/2]}{\overset{\ksu(2)}2}  \,\,   {\overset{\mathfrak{g_{2}}}2}  \,\,  [Sp(3)]$
& 2 & 0 & 0 \\ \hline

$[4^2]$ & $[3, 2^2, 1]$ & $ {\overset{\ksu(2)}2}  \,\,   \underset{[SU(2)]}{\overset{\kso(7)}3} \,\, 1  \,\,  \underset{[SU(2)]}{\overset{\mathfrak{g_{2}}}3}   $
& $\underset{[N_f = 1]}{\overset{\ksu(2)}2}  \,\,   {\overset{\ksu(3)}2}  \,\,  [SU(4)]$ & 3 & $\frac{1}{24}$ & $\frac{1}{48}$ \\ \hline

$[4^2]$ & $[3^2, 1^2]$ & $ {\overset{\ksu(2)}2}  \,\,   \underset{[SU(2)]}{\overset{\kso(7)}3} \,\, \underset{[SU(2)]}{1}  \,\,  {\overset{\ksu(3)}3}  $ & $[SU(2)]  \,\,  {\overset{\ksu(2)}2}  \,\,   {\overset{\ksu(2)}2}  \,\,  [SU(2) \times SU(2)]$ & 4 & $\frac{1}{12}$ & $\frac{1}{24}$ \\ \hline

$[5, 1^3]$ & $[2^2, 1^4]$ &
$ {\overset{\ksu(2)}2}  \,\,   \underset{[SU(2)]}{\overset{\kso(7)}3} \,\, 1  \,\,  \underset{[SU(2)^{\otimes 3}]}{\overset{\kso(8)}3}   $
&$ {\overset{\ksu(2)}2}  \,\,   {\overset{\kso(7)}2}  \,\,  [Sp(3)\times Sp(1)]$
& 0 & 0 & 0 \\ \hline

$[5, 1^3]$ & $[2^4]$ &
$ {\overset{\ksu(2)}2}  \,\,   \underset{[SU(2)]}{\overset{\kso(7)}3} \,\, 1  \,\,  \underset{[Sp(2)]}{\overset{\kso(7)}3}   $
&$ \underset{[N_f = 1/2]}{\overset{\ksu(2)}2}  \,\,   {\overset{\mathfrak{g_{2}}}2}  \,\,  [Sp(3)]$
& 1 & 0 & 0 \\ \hline

$[5, 1^3]$ & $[3, 1^5]$ &
$ {\overset{\ksu(2)}2}  \,\,   \underset{[SU(2)]}{\overset{\kso(7)}3} \,\, 1  \,\,  \underset{[Sp(2)]}{\overset{\kso(7)}3}   $
&${\overset{\ksu(2)}2}  \,\,   {\overset{\ksu(4)}2}  \,\,  [SU(6)]$
& 0 & 0 & 0 \\ \hline

$[5, 1^3]$ & $[3, 2^2, 1]$ &
$ {\overset{\ksu(2)}2}  \,\,   \underset{[SU(2)]}{\overset{\kso(7)}3} \,\, 1  \,\,  \underset{[SU(2)]}{\overset{\mathfrak{g_{2}}}3}   $
& $ \underset{[N_f = 1]}{ \overset{\ksu(2)}2}  \,\,   {\overset{\ksu(3)}2}  \,\,  [SU(4)]$
& 2 & $0$ & $0$ \\ \hline

$[5, 1^3]$ & $[3^2, 1^2]$ &
$ {\overset{\ksu(2)}2}  \,\,   \underset{[SU(2)]}{\overset{\kso(7)}3} \,\, \underset{[SU(2)]}{1}  \,\,  {\overset{\ksu(3)}3}   $
& $[SU(2)]  \,\,  {\overset{\ksu(2)}2}  \,\,   {\overset{\ksu(2)}2}  \,\,  [SU(2) \times SU(2)]$
& 4 & 0 & 0 \\ \hline

$[5, 3]$ & $[2^2, 1^4]$ &
$ {\overset{\ksu(2)}2}  \,\,   {\overset{\mathfrak{g_{2}}}3} \,\, 1  \,\,  \underset{[SU(2)^{\otimes 3}]}{\overset{\kso(8)}3}   $
&$ \underset{[N_f = 1/2]}{\overset{\ksu(2)}2}  \,\,   {\overset{\mathfrak{g_{2}}}2}  \,\,  [Sp(3)]$ & 2 & 0 & 0 \\ \hline

$[5, 3]$ & $[2^4]$ &
$ {\overset{\ksu(2)}2}  \,\,   {\overset{\mathfrak{g_{2}}}3} \,\, 1  \,\,  \underset{[Sp(2)]}{\overset{\kso(7)}3}   $
& $  \underset{[N_f = 1]}{\overset{\ksu(2)}2}  \,\,   {\overset{\ksu(3)}2}  \,\,  [SU(4)]$
& 3 & $\frac{1}{24}$ & $\frac{1}{48}$ \\ \hline

$[5, 3]$ & $[3, 1^5]$ &
$ {\overset{\ksu(2)}2}  \,\,   {\overset{\mathfrak{g_{2}}}3} \,\, 1  \,\,  \underset{[Sp(2)]}{\overset{\kso(7)}3}   $
& $ \underset{[N_f = 1]}{\overset{\ksu(2)}2}  \,\,   {\overset{\ksu(3)}2}  \,\,  [SU(4)]$
& 4 & 0 & 0 \\ \hline

$[5, 3]$ & $[3, 2^2, 1]$ &
$ {\overset{\ksu(2)}2}  \,\,   {\overset{\mathfrak{g_{2}}}3} \,\, \underset{[SU(2)]}{1}  \,\,  \underset{[SU(2)]}{\overset{\mathfrak{g_{2}}}3}   $
& $[SU(2)]  \,\,  {\overset{\ksu(2)}2}  \,\,   {\overset{\ksu(2)}2}  \,\,  [SU(2) \times SU(2)]$
& 4 &  $\frac{1}{12}$ & $\frac{1}{24}$ \\ \hline

$[5, 3]$ & $[3^2, 1^2]$ &
$ {\overset{\ksu(2)}2}  \,\,   {\overset{\mathfrak{g_{2}}}3} \,\, \underset{[SU(3)]}{1}  \,\,  {\overset{\ksu(3)}3}   $
&  $[G_2]  \,\,  {\overset{\ksu(2)}2}  \,\,   2$
& 4 & $\frac{1}{6}$ & $\frac{1}{12}$ \\ \hline \hline

$[2^2, 1^4]$ & $[2^2, 1^4]$ &
$ \underset{[SU(2)^{\otimes 3}]}{\overset{\kso(8)}3} \,\, 1  \,\,  \underset{[SU(2)^{\otimes 3}]}{\overset{\kso(8)}3}   $
& $ {\overset{\kso(8)}2} \,\,  [Sp(2) \times Sp(2) \times Sp(2)]$ & 0 & 0 & 0 \\ \hline

$[2^4]$ & $[2^2, 1^4]$ &
$ \underset{[Sp(2)]}{\overset{\kso(7)}3} \,\, 1  \,\,  \underset{[SU(2)^{\otimes 3}]}{\overset{\kso(8)}3}   $
&$ {\overset{\kso(7)}2} \,\,  [Sp(4) \times Sp(1)]$ & 1 & 0 & 0 \\ \hline

$[3, 1^5]$    & $[2^2, 1^4]$ &
$ \underset{[Sp(2)]}{\overset{\kso(7)}3} \,\, 1  \,\,  \underset{[SU(2)^{\otimes 3}]}{\overset{\kso(8)}3}   $
& $ {\overset{\kso(7)}2} \,\,  [Sp(4) \times Sp(1)]$ & 0 & 0 & 0 \\ \hline

$[2^4]$ & $[2^4]$ &
$ \underset{[Sp(2)]}{\overset{\kso(7)}3} \,\, 1  \,\,  \underset{[Sp(2)]}{\overset{\kso(7)}3}   $
&$ {\overset{\mathfrak{g_{2}}}2} \,\,  [Sp(4)]$
& 4 & 0 & 0 \\ \hline

$[3, 1^5]$    & $[2^4]$ &
$ \underset{[Sp(2)]}{\overset{\kso(7)}3} \,\, 1  \,\,  \underset{[Sp(2)]}{\overset{\kso(7)}3}   $
&$ {\overset{\mathfrak{g_{2}}}2} \,\,  [Sp(4)]$
& 2 & 0 & 0 \\ \hline

$[3, 1^5]$    & $[3, 1^5]$ &
$ \underset{[Sp(2)]}{\overset{\kso(7)}3} \,\, 1  \,\,  \underset{[Sp(2)]}{\overset{\kso(7)}3}   $
&$ {\overset{\ksu(4)}2} \,\,  [SU(8)]$ & 0 & 0 & 0 \\ \hline

$[3, 2^2, 1]$ & $[2^2, 1^4]$ &
$ \underset{[SU(2)]}{\overset{\mathfrak{g_{2}}}3} \,\, 1  \,\,  \underset{[SU(2)^{\otimes 3}]}{\overset{\kso(8)}3}   $
&$ {\overset{\mathfrak{g_{2}}}2} \,\,  [Sp(4)]$ & 2 & 0 & 0 \\ \hline

$[3, 2^2, 1]$  & $[2^4]$  &
$ \underset{[SU(2)]}{\overset{\mathfrak{g_{2}}}3} \,\, 1  \,\, \underset{[Sp(2)]}{\overset{\kso(7)}3}    $
&$ {\overset{\ksu(3)}2} \,\, [SU(6)]$
& 4 & $\frac{1}{24}$ & $\frac{1}{48}$ \\ \hline

$[3, 2^2, 1]$  & $[3, 1^5]$ &
$ \underset{[SU(2)]}{\overset{\mathfrak{g_{2}}}3} \,\, 1  \,\, \underset{[Sp(2)]}{\overset{\kso(7)}3}   $
&$ {\overset{\ksu(3)}2} \,\, [SU(6)]$ & 4 & 0 & 0 \\ \hline

$[3, 2^2, 1]$ & $[3, 2^2, 1]$ &
$ \underset{[SU(2)]}{\overset{\mathfrak{g_{2}}}3} \,\, \underset{[SU(2)]}{1}  \,\, \underset{[SU(2)]}{\overset{\mathfrak{g_{2}}}3} $
&$ {\overset{\ksu(2)}2} \,\,  [SO(7)]$ & 6 & $\frac{1}{12}$ & $\frac{1}{24}$ \\ \hline

$[3^2, 1^2]$ & $[2^2, 1^4]$ &
$ {\overset{\ksu(3)}3} \,\, 1  \,\,  \underset{[SU(2)^{\otimes 3}]}{\overset{\kso(8)}3}   $
& $ {\overset{\ksu(3)}2} \,\, [SU(6)]$  & 4 & 0 & 0 \\ \hline

$[3^2, 1^2]$  & $[2^4]$  &
$ {\overset{\ksu(3)}3} \,\, \underset{[SU(2)]}{1}  \,\, \underset{[Sp(2)]}{\overset{\kso(7)}3}   $
&$ {\overset{\ksu(2)}2} \,\, [SO(7)]$
& 6 & $\frac{1}{12}$ & $\frac{1}{24}$ \\ \hline

$[3^2, 1^2]$  & $[3, 1^5]$ &
$ {\overset{\ksu(3)}3} \,\, \underset{[SU(2)]}{1}  \,\, \underset{[Sp(2)]}{\overset{\kso(7)}3}   $
&$ {\overset{\ksu(2)}2} \,\, [SO(7)]$ & 8 & 0 & 0 \\ \hline

$[3^2, 1^2]$ & $[3, 2^2, 1]$ &
$ {\overset{\ksu(3)}3} \,\, \underset{[SU(3)]}{1}  \,\,  \underset{[SU(2)]}{\overset{\mathfrak{g_{2}}}3}  $
&$ 2 \,\,  [SU(2) \subset Sp(2)_{R}]$ & 7 & $\frac{1}{6}$ & $\frac{1}{12}$ \\ \hline


\caption{A catalog for $SO(8)$ kissing short quiver cases, their preceding longer theory, and the relevant terms for anomaly matching. The $\mathcal{O}_{L,R}$ columns correspond to the left and right deformations. Here $\Delta \alpha = \alpha_{\textrm{formal}} - \alpha_{F}$, and likewise for $\Delta \beta$. The ``Preceding Theory" column gives the theory whose length is one longer than the kissing theory, under the same pair of nilpotent orbits. The ``Theory" column gives the actual deformed short quiver theory, while the $\#I_{n}$ columns stands for the number of anomaly of neutral hypermultiplets to be added to the F-theory quiver in order to match the coefficients $\gamma$ and $\delta$ of the formal quiver. The last entry indicates that there is an $SU(2) \subset Sp(2)_R$ flavor symmetry. By this, we mean that the IR theory ends up flowing to a theory with $\mathcal{N} = (2,0)$ supersymmetry, where the R-symmetry group is $Sp(2)_{R}$. Viewed as an $\mathcal{N} = (1,0)$ SCFT, there is an $SU(2)$ flavor symmetry and an $SU(2)_R$ R-symmetry.}
\label{table:SO8tangentialShortQuiver}
\end{longtable}

\begin{longtable}{|c|c|c|c|c|c|c|}
\hline
$\mathcal{O}_L$ & $\mathcal{O}_R$ & ``Preceding Theory" & ``Kissing Theory" & $\#I_{n}$ & $\Delta \alpha$ & $\Delta \beta$\\ \hline \hline
$[9, 1]$ & $[9, 1]$ & $2  \,\,   {\overset{\mathfrak{su}(2)}2}  \,\,   {{\overset{\mathfrak{g}_2}3}} \,\, 1 \,\, {\overset{\mathfrak{so}(8)}4} \,\, 1  {\,\,{\overset{\mathfrak{g}_2}3}}  \,\, {\overset{\mathfrak{su}(2)}2}  \,\, 2$ &  $2  \,\,   \underset{[N_f = 1/2]}{\overset{\mathfrak{su}(2)}2}  \,\,   \underset{[N_f = 1]}{{\overset{\mathfrak{su}(3)}2}} \,\, \underset{[N_f = 1]}{\overset{\mathfrak{su}(3)}2}  \,\, \underset{[N_f = 1/2]}{\overset{\mathfrak{su}(2)}2}  \,\, 2$  & 1 & 0 & 0 \\ \hline \hline
$[9, 1]$ & $[7, 1^3]$ & $2 \,\, {\overset{\mathfrak{su}(2)}2} \,\,  {\overset{\mathfrak{g}_2}3} \,\, 1 \,\, {\overset{\mathfrak{so}(8)}4} \,\ 1 \,\, \underset{[SU(2)]}{{\overset{\mathfrak{so}(7)}3}} \,\, {\overset{\mathfrak{su}(2)}2}$ & $2  \,\,   \underset{[N_f = 1/2]}{\overset{\mathfrak{su}(2)}2}  \,\,   {\overset{\mathfrak{su}(3)}2}  \underset{[SU(3)]}{\,\,{\overset{\mathfrak{su}(4)}2}}  \,\, {\overset{\mathfrak{su}(2)}2}$ & 0 & 0 & 0 \\ \hline

$[9, 1]$ & $[7, 3]$ & $2 \,\, {\overset{\mathfrak{su}(2)}2} \,\,  {\overset{\mathfrak{g}_2}3} \,\, 1 \,\, {\overset{\mathfrak{so}(8)}4} \,\ 1 \,\, {\overset{\mathfrak{g}_2}3} \,\, {\overset{\mathfrak{su}(2)}2}$ &  $2  \,\,   \underset{[N_f = 1/2]}{\overset{\mathfrak{su}(2)}2}  \,\,   \underset{[N_f = 1]}{{\overset{\mathfrak{su}(3)}2}}  \,\, \underset{[N_f = 1]}{{\overset{\mathfrak{su}(3)}2}}  \,\, \underset{[N_f = 1]}{\overset{\mathfrak{su}(2)}2}$ & 1 & 0 & 0 \\ \hline \hline
$[7, 1^3]$ & $[7, 1^3]$ & ${\overset{\mathfrak{su}(2)}2}  \,\,  \underset{[SU(2)]}{{\overset{\mathfrak{so}(7)}3}} \,\, 1 \,\, {\overset{\mathfrak{so}(8)}4} \,\ 1 \,\, \underset{[SU(2)]}{{\overset{\mathfrak{so}(7)}3}} \,\, {\overset{\mathfrak{su}(2)}2}$ & ${\overset{\mathfrak{su}(2)}2}  \,\,  \underset{[SU(2)]}{{\overset{\mathfrak{su}(4)}2}}  \underset{[SU(2)]}{\,\,{\overset{\mathfrak{su}(4)}2}}  \,\,{\overset{\mathfrak{su}(2)}2}$ &0 &0 &0 \\ \hline

$[7,3]$ & $[7, 1^3]$ & ${\overset{\mathfrak{su}(2)}2}  \,\,  {\overset{\mathfrak{g}_2}3} \,\, 1 \,\, {\overset{\mathfrak{so}(8)}4} \,\ 1 \,\, \underset{[SU(2)]}{{\overset{\mathfrak{so}(7)}3}} \,\, {\overset{\mathfrak{su}(2)}2}$ & $\underset{[N_f = 1]}{\overset{\mathfrak{su}(2)}2}  \,\,   {\overset{\mathfrak{su}(3)}2}  \,\,   \underset{[SU(3)]}{{\overset{\mathfrak{su}(4)}2}}  \,\,{\overset{\mathfrak{su}(2)}2}$ &0 &0 &0 \\ \hline

$[7, 3]$ & $[7, 3]$ & ${\overset{\mathfrak{su}(2)}2}  \,\,  {\overset{\mathfrak{g}_2}3} \,\, 1 \,\, {\overset{\mathfrak{so}(8)}4} \,\ 1 \,\, {\overset{\mathfrak{g}_2}3} \,\, {\overset{\mathfrak{su}(2)}2} $ & $\underset{[N_f = 1]}{\overset{\mathfrak{su}(2)}2}  \,\,  \underset{[N_f = 1]}{{\overset{\mathfrak{su}(3)}2}} \,\, \underset{[N_f = 1]}{{\overset{\mathfrak{su}(3)}2}} \,\,\underset{[N_f = 1]}{\overset{\mathfrak{su}(2)}2}$ &1 &0 &0 \\ \hline

$[9, 1]$ & $[4^2, 1^2]$ & $2 \,\, {\overset{\mathfrak{su}(2)}2} \,\,  {\overset{\mathfrak{g}_2}3} \,\, 1 \,\, \underset{[Sp(1)]}{\overset{\mathfrak{so}(9)}4} \,\ 1 \,\, {\overset{\mathfrak{su}(3)}3}$ & $2  \,\,   {\overset{\mathfrak{su}(2)}2}  \,\,   \underset{[Sp(2)]}{{\overset{\mathfrak{g_{2}}}2}}  \,\, \underset{[N_f = 1/2]}{\overset{\mathfrak{su}(2)}2} $&1 &0 &0 \\ \hline

$[9, 1]$ & $[5, 1^5]$ & $2 \,\, {\overset{\mathfrak{su}(2)}2} \,\,  {\overset{\mathfrak{g}_2}3} \,\, 1 \,\, {\overset{\mathfrak{so}(8)}4} \,\ 1 \,\,{\overset{\mathfrak{so}(7)}3} \,\, [Sp(2)]$ & $ 2 \,\,   \underset{[N_f = 1/2]}{\overset{\mathfrak{su}(2)}2}  \,\,   {\overset{\mathfrak{su}(3)}2}  \,\,{\overset{\mathfrak{su}(4)}2}  \,\,  [SU(5)]$ & 0 & 0 & 0 \\ \hline

$[9, 1]$ & $[5, 2^2, 1]$ & $2 \,\, {\overset{\mathfrak{su}(2)}2} \,\,  {\overset{\mathfrak{g}_2}3} \,\, 1 \,\, {\overset{\mathfrak{so}(8)}4} \,\ 1 \,\, \underset{[SU(2)]}{{\overset{\mathfrak{g}_2}3}}$  & $  2 \,\,   \underset{[N_f = 1/2]}{\overset{\mathfrak{su}(2)}2}  \,\,   \underset{[N_f = 1]}{{\overset{\mathfrak{su}(3)}2}}  \,\,{\overset{\mathfrak{su}(3)}2} \,\, [SU(3)]$ & 1 & 0 & 0 \\ \hline

$[9, 1]$ & $[5, 3, 1^2]$ & $2 \,\, {\overset{\mathfrak{su}(2)}2} \,\,  {\overset{\mathfrak{g}_2}3} \,\, 1 \,\, {\overset{\mathfrak{so}(8)}4} \,\ 1 \,\, {{\overset{\mathfrak{su}(3)}3}}$ & $  2  \,\,   \underset{[N_f = 1/2]}{\overset{\mathfrak{su}(2)}2}  \,\,   \underset{[SU(2)]}{{\overset{\mathfrak{su}(3)}2}}  \,\,\underset{[N_f = 1]}{\overset{\mathfrak{su}(2)}2}$ & 2 & 0 & 0 \\ \hline\hline

$[5^2]$ & $[5^2]$ & ${\overset{\mathfrak{su}(2)}2}  \,\,  {\overset{\mathfrak{so}(7)}3} \,\, \underset{[SO(4)]}{{\overset{\mathfrak{sp}(1)}1}}  \,\, {\overset{\mathfrak{so}(7)}3} \,\, {\overset{\mathfrak{su}(2)}2} $  & $[SU(2)]  \,\,  {\overset{\mathfrak{su}(2)}2}  \,\,   {{\overset{\mathfrak{su}(2)}2}}  \,\,{\overset{\mathfrak{su}(2)}2}  \,\,   [SU(2) \times SU(2)]$ & 2 & $\frac{1}{12}$ & $\frac{1}{24}$ \\ \hline

$[7, 1^3]$ & $[4^2, 1^2]$ & ${\overset{\mathfrak{su}(2)}2}  \,\,  \underset{[SU(2)]}{{\overset{\mathfrak{so}(7)}3}} \,\, 1 \,\, \underset{[Sp(1)]}{\overset{\mathfrak{so}(9)}4} \,\ 1 \,\, {{\overset{\mathfrak{su}(3)}3}} $  & ${\overset{\mathfrak{su}(2)}2}  \,\,   \underset{[Sp(2) \times Sp(1)]}{{\overset{\mathfrak{so}(7)}2}}  \,\,{\overset{\mathfrak{su}(2)}2} $ &0 &0 &0 \\ \hline

$[7, 1^3]$ & $[5, 1^5]$ & ${\overset{\mathfrak{su}(2)}2}  \,\,  \underset{[SU(2)]}{{\overset{\mathfrak{so}(7)}3}} \,\, 1 \,\, {\overset{\mathfrak{so}(8)}4} \,\ 1 \,\, {{\overset{\mathfrak{so}(7)}3}} \,\, [Sp(2)] $  & ${\overset{\mathfrak{su}(2)}2}  \,\,   \underset{[SU(2)]}{{\overset{\mathfrak{su}(4)}2}}  \,\,{\overset{\mathfrak{su}(4)}2} \,\, [SU(4)] $ &0 &0 &0 \\ \hline

$[7, 1^3]$ & $[5, 2^2, 1]$ & ${\overset{\mathfrak{su}(2)}2}  \,\,  \underset{[SU(2)]}{{\overset{\mathfrak{so}(7)}3}} \,\, 1 \,\, {\overset{\mathfrak{so}(8)}4} \,\ 1 \,\, {{\overset{\mathfrak{g}_2}3}} \,\, [SU(2)] $ & ${\overset{\mathfrak{su}(2)}2}  \,\,   \underset{[SU(3)]}{{\overset{\mathfrak{su}(4)}2}}  \,\,{\overset{\mathfrak{su}(3)}2} \,\, [SU(2)]$ &0 &0 & 0\\ \hline

$[7, 1^3]$ & $[5, 3, 1^2]$ & ${\overset{\mathfrak{su}(2)}2}  \,\,  \underset{[SU(2)]}{{\overset{\mathfrak{so}(7)}3}} \,\, 1 \,\, {\overset{\mathfrak{so}(8)}4} \,\ 1 \,\, {{\overset{\mathfrak{su}(3)}3}} $ & ${\overset{\mathfrak{su}(2)}2}  \,\,   \underset{[SU(4)]}{{\overset{\mathfrak{su}(4)}2}}  \,\,{\overset{\mathfrak{su}(2)}2} $& 0&0 &0 \\ \hline

$[7, 3]$ & $[4^2, 1^2]$& ${\overset{\mathfrak{su}(2)}2}  \,\,  {\overset{\mathfrak{g}_2}3} \,\, 1 \,\, \underset{[Sp(1)]}{\overset{\mathfrak{so}(9)}4} \,\ 1 \,\, {{\overset{\mathfrak{su}(3)}3}} $  & $\underset{[N_f = 1/2]}{\overset{\mathfrak{su}(2)}2}  \,\,   \underset{[Sp(2)]}{{\overset{\mathfrak{g_{2}}}2}}  \,\, \underset{[N_f = 1/2]}{\overset{\mathfrak{su}(2)}2}$&1 &0&0 \\ \hline

$[7, 3]$ & $[5, 1^5]$ & ${\overset{\mathfrak{su}(2)}2}  \,\, {\overset{\mathfrak{g}_2}3} \,\, 1 \,\, {\overset{\mathfrak{so}(8)}4} \,\ 1 \,\, \underset{[Sp(2)]}{{\overset{\mathfrak{so}(7)}3}} $  & $ \underset{[N_f = 1]}{\overset{\mathfrak{su}(2)}2}  \,\,   {\overset{\mathfrak{su}(3)}2}  \,\,{\overset{\mathfrak{su}(4)}2} \,\,[SU(5)]$ & 0&0 & 0\\ \hline

$[7, 3]$ & $[5, 2^2, 1]$  & ${\overset{\mathfrak{su}(2)}2}  \,\,  {\overset{\mathfrak{g}_2}3} \,\, 1 \,\, {\overset{\mathfrak{so}(8)}4} \,\ 1 \,\, \underset{[SU(2)]}{{\overset{\mathfrak{g}_2}3}} $ & $ \underset{[N_f = 1]}{\overset{\mathfrak{su}(2)}2}  \,\,  \underset{[N_f = 1]}{{\overset{\mathfrak{su}(3)}2}}  \,\,{\overset{\mathfrak{su}(3)}2} \,\, [SU(3)] $  &1 &0 &0 \\ \hline

$[7, 3]$ & $[5, 3, 1^2]$ & ${\overset{\mathfrak{su}(2)}2}  \,\,  {\overset{\mathfrak{g}_2}3} \,\, 1 \,\, {\overset{\mathfrak{so}(8)}4} \,\ 1 \,\, {{\overset{\mathfrak{su}(3)}3}} $ & $ \underset{[N_f = 1]}{\overset{\mathfrak{su}(2)}2}  \,\,   \underset{[SU(2)]}{{\overset{\mathfrak{su}(3)}2}}  \,\, \underset{[N_f = 1]}{\overset{\mathfrak{su}(2)}2}$  &3 & 0& 0\\ \hline \hline

$[4^2, 1^2]$  & $[4^2, 1^2]$ & ${\overset{\mathfrak{su}(3)}3}  \,\, 1 \,\,  \underset{[Sp(2)]}{{\overset{\mathfrak{so}(10)}4}} \,\, 1 \,\,{\overset{\mathfrak{su}(3)}3}$  & $[SU(3)] \,\, {\overset{\mathfrak{su}(3)}2}  \,\,   {\overset{\mathfrak{su}(3)}2} \,\,[SU(3)]$ & 1& 0 & 0\\ \hline

$[5, 1^5]$ & $[4^2, 1^2]$ & $\underset{[Sp(2)]}{\overset{\mathfrak{so}(7)}3}  \,\, 1 \,\,  \underset{[Sp(1)]}{\overset{\mathfrak{so}(9)}4} \,\, 1 \,\,{\overset{\mathfrak{su}(3)}3}$ & $[Sp(3) \times Sp(1)] \,\,{\overset{\mathfrak{so}(7)}2}  \,\,   {\overset{\mathfrak{su}(2)}2}   $ &0 &0 & 0\\ \hline

$[5, 1^5]$  &$[5, 1^5]$ & $\underset{[Sp(2)]}{\overset{\mathfrak{so}(7)}3}  \,\, 1 \,\,  {\overset{\mathfrak{so}(8)}4} \,\, 1 \,\,\underset{[Sp(2)]}{\overset{\mathfrak{so}(7)}3} $ & $[SU(4)] \,\, {\overset{\mathfrak{su}(4)}2}  \,\,   {\overset{\mathfrak{su}(4)}2} \,\,[SU(4)]$  &0 &0 &0 \\ \hline

$[5, 2^2, 1]$ & $[4^2, 1^2]$ & $\underset{[SU(2)]}{\overset{\mathfrak{g}_2}3}  \,\, 1 \,\,  \underset{[Sp(1)]}{\overset{\mathfrak{so}(9)}4} \,\, 1 \,\,{\overset{\mathfrak{su}(3)}3}$  &$[Sp(3)] \,\,{\overset{\mathfrak{g_{2}}}2}  \,\,   \underset{[N_f = 1/2]}{\overset{\mathfrak{su}(2)}2}   $& 1&0 &0 \\ \hline

$[5, 2^2, 1]$ & $[5, 1^5]$  &$\underset{[SU(2)]}{\overset{\mathfrak{g}_2}3}  \,\, 1 \,\,  {\overset{\mathfrak{so}(8)}4} \,\, 1 \,\,\underset{[Sp(2)]}{\overset{\mathfrak{so}(7)}3} $ &  $[SU(2)] \,\,{\overset{\mathfrak{su}(3)}2}  \,\,   {\overset{\mathfrak{su}(4)}2} \,\,[SU(5)]$ &0 &0 &0 \\ \hline

$[5, 2^2, 1]$ & $[5, 2^2, 1]$ & $\underset{[SU(2)]}{\overset{\mathfrak{g}_2}3}  \,\, 1 \,\,  {\overset{\mathfrak{so}(8)}4} \,\, 1 \,\,\underset{[SU(2)]}{\overset{\mathfrak{g}_2}3} $ & $[SU(3)] \,\, {\overset{\mathfrak{su}(3)}2}  \,\,   {\overset{\mathfrak{su}(3)}2} \,\,[SU(3)]$ &1 &0&0 \\ \hline

$[5, 3, 1^2]$ & $[4^2, 1^2]$  &  ${\overset{\mathfrak{su}(3)}3}  \,\, 1 \,\,  \underset{[Sp(1)]}{\overset{\mathfrak{so}(9)}4} \,\, 1 \,\,{\overset{\mathfrak{su}(3)}3}$ & $[SU(4)] \,\, {\overset{\mathfrak{su}(3)}2}  \,\,   \underset{[N_f = 1]}{\overset{\mathfrak{su}(2)}2} $ &2 &0 &0 \\ \hline

$[5, 3, 1^2]$ & $[5, 1^5]$   & ${\overset{\mathfrak{su}(3)}3}  \,\, 1 \,\,  {\overset{\mathfrak{so}(8)}4} \,\, 1 \,\,\underset{[Sp(2)]}{\overset{\mathfrak{so}(7)}3} $ & ${\overset{\mathfrak{su}(2)}2}  \,\,   {\overset{\mathfrak{su}(4)}2} \,\,[SU(6)]  $  &0 &0 & 0\\ \hline

$[5, 3, 1^2]$ & $[5, 2^2, 1]$   & ${\overset{\mathfrak{su}(3)}3}  \,\, 1 \,\,  {\overset{\mathfrak{so}(8)}4} \,\, 1 \,\,\underset{[SU(2)]}{\overset{\mathfrak{g}_2}3} $  & $\underset{[N_f = 1]}{\overset{\mathfrak{su}(2)}2}  \,\,   {\overset{\mathfrak{su}(3)}2} \,\, [SU(4)] $  &2 &0&0 \\ \hline

$[5, 3, 1^2]$ & $[5, 3, 1^2]$  & ${\overset{\mathfrak{su}(3)}3}  \,\, 1 \,\,  {\overset{\mathfrak{so}(8)}4} \,\, 1 \,\,{\overset{\mathfrak{su}(3)}3} $ & $[SU(2)] \,\,{\overset{\mathfrak{su}(2)}2}  \,\,   {\overset{\mathfrak{su}(2)}2}\,\,[SU(2) \times SU(2)]$   &4 &0 &0 \\ \hline

$[5^2]$ & $[2^4, 1^2]$      & ${\overset{\mathfrak{su}(2)}2}  \,\,  {\overset{\mathfrak{so}(7)}3} \,\, \underset{[N_f = 1]}{\overset{\mathfrak{sp}(1)}1} \,\, \underset{[N_s = 1]}{\overset{\mathfrak{so}(10)}3} \,\, [Sp(2)] $ & ${\overset{\mathfrak{su}(2)}2}  \,\,{\overset{\mathfrak{so}(7)}2} \,\, [Sp(3) \times Sp(1)]$  &1  & 0 &0  \\ \hline

$[5^2]$ & $[3, 2^2, 1^3]$     & ${\overset{\mathfrak{su}(2)}2}  \,\,  {\overset{\mathfrak{so}(7)}3} \,\, \underset{[SO(3)]}{\overset{\mathfrak{sp}(1)}1} \,\, \underset{[Sp(1)\times Sp(1)]}{\overset{\mathfrak{so}(9)}3} $ &  $\underset{[N_f = 1/2]}{\overset{\mathfrak{su}(2)}2}  \,\,{\overset{\mathfrak{g_{2}}}2} \,\, [Sp(3)]$  &2  &0 &  0\\ \hline

$[5^2]$ & $[3^2, 1^4]$      & ${\overset{\mathfrak{su}(2)}2}  \,\,  {\overset{\mathfrak{so}(7)}3} \,\, \underset{[SO(4)]}{\overset{\mathfrak{sp}(1)}1} \,\, \underset{[Sp(1) \times Sp(1)]}{\overset{\mathfrak{so}(8)}3} $ &  $\underset{[N_f = 1]}{\overset{\mathfrak{su}(2)}2}  \,\,   {\overset{\mathfrak{su}(3)}2} \,\, [SU(4)] $  &4  &0  & 0 \\ \hline

$[5^2]$ & $[3^2,2^2]$      &${\overset{\mathfrak{su}(2)}2}  \,\,  {\overset{\mathfrak{so}(7)}3} \,\, \underset{[SO(4)]}{\overset{\mathfrak{sp}(1)}1} \,\, \underset{[Sp(2)]}{\overset{\mathfrak{so}(7)}3} $&  $[SU(2)] \,\,{\overset{\mathfrak{su}(2)}2}  \,\,   {\overset{\mathfrak{su}(2)}2}\,\,[SU(2) \times SU(2)]$ & 4&$\frac{1}{12}$ &$\frac{1}{24}$  \\ \hline

$[5^2]$ & $[3^3, 1]$        &${\overset{\mathfrak{su}(2)}2}  \,\,  {\overset{\mathfrak{so}(7)}3} \,\, \underset{{[SO(5)]}}{\overset{\mathfrak{sp}(1)}1} \,\, {\overset{\mathfrak{g}_2}3} $  &  $[G_2] \,\, {\overset{\mathfrak{su}(2)}2} \,\,   2 $  &4 &$\frac{1}{6}$ &$\frac{1}{12}$   \\ \hline \hline

$[2^4, 1^2]$& $[2^4, 1^2]$& $[Sp(2)] \,\, \underset{[N_s = 1]}{\overset{\mathfrak{so}(10)}3}  \,\, {\overset{\mathfrak{sp}(1)}1} \,\,  \underset{[N_s = 1]}{\overset{\mathfrak{so}(10)}3} \,\, [Sp(2)] $ & $ {\overset{\mathfrak{so}(10)}2}  \,\,  [Sp(4) \times SU(2)]$ & 0 & 0  & 0 \\ \hline

$[3,2^2,1^3]$& $[2^4, 1^2]$& $ \underset{[Sp(1)\times Sp(1)]}{\overset{\mathfrak{so}(9)}3}  \,\, \underset{[N_f = 1/2]}{\overset{\mathfrak{sp}(1)}1} \,\,  \underset{[N_s = 1]}{\overset{\mathfrak{so}(10)}3} \,\, [Sp(2)]$ & $ {\overset{\mathfrak{so}(9)}2} \,\, [Sp(3) \times Sp(2)] $ & 0 & 0 & 0\\ \hline

$[3,2^2,1^3]$&$[3, 2^2, 1^3]$& $ \underset{[Sp(1)\times Sp(1)]}{\overset{\mathfrak{so}(9)}3}  \,\, \underset{[N_f = 1]}{\overset{\mathfrak{sp}(1)}1} \,\,  \underset{[Sp(1)\times Sp(1)]}{\overset{\mathfrak{so}(9)}3} $ & $ {\overset{\mathfrak{so}(8)}2} \,\, [Sp(2) \times Sp(2) \times Sp(2)]  $ & 0 & 0 & 0\\ \hline

$[3^2, 1^4]$& $[2^4, 1^2]$& $ \underset{[Sp(1) \times Sp(1)]}{\overset{\mathfrak{so}(8)}3}  \,\, \underset{[N_f = 1]}{\overset{\mathfrak{sp}(1)}1} \,\,  \underset{[N_s = 1]}{\overset{\mathfrak{so}(10)}3} \,\, [Sp(2)]$ & $ {\overset{\mathfrak{so}(8)}2}\,\, [Sp(2) \times Sp(2) \times Sp(2)] $ & 0 & 0 & 0\\ \hline

$[3^2, 1^4]$& $[3, 2^2, 1^3]$& $ \underset{[Sp(1) \times Sp(1)]}{\overset{\mathfrak{so}(8)}3}  \,\, \underset{[SO(3)]}{\overset{\mathfrak{sp}(1)}1} \,\,  \underset{[Sp(1) \times Sp(1)]}{\overset{\mathfrak{so}(9)}3} $ & $ {\overset{\mathfrak{so}(7)}2} \,\, [Sp(4) \times Sp(1) ]$ & 0 & 0 & 0\\ \hline

$[3^2, 1^4]$& $[3^2, 1^4]$& $ \underset{[Sp(1) \times Sp(1)]}{\overset{\mathfrak{so}(8)}3}  \,\, \underset{[SO(4)]}{\overset{\mathfrak{sp}(1)}1} \,\,  \underset{[Sp(1) \times Sp(1)]}{\overset{\mathfrak{so}(8)}3} $ & $ {\overset{\mathfrak{su}(4)}2} \,\, [SU(8)]$ & 0 & 0 & 0\\ \hline

$[3^2,2^2]$& $[2^4, 1^2]$& $ \underset{[Sp(1)]}{\overset{\mathfrak{so}(7)}3}  \,\, \underset{[N_f = 1]}{\overset{\mathfrak{sp}(1)}1} \,\,  \underset{[N_s = 1]}{\overset{\mathfrak{so}(10)}3} \,\, [Sp(2)] $ & $ {\overset{\mathfrak{so}(7)}2} \,\, [Sp(4) \times Sp(1) ] $ & 1 & 0 & 0\\ \hline

$[3^2,2^2]$& $[3, 2^2, 1^3]$& $ \underset{[Sp(1)]}{\overset{\mathfrak{so}(7)}3}  \,\, \underset{[SO(3)]}{\overset{\mathfrak{sp}(1)}1} \,\,  \underset{[Sp(1) \times Sp(1)]}{\overset{\mathfrak{so}(9)}3} $ & $ {\overset{\mathfrak{g_{2}}}2} \,\, [Sp(4)] $ &2& 0 & 0\\ \hline

$[3^2,2^2]$& $[3^2, 1^4]$& $ \underset{[Sp(1)]}{\overset{\mathfrak{so}(7)}3}  \,\, \underset{[SO(4)]}{\overset{\mathfrak{sp}(1)}1} \,\,  \underset{[Sp(1) \times Sp(1)]}{\overset{\mathfrak{so}(8)}3} $ & $ {\overset{\mathfrak{su}(3)}2} \,\, [SU(6)]$ & 4 & 0 & 0\\ \hline

$[3^2,2^2]$& $[3^2, 2^2]$& $ \underset{[Sp(1)]}{\overset{\mathfrak{so}(7)}3}  \,\, \underset{[SO(4)]}{\overset{\mathfrak{sp}(1)}1} \,\,  \underset{[Sp(1)]}{\overset{\mathfrak{so}(7)}3} $ & $ {\overset{\mathfrak{su}(2)}2} \,\, [SO(7)]$ & 6 & $\frac{1}{12}$ & $\frac{1}{24}$\\ \hline

$[3^3,  1]$& $[2^4, 1^2]$& $ {\overset{\mathfrak{g}_2}3}  \,\, \underset{{[SO(3)]}}{\overset{\mathfrak{sp}(1)}1} \,\,  \underset{[N_s = 1]}{\overset{\mathfrak{so}(10)}3} \,\, [Sp(2)] $ & $ {\overset{\mathfrak{g_{2}}}2} \,\, [Sp(4)]$ & 2 & 0 & 0\\ \hline

$[3^3,  1]$& $[3,2^2,1^3]$& $ {\overset{\mathfrak{g}_2}3}  \,\, \underset{{[SO(4)]}}{\overset{\mathfrak{sp}(1)}1} \,\,  \underset{[Sp(1) \times Sp(1)]}{\overset{\mathfrak{so}(9)}3} $ & $ {\overset{\mathfrak{su}(3)}2} \,\, [SU(6)]$ & 4 & 0 & 0\\ \hline

$[3^3,  1]$& $[3^2, 1^4]$& $ {\overset{\mathfrak{g}_2}3}  \,\, \underset{{[SO(5)]}}{\overset{\mathfrak{sp}(1)}1} \,\,  \underset{[Sp(1) \times Sp(1)]}{\overset{\mathfrak{so}(8)}3} $ & $ {\overset{\mathfrak{su}(2)}2} \,\, [SO(7)]$ & 8 & 0 & 0\\ \hline

$[3^3,  1]$& $[3^2, 2^2]$& $ {\overset{\mathfrak{g}_2}3}  \,\underset{{[SO(5)]}}{\overset{\mathfrak{sp}(1)}1} \,\,  \underset{[Sp(1)]}{\overset{\mathfrak{so}(7)}3} $ & $ 2 \,\, [SU(2) \subset Sp(2)_{R}]$ & 7 & $\frac{1}{6}$ & $\frac{1}{12}$\\ \hline


%

\caption{$SO(10)$ short quiver tangential cases, in parallel to table \ref{table:SO8tangentialShortQuiver}. See table \ref{table:SO8tangentialShortQuiver} for conventions and notation.}
\label{table:SO10TangentshortQuiver}
\end{longtable}

\section{Generators of $E_{6,7,8}$}

\label{app:ExpMat} In this section we list the generators $X_i$ and $Y_i$ for the exceptional
algebras $E_{6,7,8}$ in the basis used throughout this paper. All
other generators can be obtained from appropriate commutators.

The six positive simple roots of $E_{6}$ are associated with:
\begin{align}
X_{1}  &  = E_{1,2}+E_{12,13}+E_{15,16}+E_{17,18}+E_{19,20}+E_{21,22}%
,\nonumber\\
X_{2}  &  = E_{4,6}+E_{5,8}+E_{7,9}+E_{19,21}+E_{20,22}+E_{23,24},\nonumber\\
X_{3}  &  = E_{2,3}+E_{10,12}+E_{11,15}+E_{14,17}+E_{20,23}+E_{22,24}%
,\nonumber\\
X_{4}  &  = E_{3,4}+E_{8,10}+E_{9,11}+E_{17,19}+E_{18,20}+E_{24,25}%
,\nonumber\\
X_{5}  &  = E_{4,5}+E_{6,8}+E_{11,14}+E_{15,17}+E_{16,18}+E_{25,26}%
,\nonumber\\
X_{6}  &  = E_{5,7}+E_{8,9}+E_{10,11}+E_{12,15}+E_{13,16}+E_{26,27}.
\end{align}
The corresponding negative roots are $Y_{i} = X_{i}^{T}$ and Cartans $H_{i} =
[X_{i},Y_{i}]$.

The seven positive simple roots of $E_{7}$ are taken to be:
\begin{align}
X_{1}  &  = E_{7,8}+E_{9,10}+E_{11,12}+E_{13,14}+E_{16,17}+E_{19,20}%
+E_{37,38}+E_{40,41}+E_{43,44}+E_{45,46}+E_{47,48}+E_{49,50},\nonumber\\
X_{2}  &  = E_{5,6}+E_{7,9}+E_{8,10}+E_{22,25}+E_{24,28}+E_{26,30}%
+E_{27,31}+E_{29,33}+E_{32,35}+E_{47,49}+E_{48,50}+E_{51,52},\nonumber\\
X_{3}  &  = E_{5,7}+E_{6,9}+E_{12,15}+E_{14,18}+E_{17,21}+E_{20,23}%
+E_{34,37}+E_{36,40}+E_{39,43}+E_{42,45}+E_{48,51}+E_{50,52},\nonumber\\
X_{4}  &  = E_{4,5}+E_{9,11}+E_{10,12}+E_{18,22}+E_{21,24}+E_{23,26}%
+E_{31,34}+E_{33,36}+E_{35,39}+E_{45,47}+E_{46,48}+E_{52,53},\nonumber\\
X_{5}  &  = E_{3,4}+E_{11,13}+E_{12,14}+E_{15,18}+E_{24,27}+E_{26,29}%
+E_{28,31}+E_{30,33}+E_{39,42}+E_{43,45}+E_{44,46}+E_{53,54},\nonumber\\
X_{6}  &  = E_{2,3}+E_{13,16}+E_{14,17}+E_{18,21}+E_{22,24}+E_{25,28}%
+E_{29,32}+E_{33,35}+E_{36,39}+E_{40,43}+E_{41,44}+E_{54,55},\nonumber\\
X_{7}  &  = E_{1,2}+E_{16,19}+E_{17,20}+E_{21,23}+E_{24,26}+E_{27,29}%
+E_{28,30}+E_{31,33}+E_{34,36}+E_{37,40}+E_{38,41}+E_{55,56}.
\end{align}
Again corresponding negative roots are $Y_{i} = X_{i}^{T}$ and Cartans $H_{i}
= [X_{i},Y_{i}]$.

Finally, the eight positive simple roots of $E_{8}$ are taken to be:
\begin{align}
X_{1}  &  = E_{8,9}+E_{10,11}+E_{12,13}+E_{14,15}+E_{17,18}+E_{20,21}%
+E_{24,25}+E_{46,47}+E_{52,53}+E_{57,59}+E_{58,60}+E_{63,65}\nonumber\\
&  +E_{64,66}+E_{68,71}+E_{69,72}+E_{70,73}+E_{75,78}+E_{76,79}+E_{77,80}%
+E_{82,85}+E_{83,86}+E_{84,87}+E_{90,92}+E_{91,93}\nonumber\\
&  +E_{97,99}+E_{98,100}+E_{105,106}+E_{112,113}+E_{120,121}+2 E_{121,129}%
-E_{122,129}+E_{136,137}+E_{143,144}+E_{149,151}\nonumber\\
&  +E_{150,152}+E_{156,158}+E_{157,159}+E_{162,165}+E_{163,166}+E_{164,167}%
+E_{169,172}+E_{170,173}+E_{171,174}+E_{176,179}\nonumber\\
&  +E_{177,180}+E_{178,181}+E_{183,185}+E_{184,186}+E_{189,191}+E_{190,192}%
+E_{196,197}+E_{202,203}+E_{224,225}+E_{228,229}\nonumber\\
&  +E_{231,232}+E_{234,235}+E_{236,237}+E_{238,239}+E_{240,241},\nonumber\\
X_{2}  &  = -E_{6,7}-E_{8,10}-E_{9,11}-E_{23,28}-E_{27,32}-E_{30,35}%
-E_{31,36}-E_{33,39}-E_{34,40}-E_{37,43}-E_{38,44}-E_{42,49}\nonumber\\
&  -E_{48,54}-E_{70,77}-E_{73,80}-E_{76,84}-E_{79,87}-E_{81,89}-E_{83,91}%
-E_{86,93}-E_{88,95}-E_{90,98}-E_{92,100}-E_{94,102}\nonumber\\
&  -E_{97,105}-E_{99,106}-E_{101,108}-E_{107,114}+E_{115,128}-E_{123,134}+2
E_{128,134}-E_{135,142}-E_{141,148}-E_{143,150}\nonumber\\
&  -E_{144,152}-E_{147,155}-E_{149,157}-E_{151,159}-E_{154,161}-E_{156,163}%
-E_{158,166}-E_{160,168}-E_{162,170}-E_{165,173}\nonumber\\
&  -E_{169,176}-E_{172,179}-E_{195,201}-E_{200,207}-E_{205,211}-E_{206,212}%
-E_{209,215}-E_{210,216}-E_{213,218}\nonumber\\
&  -E_{214,219}-E_{217,222}-E_{221,226}-E_{238,240}-E_{239,241}-E_{242,243}%
,\nonumber\\
X_{3}  &  = -E_{6,8}-E_{7,10}-E_{13,16}-E_{15,19}-E_{18,22}-E_{21,26}%
-E_{25,29}-E_{41,46}-E_{45,52}-E_{50,57}-E_{51,58}-E_{55,63}\nonumber\\
&  -E_{56,64}-E_{61,68}-E_{62,69}-E_{67,75}-E_{73,81}-E_{74,82}-E_{79,88}%
-E_{80,89}-E_{86,94}-E_{87,95}-E_{92,101}-E_{93,102}\nonumber\\
&  -E_{99,107}-E_{100,108}-E_{106,114}-E_{112,120}+E_{113,122}-E_{121,136}+2
E_{122,136}-E_{123,136}-E_{129,137}-E_{135,143}\nonumber\\
&  -E_{141,149}-E_{142,150}-E_{147,156}-E_{148,157}-E_{154,162}-E_{155,163}%
-E_{160,169}-E_{161,170}-E_{167,175}-E_{168,176}\nonumber\\
&  -E_{174,182}-E_{180,187}-E_{181,188}-E_{185,193}-E_{186,194}-E_{191,198}%
-E_{192,199}-E_{197,204}-E_{203,208}-E_{220,224}\nonumber\\
&  -E_{223,228}-E_{227,231}-E_{230,234}-E_{233,236}-E_{239,242}-E_{241,243}%
,\nonumber\\
X_{4}  &  = E_{5,6}+E_{10,12}+E_{11,13}+E_{19,23}+E_{22,27}+E_{26,30}%
+E_{29,33}+E_{36,41}+E_{40,45}+E_{43,50}+E_{44,51}+E_{49,55}+E_{54,61}%
\nonumber\\
&  +E_{64,70}+E_{66,73}+E_{69,76}+E_{72,79}+E_{75,83}+E_{78,86}+E_{82,90}%
+E_{85,92}+E_{89,96}+E_{95,103}+E_{102,109}+E_{105,112}\nonumber\\
&  +E_{106,113}+E_{107,115}+E_{108,116}+E_{114,123}-E_{122,135}+2
E_{123,135}-E_{124,135}-E_{128,135}+E_{133,141}+E_{134,142}\nonumber\\
&  +E_{136,143}+E_{137,144}+E_{140,147}+E_{146,154}+E_{153,160}+E_{157,164}%
+E_{159,167}+E_{163,171}+E_{166,174}+E_{170,177}\nonumber\\
&  +E_{173,180}+E_{176,183}+E_{179,185}+E_{188,195}+E_{194,200}+E_{198,205}%
+E_{199,206}+E_{204,209}+E_{208,213}+E_{216,220}\nonumber\\
&  +E_{219,223}+E_{222,227}+E_{226,230}+E_{236,238}+E_{237,239}+E_{243,244}%
,\nonumber\\
X_{5}  &  = -E_{4,5}-E_{12,14}-E_{13,15}-E_{16,19}-E_{27,31}-E_{30,34}%
-E_{32,36}-E_{33,37}-E_{35,40}-E_{39,43}-E_{51,56}-E_{55,62}\nonumber\\
&  -E_{58,64}-E_{60,66}-E_{61,67}-E_{63,69}-E_{65,72}-E_{68,75}-E_{71,78}%
-E_{90,97}-E_{92,99}-E_{96,104}-E_{98,105}-E_{100,106}\nonumber\\
&  -E_{101,107}-E_{103,110}-E_{108,114}-E_{109,117}+E_{116,124}-E_{123,133}+2
E_{124,133}-E_{125,133}-E_{132,140}-E_{135,141}\nonumber\\
&  -E_{139,146}-E_{142,148}-E_{143,149}-E_{144,151}-E_{145,153}-E_{150,157}%
-E_{152,159}-E_{171,178}-E_{174,181}-E_{177,184}\nonumber\\
&  -E_{180,186}-E_{182,188}-E_{183,189}-E_{185,191}-E_{187,194}-E_{193,198}%
-E_{206,210}-E_{209,214}-E_{212,216}-E_{213,217}\nonumber\\
&  -E_{215,219}-E_{218,222}-E_{230,233}-E_{234,236}-E_{235,237}-E_{244,245}%
,\nonumber\\
X_{6}  &  = E_{3,4}+E_{14,17}+E_{15,18}+E_{19,22}+E_{23,27}+E_{28,32}%
+E_{34,38}+E_{37,42}+E_{40,44}+E_{43,49}+E_{45,51}+E_{50,55}\nonumber\\
&  +E_{52,58}+E_{53,60}+E_{57,63}+E_{59,65}+E_{67,74}+E_{75,82}+E_{78,85}%
+E_{83,90}+E_{86,92}+E_{91,98}+E_{93,100}+E_{94,101}\nonumber\\
&  +E_{102,108}+E_{104,111}+E_{109,116}+E_{110,118}+E_{117,125}-E_{124,132}+2
E_{125,132}-E_{126,132}+E_{131,139}+E_{133,140}\nonumber\\
&  +E_{138,145}+E_{141,147}+E_{148,155}+E_{149,156}+E_{151,158}+E_{157,163}%
+E_{159,166}+E_{164,171}+E_{167,174}+E_{175,182}\nonumber\\
&  +E_{184,190}+E_{186,192}+E_{189,196}+E_{191,197}+E_{194,199}+E_{198,204}%
+E_{200,206}+E_{205,209}+E_{207,212}+E_{211,215}\nonumber\\
&  +E_{217,221}+E_{222,226}+E_{227,230}+E_{231,234}+E_{232,235}+E_{245,246}%
,\nonumber\\
X_{7}  &  = -E_{2,3}-E_{17,20}-E_{18,21}-E_{22,26}-E_{27,30}-E_{31,34}%
-E_{32,35}-E_{36,40}-E_{41,45}-E_{42,48}-E_{46,52}-E_{47,53}\nonumber\\
&  -E_{49,54}-E_{55,61}-E_{62,67}-E_{63,68}-E_{65,71}-E_{69,75}-E_{72,78}%
-E_{76,83}-E_{79,86}-E_{84,91}-E_{87,93}-E_{88,94}\nonumber\\
&  -E_{95,102}-E_{103,109}-E_{110,117}-E_{111,119}+E_{118,126}-E_{125,131}+2
E_{126,131}-E_{127,131}-E_{130,138}-E_{132,139}\nonumber\\
&  -E_{140,146}-E_{147,154}-E_{155,161}-E_{156,162}-E_{158,165}-E_{163,170}%
-E_{166,173}-E_{171,177}-E_{174,180}-E_{178,184}\nonumber\\
&  -E_{181,186}-E_{182,187}-E_{188,194}-E_{195,200}-E_{196,202}-E_{197,203}%
-E_{201,207}-E_{204,208}-E_{209,213}-E_{214,217}\nonumber\\
&  -E_{215,218}-E_{219,222}-E_{223,227}-E_{228,231}-E_{229,232}-E_{246,247}%
,\nonumber\\
X_{8}  &  = E_{1,2}+E_{20,24}+E_{21,25}+E_{26,29}+E_{30,33}+E_{34,37}%
+E_{35,39}+E_{38,42}+E_{40,43}+E_{44,49}+E_{45,50}+E_{51,55}\nonumber\\
&  +E_{52,57}+E_{53,59}+E_{56,62}+E_{58,63}+E_{60,65}+E_{64,69}+E_{66,72}%
+E_{70,76}+E_{73,79}+E_{77,84}+E_{80,87}+E_{81,88}\nonumber\\
&  +E_{89,95}+E_{96,103}+E_{104,110}+E_{111,118}+E_{119,127}-E_{126,130}+2
E_{127,130}+E_{131,138}+E_{139,145}+E_{146,153}\nonumber\\
&  +E_{154,160}+E_{161,168}+E_{162,169}+E_{165,172}+E_{170,176}+E_{173,179}%
+E_{177,183}+E_{180,185}+E_{184,189}+E_{186,191}\nonumber\\
&  +E_{187,193}+E_{190,196}+E_{192,197}+E_{194,198}+E_{199,204}+E_{200,205}%
+E_{206,209}+E_{207,211}+E_{210,214}+E_{212,215}\nonumber\\
&  +E_{216,219}+E_{220,223}+E_{224,228}+E_{225,229}+E_{247,248}.
\end{align}
The corresponding negative roots are almost the transpose of these positive
roots:
\begin{align}
Y_{1}  &  = E_{9,8}+E_{11,10}+E_{13,12}+E_{15,14}+E_{18,17}+E_{21,20}%
+E_{25,24}+E_{47,46}+E_{53,52}+E_{59,57}+E_{60,58}+E_{65,63}\nonumber\\
&  +E_{66,64}+E_{71,68}+E_{72,69}+E_{73,70}+E_{78,75}+E_{79,76}+E_{80,77}%
+E_{85,82}+E_{86,83}+E_{87,84}+E_{92,90}+E_{93,91}\nonumber\\
&  +E_{99,97}+E_{100,98}+E_{106,105}+E_{113,112}+2 E_{121,120}-E_{122,120}%
+E_{129,121}+E_{137,136}+E_{144,143}+E_{151,149}\nonumber\\
&  +E_{152,150}+E_{158,156}+E_{159,157}+E_{165,162}+E_{166,163}+E_{167,164}%
+E_{172,169}+E_{173,170}+E_{174,171}+E_{179,176}\nonumber\\
&  +E_{180,177}+E_{181,178}+E_{185,183}+E_{186,184}+E_{191,189}+E_{192,190}%
+E_{197,196}+E_{203,202}+E_{225,224}+E_{229,228}\nonumber\\
&  +E_{232,231}+E_{235,234}+E_{237,236}+E_{239,238}+E_{241,240},\nonumber\\
Y_{2}  &  = -E_{7,6}-E_{10,8}-E_{11,9}-E_{28,23}-E_{32,27}-E_{35,30}%
-E_{36,31}-E_{39,33}-E_{40,34}-E_{43,37}-E_{44,38}-E_{49,42}\nonumber\\
&  -E_{54,48}-E_{77,70}-E_{80,73}-E_{84,76}-E_{87,79}-E_{89,81}-E_{91,83}%
-E_{93,86}-E_{95,88}-E_{98,90}-E_{100,92}-E_{102,94}\nonumber\\
&  -E_{105,97}-E_{106,99}-E_{108,101}-E_{114,107}-E_{123,115}+2 E_{128,115}%
+E_{134,128}-E_{142,135}-E_{148,141}-E_{150,143}\nonumber\\
&  -E_{152,144}-E_{155,147}-E_{157,149}-E_{159,151}-E_{161,154}-E_{163,156}%
-E_{166,158}-E_{168,160}-E_{170,162}-E_{173,165}\nonumber\\
&  -E_{176,169}-E_{179,172}-E_{201,195}-E_{207,200}-E_{211,205}-E_{212,206}%
-E_{215,209}-E_{216,210}-E_{218,213}-E_{219,214}\nonumber\\
&  -E_{222,217}-E_{226,221}-E_{240,238}-E_{241,239}-E_{243,242},\nonumber\\
Y_{3}  &  = -E_{8,6}-E_{10,7}-E_{16,13}-E_{19,15}-E_{22,18}-E_{26,21}%
-E_{29,25}-E_{46,41}-E_{52,45}-E_{57,50}-E_{58,51}-E_{63,55}\nonumber\\
&  -E_{64,56}-E_{68,61}-E_{69,62}-E_{75,67}-E_{81,73}-E_{82,74}-E_{88,79}%
-E_{89,80}-E_{94,86}-E_{95,87}-E_{101,92}-E_{102,93}\nonumber\\
&  -E_{107,99}-E_{108,100}-E_{114,106}-E_{120,112}-E_{121,113}+2
E_{122,113}-E_{123,113}+E_{136,122}-E_{137,129}-E_{143,135}\nonumber\\
&  -E_{149,141}-E_{150,142}-E_{156,147}-E_{157,148}-E_{162,154}-E_{163,155}%
-E_{169,160}-E_{170,161}-E_{175,167}-E_{176,168}\nonumber\\
&  -E_{182,174}-E_{187,180}-E_{188,181}-E_{193,185}-E_{194,186}-E_{198,191}%
-E_{199,192}-E_{204,197}-E_{208,203}-E_{224,220}\nonumber\\
&  -E_{228,223}-E_{231,227}-E_{234,230}-E_{236,233}-E_{242,239}-E_{243,241}%
,\nonumber\\
Y_{4}  &  = E_{6,5}+E_{12,10}+E_{13,11}+E_{23,19}+E_{27,22}+E_{30,26}%
+E_{33,29}+E_{41,36}+E_{45,40}+E_{50,43}+E_{51,44}+E_{55,49}+E_{61,54}%
\nonumber\\
&  +E_{70,64}+E_{73,66}+E_{76,69}+E_{79,72}+E_{83,75}+E_{86,78}+E_{90,82}%
+E_{92,85}+E_{96,89}+E_{103,95}+E_{109,102}+E_{112,105}\nonumber\\
&  +E_{113,106}+E_{115,107}+E_{116,108}-E_{122,114}+2 E_{123,114}%
-E_{124,114}-E_{128,114}+E_{135,123}+E_{141,133}+E_{142,134}\nonumber\\
&  +E_{143,136}+E_{144,137}+E_{147,140}+E_{154,146}+E_{160,153}+E_{164,157}%
+E_{167,159}+E_{171,163}+E_{174,166}+E_{177,170}\nonumber\\
&  +E_{180,173}+E_{183,176}+E_{185,179}+E_{195,188}+E_{200,194}+E_{205,198}%
+E_{206,199}+E_{209,204}+E_{213,208}+E_{220,216}\nonumber\\
&  +E_{223,219}+E_{227,222}+E_{230,226}+E_{238,236}+E_{239,237}+E_{244,243}%
,\nonumber\\
Y_{5}  &  = -E_{5,4}-E_{14,12}-E_{15,13}-E_{19,16}-E_{31,27}-E_{34,30}%
-E_{36,32}-E_{37,33}-E_{40,35}-E_{43,39}-E_{56,51}-E_{62,55}\nonumber\\
&  -E_{64,58}-E_{66,60}-E_{67,61}-E_{69,63}-E_{72,65}-E_{75,68}-E_{78,71}%
-E_{97,90}-E_{99,92}-E_{104,96}-E_{105,98}-E_{106,100}\nonumber\\
&  -E_{107,101}-E_{110,103}-E_{114,108}-E_{117,109}-E_{123,116}+2
E_{124,116}-E_{125,116}+E_{133,124}-E_{140,132}-E_{141,135}\nonumber\\
&  -E_{146,139}-E_{148,142}-E_{149,143}-E_{151,144}-E_{153,145}-E_{157,150}%
-E_{159,152}-E_{178,171}-E_{181,174}-E_{184,177}\nonumber\\
&  -E_{186,180}-E_{188,182}-E_{189,183}-E_{191,185}-E_{194,187}-E_{198,193}%
-E_{210,206}-E_{214,209}-E_{216,212}-E_{217,213}\nonumber\\
&  -E_{219,215}-E_{222,218}-E_{233,230}-E_{236,234}-E_{237,235}-E_{245,244}%
,\nonumber\\
Y_{6}  &  = E_{4,3}+E_{17,14}+E_{18,15}+E_{22,19}+E_{27,23}+E_{32,28}%
+E_{38,34}+E_{42,37}+E_{44,40}+E_{49,43}+E_{51,45}+E_{55,50}\nonumber\\
&  +E_{58,52}+E_{60,53}+E_{63,57}+E_{65,59}+E_{74,67}+E_{82,75}+E_{85,78}%
+E_{90,83}+E_{92,86}+E_{98,91}+E_{100,93}+E_{101,94}\nonumber\\
&  +E_{108,102}+E_{111,104}+E_{116,109}+E_{118,110}-E_{124,117}+2
E_{125,117}-E_{126,117}+E_{132,125}+E_{139,131}+E_{140,133}\nonumber\\
&  +E_{145,138}+E_{147,141}+E_{155,148}+E_{156,149}+E_{158,151}+E_{163,157}%
+E_{166,159}+E_{171,164}+E_{174,167}+E_{182,175}\nonumber\\
&  +E_{190,184}+E_{192,186}+E_{196,189}+E_{197,191}+E_{199,194}+E_{204,198}%
+E_{206,200}+E_{209,205}+E_{212,207}+E_{215,211}\nonumber\\
&  +E_{221,217}+E_{226,222}+E_{230,227}+E_{234,231}+E_{235,232}+E_{246,245}%
,\nonumber\\
Y_{7}  &  = -E_{3,2}-E_{20,17}-E_{21,18}-E_{26,22}-E_{30,27}-E_{34,31}%
-E_{35,32}-E_{40,36}-E_{45,41}-E_{48,42}-E_{52,46}-E_{53,47}\nonumber\\
&  -E_{54,49}-E_{61,55}-E_{67,62}-E_{68,63}-E_{71,65}-E_{75,69}-E_{78,72}%
-E_{83,76}-E_{86,79}-E_{91,84}-E_{93,87}-E_{94,88}\nonumber\\
&  -E_{102,95}-E_{109,103}-E_{117,110}-E_{119,111}-E_{125,118}+2
E_{126,118}-E_{127,118}+E_{131,126}-E_{138,130}-E_{139,132}\nonumber\\
&  -E_{146,140}-E_{154,147}-E_{161,155}-E_{162,156}-E_{165,158}-E_{170,163}%
-E_{173,166}-E_{177,171}-E_{180,174}-E_{184,178}\nonumber\\
&  -E_{186,181}-E_{187,182}-E_{194,188}-E_{200,195}-E_{202,196}-E_{203,197}%
-E_{207,201}-E_{208,204}-E_{213,209}-E_{217,214}\nonumber\\
&  -E_{218,215}-E_{222,219}-E_{227,223}-E_{231,228}-E_{232,229}-E_{247,246}%
,\nonumber\\
Y_{8}  &  = E_{2,1}+E_{24,20}+E_{25,21}+E_{29,26}+E_{33,30}+E_{37,34}%
+E_{39,35}+E_{42,38}+E_{43,40}+E_{49,44}+E_{50,45}+E_{55,51}\nonumber\\
&  +E_{57,52}+E_{59,53}+E_{62,56}+E_{63,58}+E_{65,60}+E_{69,64}+E_{72,66}%
+E_{76,70}+E_{79,73}+E_{84,77}+E_{87,80}+E_{88,81}\nonumber\\
&  +E_{95,89}+E_{103,96}+E_{110,104}+E_{118,111}-E_{126,119}+2 E_{127,119}%
+E_{130,127}+E_{138,131}+E_{145,139}+E_{153,146}\nonumber\\
&  +E_{160,154}+E_{168,161}+E_{169,162}+E_{172,165}+E_{176,170}+E_{179,173}%
+E_{183,177}+E_{185,180}+E_{189,184}+E_{191,186}\nonumber\\
&  +E_{193,187}+E_{196,190}+E_{197,192}+E_{198,194}+E_{204,199}+E_{205,200}%
+E_{209,206}+E_{211,207}+E_{214,210}+E_{215,212}\nonumber\\
&  +E_{219,216}+E_{223,220}+E_{228,224}+E_{229,225}+E_{248,247}.
\end{align}


\newpage\FloatBarrier
\providecommand{\href}[2]{#2}\begingroup\raggedright\endgroup

\end{document}